\documentclass[
 aip,
 amsmath,amssymb,
 preprint,%
]{revtex4-2}

\usepackage{graphicx}
\usepackage{dcolumn}
\usepackage{bm}
\usepackage[paperwidth=210mm,paperheight=297mm,centering,hmargin=1cm,vmargin=2.5cm]{geometry} 
\usepackage[utf8]{inputenc}
\usepackage[T1]{fontenc}
\usepackage{mathptmx}

\usepackage{amsmath}
\usepackage{mathtools, cuted}
\usepackage{lineno,hyperref}
\usepackage{xcolor}
\modulolinenumbers[5]

\usepackage{blindtext}

\newcommand{\ket}[1]{\left|#1\right\rangle}
\newcommand{\bra}[1]{\left\langle#1\right|}
\newcommand{\braket}[2]{\left\langle#1\middle|#2\right\rangle}

\begin{document}
\title{Analytical Solutions for $N$-Electron Interacting System Confined in Graph of Coupled Electrostatic Semiconductor and Superconducting Quantum Dots in Tight-Binding Model with Focus on Quantum Information Processing }

\author{Krzysztof Pomorski$^{1}$}   \email[Corresponding author:]{kdvpomorski@gmail.com}
 \affiliation{
  1:University College Dublin-School of Electrical and Electronic Engineering, Ireland, \\
  2:Quantum Hardware Systems, www.quantumhardwaresystems.com
}
\author{Robert Bogdan Staszewski$^{1}$} 
 \affiliation{
  1:University College Dublin-School of Electrical and Electronic Engineering, Ireland, \\
}

\begin{abstract}
Analytical solutions for a tight-binding model are presented for a position-based qubit and $N$ interacting qubits realized by quasi-one-dimensional network of coupled quantum dots expressed by connected or disconnected graphs of any topology in 2 and 3 dimensions where one electron is presented at each separated graphs. Electron(s) quantum dynamic state is described under various electromagnetic circumstances with an omission spin degree-of-freedom. The action of Hadamard and phase rotating gate is given by analytical formulas derived and formulated for any case of physical field evolution preserving the occupancy of two-energy level system. The procedure for heating up and cooling down of the quantum state placed in position based qubit is described. The interaction of position-based qubit with electromagnetic cavity is described. In particular non-local communication between position based qubits is given. It opens the perspective of implementation of quantum internet among electrostatic CMOS quantum computers (quantum chips). The interface between superconducting Josephson junction and semiconductor position-based qubit implemented in coupled semiconductor q-dots is described such that it can be the base for electrostatic interface between superconducting and semiconductor quantum computer. Modification of Andreev Bound State in Josephson junction by the presence of semiconductor qubit in its proximity and electrostatic interaction with superconducting qubit is spotted by the minimalistic tight-binding model. The obtained results allow in creating interface between semiconductor quantum computer and superconducting quantum computer. They open the perspective of construction of QISKIT like software that will describe both types of quantum computers as wel\textbf{l as their interface. }
\end{abstract}



\maketitle
\onecolumngrid
\newpage
\tableofcontents
\newpage
\section{Introduction to recent trends in Q-Technologies}
Quantum technology opens the gate for quantum computation and quantum sensing as well as quantum communication. Also in the nearest perspective one shall consider quantum Artificial Intelligence as extension of classical Artificial Intelligence. Because of high technical cost of implementation of quantum technologies one shall think about usage of both classical and quantum technologies at one chip what is possible in FD SOI CMOS technology that currently manufactures transistors with 3nm of channel length.
The quantum mechanics offers the superposition of states and massive parallelism as well as non-local correlations that are non-present in classical world perceived by us. However these phenomena occurs only in special time scale and under specific thermodynamic conditions in the case of special geometries and confining potentials. Basically the quantum system needs to be maximally decoupled from the world to keep its unique quantum features. On the other hand we need to be able to interact with quantum system relatively quickly what brings the need for not so small interaction of qubit with classical or semiclassical interface via specific channels. At the same time we would expect the quantum technology to be highly reproducible in large scale, compact and having an easy interface with already existing technologies mostly working at room temperature. Basically ideal candidate for qubit does not exist and we have to make trade-off between certain technical parameters. The first option is to chose the system that is maximally decoupled from external world so we arrive to the idea of ion traps. We are placing atomic ions in almost ideal vacuum and we trap them by strong magnetic and electric fields. Maxwells equations does not allow for complicated topologies of EM confinement field affecting ion positions and thus we are limited to the case of ions on one line as it is indicated by many experimentalist. However every time we are about to use quantum ionic processor we need to cool down and set the ions in certain positions what makes structure to be practically not adjustable for large scales. However the decoherence times are more than promising since $T_1$ and $T_2$ time is in range of seconds what makes it bigger by 4 orders magnitude than any other quantum technology available so far. This makes ion trap to be excellent quantum sensors.
\onecolumngrid
On the other hand we can think about use of electron or electron spin to represent the state of qubit. So far the electron is most successful carrier of classical information. Thus we need to use it on the level of qubit implementation in semiconductor or in superconductor. In the natural way we arrive to the electrostatic qubit in semiconductor, where presence of single electron corresponds to logical $\ket{1}$ and its lack to
$\ket{0}$ (Fujisawa \cite{Fujisawa}, Petta \cite{Petta} ) or to superconducting Cooper pair box.  However electron-electron interaction is quite strong as in comparison with spin to spin interaction. The strength of the interaction preimposes the decoherence time since the stronger is the interaction the smaller is the decoherence time. At the same time big quantum information density is usually leading to higher decoherence times since qubit-qubit interaction is more prominent.

Every qubit assembles can be described by the following Hamiltonian operator:

\begin{eqnarray*}
\hat{H}_t=\textcolor{blue}{H_{[Q_0]}}+([H_{Q}-H_{Q_0}])_{[Q \setminus Q_0]}+H_{[Q-Env]}+H_{[Q-Q]} + H_{[Env]}=   \nonumber \\ \sum_{l=1}^{N_{qbits}} ( \textcolor{blue}{ E_{e,l}(t)|e_l(t)><e_l(t)|+E_{g,l}(t)|g_l(t)><g_l(t)|}+ \nonumber \\
+\textcolor{green}{T_{g \rightarrow e,l}(t)|e_l(t)><g_l(t)|+T_{e  \rightarrow g,l}(t)|g_l(t)><e_l(t)|)_{[Q_0]}} \nonumber \\ \textcolor{gray}{ + \sum_{l=1}^{N_{qbits}} ( \sum_{s1_{l}={3}}^{+\infty} ( E_{s1,l}(t)|s1_l(t)><s1_l(t)|+\sum_{s2_{l}=3,s2_l \neq s1_l}^{+\infty} T_{s1_l \rightarrow s2_l,l}(t)|s1_l(t)><s2_l(t)|)+ } \nonumber \\  \textcolor{red}{ +T_{e_{s_l} \rightarrow s3_l,l}(t)|e_{s_l}(t)><s3_l(t)|+T_{s3_l \rightarrow e_{s_l},l}(t)|s3(t)_l><e(t)_{s_l}|+T_{g_{s_l} \rightarrow s3_l,l}(t)|g_{s_l}(t)><s3_l(t)| }+ \nonumber \\ + \textcolor{red}{ T_{s3_l \rightarrow g_{s_l},l}(t)|s3(t)_l><g(t)_{s_l}|)_{ [Q \setminus Q_0] } } +  \nonumber \\ +( \sum_{i=1}^{+\infty} \sum_{l=1}^{N_{qbits}}  \sum_{s1_l={(\textcolor{red}{g,e},..)}}^{+\infty} \sum_{s2_l={( \textcolor{red}{g,e},..)}}^{+\infty}  \textcolor{red}{ U_{3}(s1_l,s2_l,i,t)|s1_l(t),i(t)><s2_l(t),i(t)| )_{[Q-Env]}}  + \nonumber \\ \textcolor{orange}{ + ( \sum_{l=1}^{N_{qbits}} \sum_{k=1, k \neq l}^{N_{qbits}}\sum_{s1_k={(g,e, ..)}}^{+\infty}\sum_{s2_l={(g,e,..)}}^{+\infty} U_4(s1_k,s2_l,t)|s1_k(t),s2_l(t)><s1_k(t),s2_l(t)|  )_{[Q-Q]} + }  \nonumber \\   \textcolor{red}{ +(\sum_{i=1}^{+\infty}E_{i}(t)|i(t)><i(t)|)_{[Env]}. } 
\end{eqnarray*}
The given Hamiltonian is describing quantum system embedded in external environment (external world) and it has terms $H_{Q_0}, H_{Q \setminus Q_0}, H_{Q-Env}, H_{Q-Q}, H_{Env}$. In particular we have idealistic mathematical model of qubit that is isolated from external world and denoted by $H_{Q_0}$ (blue color). Next Hamiltonian term $H_{Q \setminus Q_0}$ (green color) describes Hamiltonian setting qubit state and Hamiltonian term capable of qubit readout. However it is not suprsing that Hamiltonian term responsbile for qubit setting and reading can also contribute to its decoherence. The Hamiltonian terms describing the decoherence are due to qubit-qubit interaction and due to qubit-enviroment interaction  (red and orange color).  Usually we drop the last $H_{env}$ term since we assume that environment has infinite size and has well-defined thermodynamical state that cannot be changed by the small size and finite quantum system Q that implements qubits.
The value of $E_e$ and $E_g$ is determined by the qubit confinement potential, while functions $f_1(t)$ and $f_2(t)$ give us the ways to implement qubit setting mechanism and qubit reading mechanism by means of time-dependent Hamiltonians that are driven by external biasing circuit [qubit controlling circuit]. Formally we recognize that qubit assembly is the many body system with certain desired degrees of freedom (in general the available number of degrees of freedom is much bigger than desired and it is a function of given quantum technology) that are controlled by given quantum technology placed in special thermodynamic conditions and given qubit implementation scheme. Study of the quantum system Q being assemble of interacting qubits embedded in external world denoted as environment is fundamental study and as much important for fundamental science as for technology.

Basically the thermodynamics is against preservation of information stored in qubit since entropy is increasing with time. The difference between energy levels of ground state (g) and excited state (e) is tiny and can be directly evaluated from Schrodinger equation. Once the excited level is occupied to certain extent it is in metastable state and tends to decay into ground state g. This decay time in case lack of external perturbations is shorter for the case of systems with bigger difference between excited and ground state. Because of this decay quantum state needs to be refreshed all the times to maintain its content (IBM Q-Experience provides superconducting Josephson junction qubits of 100 mikroseconds coherence time) . The biggest danger to qubit coherence is energy of surrounding environment that is expressed especially by $H_{env}$ last Hamiltonian term and by $H_{env-Q}$. Moderate decoherence to qubit state is by qubit-qubit interaction $H_{Q-Q}$ that is potential factor limiting the maximum density of quantum logic.

\begin{figure}
\centering
\includegraphics[scale=0.4]{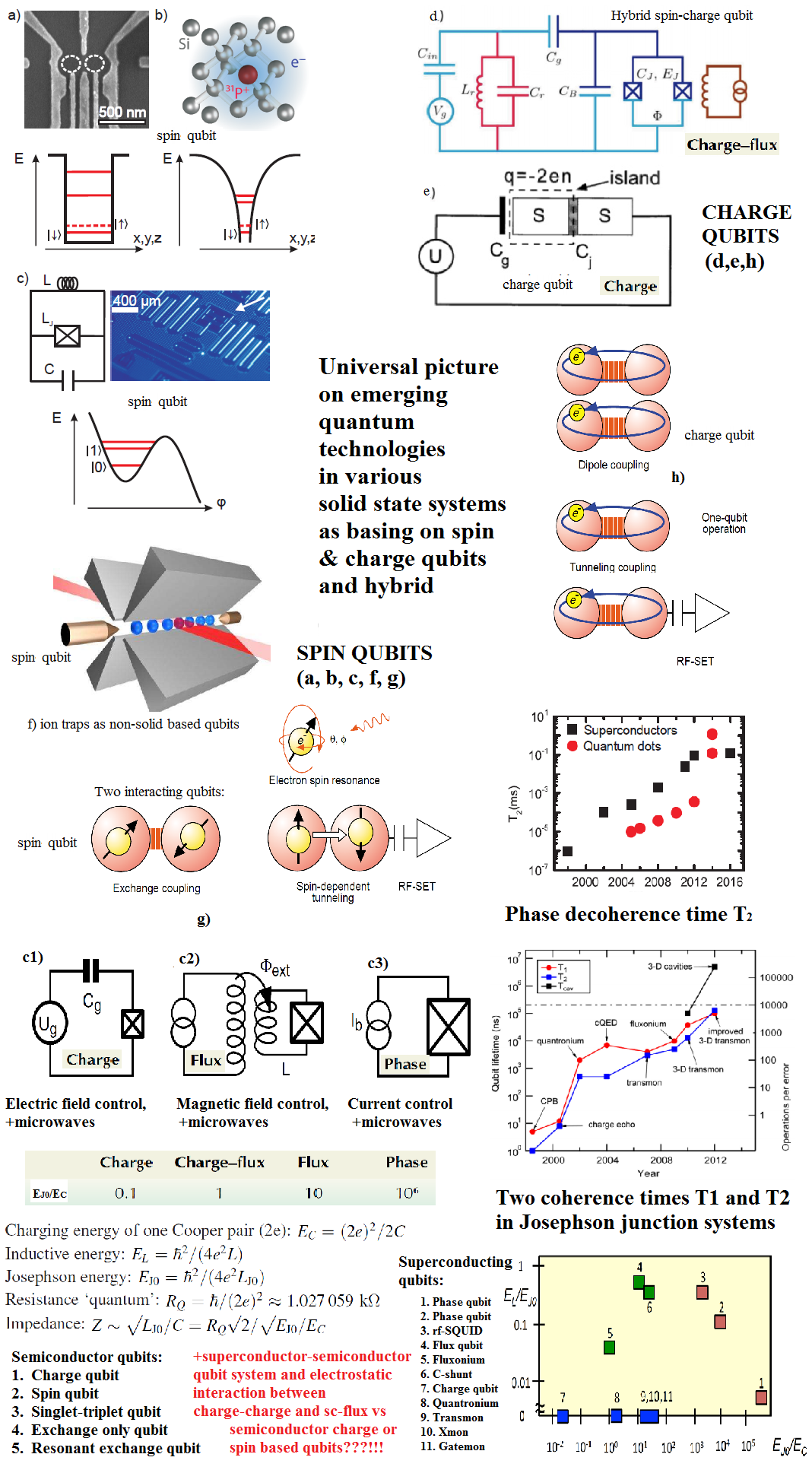} 
\caption{Summary of quantum information technologies [
From "Solid-state qubits",M. Fernando Gonzalez-Zalba, Arxiv: 1801.06722], one- and two-qubit operations and single-shot measurement, for
(a) a charge qubit in a double quantum dot and (b) a spin qubit in a single quantum dot. [From "Quantum Information Technology based
on Single Electron Dynamics", T.Fujsawa, NTT Technical Review, Vol. 1 No. 3,2003]}
\label{QTtechnologies}
\end{figure}

Zoo of existing quantum technologies is growing. However still there exists two fundamental representation of q-information in spin of electron or Cooper pair and in electric charge as it is depicted in Fig.\ref{QTtechnologies}.

Currently there exist various paradigms for quantum computation. The most common is by the use of quasiparticle that is trapped in effective field that builds up the quantization of the energetic levels.

Paradigms existing currently assume that the quantum system shall be controlled either by electric or magnetic field factor or by combined magnetic and electric field that is generated by the controlling circuit. 
The good example are superconducting Cooper pair box (JJ-Josephson junction controlled by external capacitor), flux-qubit JJ (JJ controlled by external solenoid), phase JJ qubit (controlled by biasing electric current) and transmon qubit (controlled both by solenoid and capacitor).
Indeed very recent progress was done very much up to electrical control of various types of technologies. The experiments conducted in 2003 by Fujisawa \cite{Fujisawa} have revealed significant charge noise problem and contributed to the change of the dominant paradigm in development of quantum circuits that was about shift from electric to magnetic field control and later electromagnetic control what is greatly expressed in superconducting technologies by common use of transmon superconducting qubit (and transmon like qubits: Xmon, etc). The details are specified in the attached table \ref{table_qt}.

\begin{table}[ht]
\centering
\begin{tabular}{ |p{5cm}||p{1.8cm}|p{1.9cm}|p{1.9cm}|  }
 \hline
 \multicolumn{4}{|c|}{ \texttt{Comparison of dominant quantum technologies} } \\
 \hline
\textbf{Quantum Technology
[S-spin or C-charge like] qbits} & \textbf{Scalability} & \textbf{Coherence time $T_1$ } & \textbf{Coherence time $T_2$} \\
 \hline
Ion Traps [S]  & Relatively low    & $ >10^{10} \mu$s !   & $ >10^{6} \mu$s   ! \\
 \hline
Semiconductor qubits: &  High   & $  \sim 1-10ns $   & $  \sim 1-10ns $   \\
 $\rightarrow$charge qubit [C] & High & $ 7 ns$ &  $250ps$ \\
 $\rightarrow$spin qubit [S]    & High & 59 ns &  59ns \\
 $\rightarrow$spin singlet-triplet qubit [S]    &  &  &   \\
 $\rightarrow$spin exchange qubit [S]    & High &  &  19\\
 $\rightarrow$spin resonant exchange qbit [S]  &   High  & 0 & $19 \mu s $  \\
 $\rightarrow$spin-charge qbit [S-C qbit]&   High  &  & 80ns\\
 \hline
 Josephson junction qubits: & Moderate  & $ 0.1-100\mu s $   & $ 0.1-100\mu s $   \\
 $\rightarrow$ Cooper pair box [C] & Moderate & 2 $\mu$s   & 2  $\mu$s \\
 $\rightarrow$ Flux qubit [S]      & Moderate  &  4.6 $\mu$s   & 1.2 $\mu$s   \\
 $\rightarrow$ Phase qubit & Moderate  & 0.5 $\mu$ s & 0.3 $\mu$ s \\
 $\rightarrow$ 3D Transmon [S-C] & High  & $ > 100 \mu s $ &  $ > 140\mu s$ \\
 $\rightarrow$ 2 D Transmon [S-C] & Moderate  & 50 $\mu s $ & 20 $\mu s $ \\
 $\rightarrow$ Fluxm [S-C] & Moderate  & 1000 $\mu s $  & $ >10 \mu s $ \\
 $\rightarrow$ C-shunt [S-C] & Moderate  & 55 $\mu$s  & 40 $\mu$s  \\
 $\rightarrow$ Xmon [S-C] & Moderate  & 50 $\mu$s & 20 $\mu$s \\
 $\rightarrow$ Gatemon [S-C] & Moderate  & 5.3 $\mu$s & 3.7 $\mu$s  \\
 \hline
\end{tabular}
\caption{Quick overview on quantum technologies}
\label{table_qt}
\end{table}



\begin{figure}
\centering
\includegraphics[scale=0.5]{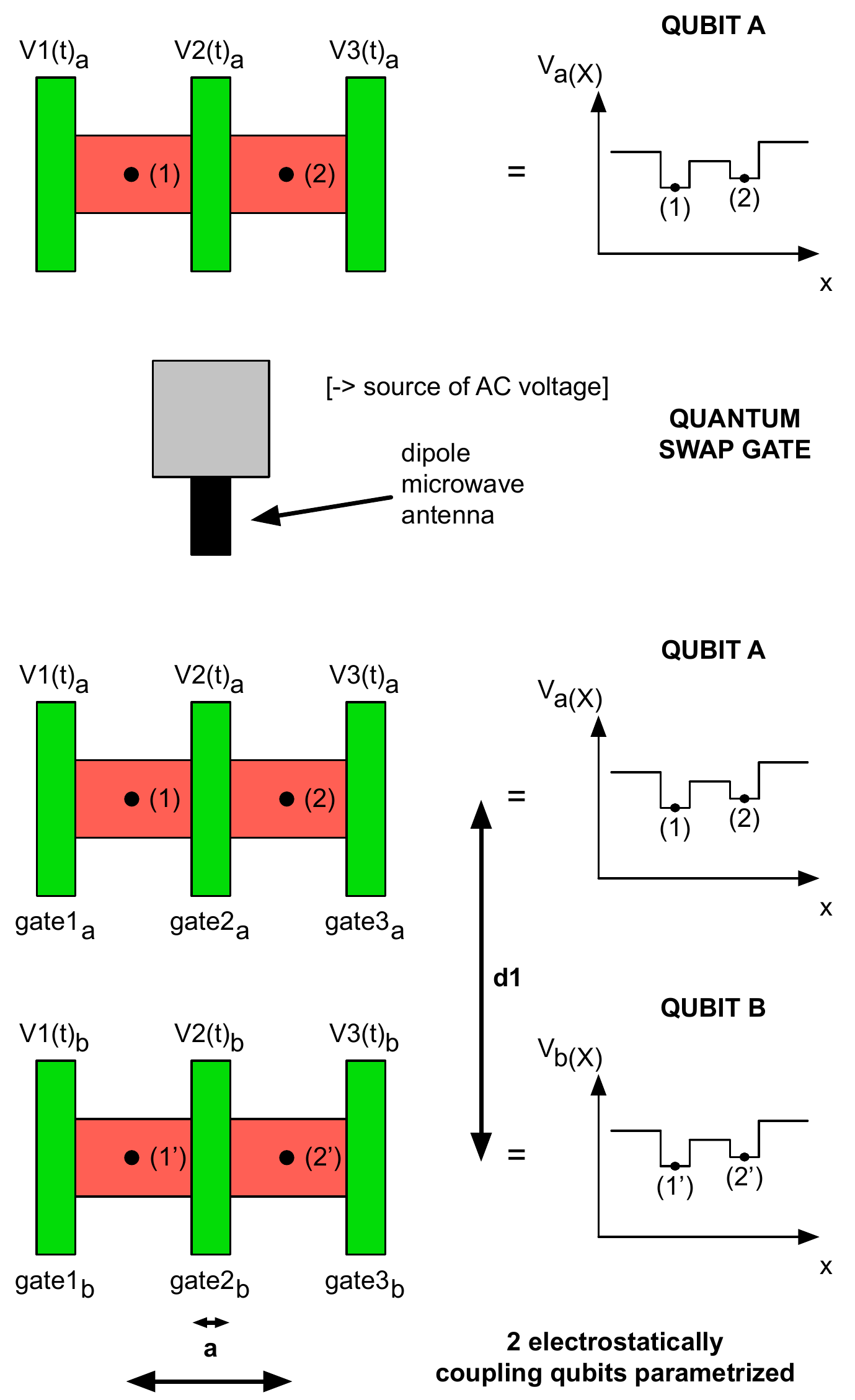} 
\caption{[Left]: Electrostatic position-based qubit implemented in CMOS technology \cite{Pomorski_spie}. [Upper Left]: Simplistic representation by particle localized in two regions of space denoted by nodes (1) and (2); [Lower Left]: Case of two electrostatically interacting qubits implementing quantum swap gate. Quantum dynamics are parameterized by presence of electrons at nodes 1, 2, 1' and 2'.}
\label{fig:central}
\end{figure}
\section{Description of position based-qubit in tight-binding model}
We refer to the physical situation from Fig.\ref{fig:central} and we consider position based-qubit in tight-binding model \cite{SEL} and its the Hamiltonian of this system is given as
\begin{eqnarray}
\label{simplematrix}
\hat{H}(t)=
\begin{pmatrix}
E_{p1}(t) & t_{s12}(t) \\
t_{s12}^{\dag}(t) & E_{p2}(t)
\end{pmatrix}_{[x=(x_1,x_2)]}= 
(E_1(t)\ket{E_1}_t \bra{E_1}_t+E_2(t)\ket{E_2}\bra{E_2})_{[E=(E_1,E_2)]}.
\end{eqnarray}
The Hamiltonian $\hat{H}(t)$ eigenenergies $E_1(t)$ and $E_2(t)$ with $E_2(t)>E_1(t)$ are given as
\begin{eqnarray}
E_1(t)= \left(-\sqrt{\frac{(E_{p1}(t)-E_{p2}(t))^2}{4}+|t_{s12}(t)|^2}+\frac{E_{p1}(t)+E_{p2}(t)}{2}\right), \nonumber \\
E_2(t)= \left(+\sqrt{\frac{(E_{p1}(t)-E_{p2}(t))^2}{4}+|t_{s12}(t)|^2}+\frac{E_{p1}(t)+E_{p2}(t)}{2}\right),
\end{eqnarray}
and energy eigenstates $\ket{E_1(t)}$ and $\ket{E_2(t)}$ have the following form
\begin{eqnarray}
\ket{E_1,t}=
\begin{pmatrix}
\frac{(E_{p2}(t)-E_{p1}(t))+\sqrt{\frac{(E_{p2}(t)-E_{p1}(t))^2}{2}+|t_{s12}(t)|^2}}{-i t_{sr}(t)+t_{si}(t)} \\
-1
\end{pmatrix},  \nonumber \\
\ket{E_2,t}=
\begin{pmatrix}
\frac{-(E_{p2}(t)-E_{p1}(t))+\sqrt{\frac{(E_{p2}(t)-E_{p1}(t))^2}{2}+|t_{s12}(t)|^2}}{t_{sr}(t) - i t_{si}(t)} \\
1
\end{pmatrix}.
\end{eqnarray}
This Hamiltonian gives a description of two coupled quantum wells as depicted in Fig.2.
In such situation we have real-valued functions $E_{p1}(t)$, $E_{p2}(t)$ and complex-valued functions $t_{s12}(t)=t_s(t)=t_{sr}(t)+i t_{si}(t)$ and $t_{s21}(t)=t_{s12}^{*}(t)$, what is equivalent to the knowledge of four real valued time-dependent continuous or discontinues functions $E_{p1}(t)$, $E_{p1}(2)$ , $t_{sr}(t)$ and $t_{si}(t)$. The quantum state is a superposition of state localized at node 1 and 2 and therefore is given as
\begin{equation}
\ket{\psi}_{[x]}=\alpha(t)\ket{1,0}_x+\beta(t)\ket{0,1}_x=
\alpha(t)
\begin{pmatrix}
1 \\
0 \\
\end{pmatrix}
+
\beta(t)
\begin{pmatrix}
0 \\
1 \\
\end{pmatrix} ,
\end{equation}
where $|\alpha(t)|^2$ ($|\beta(t)|^2$) is probability of finding particle at node 1(2) respectively, which brings $|\alpha(t)|^2+|\beta(t)|^2=1$ and obviously $\bra{1,0}_x\ket{|1,0}_x=1=\bra{0,1}_x\ket{|0,1}_x$ and $\bra{1,0}_x\ket{|0,1}_x=0=\bra{0,1}_x\ket{|1,0}_x$. In Schr\"odinger formalism, states $\ket{1,0}_x$ and $\ket{0,1}_x$ are Wannier functions that are parameterized by position $x$. We work in tight-binding approximation and quantum state evolution with time as given by
\begin{equation}
 i \hbar \frac{d}{dt}\ket{\psi(t)}=\hat{H}(t)\ket{\psi(t)}=E(t)\ket{\psi(t)}.
\end{equation}
The last equation has an analytic solution
\begin{equation}
\ket{\psi(t)}=e^{\frac{1}{i \hbar}\int_{t_0}^{t}\hat{H}(t_1)dt_1}\ket{\psi(t_0)}=e^{\frac{1}{i \hbar}\int_{t_0}^{t}\hat{H}(t_1)dt_1}
\begin{pmatrix}
\alpha(0) \\
\beta(0) \\
\end{pmatrix}
\end{equation}
and in quantum density matrix theory we obtain
\begin{eqnarray}
\hat{\rho}(t)=\hat{\rho}^{\dag}(t)=\ket{\psi(t)}\bra{\psi(t)}=\nonumber \\ =\hat{U}(t,t_0)\hat{\rho}(t_0)\hat{U}(t,t_0)^{-1}= \nonumber \\
=e^{\frac{1}{i \hbar}\int_{t_0}^{t}\hat{H}(t_1)dt_1}(\ket{\psi(t_0)}\bra{\psi(t_0)})e^{-\frac{1}{i \hbar}\int_{t_0}^{t}\hat{H}(t_1)dt_1}= \nonumber \\
=e^{\frac{1}{i \hbar}\int_{t_0}^{t}\hat{H}(t_1)dt_1}
\bigg(
\begin{pmatrix}
\alpha(0) \\
\beta(0) \\
\end{pmatrix} 
\begin{pmatrix}
\alpha^{*}(0) & \beta^{*}(0) \\
\end{pmatrix}
\bigg)e^{-\frac{\int_{t_0}^{t}\hat{H}(t_1)dt_1}{i \hbar}}= \nonumber \\
=\hat{U}(t,t_0)  
\begin{pmatrix}
|\alpha(0)|^2 & \alpha(0)\beta^{*}(0)  \\
\beta(0)\alpha(0)^{*} &  |\beta(0)|^2 \\
\end{pmatrix} \hat{U}(t,t_0)^{\dag}.
\end{eqnarray}

Having Hermitian matrix $\hat{A}$ with real-valued coefficients $a_{11}(t)$, $a_{22}(t)$, $a_{12r}(t)$, $a_{12i}(t)$  and Pauli matrices $\sigma_1$, $\sigma_2$, $\sigma_3$, $\sigma_0=\hat{I}_{2 by 2}$ we observe that
\begin{eqnarray}
\hat{A}_{2 \times 2}=
\begin{pmatrix}
 a_{11} & a_{12r}+ia_{12i} \\
 a_{12r}-ia_{12i} & a_{22}
\end{pmatrix}, = \nonumber \\
=a_{12r}\sigma_1 -a_{12i}\sigma_2+\frac{1}{2}(a_{11}-a_{22})\sigma_3+\frac{1}{2}(a_{11}+a_{22})\sigma_0.
\end{eqnarray}
and for $\hat{A}_{2N \times 2N}=\Sigma_{k_1,k_2,..,k_N} b_{k_1,k_2,..,k_N}(\sigma_{k_1}\times \sigma_{k_2} \times .. \times  \sigma_{k_N})$ we obtain the unique matrix decomposition in terms of Pauli matrix tensor products, where $k_i=0,..,3$.
Using the above property for matrix of size 2$\times$2 we obtain $e^{\frac{1}{i\hbar}\int_{t_0}^{t}\hat{H}(t_1)dt_1}=\hat{U}(t,t_0), $ and assuming $E_{p1}(t)=E_{p2}(t)=E_{p}(t)$ and we  are given matrix $e^{\frac{1}{i\hbar}\int_{t_0}^{t}\hat{H}(t_1)dt_1}= 
 $ 
\begin{eqnarray}
\begin{pmatrix}
e^{\frac{-i \int_{t_0}^{t}E_p(t')dt'}{\hbar}}ch\left(\frac{\sqrt{-\int_{t_0}^{t}(|t_{s}(t')|^2)dt'}}{\hbar}\right) & \frac{e^{\frac{-i \int_{t_0}^{t}E_{p}(t')dt'}{\hbar}} (\int_{t_0}^{t}(t_{s}^{*}(t'))dt')sh\left(\frac{\sqrt{-\int_{t_0}^{t}|t_{s}(t')|^2)}}{\hbar}\right)}{\sqrt{-\int_{t_0}^{t}((t_{si}(t')^2+t_{sr}(t'))^2)dt'}}  \\
\frac{e^{\frac{-i \int_{t_0}^{t}E_{p}(t')dt'}{\hbar}} (\int_{t_0}^{t}(-t_{s}(t'))dt')sh\left(\frac{\sqrt{-\int_{t_0}^{t}|t_{s}(t')|^2dt'}}{\hbar}\right)}{\sqrt{-\int_{t_0}^{t}((t_{si}(t')^2+t_{sr}(t'))^2)dt'}} & e^{\frac{-i\int_{t_0}^{t}E_p(t')dt'}{\hbar}}ch\left(\frac{\sqrt{-\int_{t_0}^{t}(|t_{s}(t')|^2)dt'}}{\hbar}\right)  \end{pmatrix}, \nonumber \\
\end{eqnarray}
where $sh$(.) and $ch$(.) are $\sinh$ and $\cosh$ hyperbolic functions, where $|t_s(t)|^2=|t_{sr}(t)|^2+|t_{si}(t)|^2$. This matrix is unitary so $\hat{U}^{\dag}(t,t_0)=\hat{U}^{-1}(t,t_0)$.
At the very end we will also consider more general case when $E_{p1}(t) \neq E_{p2}(t)$. At first let us consider the case of two localized states in the left and right quantum well so there is no hopping which implies $t_s=0$.
In such case the evolution matrix $\hat{U}(t,t_0)$ is unitarian and has the following form
\begin{eqnarray}
 \hat{U}(t,t_0)=e^{\frac{1}{i\hbar}\int_{t_0}^{t}\hat{H}(t_1)dt_1}=
\begin{pmatrix}
e^{\frac{-i \int_{t_0}^{t}E_{p1}(t')dt'}{\hbar}} & 0 \\
0 & e^{\frac{-i \int_{t_0}^{t}E_{p2}(t')dt'}{\hbar}}
\end{pmatrix},
\end{eqnarray}
what implies that left and right quantum dot are two disconnected physical systems subjected to its own evolution with time. However since one electron is distributed between those physical systems the measurement conducted on the left quantum dot will have its immediate effect on the right quantum dot. Another extreme example is the situation when hopping energy is considerably bigger than localization energy. In such case we set $E_{p1}=E_{p2}=0$ and in case of non-zero hopping terms we obtain
\begin{eqnarray}
 \hat{U}(t,t_0)=e^{\frac{1}{i\hbar}\int_{t_0}^{t}\hat{H}(t_1)dt_1}= 
\begin{pmatrix}
ch\left(\frac{\sqrt{-\int_{t_0}^{t}(|t_{s}(t')|^2)dt'}}{\hbar}\right) & \frac{ (\int_{t_0}^{t}(t_{s}^{*}(t'))dt')sh\left(\frac{\sqrt{-\int_{t_0}^{t}|t_{s}(t')|^2)}}{\hbar}\right)}{\sqrt{-\int_{t_0}^{t}((t_{si}(t')^2+t_{sr}(t'))^2)dt'}}\\
\frac{ (\int_{t_0}^{t}(-t_{s}(t'))dt')sh\left(\frac{\sqrt{-\int_{t_0}^{t}|t_{s}(t')|^2dt'}}{\hbar}\right)}{\sqrt{-\int_{t_0}^{t}((t_{si}(t')^2+t_{sr}(t'))^2)dt'}} & ch\left(\frac{\sqrt{-\int_{t_0}^{t}(|t_{s}(t')|^2)dt'}}{\hbar}\right)
\end{pmatrix}, \nonumber \\
\end{eqnarray}
Now it is time to move to most general situation of $E_{p1} \neq E_{p2}$, $t_{sr}, t_{si} \neq 0$. We have 4 elements of evolution matrix given as
$ \hat{U}(t,t_0)=e^{\frac{1}{i\hbar}\int_{t_0}^{t}\hat{H}(t_1)dt_1}= 
\begin{pmatrix}
 U(t,t_0)_{1,1} & U(t,t_0)_{1,2} \nonumber \\
 U(t,t_0)_{2,1}= U(t,t_0)_{1,2}^{*} & U(t,t_0)_{2,2}
\end{pmatrix}.$
\begin{eqnarray}
 U(t,t_0)_{1,1}=
\frac{\exp \left(-  \frac{\sqrt{-\hbar^2 \left(|\int_{t_0}^{t}dt'(E_{p1}(t')-E_{p2}(t'))|^2+4 \left(|\int_{t_0}^{t}dt't_{si}(t')|^2+|\int_{t_0}^{t}dt't_{sr}(t')|^2\right)\right)}+i \hbar \int_{t_0}^{t}dt'(E_{p1}(t')+E_{p2}(t'))}{2 \hbar^2}\right)}{2 \hbar
   \left( ( \int_{t_0}^{t}dt'(E_{p1}(t')-E_{p2}(t')))^2+4 \left(|\int_{t_0}^{t}dt't_{si}(t')|^2+|\int_{t_0}^{t}dt't_{sr}(t')|^2\right)\right)} \times \nonumber \\
\times  \Bigg[-i (\int_{t_0}^{t}dt'E_{p1}(t')) \sqrt{-\hbar^2 \left(|\int_{t_0}^{t}dt'(E_{p1}(t')-E_{p2}(t'))|^2+ 4 \left(|\int_{t_0}^{t}dt't_{si}(t')|^2+|\int_{t_0}^{t}dt't_{sr}(t')|^2\right)\right)}+ \nonumber \\  +\hbar \left(|\int_{t_0}^{t}dt'(E_{p1}(t')-E_{p2}(t'))|^2+4
   \left(|\int_{t_0}^{t}dt't_{si}(t')|^2+|\int_{t_0}^{t}dt't_{sr}(t')|^2\right)\right) \times \nonumber \\ e^{\frac{\sqrt{-\hbar^2 \left(|\int_{t_0}^{t}dt'(E_{p1}(t')-E_{p2}(t'))|^2+4 \left(|\int_{t_0}^{t}dt't_{si}(t')|^2+|\int_{t_0}^{t}dt't_{sr}(t')|^2\right)\right)}}{\hbar^2}}+ \nonumber\\  + \left(
   \left(( \int_{t_0}^{t}dt'(E_{p1}(t')-E_{p2}(t')))^2+4 \left(|\int_{t_0}^{t}dt't_{si}(t')|^2+|\int_{t_0}^{t}dt't_{sr}(t')|^2\right)\right) \right) + \nonumber \\
+i (\int_{t_0}^{t}dt'E_{p1}(t')) e^{\frac{\sqrt{-h^2 \left(|\int_{t_0}^{t}dt'(E_{p1}(t')-E_{p2}(t'))|^2+4 \left(|\int_{t_0}^{t}dt't_{si}(t')|^2+|\int_{t_0}^{t}dt't_{sr}(t')|^2\right)\right)}}{\hbar^2}} \times \nonumber \\ \sqrt{-\hbar^2 \left(|\int_{t_0}^{t}dt'(E_{p1}(t')-E_{p2}(t'))|^2+4
   \left(|\int_{t_0}^{t}dt't_{si}(t')|^2+|\int_{t_0}^{t}dt't_{sr}(t')|^2\right)\right)} \nonumber \\ -i (\int_{t_0}^{t}dt'E_{p2}(t')) e^{\frac{\sqrt{-\hbar^2 \left(|\int_{t_0}^{t}dt'(E_{p1}(t')-E_{p2}(t'))|^2+4 \left(|\int_{t_0}^{t}dt't_{si}(t')|^2+|\int_{t_0}^{t}dt't_{sr}(t')|^2\right)\right)}}{\hbar^2}}\times \nonumber \\
   \sqrt{-\hbar^2 \left(|\int_{t_0}^{t}dt'(E_{p1}(t')-E_{p2}(t'))|^2+4 \left(|\int_{t_0}^{t}dt't_{si}(t')|^2+|\int_{t_0}^{t}dt't_{sr}(t')|^2\right)\right)}+ \nonumber \\ + i (\int_{t_0}^{t}dt' E_{p2}(t')) \sqrt{-\hbar^2 \left(|\int_{t_0}^{t}dt'(E_{p1}(t')-E_{p2}(t'))|^2+4
   \left(|\int_{t_0}^{t}dt't_{si}(t')|^2+|\int_{t_0}^{t}dt't_{sr}(t')|^2\right) \right)} \Bigg].
\end{eqnarray}

\begin{eqnarray}
U(t,t_0)_{1,2}=\nonumber \\
\frac{2 \hbar (\int_{t_0}^{t}dt'(t_{si}(t')-i t_{sr}(t'))) e^{-\frac{i \int_{t_0}^{t}dt'(E_{p1}(t')+E_{p2}(t'))}{2 \hbar}} \sinh \left(\frac{\sqrt{-\hbar^2 \left(|\int_{t_0}^{t}dt'(E_{p1}(t')-E_{p2}(t'))|^2+4
   \left(|\int_{t_0}^{t}dt't_{si}(t')|^2+|\int_{t_0}^{t}dt't_{sr}(t')|^2\right)\right)}}{2 h^2}\right)}{\sqrt{-\hbar^2 \left(|\int_{t_0}^{t}dt'(E_{p1}(t')-E_{p2}(t'))|^2+4 \left(|\int_{t_0}^{t}dt't_{si}(t')|^2+|\int_{t_0}^{t}dt't_{sr}(t')|^2\right)\right)}}= \nonumber \\
   =U(t,t_0)_{2,1}^{*}. \nonumber \\
\end{eqnarray}

\begin{eqnarray}
 U(t,t_0)_{2,2}=
\frac{\exp \left(-  \frac{\sqrt{-\hbar^2 \left(|\int_{t_0}^{t}dt'(E_{p1}(t')-E_{p2}(t'))|^2+4 \left(|\int_{t_0}^{t}dt't_{si}(t')|^2+|\int_{t_0}^{t}dt't_{sr}(t')|^2\right)\right)}-i \hbar \int_{t_0}^{t}dt'(E_{p1}(t')+E_{p2}(t'))}{2 \hbar^2}\right)}{2 \hbar
   \left( ( \int_{t_0}^{t}dt'(E_{p1}(t')-E_{p2}(t')))^2+4 \left(|\int_{t_0}^{t}dt't_{si}(t')|^2+|\int_{t_0}^{t}dt't_{sr}(t')|^2\right)\right)} \times \nonumber \\
\times  \Bigg[+i (\int_{t_0}^{t}dt'E_{p1}(t')) \sqrt{-\hbar^2 \left(|\int_{t_0}^{t}dt'(E_{p1}(t')-E_{p2}(t'))|^2+ 4 \left(|\int_{t_0}^{t}dt't_{si}(t')|^2+|\int_{t_0}^{t}dt't_{sr}(t')|^2\right)\right)}+ \nonumber \\  +\hbar \left(|\int_{t_0}^{t}dt'(E_{p1}(t')-E_{p2}(t'))|^2+4
   \left(|\int_{t_0}^{t}dt't_{si}(t')|^2+|\int_{t_0}^{t}dt't_{sr}(t')|^2\right)\right) \times \nonumber \\ e^{\frac{\sqrt{-\hbar^2 \left(|\int_{t_0}^{t}dt'(E_{p1}(t')-E_{p2}(t'))|^2+4 \left(|\int_{t_0}^{t}dt't_{si}(t')|^2+|\int_{t_0}^{t}dt't_{sr}(t')|^2\right)\right)}}{\hbar^2}}+ \nonumber\\  + \left(
   \left(( \int_{t_0}^{t}dt'(E_{p1}(t')-E_{p2}(t')))^2+4 \left(|\int_{t_0}^{t}dt't_{si}(t')|^2+|\int_{t_0}^{t}dt't_{sr}(t')|^2\right)\right) \right) + \nonumber \\
-i (\int_{t_0}^{t}dt'E_{p1}(t')) e^{\frac{\sqrt{-\hbar^2 \left(|\int_{t_0}^{t}dt'(E_{p1}(t')-E_{p2}(t'))|^2+4 \left(|\int_{t_0}^{t}dt't_{si}(t')|^2+|\int_{t_0}^{t}dt't_{sr}(t')|^2\right)\right)}}{\hbar^2}} \times \nonumber \\ \sqrt{-\hbar^2 \left(|\int_{t_0}^{t}dt'(E_{p1}(t')-E_{p2}(t'))|^2+4
   \left(|\int_{t_0}^{t}dt't_{si}(t')|^2+|\int_{t_0}^{t}dt't_{sr}(t')|^2\right)\right)} \nonumber \\ +i (\int_{t_0}^{t}dt'E_{p2}(t')) e^{\frac{\sqrt{-\hbar^2 \left(|\int_{t_0}^{t}dt'(E_{p1}(t')-E_{p2}(t'))|^2+4 \left(|\int_{t_0}^{t}dt't_{si}(t')|^2+|\int_{t_0}^{t}dt't_{sr}(t')|^2\right)\right)}}{\hbar^2}} \times \nonumber \\
   \sqrt{-\hbar^2 \left(|\int_{t_0}^{t}dt'(E_{p1}(t')-E_{p2}(t'))|^2+4 \left(|\int_{t_0}^{t}dt't_{si}(t')|^2+|\int_{t_0}^{t}dt't_{sr}(t')|^2\right)\right)}+ \nonumber \\ - i (\int_{t_0}^{t}dt' E_{p2}(t')) \sqrt{-\hbar^2 \left(|\int_{t_0}^{t}dt'(E_{p1}(t')-E_{p2}(t'))|^2+4
   \left(|\int_{t_0}^{t}dt't_{si}(t')|^2+|\int_{t_0}^{t}dt't_{sr}(t')|^2\right)\right)} \Bigg].
\end{eqnarray}

We recognize that more efficient mathematical representation of qubit evolution with time is by introducing 4 quantities that are real valued functions of the form:
\begin{eqnarray*}
EP_{1}(t)=EP[E_{p1}]_{t}=\int_{t_0}^{t}dt'E_{p1}(t'), EP_{2}(t)=EP[E_{p2}]_{t}=\int_{t_0}^{t}dt'E_{p2}(t'), \nonumber \\
TR(t)=TR[t_{sr}]_t=\int_{t_0}^{t}dt't_{sr}(t'),
TI(t)=TI[t_{si}]_t=\int_{t_0}^{t}dt't_{si}(t').
\end{eqnarray*}
It shall be underlined that $E_{p1}(t')$, $E_{p2}(t')$,$t_{sr}(t')$ and $t_{si}(t')$ can be continuous or discontinuous real valued functions of finite value of any dependence and that $EP[.]$, $TR[.]$ and $TI[.]$ are functionals of Hamiltonian parameters. Usually in case of nano-circuit their range of values and time-dependence is limited but can be extended with more advanced engineering and circuit topology. It can be carefully
examined if one moves from Schroedinger to tight-binding formalism so value $E_{p1}$ is associated with energy of particle localized at node 1 and $E_{p2}$ is associated with energy of particle localized at node 2, while $t_s$ is measure of energy that can be transported
between node 1 and 2 that takes places during particle movement. $t_s$ can also be measured by the delocalized energy between 2 nodes. Therefore highly energetic particle moving across nanostructure of q-wells shall have high value of $t_s$ and low value of $E_{p1}$ and
$E_{p2}$ so ballistic transport takes place. On another hand slowly moving particle participating in diffusive transport between one q-well and neighbouring q-well is strongly localized so $E_{p1}, E_{p2} >> |t_s|$.
\section{Action of phase rotating gate described analytically}

Let us consider the situation of single qubit from Fig.\ref{fig:central} when we assume the following dependencies: $E_{p1}(t)=E_{p2}(t)=E_p=constant$ and $t_{s12}(t)=t_{s21}(t)=t_s(t)=constant_1$.
In such we have two time-independent eigenenergies $E_1=E_p-t_s$ and $E_p+t_s$. For simplicity we assume $(\alpha(0)\in R),(\beta(0)\in R)$ .The probability of finding electron at node 1 is given by angle $\Theta$ at Bloch sphere expressed as
\begin{eqnarray}
P_1(t)=|\alpha(t)|^2=\frac{1}{2}((|\alpha(0)|^2+|\beta(0)|^2)+\frac{1}{2}(|\alpha(0)|^2-|\beta(0)|^2)\cos((\frac{E_2-E_1)t}{\hbar}))=\cos(\Theta(t))^2, \nonumber \\
P_2(t)=|\alpha(t)|^2=\frac{1}{2}((|\alpha(0)|^2+|\beta(0)|^2)-\frac{1}{2}(|\alpha(0)|^2-|\beta(0)|^2)\cos((\frac{E_2-E_1)t}{\hbar}))=\sin(\Theta(t))^2,
\end{eqnarray} and it oscillates periodically with frequency proportional to distance between energetic levels $E_2$ and $E_1$ and is given as $\omega_0=\frac{E_2-E_1}{\hbar}$. Therefore the same occupancy at node is repeating with periodic time $t_d=n\frac{2\pi \hbar}{E_2-E_1}$ for integer n. Obviously probability of finding of particle at node 2 is $P_2=1-P_1$. The phase difference between wavefunctions at node 1 and 2 is denoted as $\phi(t)$ and can be expressed analytically by formula
\begin{eqnarray}
\label{qq}
-\phi(t)=ASin\left[\frac{\sin(\frac{E_1 t}{\hbar})(|\alpha(0)|^2-|\beta(0)|^2)+\sin(\frac{E_2 t}{\hbar})(|\alpha(0)|^2+|\beta(0)|^2)}{\cos(\frac{E_1 t}{\hbar})(|\alpha(0)|^2-|\beta(0)|^2)+\cos(\frac{E_2 t}{\hbar})(|\alpha(0)|^2+|\beta(0)|^2)}\right] \nonumber  \\
=ASin\left[ \frac{\frac{1}{2i}(\exp(i\frac{E_1 t}{\hbar})-\exp(-i\frac{E_1 t}{\hbar}))(|\alpha(0)|^2-|\beta(0)|^2)+\frac{1}{2i}(\exp(i\frac{E_2 t}{\hbar})-\exp(-i\frac{E_2 t}{\hbar}))(|\alpha(0)|^2+|\beta(0)|^2)}{\frac{1}{2}(\exp(i\frac{E_1 t}{\hbar})+\exp(-i\frac{E_1 t}{\hbar})))(|\alpha(0)|^2-|\beta(0)|^2)+\frac{1}{2}(\exp(i\frac{E_2 t}{\hbar})+\exp(-i\frac{E_2 t}{\hbar}))
(|\alpha(0)|^2+|\beta(0)|^2)}  \right] \nonumber \\
=ASin\left[ \frac{\frac{1}{2i}(1-\exp(-i\frac{2E_1 t}{\hbar}))(|\alpha(0)|^2-|\beta(0)|^2)+\frac{1}{2i}(\exp(i\frac{(E_2 -E_1)t}{\hbar})-\exp(-i\frac{(E_2 +E_1)t}{\hbar}))}{\frac{1}{2}(1+\exp(-i\frac{2E_1 t}{\hbar})))(|\alpha(0)|^2-|\beta(0)|^2)+\frac{1}{2}(\exp(i\frac{(E_2-E_1) t}{\hbar})+\exp(-i\frac{(E_1+E_2 )t}{\hbar}))
} \right ]= \nonumber \\
=ASin\left[ \frac{\frac{1}{2i}(1-\exp(-i\frac{2E_1 t}{\hbar}))(|\alpha(0)|^2-|\beta(0)|^2)+\frac{1}{2i}(  \cos(\frac{(E_2-E_1) t}{\hbar})+i\sin(\frac{(E_2-E_1) t}{\hbar}))-\exp(-i\frac{(E_2 +E_1)t}{\hbar}))}{\frac{1}{2}(1+\exp(-i\frac{2E_1 t}{\hbar})))(|\alpha(0)|^2-|\beta(0)|^2)+\frac{1}{2}(\cos(\frac{(E_2-E_1) t}{\hbar})+i\sin(\frac{(E_2-E_1) t}{\hbar})+\exp(-i\frac{(E_1+E_2 )t}{\hbar}))
}  \right]= \nonumber \\
ASin\left[ \frac{(1-e^{-i\frac{2E_1 t}{\hbar}})(|\alpha(0)|^2-|\beta(0)|^2)+\left(\frac{\cos(\Theta(t))^2-\frac{1}{2}}{\frac{1}{2}(|\alpha(0)|^2-|\beta(0)|^2)}+ i |\sqrt{1-(\frac{\cos(\Theta(t))^2-\frac{1}{2}}{\frac{1}{2}(|\alpha(0)|^2-|\beta(0)|^2)})^2}|s_{\sin(\frac{(E_2-E_1) t}{\hbar})} -e^{-i\frac{(E_2 +E_1)t}{\hbar}}\right)}{i(1+e^{-i\frac{2E_1 t}{\hbar}}))(|\alpha(0)|^2-|\beta(0)|^2)+i\left(\frac{\cos(\Theta(t))^2-\frac{1}{2}}{\frac{1}{2}(|\alpha(0)|^2-|\beta(0)|^2)} +  i |\sqrt{1-(\frac{\cos(\Theta(t))^2-\frac{1}{2}}{\frac{1}{2}(|\alpha(0)|^2-|\beta(0)|^2)})^2}|s_{\sin(\frac{(E_2-E_1) t}{\hbar})} +e^{-i\frac{(E_1+E_2 )t}{\hbar}}\right)
}  \right]  \nonumber \\
\end{eqnarray}
We recognize that three frequencies are involved  $\omega_1=\frac{E_1}{\hbar}$, $\omega_{21m}=\frac{E_2-E_1}{\hbar}$,$\omega_{21p}=\frac{E_2+E_1}{\hbar}$  in the dynamics of phase difference of quantum state between nodes 2 and 1. We are using sign function as
$Sign(\sin(\frac{(E_2-E_1) t}{\hbar})))=s_{(\sin(\frac{(E_2-E_1) t}{\hbar}))}$ so it has 1 and -1 values for positive and negative values of $\sin\frac{(E_2-E_1) t}{\hbar}$ and 0 otherwise.
More phase difference across position based qubit between nodes 1 and 2 is codependent on the occupancy of the left and right node as given by last equation  in the case of time-independent Hamiltonian. Such situation is not taking place in most conventional qubits using energy eigenbases to encode information but takes place in position based semiconductor qubit. The ideal phase rotating gate implemented in position based qubit brings desired phase difference between wavefunctions at nodes 2 and 1 is not changing the occupancy of node 1 and 2. If we want to keep the occupancy from time t=0 we need to consider times $t_d=n\frac{2\pi \hbar}{E_2-E_1}$. At time t=0 phase difference was assumed to be 0.
\section{Action of Hadamard gate in position qubit}

The Hadamard gate is able to conduct the following unitary transformation on quantum state $\ket{\psi(t)}$ and is given as
\begin{equation}
U_{Hadamard}=
\begin{pmatrix}
1 & 1 \\
1 & -1
\end{pmatrix}
. 
\end{equation}
It has property $U_{Hadamard}^{\dag}=U_{Hadamard}$ and $U_{Hadamard}U_{Hadamard}^{\dag}=1$ so double action of Hadamard gate
gives $U_{Hadamard}U_{Hadamard}^{-1}=1$.

Let us concentrate on the position dependent qubit with time-independent parameters $E_{p1},E_{p2}=E_{p1}=E_p,t_s\in R$.
In such case we obtain following eigenenrgies $E_1=E_p-t_s$ and $E_2=E_p+t_s$.
From simple calculations we can notice that two eigenergies $E_1=E_p-t_s$ and $E_1=E_p+t_s$ have corresponding eigenstates
\begin{equation}
\ket{E_1}=\frac{1}{\sqrt{2}}(\ket{1,0}_x-\ket{0,1}_x),\ket{E_2}=\frac{1}{\sqrt{2}}(\ket{1,0}_x+\ket{0,1}_x),
\label{eq:Hadamard}
\end{equation}
that are orthonormal so $\braket{1,0}{1,0}=\braket{0,1}{0,1}=1$ and $\braket{1,0}{0,1}=\braket{0,1}{1,0}=0$.
At the same time $\braket{E_1}{E_1}=\braket{E_2}{E_2}=1$ and $\braket{E_1}{E_2}=\braket{E_2}{E_1}=0$.
We recognize that formula \ref{eq:Hadamard} can be written in the compact form as
\begin{eqnarray}
\begin{pmatrix}
\ket{E_2} \\
\ket{E_1}
\end{pmatrix}=
\frac{1}{\sqrt{2}}
\begin{pmatrix}
1 & 1 \\
1 & -1
\end{pmatrix}
\begin{pmatrix}
\ket{1,0}_x \\
\ket{0,1}_x
\end{pmatrix}
=\hat{U}_{Hadamard}
\begin{pmatrix}
\ket{1,0}_x \\
\ket{0,1}_x
\end{pmatrix}, \nonumber \\
\begin{pmatrix}
\ket{1,0}_x \\
\ket{0,1}_x
\end{pmatrix}=
\frac{1}{\sqrt{2}}
\begin{pmatrix}
1 & 1 \\
1 & -1
\end{pmatrix}
\begin{pmatrix}
\ket{E_2} \\
\ket{E_1}
\end{pmatrix}
=\hat{U}_{Hadamard}
\begin{pmatrix}
\ket{E_2} \\
\ket{E_1}
\end{pmatrix}.
\end{eqnarray}
We recognize that quantum transformation is naturally encoded in transformation from position quantum system eigenbases into energy eigenbases.
Quantum logical 0 can be spanned (represented) by state $\ket{1,0}_x=\ket{0}_L$ (presence of electron in qubit on the left side in Fig.\ref{fig:central}) and quantum logical 1 can be spanned (represented) by the state $\ket{0,1}_x=\ket{1}_R$ (presence of electron in qubit on the right side).
Therefore qubit state shall be defined by
\begin{eqnarray}
\ket{\psi_t}=\alpha(t)\ket{1,0}_x+\beta\ket{0,1}_x=e^{iPh(\alpha(t))=i \xi(t)}(|\alpha(t)|\ket{1,0}_x  
+e^{Ph(\beta(t))-Ph(\alpha(t))}\beta(t)\ket{0,1}_x)= \nonumber \\
=e^{i \xi(t)}(|\alpha(t)|\ket{1,0}_x+e^{i\phi(t)}\beta \ket{0,1}_x).
\end{eqnarray}
Action of Hadamard gate requires
\begin{eqnarray}
\ket{0}_L =\ket{1,0}_x \rightarrow \frac{1}{\sqrt{2}}(\ket{1,0}_x+\ket{0,1}_x)=\frac{1}{\sqrt{2}}(\ket{1}_L+\ket{2}_L), 
\ket{0}_R =\ket{0,1}_x \rightarrow \frac{1}{\sqrt{2}}(\ket{1,0}_x-\ket{0,1}_x)=\frac{1}{\sqrt{2}}(\ket{1}_L-\ket{2}_L).  \nonumber \\ 
\end{eqnarray}
that is heating up (left transition from occupancy of two energetic levels expressed by quantum state $\ket{1,0}_x$ to occupancy of $E_2$ level given by quantum state $\frac{1}{\sqrt{2}}(\ket{1,0}_x+\ket{0,1}_x)$) or cooling down (right transition from occupancy of 2 energetic levels expressed by quantum state $\ket{0,1}_x$ to the occupancy of ground state $E_1$ given by quantum state $\frac{1}{\sqrt{2}}(\ket{1}_L-\ket{2}_L)$) of quantum state in 2 energy level system. We recognize that quantum logical 0 or presence of state (electron) in left well is achieved when there is equal occupancy (given by $c_{E1}$) of energetic level $E_1$ and  $E_2$ so $|c_{E1}(t)|^2=|c_{E2}(t)|^2$. The scheme how to change the complete occupancy of energetic level $E_1$ into full occupancy of energetic level $E_2$ is given by formula \ref{example} that is associated with time-dependent Hamiltonian applied to position based qubit.
The quantum state is given as
\begin{eqnarray*}
 \ket{\psi(t)}=\frac{1}{\sqrt{2}}[(c_{E1}(t)(\ket{1,0}_x-\ket{0,1}_x))+(c_{E2}(t)(\ket{1,0}_x+\ket{0,1}_x))]= \nonumber \\
\frac{1}{\sqrt{2}}[e^{\frac{1}{\hbar}}(e^{\frac{1}{\hbar i}(t-t_0)E_1}c_{E1}(t_0)(\ket{1,0}_x-\ket{0,1}_x))+
(e^{\frac{1}{\hbar i}(t-t_0)E_2(t-t_0)}c_{E2}(t_0)(\ket{1,0}_x+\ket{0,1}_x))]= \nonumber \\
\frac{1}{\sqrt{2}}[((+e^{\frac{1}{\hbar i}(t-t_0)E_1}c_{E1}(t_0)+(e^{\frac{1}{\hbar i}(t-t_0)E_2(t-t_0)}c_{E2}(t_0))\ket{1,0}_x+ \nonumber \\
((-e^{\frac{1}{\hbar i}(t-t_0)E_1}c_{E1}(t_0)+(e^{\frac{1}{\hbar i}(t-t_0)E_2(t-t_0)}c_{E2}(t_0))\ket{0,1}_x].
\end{eqnarray*}
Such state will evolve after characteristic time from
logical state $\ket{0}_L$ into quantum logical $\ket{1}_L$ and later into $\ket{0}_L$ and so on.
We can also set logical quantum state in position space parameterized by x and  we can read the results of Hadamard operation action in energy space or reversely. Engineers have the choice of setting qubit state in a position space (what is more intuitive if one aims to obtain high integration circuits) or in energy space. By setting the quantum state in position space (as by injecting electron from left side into left well of qubit) one needs to read it by energy space or reversely. Reading the quantum state after Hadamard operation (or any other quantum operation) in energy space requires either spectroscopy of occupation of energy levels which basically means that we need to use microwaves in order to populate or depopulate given energy level(s). Alternative method for reading the qubit state after Hadamard operation (or any other quantum operation)is determination the state of neighbouring qubit that interacts with measured qubit in electrostatic way as it is depicted in the right side of Fig.\ref{fig:central}. The determination of occupancy of energy level $E_1$ and $E_2$ will give us
the information on the qubit state after Hadamard operation (so presence of at least 2 energy levels in physical system is the requirement) and formally we have
\begin{eqnarray}
\ket{\psi}_{output}=c_{E1}\ket{E_1}+c_{E2}\ket{E_2}=c_{E1}\ket{0}_{L-output}+c_{E2}\ket{1}_{L-output}=
\hat{U}_{Hadamard}(\alpha\ket{0}_{L-input}+\beta\ket{1}_{L-input}).
\end{eqnarray}
\section{Rabi oscillations in general case for 2 energy level system}
In general case during heating up of q-state or during cooling down of q-state we need to consider the Hamiltonian as
$H=E_1\ket{E_1}\bra{E_1}+E_2\ket{E_2}\bra{E_2}+f_1(t)\ket{E_2}\bra{E_1}+f_2(t)\ket{E_1}\bra{E_2}$. If we want to have time-dependent only $E_1(t)$ and only$ E_2(t)$ states we need to consider
$H=E_1\ket{E_1(t)}\bra{E_1(t)}+E_2\ket{E_2(t)}\bra{E_2(t)}+f_1(t)\ket{E_2(t)}\bra{E_1(t)}+f_2(t)\ket{E_1(t)}\bra{E_2(t)}$.
Let us see the dynamics of quantum states with time so we have $f_1(t),f_2(t)=0$ for $t<=0$ and constant non-zero otherwise ($f_1(t)=f_1=const_1,f_2(t)=f_2=const_2$) so one obtains the equation
\begin{eqnarray}
+\hbar i \frac{d}{dt}c_{E1}(t) =(c_{E1}(t)E_1+f_2(t) c_{E2}(t)), 
+\hbar i \frac{d}{dt}c_{E2}(t) =(c_{E2}(t) E_2+f_1(t) c_{E1}(t)).
\end{eqnarray}
From first equation we have $\frac{1}{f_2(t)}(+\hbar i \frac{d}{dt}c_{E1}(t)(t)-E_1 c_{E1}(t)(t))=c_{E2}(t)$ and we obtain the second equation
\begin{eqnarray}
+\hbar i \frac{d}{dt} (\frac{1}{f_2(t)}(+\hbar i \frac{d}{dt}c_{E1}(t)-E_1c_{E1}(t))) 
=( \frac{1}{f_2(t)}(+\hbar i \frac{d}{dt}c_{E1}(t)-E_1c_{E1}(t)) ) E_2 + f_1(t)c_{E1}(t).
\end{eqnarray}
which gives,
\begin{eqnarray}
\frac{d}{dt} (\frac{1}{f_2(t)}(+\hbar i \frac{d}{dt}c_{E1}(t)-E_1c_{E1}(t)(t))) = \nonumber \\
=-\frac{df_2}{dt}\frac{1}{f_2^2(t)}(+\hbar i \frac{d}{dt}c_{E1}(t)-E_1c_{E1}(t)(t)) +(\frac{1}{f_2(t)}(+\hbar i \frac{d^2}{dt^2}c_{E1}(t)(t)-E_1\frac{d}{dt}c_{E1}(t))
\nonumber \\
=\frac{1}{i \hbar}( \frac{1}{f_2(t)}(+\hbar i \frac{d}{dt}c_{E1}(t)-E_1c_{E1}(t)(t)) ) E_2 + \frac{1}{i \hbar} f_1(t)c_{E1}(t)(t).
\end{eqnarray}
and it implies
\begin{eqnarray}
\frac{d^2}{dt^2}c_{E2}(t)\frac{\hbar i}{f_2(t)}+\frac{d}{dt}c_{E2}(t)[-\frac{df_2}{dt}\frac{\hbar i}{f_2^2(t)}-\frac{(E_1+E_2)}{f_2(t)}]+c_{E2}(t)[\frac{E_1}{i \hbar} \frac{E_2}{f_2(t)}+\frac{df_2}{dt}\frac{E_1}{f_2(t)^2}-\frac{1}{\hbar i}f_1(t)]=0.
\end{eqnarray}
After multiplication by $\frac{f_2(t)}{\hbar i}$ the last equation gives
\begin{eqnarray}
\frac{d^2}{dt^2}c_{E1}(t)+\frac{d}{dt}c_{E1}(t)[-\frac{df_2}{dt}\frac{1}{f_2(t)}+i\frac{(E_1+E_2)}{\hbar}]+\beta(t)[-\frac{E_1E_2}{ \hbar^2} - \frac{i}{\hbar} \frac{df_2}{dt}\frac{E_1}{f_2(t)}+\frac{1}{\hbar^2} f_1(t)f_2(t)]=0.
\end{eqnarray}
In analogical way we obtain
\begin{eqnarray}
\frac{d^2}{dt^2}c_{E2}(t)+\frac{d}{dt}c_{E2}(t)[-\frac{df_1}{dt}\frac{1}{f_1(t)}+i\frac{(E_1+E_2)}{\hbar}]+\beta(t)[-\frac{E_1E_2}{ \hbar^2} - \frac{i}{\hbar} \frac{df_1}{dt}\frac{E_2}{f_1(t)}+\frac{1}{\hbar^2} f_1(t)f_2(t)]=0.
\end{eqnarray}
Boundary conditions are given as
\begin{eqnarray}
i \hbar \frac{d}{dt}c_{E1}(t_0^{+}) = E_1 c_{E2}(t_0^{+}) + f_2(t_0^{+}) c_{E1}(t_0),
i \hbar \frac{d}{dt}c_{E2}(t_0^{+}) = E_2 c_{E1}(t_0^{+}) + f_1(t_0^{+}) c_{E2}(t_0), \nonumber  \\
c_{E2}(t_0^{+})=c_{E2}(t_0),
c_{E1}(t_0^{+})=c_{E1}(t_0) .
\end{eqnarray}
From later considerations it turns out that $f_1(t)^{*}=f_2(t)$ so $f_1(t)=f_a(t)+i f_b(t)$ and $f_2(t)=f_a(t)- i f_b(t)$, where $f_a(t)$ and $f_b(t)$ are real valued functions. Therefore we can write the equations of motion as
\begin{eqnarray}
\frac{d^2}{dt^2}c_{E1}(t)(t)+\frac{d}{dt}c_{E1}(t)[-\frac{df_2}{dt}\frac{1}{f_2(t)}+i\frac{(E_1+E_2)}{\hbar}]+c_{E1}(t_0)[-\frac{E_1E_2}{ \hbar^2} - \frac{i}{\hbar} \frac{df_2}{dt}\frac{E_1}{f_2(t)}+\frac{1}{\hbar^2} (f_a(t)^2+f_b(t)^2)=0. \nonumber \\
\end{eqnarray}
In analogical way we obtain
\begin{eqnarray}
\frac{d^2}{dt^2}c_{E2}(t_0)(t)+\frac{d}{dt}c_{E2}(t_0)[-\frac{df_1}{dt}\frac{1}{f_1(t)}+i\frac{(E_1+E_2)}{\hbar}]+c_{E1}(t)[-\frac{E_1E_2}{ \hbar^2} - \frac{i}{\hbar} \frac{df_1}{dt}\frac{E_2}{f_1(t)}+\frac{1}{\hbar^2}  (f_a(t)^2+f_b(t)^2)]=0. \nonumber \\
\end{eqnarray}
Boundary conditions are given as
\begin{eqnarray}
i \hbar \frac{d}{dt}c_{E2}(t_0^{+})=E_1 c_{E2}(t_0^{+}) + (f_a(t_0)-i f_b(t_0))c_{E1}(t_0), \nonumber  \\
i \hbar \frac{d}{dt}c_{E1}(t_0)(t_0^{+})= E_2 c_{E1}(t_0)(t_0^{+})  + (f_a(t_0)+i f_b(t_0)c_{E2}(t_0), \nonumber  \\
c_{E2}(t_0^{+})=c_{E2}(t_0), 
c_{E1}(t_0^{+})=c_{E1}(t_0) .
\end{eqnarray}
Very special case is when
$f_1(t)=a \exp(ct) + i b \exp(ct), f_2(t)=a \exp(ct) - i b \exp(ct)$,
 where c, a and b are real valued. 
In such cases we obtain the equations for the occupancy of energy state $E_1$ and $E_2$ expressed as
\begin{eqnarray}
\label{special1q}
\frac{d^2}{dt^2}c_{E2}(t)+\frac{d}{dt}c_{E2}(t)(t)[-c+i\frac{(E_1+E_2)}{\hbar}]+c_{E2}(t)(t)[-\frac{E_1E_2}{ \hbar^2} - \frac{i}{\hbar}E_1 c +\frac{1}{\hbar^2}(a^2+b^2)exp(2ct)]=0.
\end{eqnarray}
First case is $c=0,\hbar=1$  and solution is 
\begin{eqnarray}
c_{E1}(t) =
 e^{-\frac{1}{2} i (E_1 + E_2 - i \sqrt{-4 a^2 - 4 b^2 - E_1^2 + 2 E_1 E_2 - E_2^2}) t}g_1+   e^{\frac{1}{2}(- i(E_1 + E_2) + \sqrt{-4 a^2 - 4 b^2 - E_1^2 + 2 E_1E_2 - E_2^2}) t} g_2,
\end{eqnarray}
where $g_1$ and $g_1$ are complex values.
Having non-zero c we obtain solutions 

\begin{eqnarray}
c_{E1}(t)= c_1 \exp \left(\frac{1}{2} t \left(-\sqrt{-4 a^2 e^{2 ct}-4 b^2 e^{2 ct}+c^2-2 i c E_1+2 i c E_2- E_1^2+2 E_1
   E_2-E_2^2}+c-i E_1-i E_2\right)\right)+  \nonumber \\
c_2 \exp \left(\frac{1}{2} t \left(\sqrt{-4 a^2 e^{2 ct}-4 b^2 e^{2 ct}+c^2-2 i c E_1+2 i c
   E_2-E_1^2+2 E_1 E_2-E_2^2}+c-i E_1-i E_2\right)\right), \nonumber \\
\end{eqnarray}
\begin{eqnarray}
c_{E2}(t) = c_1 \exp \left(\frac{1}{2} t \left(-\sqrt{-4 a^2 e^{2 ct}-4 b^2 e^{2 ct}+c^2+2 i c E_1-2 i c E_2-E_1^2+2 E_1
   E_2-E_2^2}+c-i E_1-i E_2 \right)\right)+     \nonumber \\     c_2 \exp \left(\frac{1}{2} t \left(\sqrt{-4 a^2 e^{2 ct}-4 b^2 e^{2 ct}+c^2+2 i c E_1-2 i c
   E_2-E_1^2+2 E_1E_2-E_2^2}+c-i E_1-i E_2\right)\right).\nonumber \\
\end{eqnarray}
The simplified case of last formula can be given as
\begin{eqnarray}
\label{example}
c_{E2}(t) =
-g_4\exp \left(-\frac{1}{2} i t \left(-i \sqrt{-E1^2+2 E_1 E_2-\text{E2}^2-4}+E_1+E_2\right)\right) \nonumber \\
\left(-1+\exp \left(\frac{1}{2} i t
   \left(-i \sqrt{-E_1^2+2 E_1 E_2-E_2^2-4}+E_1+E_2\right)+\frac{1}{2} t \left(\sqrt{-E_1^2+2 E_1 E_2-E_2^2-4}-i
   (E_1+E_2)\right)\right)\right) \nonumber \\
\end{eqnarray}
and the numerical example of its dependence on time is depicted in Fig.\,\ref{fig:spectra}, where initially energy level $E_1$ was completely populated and with time the full population of energy level $E_2$ was achieved while energy level $E_1$ was completely depopulated. Such dependence can be used for example in the action of Hadamard gate implemented in electrostatic position dependent qubit.  If $f_1(t)$ and $f_2(t)$ functions have small values one can assume $\ket{E_1}=\frac{1}{\sqrt{2}}(\ket{1,0}_x-\ket{0,1}_x)$ and $\ket{E_2}=\frac{1}{\sqrt{2}}(\ket{1,0}_x+\ket{0,1}_x)$ and
\begin{equation}
\hat{H}(t)_x=
\begin{pmatrix}
E_p & t_s \\
t_s^{*} & E_p
\end{pmatrix}+\frac{1}{2}
\begin{pmatrix}
+f_1(t)+f_2(t) & -f_1(t)+f_2(t) \\
+f_1(t)-f_2(t) & -(f_1(t)+f_2(t))
\end{pmatrix}.
\end{equation}
Hermicity of last Hamiltonian requires that $f_1(t)=f_2(t)^{*}$.
\normalsize

\section{Extension of 2-energy tight binding model into N energetic levels for position based qubit in arbitrary electromagnetic enviroment}

\begin{figure}
\includegraphics[scale=0.3]{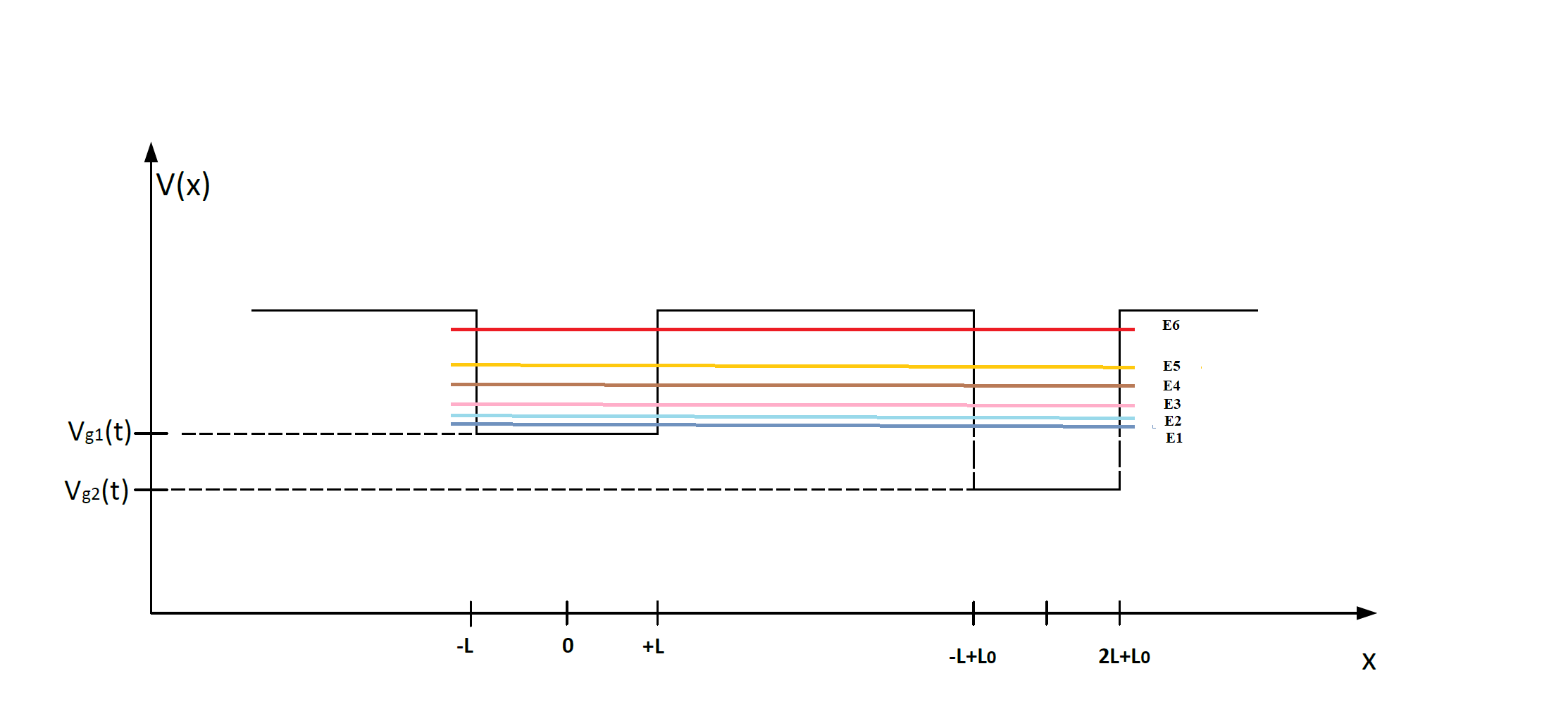} 
\caption{Case of position based qubit with N=6 energetic levels and unoccupied ground state.}
\end{figure}

Pictures presented before as in equation \ref{simplematrix} with N=2 energetic levels  can be easily extended for arbitrary number of energy levels $E_1<E_2<..<E_{2N_1=N}$ what is valid in time-independent case.
It is worth mentioning that very last chain of inequalities between time depedent eigenenergies does not need to be always valid in the general case of time-dependent Hamiltonian.
In most general case we have $N=2N_1$ energetic levels among 2 coupled quantum wells controlled electrostatically. Quite obviously we are omitting continuum spectrum of eigenenergies and we only concentrate on the system with electrons confiment by some effective potential. It requires introduction of $2N_1$ orthogonal Wannier functionl bases such that $\ket{x_1}_1,..,\ket{x_1}_{N_1}, \ket{x_2}_1,..,\ket{x_2}_{N_1}$=($\ket{1,0}_{E1-E2},$.., \newline
$\ket{1,0}_{E_{{N_1}-1}-E_{N_1}}, \ket{0,1}_{E1-E2},..,\ket{0,1}_{E_{N_1-1},E_{N_1}}$) and such that $\bra{x_1}_k (\ket{x_2}_m )=0$ for any m different than k. In such case the quantum state for $N_1=3$ ($N=2N_1$) is described as
\begin{eqnarray}
\ket{\psi}(t)=\gamma_{E_1-E_2,p1}(t)\ket{x_1}_{E_1,E_2}+\gamma_{E_3-E_4,p1}(t) \ket{x_1}_{E_3,E_4}+\gamma_{E_5-E_6,p1}(t)\ket{x_1}_{E_5,E_6} + \nonumber \\ \gamma_{E_5-E_6,p2}(t)\ket{x_2}_{E_5-E_6}+\gamma_{E_3-E_4,p2}(t)\ket{x_2}_{E_3-E_4} + \gamma_{E_1,p2}(t) \ket{x_2}_{E_1-E_2}= \nonumber \\
=\frac{1}{\sqrt{N}}[\gamma_{E_1-E_2,p1}(t)
\begin{pmatrix}
1 \\
0 \\
0 \\
0 \\
0 \\
0  \\
\end{pmatrix}
+\gamma_{E_3-E_4,p1}(t)
\begin{pmatrix}
0 \\
1 \\
0 \\
0 \\
0 \\
0  \\
\end{pmatrix}
+
..+
\gamma_{E_1-E_2,p2}(t)
\begin{pmatrix}
0 \\
0 \\
0 \\
0 \\
0 \\
1  \\
\end{pmatrix}
]=
\begin{pmatrix}
\gamma_{E_1-E_2,p1}(t) \\
\gamma_{E_3-E_4,p1}(t) \\
\gamma_{E_5-E_6,p1}(t) \\
\gamma_{E_5-E_6,p2}(t) \\
\gamma_{E_3-E_4,p2}(t) \\
\gamma_{E_1-E_2,p2}(t)  \\
\end{pmatrix}.
\end{eqnarray}
The probability of presence of electron at node 1 is $P_1(t)=|\gamma_{E_1-E_2,p1}(t)+\gamma_{E_3-E_4,p1}(t)+\gamma_{E_5-E_6,p1}(t)|^2$ and the probability of presence of electrone at node 2 is $P_2(t)=|\gamma_{E_1-E_2,p2}(t)+\gamma_{E_3-E_4,p2}(t)+\gamma_{E_5-E_6,p2}(t)|^2$. The act of measurement on position based qubit is represented by the operator
\begin{equation}
P_{Left}=\ket{1,0}_{E_1,E_2}\bra{1,0}_{E_1,E_2}+\ket{1,0}_{E_3,E_4}\bra{1,0}_{E_3,E_4}+\ket{1,0}_{E_5,E_6}\bra{1,0}_{E_5,E_6},
\end{equation}
\begin{equation}
P_{Right}=\ket{0,1}_{E_1,E_2}\bra{0,1}_{E_1,E_2}+\ket{0,1}_{E_3,E_4}\bra{0,1}_{E_3,E_4}+\ket{0,1}_{E_5,E_6}\bra{0,1}_{E_5,E_6}.
\end{equation}

Let us review the Hamiltonian describing system with $N=2N_1$ energy levels. Essientially we have $2N_1$ coefficients describing energy localized at 2 nodes $E_{p1,1},E_{p1,2},..,E_{p1,N_1},$ $E_{p2,1},E_{p2,2},..,E_{p2,N_1}$, so we are dealing with $E_{pu,m}$ coefficients, where m=1..$N_1$, pu is 1 or 2 and we have taken into account existence of all $N=2N_1$ energetic levels. Let us set $N_1=3$ and in such case the quantum state Hamiltonia in the case of lack of transition between energetic levels corresponding to Fig.4. can be written as
\begin{eqnarray}
\hat{H}=
\begin{pmatrix}
E_{1,p1} & 0 & 0 & 0 & 0  & t_{1,p1 \rightarrow p2} \\
0 & E_{2,p1}  & 0 & 0 & t_{2,p1 \rightarrow p2} & 0 \\
0 & 0 & E_{3,p1}  & t_{3,p1 \rightarrow p2} & 0 & 0  \\
0 & 0 &  t_{3,p2 \rightarrow p1} & E_{3,p2}  & 0 & 0 \\
0 & t_{2,p2 \rightarrow p1} & 0 & 0 & E_{2,p2}  & 0 \\
t_{1,p2 \rightarrow p1} & 0 & 0 & 0 & 0 & E_{1,p2} \\
\end{pmatrix}_x= \nonumber \\
=E_{1,t}\ket{E_{1,t}}\bra{E_{1,t}}+E_{2,t}\ket{E_{2,t}}\bra{E_{2,t}}+E_{3,t}\ket{E_{3,t}}\bra{E_{3,t}}+E_{4,t}\ket{E_{4,t}}\bra{E_{4,t}}+E_{5,t}\ket{E_{5,t}}\bra{E_{5,t}}+ E_{6,t}\ket{E_{6,t}}\bra{E_{6,t}}. \nonumber \\
\end{eqnarray}

\begin{figure}
\label{CMOSqubitQ}
\includegraphics[scale=1.2]{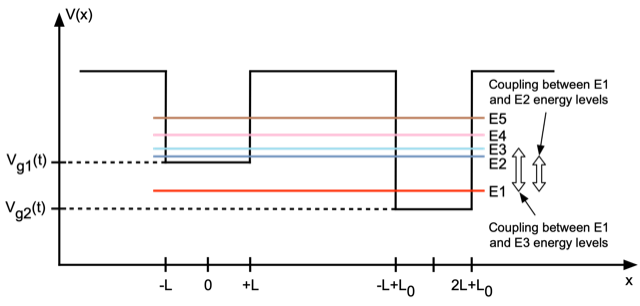}    
\caption{Position based qubit with 5 energetic levels, two-different potential minima and one occupied localized state.}  
\end{figure}

It is important to mention that in the case of lack of time-dependent Hamiltonian having any among frequency components $\frac{E_k-E_l}{\hbar}$ for $k \neq l$ such that $(k,l)=1..6$ there is no possibility for the occurence of resonant state and change of probability of
occupancy among different energetic levels. In such case \newline $(\ket{1,0}_{E_1,E_2}\bra{1,0}_{E_1,E_2})(\ket{1,0}_{E_3,E_4}\bra{1,0}_{E_3,E_4})=0$. However it is not true if there exists resonant state and if for example Hamiltonian consists following non-zero components with frequencies $(\frac{E_1-E_3}{\hbar},\frac{E_1-E_4}{\hbar},\frac{E_2-E_3}{\hbar},\frac{E_2-E_4}{\hbar})$.


Now we are moving towards the situation of system with position based qubit with 5 energetic levels, two-different potential minima and one occupied localized state on the right side as depicted in Fig.5. We have Hamiltonian of the form
\begin{equation}
\hat{H}=
\begin{pmatrix}
E_{2,p1}                                        & 0                                      & 0                                     & t_{2,p1 \rightarrow p2} & 0               \\ 
0                                                     & E_{3,p1}                         & t_{3,p1 \rightarrow p2} & 0                                     & 0               \\ 
0                                                     & t_{3,p2 \rightarrow p1} & E_{3,p2}                         & 0                                     & 0               \\
t_{2,p2 \rightarrow p1}                 & 0                                     & 0                                     & E_{2,p2}                         & 0               \\ 
0                                                     & 0                                     & 0                                     & 0                                     & E_{1,p1}   \\ 
\end{pmatrix}=E_1(t)\ket{E_1(t)}\bra{E_1(t)}+..+E_5(t)\ket{E_5(t)}\bra{E_5(t)}
\end{equation}
with corresponding quantum state given as
\begin{eqnarray}
\ket{\psi,t}_x=
\gamma_{E5,E4,p1}(t)\ket{1,0}_{E_5,E_4}+\gamma_{E_3,E_2,p1}(t)\ket{1,0}_{E_3,E_2}+\gamma_{E5,E4,p2}(t)\ket{0,1}_{E_5,E_4}+\gamma_{E3,E2,p2}(t)\ket{1,0}_{E_3,E_2}+ \nonumber \\
+\gamma_{E1,p2}(t)\ket{0,1}_{E_1}=
\begin{pmatrix}
\gamma_{E_5,E_4,p1}(t) \\
\gamma_{E_3,E_2,p1}(t) \\
\gamma_{E_3,E_2,p2}(t) \\
\gamma_{E_5,E_4,p2}(t) \\
\gamma_{E_1,p2}(t)
\end{pmatrix}_x .
\end{eqnarray}
The energetic states parametrized by $E_5,E_4$ or $E_3,E_2$ can move freely between node 1 and 2 so they are delocalized while the state numerated by $E_1$ is the particular localized ground state.
Specified Hamiltonian structure implies that the ground state cannot be moved to excited states  and reversely excited states cannot be moved into ground state .

The coupling between ground state and first excited state at node 2 occurs in the case of modified Hamiltonian of the following form as
\begin{eqnarray}
\label{modifiedH}
\hat{H}=
\begin{pmatrix}
E_{2,p1}                                        & 0                                      & 0                                     & t_{2,p1 \rightarrow p2} & 0               \\ 
0                                                     & E_{3,p1}                         & t_{3,p1 \rightarrow p2} & 0                                     & 0               \\ 
0                                                     & t_{3,p2 \rightarrow p1} & E_{3,p2}                         & 0                                     & 0               \\
t_{2,p2 \rightarrow p1}                 & 0                                     & 0                                     & E_{2,p2}                         & t_{1 \rightarrow 2,p2 \rightarrow p2}               \\ 
0                                                     & 0                                     & 0                                     & t_{2 \rightarrow 1,p2 \rightarrow p2}                                     & E_{1,p1}   \\ 
\end{pmatrix}=E_1(t)\ket{E_1(t)}\bra{E_1(t)}+E_2(t)\ket{E_2(t)}\bra{E_2(t)}+.. \nonumber \\
+E_5(t)\ket{E_5(t)}\bra{E_6(t)}+f_1(t)\ket{E_2}\bra{E_1}+f_2(t)\ket{E_1}\bra{E_2}+f_3(t)\ket{E_3}\bra{E_1}+f_4(t)\ket{E_1}\bra{E_3}. \nonumber \\
\end{eqnarray}
In a particular state it is allowed for the wave-packet in the right-well to undergoe transition from energetic state $E_1$ to $E_2$ and $E_3$ and reversely. A better picture can be obtained from Schroedinger equation.
Last Hamiltonian implies presence of time-dependent component in matrix that has $\omega_{21}=\frac{E_2-E_1}{\hbar}$ and $\omega_{31}=\frac{E_3-E_1}{\hbar}$ frequency components.

In such case the projectors
$(\ket{0,1}_{E_1,E_2}\bra{0,1}_{E_1,E_2})(\ket{0,1}_{E_1,E_3}\bra{0,1}_{E_3,E_1})$ are different from zero because of existence of resonant states characterized by frequencies $\omega_{21}$ and $\omega_{31}$.
Now we are moving from position based Hamiltonian representation into energy based that is by identity transformation
\begin{eqnarray*}
\hat{H}(t)=
\begin{pmatrix}
E_5 & 0 & 0 & 0 & 0  \\
0 & E_4 & 0 & 0 & 0  \\
0 & 0 & E_3 & 0 & 0  \\
0 & 0 & 0 & E_2 & 0  \\
0 & 0 & 0 & 0 & E_1  \\
\end{pmatrix}
\begin{pmatrix}
\frac{1}{E_5} & 0 & 0 & 0 & 0 \\
0 & \frac{1}{E_4} & 0 & 0 & 0  \\
0 & 0 & \frac{1}{E_3} & 0 & 0  \\
0 & 0 & 0 & \frac{1}{E_2} & 0 \\
0 & 0 & 0 & 0 & \frac{1}{E_1} \\
\end{pmatrix}
\begin{pmatrix}
E_{2,p1}                                        & 0                                      & 0                                     & t_{2,p1 \rightarrow p2} & 0               \\ 
0                                                     & E_{3,p1}                         & t_{3,p1 \rightarrow p2} & 0                                     & 0               \\ 
0                                                     & t_{3,p2 \rightarrow p1} & E_{3,p2}                         & 0                                     & 0               \\
t_{2,p2 \rightarrow p1}                 & 0                                     & 0                                     & E_{2,p2}                         & t_{1 \rightarrow 2,p2 \rightarrow p2}               \\ 
0                                                     & 0                                     & 0                                     & t_{2 \rightarrow 1,p2 \rightarrow p2}                                     & E_{1,p1}   \\ 
\end{pmatrix} = \nonumber \\ =
\begin{pmatrix}
E_5 & 0 & 0 & 0 & 0  \\
0 & E_4 & 0 & 0 & 0  \\
0 & 0 & E_3 & 0 & 0  \\
0 & 0 & 0 & E_2 & 0  \\
0 & 0 & 0 & 0 & E_1  \\
\end{pmatrix}
\begin{pmatrix}
\frac{E_{2,p1}}{E_5}                                        & 0                                      & 0                                     & \frac{t_{2,p1 \rightarrow p2}}{E_5} & 0               \\ 
0                                                     & \frac{E_{3,p1}}{E_4}                         & \frac{t_{3,p1 \rightarrow p2}}{E_4} & 0                                     & 0               \\ 
0                                                     & \frac{t_{3,p2 \rightarrow p1}}{E_3} & \frac{E_{3,p2}}{E_3}                         & 0                                     & 0               \\
\frac{t_{2,p2 \rightarrow p1}}{E_2}                 & 0                                     & 0                                     & \frac{E_{2,p2}}{E_2}                         & \frac{t_{1 \rightarrow 2,p2 \rightarrow p2}}{E_2}               \\ 
0                                                     & 0                                     & 0                                     & \frac{t_{2 \rightarrow 1,p2 \rightarrow p2}}{E_1}                                     &\frac{ E_{1,p1}}{E_1}   \\ 
\end{pmatrix}
\end{eqnarray*}
Now we need to specify the energy eigenstates introducing $\hat{E}=diag(E_5,E_4,E_3,E_2,E_1)$ and we obtain $\hat{E}$ acting on
\begin{eqnarray}
\begin{pmatrix}
\frac{E_{2,p1}}{E_5}                                        & 0                                      & 0                                     & \frac{t_{2,p1 \rightarrow p2}}{E_5} & 0               \\ 
0                                                     & \frac{E_{3,p1}}{E_4}                         & \frac{t_{3,p1 \rightarrow p2}}{E_4} & 0                                     & 0               \\ 
0                                                     & \frac{t_{3,p2 \rightarrow p1}}{E_3} & \frac{E_{3,p2}}{E_3}                         & 0                                     & 0               \\
\frac{t_{2,p2 \rightarrow p1}}{E_2}                 & 0                                     & 0                                     & \frac{E_{2,p2}}{E_2}                         & \frac{t_{1 \rightarrow 2,p2 \rightarrow p2}}{E_2}               \\ 
0                                                     & 0                                     & 0                                     & \frac{t_{2 \rightarrow 1,p2 \rightarrow p2}}{E_1}                                     &\frac{ E_{1,p1}}{E_1}   \\ 
\end{pmatrix}
\begin{pmatrix}
\gamma_{E_3,E_2,p1} \\
\gamma_{E_4,E_5,p1} \\
\gamma_{E_4,E_5,p2} \\
\gamma_{E_3,E_2,p2} \\
\gamma_{E_1,p2}
\end{pmatrix}_x =\hat{E}
\begin{pmatrix}
\gamma_{E_3,E_2,p1}(t)\frac{E_{2,p1}}{E_5} + \frac{t_{2,p1 \rightarrow p2}}{E_5}\gamma_{E_3,E_2,p2}(t) \\
\gamma_{E_4,E_5,p1}(t)\frac{E_{3,p1}}{E_4}  + \frac{t_{3,p1 \rightarrow p2}}{E_4}\gamma_{E_4,E_5,p2}(t) \\
\gamma_{E_4,E_5,p2}(t)\frac{E_{3,p2}}{E_3} + \gamma_{E_4,E_5,p1}(t)\frac{t_{3,p2 \rightarrow p1}}{E_3}  \\
\frac{E_{2,p2} \gamma_{E_3,E_2,p2}}{E_2}+\frac{\gamma_{E_1,p2} t_{1 \rightarrow 2,p2 \rightarrow p2}}{E_2}+\frac{\gamma_{E_3,E_2,p1} t_{2,p2 \rightarrow p1}}{E_2} \\
\frac{t_{2 \rightarrow 1,p2 \rightarrow p2}}{E_1}\gamma_{E_2,E_3,p2}(t)+\gamma_{E1,p2}(t)\frac{ E_{1,p1}}{E_1}
\end{pmatrix}_x \nonumber
\end{eqnarray}
\begin{eqnarray}
=
\begin{pmatrix}
\gamma_{E_2,E_3,p1}(t)E_{2,p1} + t_{2,p1 \rightarrow p2}\gamma_{E_2,E_3,p2}(t) \\
0 \\
0 \\
0 \\
0 \\
\end{pmatrix}_{E_5}
+
\begin{pmatrix}
0 \\
\gamma_{E_4,E_5,p1}(t)E_{3,p1}  + t_{3,p1 \rightarrow p2}\gamma_{E_4,E_5,p2}(t) \\
0 \\
0 \\
0 \\
\end{pmatrix}_{E_4}
+
\end{eqnarray}
\begin{eqnarray}
\begin{pmatrix}
0 \\
0 \\
\gamma_{E_4,E_5,p2}(t)E_{3,p2} + \gamma_{E_4,E_5,p1}(t)t_{3,p2 \rightarrow p1}  \\
0 \\
0 \\
\end{pmatrix}_{E_3}
+
\begin{pmatrix}
0 \\
0 \\
0 \\
E_{2,p2} \gamma_{E_2,E_3,p2}+\gamma_{E_1,p2} t_{1 \rightarrow 2,p2 \rightarrow p2}+\gamma_{E_2,E_3,p1} t_{2,p2 \rightarrow p1} \\
0
\end{pmatrix}_{E_2}
+ \nonumber \\ +
\begin{pmatrix}
0 \\
0 \\
0 \\
0 \\
t_{2 \rightarrow 1,p2 \rightarrow p2}\gamma_{E_2,E_3,p2}(t)+\gamma_{E1,p2}(t) E_{1,p1}
\end{pmatrix}_{E_1} \nonumber.
\end{eqnarray}
It is noticable to recognize that the ground state eigenvector from localized state was converted into delocalized state by the presence of non-zero $\gamma_{E1,p2}(t) E_{1,p1}$ term in the Hamiltonian .
\begin{eqnarray}
\begin{pmatrix}
0 \\
0 \\
0 \\
0 \\
\gamma_{E1,p2}(t) E_{1,p1}
\end{pmatrix}_{E_1} \nonumber.
\rightarrow
\begin{pmatrix}
0 \\
0 \\
0 \\
0 \\
t_{2 \rightarrow 1,p2 \rightarrow p2}\gamma_{E_5,E_4,p2}(t)+\gamma_{E1,p2}(t) E_{1,p1}
\end{pmatrix}_{E_1} \nonumber.
\end{eqnarray}
Also second energergy level eigenvector was changed.
\begin{eqnarray}
\begin{pmatrix}
0 \\
0 \\
0 \\
E_{2,p2} \gamma_{E_5,E_4,p2}+\gamma_{E_5,E_4,p1} t_{2,p2 \rightarrow p1} \\
0
\end{pmatrix}_{E_2}
\rightarrow
\begin{pmatrix}
0 \\
0 \\
0 \\
(E_{2,p2} \gamma_{E_5,E_4,p2}+\gamma_{E_5,E_4,p1} t_{2,p2 \rightarrow p1}) +\gamma_{E_1,p2} t_{1 \rightarrow 2,p2 \rightarrow p2} \\
0
\end{pmatrix}_{E_2}.
\end{eqnarray}
The element $t_{2 \rightarrow 1,p2 \rightarrow p2}$ is responsible for heating up or cooling down of the localized state.
We notice that all other eigenenergy vectors were not changed by the presence of non-zero elements $t_{2 \rightarrow 1,p2 \rightarrow p2}=t_{1 \rightarrow 2,p2 \rightarrow p2}^{*}$ in the Hamiltonian \ref{modifiedH}.

It may occur that potential minima (bottom) in position based qubit can have arbitrary depth so more than one eigenenergy state can be localized. The number of localized states can be arbitrary big both on the left and the right side. In considered example we have only localized on the right state.  Localized states can be heated up or cool down so one localized state is transfering into
another localized state in the same quantum well. 
In general k states (as k =2 in reference to the matrix \ref{massiveL}) can be localized on the right side among k+m all energetic states (where m=4 is number of delocalized eigenenergy states) so total number of Hamiltonian eigenenergy state k+m is 4+2=6.


\begin{eqnarray}
\label{massiveL}
\hat{H}=
\begin{pmatrix}
E_{2,p1}                                        & 0                                      & 0                                     & t_{2,p1 \rightarrow p2}                          & 0                                                                 & 0  \\ 
0                                                     & E_{3,p1}                         & t_{3,p1 \rightarrow p2} & 0                                                              & 0                                                                 & 0 \\ 
0                                                     & t_{3,p2 \rightarrow p1} & E_{3,p2}                         & 0                                                              & 0                                                                 & 0 \\ 
t_{2,p2 \rightarrow p1}                 & 0                                     & 0                                     & E_{2,p2}                                                  & t_{1 \rightarrow 2,p2 \rightarrow p2}       &  t_{0 \rightarrow 2,p2 \rightarrow p2}  \\ 
0                                                     & 0                                     & 0                                     & t_{2 \rightarrow 1,p2 \rightarrow p2}     & \textcolor{red}{ E_{1,p2}}                       & \textcolor{red}{t_{0 \rightarrow 1 ,p2 \rightarrow p2}} \\ 
0                                                     & 0                                     & 0                                     &  t_{2 \rightarrow 0 ,p2 \rightarrow p2}                                                               & \textcolor{red}{ t_{1 \rightarrow 0 ,p2 \rightarrow p2}} &\textcolor{red}{ E_{0,p2}} \\ 
\end{pmatrix}
\end{eqnarray}

We recognize that term the $ t_{1 \rightarrow 0 ,p2 \rightarrow p2}$ is able to heat up and cool down the localized q-state between 0 and 1 energetic level in q-well p2 and term $ t_{2 \rightarrow 0 ,p2 \rightarrow p2}$ is describing interaction between 0 and 2 energy level in q-well p2, while
term $t_{2 \rightarrow 1,p2 \rightarrow p2}$ describes the interaction between 1st and 2nd energetic level in second quantum well p2.


Now to describe the situation of 3 localized states in the left well (associated with matrix coefficients in green) and 2 localized states in the right wells (associated with matrix coefficients in red) and 4 states that are delocalized so we are dealing with matrix of 9 states.
\begin{eqnarray}
\label{preuniversal}
\hat{H}=
\begin{pmatrix}
\textcolor{green}{E_{-1,p1}} & \textcolor{green}{t_{0 \rightarrow -1,p1 \rightarrow p1}} & \textcolor{green}{t_{1 \rightarrow -1,p1 \rightarrow p1}} & 0 & 0 & 0 & 0 & 0 & \textcolor{brown}{t_{0 \rightarrow -1,p1 \rightarrow p2}} \\
\textcolor{green}{t_{-1 \rightarrow 0,p1 \rightarrow p1}} & \textcolor{green}{E_{0,p1}} & \textcolor{green}{t_{1 \rightarrow 0 ,p1 \rightarrow p1}} & 0 & 0 & 0 & 0 & 0 & 0 \\
\textcolor{green}{t_{-1 \rightarrow 1,p1 \rightarrow p1}} & \textcolor{green}{t_{0 \rightarrow 1,p1 \rightarrow p1}} & \textcolor{green}{E_{1,p1}} & \textcolor{orange}{ t_{2 \rightarrow 1,p1 \rightarrow p1}} & 0 & 0 & 0 & 0 & 0 \\
0 & 0 & \textcolor{orange}{t_{1 \rightarrow 2,p1 \rightarrow p1}} & E_{2,p1}                                        & 0                                      & 0                                     & t_{2,p1 \rightarrow p2} & 0            & 0  \\ 
0 & 0 & 0 & 0                                                     & E_{3,p1}                         & t_{3,p1 \rightarrow p2} & 0                                     & 0            & 0 \\ 
0 & 0 & 0 & 0                                                     & t_{3,p2 \rightarrow p1} & E_{3,p2}                         & 0                                     & 0            & 0 \\ 
0 & 0 & 0 & t_{2,p2 \rightarrow p1}                 & 0                                     & 0                                     & E_{2,p2}                         & t_{1 \rightarrow 2,p2 \rightarrow p2}       &   t_{0 \rightarrow 2,p2 \rightarrow p2}   \\ 
0 & 0 & 0 & 0                                                     & 0                                     & 0                                      & t_{2 \rightarrow 1,p2 \rightarrow p2}                                     & \textcolor{red}{ E_{1,p2}} & \textcolor{red}{t_{0 \rightarrow 1 ,p2 \rightarrow p2}} \\ 
\textcolor{brown}{t_{-1 \rightarrow 0,p2 \rightarrow p1}} & 0 & 0 & 0                                                     & 0                                     & 0                                     &  t_{2 \rightarrow 0,p2 \rightarrow p2}                                     & \textcolor{red}{ t_{1 \rightarrow 0 ,p2 \rightarrow p2}} &\textcolor{red}{ E_{0,p2}} \\ 
\end{pmatrix} \nonumber
\end{eqnarray}

Heating up and cooling down of the localized quantum state in the left q-well is controlled by Hamiltonian coeffcients $ t_{0 \rightarrow -1,p1 \rightarrow p1}$, $t_{1 \rightarrow 0 ,p1 \rightarrow p1}$, $t_{1 \rightarrow -1,p1 \rightarrow p1}$ and its conjugate counterparts
$ t_{-1 \rightarrow 0,p1 \rightarrow p1}$, $t_{0 \rightarrow 1 ,p1 \rightarrow p1}$, $t_{-1 \rightarrow 1,p1 \rightarrow p1}$. Moving delocalized q-state in the left q-well p1 into delocalized q-state in the left p2 well is by non-zero $t_{1 \rightarrow 2,p1 \rightarrow p1}$ and its conjugate $t_{2 \rightarrow 1,p1 \rightarrow p1}$ in orange color. From the point of view of q-mechanics it is also possible to transfer one q-state localized in the left q-well into the q-state localized in the right q-well. It is achieved by the non-zero coefficient $t_{0 \rightarrow -1,p1 \rightarrow p2}$ and its conjugate $t_{-1 \rightarrow 0,p2 \rightarrow p1}$ in brown color. All these transfer between states of different energies requires microwave field or AC voltage components. In case of matrix 9 by 9 we can spot $(9^2-9)/2$ processes of transfer
from one energetic state into another energetic state in the same q-well or into opposite q-well. In general for a N by N matrix one has $(N^2-N)/2$ such processes.   More detailed knowledge about this processes might be only extracted from Schroediger formalism in 1, 2 or 3 dimensions. In most general case in the case of system with 9 energetic levels are depicted in Fig.6.

\begin{figure}
\centering
\includegraphics[scale=0.5]{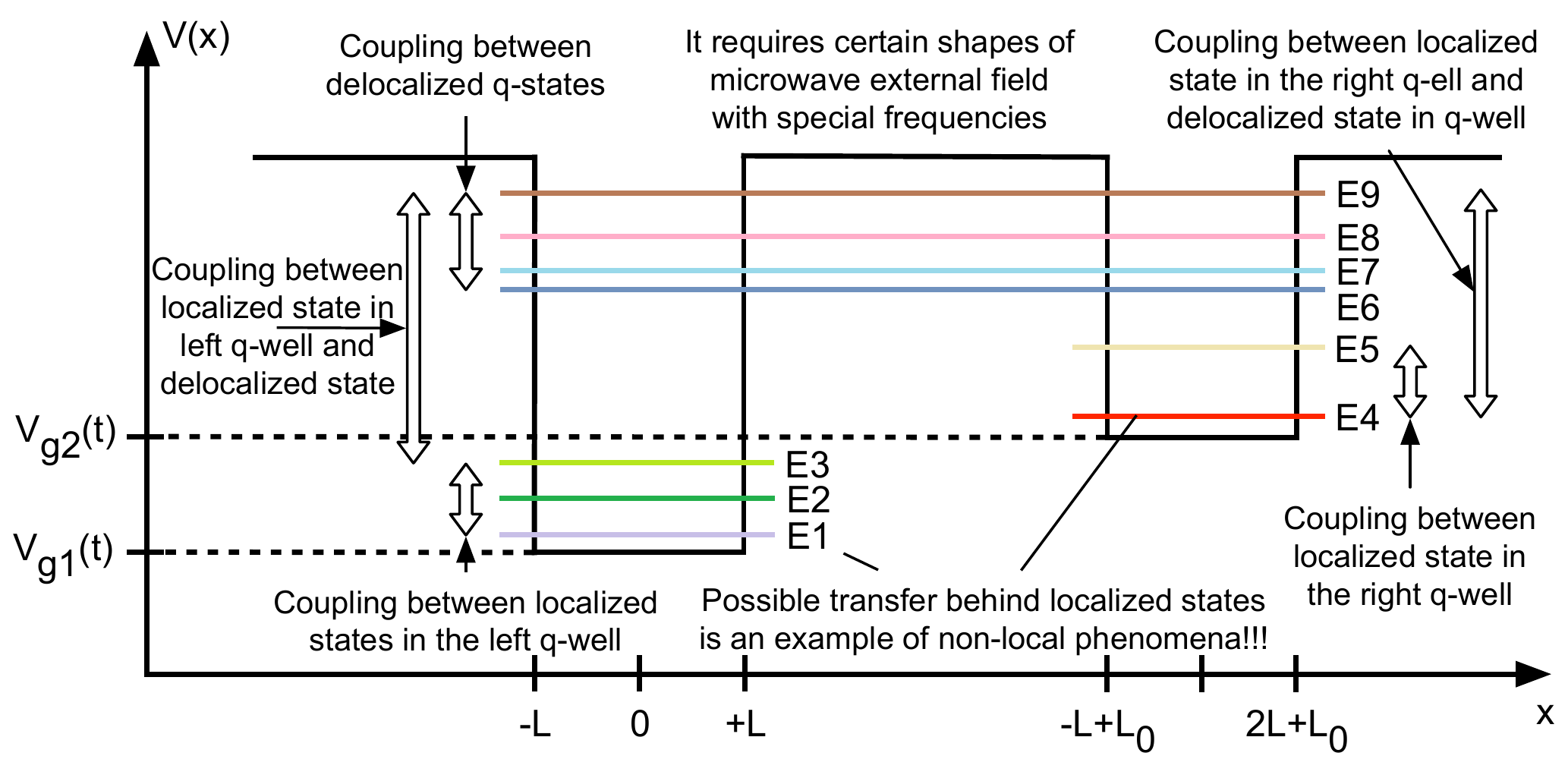}
\caption{All possible quantum processes in the system of 2 coupled q-dots in the case of various microwave fields: transitions between delocalized eigen energetic levels (P1), transitions between left localized eigen energies (P2), transitions between right localized eigen energy states (P3), transitions between left and right delocalized eigen energy states (P4), transitions between left localized q-states and delocalized q-states (P5), transitions between right localized q-states and delocalized q-states (P6). One can also distinguish process on injection of electron from outside to 2-qwell sytem (P7) and process of ejection of electron from 2-qwell system to the outside (P8).Six processes P1-P6 are described by the Hamiltonian \ref{universal} and its precurson Hamiltonian \ref{preuniversal}.}
\end{figure}

\normalsize
Now we are describing the most general situation for the system preserving  6 energy levels where position of potential minima and maxima can change in time so localized states can change into delocalized or reversely. It is thus describing the system is placed in outside time-dependent electromagnetic field of any dependence so the matrix of position-based qubit $\hat{H}(t)$ can be written as
\begin{equation}
\label{universal}
\hat{H}(t)=
\begin{pmatrix}
E_{1,p1} & t_{2 \rightarrow 1,p1 \rightarrow p1} & t_{3 \rightarrow 1,p1 \rightarrow p1} & t_{3 \rightarrow 1, p2 \rightarrow p1} & t_{2 \rightarrow 1, p2 \rightarrow p1}  & t_{1,p2 \rightarrow p1} \\
t_{1 \rightarrow 2,p1 \rightarrow p1} & E_{2,p1}  & t_{3 \rightarrow 2, p1 \rightarrow p1} & t_{3 \rightarrow 2, p2 \rightarrow p1} & t_{2,p2 \rightarrow p1} & t_{1 \rightarrow 2,p2 \rightarrow p1} \\
t_{1 \rightarrow 3,p1 \rightarrow p1} & t_{2 \rightarrow 3, p1 \rightarrow p1} & E_{3,p1}  & t_{3,p2 \rightarrow p1} & t_{2 \rightarrow 3,p2 \rightarrow p1} & t_{1 \rightarrow 3,p2 \rightarrow p1}  \\
t_{1 \rightarrow 3,p1 \rightarrow p2} & t_{2 \rightarrow 3, p1 \rightarrow p2} &  t_{3,p1 \rightarrow p2} & E_{3,p2}  & t_{2\rightarrow 3, p2 \rightarrow p2} & t_{1 \rightarrow 3,p2 \rightarrow p2} \\
t_{1 \rightarrow 2,p1 \rightarrow p2} & t_{2,p1 \rightarrow p2} &  t_{3 \rightarrow 2,p1 \rightarrow p2} & t_{3 \rightarrow 2,p2 \rightarrow p2} & E_{2,p2}  &  t_{1 \rightarrow 2,p2 \rightarrow p2} \\
t_{1,p1 \rightarrow p2} & t_{2 \rightarrow 1,p1 \rightarrow p2} & t_{3 \rightarrow 1,p1 \rightarrow p2} &  t_{3 \rightarrow 1,p2 \rightarrow p2} & t_{2 \rightarrow 1,p2 \rightarrow p2} & E_{1,p2} \\
\end{pmatrix}_x .
\end{equation}
Such matrix is Hermitian so $t_{k \rightarrow s ,pk \rightarrow p_l}^{*}=t_{k \rightarrow s ,pk \rightarrow p_l}^{*}$ for k and s among $1$, $2$ and 3 and $p_k$ and $p_l$ having value $p_1$ (presence of electron in left quantum well) or $p_2$ (presence of electron in right quantum well) and having real-valued diagonal elements.  The meaning of non-diagonal coefficients is non-trivial.

In the general case the eigenvalues of described matrix cannot be determined analytically unless there are some preimposed symmetries as for example $E_{k,p1}$=$E_{k,p2}$ for k=1,2 and 3
and in such case eigenvalues are determined by the roots of polynomial of 3rd order in an analytical way. Final reasoning can be conducted also for the system with 8 energetic levels when one deals with roots of polynomial of 4th order.
By proper electromagnetic engineering the system with 6 energetic levels can be controlled by $((36-6)/2)+6=15+6=21$ time dependent parameters.
In most general case the system of position based qubit having 2 coupled quantum dots with 6 energy levels can be parametrized by 36 real valued functions that are time-dependent. Quite obviously the same system with 2N energetic levels can be parametrized by
$(2N)^2$ real valued functions under the assumption that occupancy of electron is distributed among 2N energetic levels. We introduce the notation $\gamma_{1,p1}=\gamma_{E1-E2,p1},$ $\gamma_{2,p1}=\gamma_{E3-E4,p1},$ $\gamma_{3,p1}=\gamma_{E5-E6,p1},$
$\gamma_{3,p2}=\gamma_{E5-E6,p2},$ $\gamma_{2,p2}=\gamma_{E3-E4,p2},$ $\gamma_{1,p2}=\gamma_{E1-E2,p2}$.
The last matrix can be written in energy bases by using the last matrix \label{GeneralMatrix1} of Hamiltonian with identity $\hat{H}(t)\ket{\psi}(t)=$
\small
\begin{eqnarray*}
\label{GeneralMatrix1}
\begin{pmatrix}
E_1 & 0 & 0 & 0 & 0 & 0 \\
0 & E_2 & 0 & 0 & 0 & 0 \\
0 & 0 & E_3 & 0 & 0 & 0 \\
0 & 0 & 0 & E_4 & 0 & 0 \\
0 & 0 & 0 & 0 & E_5 & 0 \\
0 & 0 & 0 & 0 & 0 & E_6 \\
\end{pmatrix} 
\begin{pmatrix}
\frac{1}{E_1} & 0 & 0 & 0 & 0 & 0 \\
0 & \frac{1}{E_2} & 0 & 0 & 0 & 0 \\
0 & 0 & \frac{1}{E_3} & 0 & 0 & 0 \\
0 & 0 & 0 & \frac{1}{E_4} & 0 & 0 \\
0 & 0 & 0 & 0 & \frac{1}{E_5} & 0 \\
0 & 0 & 0 & 0 & 0 & \frac{1}{E_6} \\
\end{pmatrix} 
\begin{pmatrix}
E_{1,p1} & t_{2 \rightarrow 1,p1 \rightarrow p1} & t_{3 \rightarrow 1,p1 \rightarrow p1} & t_{3 \rightarrow 1, p2 \rightarrow p1} & t_{2 \rightarrow 1, p2 \rightarrow p1}  & t_{1,p2 \rightarrow p1} \\
t_{1 \rightarrow 2,p1 \rightarrow p1} & E_{2,p1}  & t_{3 \rightarrow 2, p1 \rightarrow p1} & t_{3 \rightarrow 2, p2 \rightarrow p1} & t_{2,p2 \rightarrow p1} & t_{1 \rightarrow 2,p2 \rightarrow p1} \\
t_{1 \rightarrow 3,p1 \rightarrow p1} & t_{2 \rightarrow 3, p1 \rightarrow p1} & E_{3,p1}  & t_{3,p2 \rightarrow p1} & t_{2 \rightarrow 3,p2 \rightarrow p1} & t_{1 \rightarrow 3,p2 \rightarrow p1}  \\
t_{1 \rightarrow 3,p1 \rightarrow p2} & t_{2 \rightarrow 3, p1 \rightarrow p2} &  t_{3,p1 \rightarrow p2} & E_{3,p2}  & t_{2\rightarrow 3, p2 \rightarrow p2} & t_{1 \rightarrow 3,p2 \rightarrow p2} \\
t_{1 \rightarrow 2,p1 \rightarrow p2} & t_{2,p1 \rightarrow p2} &  t_{3 \rightarrow 2,p1 \rightarrow p2} & t_{3 \rightarrow 2,p2 \rightarrow p2} & E_{2,p2}  &  t_{1 \rightarrow 2,p2 \rightarrow p2} \\
t_{1,p1 \rightarrow p2} & t_{2 \rightarrow 1,p1 \rightarrow p2} & t_{3 \rightarrow 1,p1 \rightarrow p2} &  t_{3 \rightarrow 1,p2 \rightarrow p2} & t_{2 \rightarrow 1,p2 \rightarrow p2} & E_{1,p2} \\
\end{pmatrix}_x \nonumber \\ 
\times
\begin{pmatrix}
\gamma_{E1-E2,p1}\\
\gamma_{E3-E4,p1}\\
\gamma_{E5-E6,p1}\\
\gamma_{E5-E6,p2}\\
\gamma_{E3-E4,p2}\\
\gamma_{E1-E2,p2}\\
\end{pmatrix}.
=\begin{pmatrix}
E_1 & 0 & 0 & 0 & 0 & 0 \\
0 & E_2 & 0 & 0 & 0 & 0 \\
0 & 0 & E_3 & 0 & 0 & 0 \\
0 & 0 & 0 & E_4 & 0 & 0 \\
0 & 0 & 0 & 0 & E_5 & 0 \\
0 & 0 & 0 & 0 & 0 & E_6 \\
\end{pmatrix}
\begin{pmatrix}
\frac{E_{1,p1}}{E_1} & \frac{t_{2 \rightarrow 1,p1 \rightarrow p1}}{E1} & \frac{t_{3 \rightarrow 1,p1 \rightarrow p1}}{E_1} & \frac{t_{3 \rightarrow 1, p2 \rightarrow p1}}{E_1} & \frac{t_{2 \rightarrow 1, p2 \rightarrow p1}}{E_1}  & \frac{t_{1,p2 \rightarrow p1}}{E_1} \\
\frac{t_{1 \rightarrow 2,p1 \rightarrow p1}}{E_2} & \frac{E_{2,p1}}{E_2}  & \frac{t_{3 \rightarrow 2, p1 \rightarrow p1}}{E_2} & \frac{t_{3 \rightarrow 2, p2 \rightarrow p1}}{E_2} & \frac{t_{2,p2 \rightarrow p1}}{E_2} & \frac{t_{1 \rightarrow 2,p2 \rightarrow p1}}{E_2} \\
\frac{t_{1 \rightarrow 3,p1 \rightarrow p1}}{E_3} & \frac{t_{2 \rightarrow 3, p1 \rightarrow p1}}{E_3} & \frac{E_{3,p1}}{E_3}  & \frac{t_{3,p2 \rightarrow p1}}{E_3} & \frac{t_{2 \rightarrow 3,p2 \rightarrow p1}}{E_3} & \frac{t_{1 \rightarrow 3,p2 \rightarrow p1} }{E_3} \\
\frac{t_{1 \rightarrow 3,p1 \rightarrow p2}}{E_4} & \frac{t_{2 \rightarrow 3, p1 \rightarrow p2}}{E_4} & \frac{t_{3,p1 \rightarrow p2}}{E_4} & \frac{E_{3,p2}}{E_4}  & \frac{t_{2\rightarrow 3, p2 \rightarrow p2}}{E_4} & \frac{t_{1 \rightarrow 3,p2 \rightarrow p2}}{E_4} \\
\frac{t_{1 \rightarrow 2,p1 \rightarrow p2}}{E_5} & \frac{t_{2,p1 \rightarrow p2}}{E_5} &  \frac{t_{3 \rightarrow 2,p1 \rightarrow p2}}{E_5} & \frac{t_{3 \rightarrow 2,p2 \rightarrow p2}}{E_5} & \frac{E_{2,p2}}{E_5}  &  \frac{t_{1 \rightarrow 2,p2 \rightarrow p2}}{E_5} \\
\frac{t_{1,p1 \rightarrow p2}}{E_6} & \frac{t_{2 \rightarrow 1,p1 \rightarrow p2}}{E_6} & \frac{t_{3 \rightarrow 1,p1 \rightarrow p2}}{E_6} &  \frac{t_{3 \rightarrow 1,p2 \rightarrow p2}}{E_6} & \frac{t_{2 \rightarrow 1,p2 \rightarrow p2}}{E_6} & \frac{E_{1,p2}}{E_6} \\
\end{pmatrix}_x
\begin{pmatrix}
\gamma_{E1-E2,p1}\\
\gamma_{E3-E4,p1}\\
\gamma_{E5-E6,p1}\\
\gamma_{E5-E6,p2}\\
\gamma_{E3-E4,p2}\\
\gamma_{E1-E2,p2}\\
\end{pmatrix}_x=
\end{eqnarray*}
\begin{eqnarray*}
=\hat{E}
\begin{pmatrix}
\frac{E_{1,p1}}{E_1}\gamma_{1,p1}(t) + \frac{t_{2 \rightarrow 1,p1 \rightarrow p1}}{E1}\gamma_{2,p1}(t) + \frac{t_{3 \rightarrow 1,p1 \rightarrow p1}}{E_1}\gamma_{3,p1}(t) + \frac{t_{3 \rightarrow 1, p2 \rightarrow p1}}{E_1}\gamma_{3,p2}(t) + \frac{t_{2 \rightarrow 1, p2 \rightarrow p1}}{E_1}\gamma_{2,p2}(t)  + \frac{t_{1,p2 \rightarrow p1}}{E_1}\gamma_{1,p2}(t) \\
\frac{t_{1 \rightarrow 2,p1 \rightarrow p1}}{E_2}\gamma_{E1,p1}(t) + \frac{E_{2,p1}}{E_2}\gamma_{E2,p1}(t)  + \frac{t_{3 \rightarrow 2, p1 \rightarrow p1}}{E_2}\gamma_{E3,p1}(t) + \frac{t_{3 \rightarrow 2, p2 \rightarrow p1}}{E_2}\gamma_{E3,p2}(t) + \frac{t_{2,p2 \rightarrow p1}}{E_2}\gamma_{2,p2}(t) + \frac{t_{1 \rightarrow 2,p2 \rightarrow p1}}{E_2}\gamma_{1,p2}(t) \\
\frac{t_{1 \rightarrow 3,p1 \rightarrow p1}}{E_3}\gamma_{1,p1}(t) + \frac{t_{2 \rightarrow 3, p1 \rightarrow p1}}{E_3}\gamma_{2,p1}(t) + \frac{E_{3,p1}}{E_3}\gamma_{3,p1}(t)  + \frac{t_{3,p2 \rightarrow p1}}{E_3}\gamma_{3,p2}(t) + \frac{t_{2 \rightarrow 3,p2 \rightarrow p1}}{E_3}\gamma_{2,p2}(t) + \frac{t_{1 \rightarrow 3,p2 \rightarrow p1} }{E_3}\gamma_{1,p2}(t) \\
\frac{t_{1 \rightarrow 3,p1 \rightarrow p2}}{E_4}\gamma_{1,p1}(t) + \frac{t_{2 \rightarrow 3, p1 \rightarrow p2}}{E_4}\gamma_{2,p1}(t) + \frac{t_{3,p1 \rightarrow p2}}{E_4}\gamma_{3,p1}(t) + \frac{E_{3,p2}}{E_4}\gamma_{3,p2}(t)  + \frac{t_{2\rightarrow 3, p2 \rightarrow p2}}{E_4}\gamma_{2,p2}(t) + \frac{t_{1 \rightarrow 3,p2 \rightarrow p2}}{E_4}\gamma_{1,p2}(t) \\
\frac{t_{1 \rightarrow 2,p1 \rightarrow p2}}{E_5}\gamma_{1,p1}(t) + \frac{t_{2,p1 \rightarrow p2}}{E_5}\gamma_{2,p1}(t) +  \frac{t_{3 \rightarrow 2,p1 \rightarrow p2}}{E_5}\gamma_{3,p1}(t) + \frac{t_{3 \rightarrow 2,p2 \rightarrow p2}}{E_5}\gamma_{3,p2}(t) + \frac{E_{2,p2}}{E_5}\gamma_{2,p2}(t)  +  \frac{t_{1 \rightarrow 2,p2 \rightarrow p2}}{E_5}\gamma_{1,p2}(t) \\
\frac{t_{1,p1 \rightarrow p2}}{E_6}\gamma_{1,p1}(t) + \frac{t_{2 \rightarrow 1,p1 \rightarrow p2}}{E_6}\gamma_{2,p1}(t) + \frac{t_{3 \rightarrow 1,p1 \rightarrow p2}}{E_6}\gamma_{3,p1}(t) +  \frac{t_{3 \rightarrow 1,p2 \rightarrow p2}}{E_6}\gamma_{3,p2}(t) + \frac{t_{2 \rightarrow 1,p2 \rightarrow p2}}{E_6}\gamma_{2,p2}(t) + \frac{E_{1,p2}}{E_6}\gamma_{1,p2}(t) \\
\end{pmatrix}_E
\end{eqnarray*}
\begin{eqnarray*}
=E_1
\begin{pmatrix}
\frac{E_{1,p1}}{E_1}\gamma_{1,p1} + \frac{t_{2 \rightarrow 1,p1 \rightarrow p1}}{E1}\gamma_{2,p1} + \frac{t_{3 \rightarrow 1,p1 \rightarrow p1}}{E_1}\gamma_{3,p1}(t) + \frac{t_{3 \rightarrow 1, p2 \rightarrow p1}}{E_1}\gamma_{3,p2}(t) + \frac{t_{2 \rightarrow 1, p2 \rightarrow p1}}{E_1}\gamma_{2,p2}(t)  + \frac{t_{1,p2 \rightarrow p1}}{E_1}\gamma_{1,p2}(t) \\
0 \\
0 \\ 
0 \\ 
0 \\ 
0  \\ 
\end{pmatrix}_{E}
+\nonumber \\
+E_2
\begin{pmatrix}
0 \\ 
\frac{t_{1 \rightarrow 2,p1 \rightarrow p1}}{E_2}\gamma_{1,p1}(t) + \frac{E_{2,p1}}{E_2}\gamma_{2,p1}(t)  + \frac{t_{3 \rightarrow 2, p1 \rightarrow p1}}{E_2}\gamma_{3,p1}(t) + \frac{t_{3 \rightarrow 2, p2 \rightarrow p1}}{E_2}\gamma_{3,p2}(t) + \frac{t_{2,p2 \rightarrow p1}}{E_2}\gamma_{2,p2}(t) + \frac{t_{1 \rightarrow 2,p2 \rightarrow p1}}{E_2}\gamma_{1,p2}(t) \\
0 \\ 
0 \\ 
0 \\ 
0 \\ 
\end{pmatrix}_{E} + \nonumber \\
+E_3
\begin{pmatrix}
0 \\
0 \\ 
\frac{t_{1 \rightarrow 3,p1 \rightarrow p1}}{E_3}\gamma_{1,p1}(t) + \frac{t_{2 \rightarrow 3, p1 \rightarrow p1}}{E_3}\gamma_{2,p1}(t) + \frac{E_{3,p1}}{E_3}\gamma_{3,p1}(t)  + \frac{t_{3,p2 \rightarrow p1}}{E_3}\gamma_{3,p2}(t) + \frac{t_{2 \rightarrow 3,p2 \rightarrow p1}}{E_3}\gamma_{2,p2}(t) + \frac{t_{1 \rightarrow 3,p2 \rightarrow p1} }{E_3}\gamma_{1,p2}(t) \\
0 \\ 
0 \\ 
0 \\ 
\end{pmatrix}_{E} + \nonumber \\
+E_4
\begin{pmatrix}
0 \\ 
0 \\ 
0 \\ 
\frac{t_{1 \rightarrow 3,p1 \rightarrow p2}}{E_4}\gamma_{1,p1}(t) + \frac{t_{2 \rightarrow 3, p1 \rightarrow p2}}{E_4}\gamma_{2,p1}(t) + \frac{t_{3,p1 \rightarrow p2}}{E_4}\gamma_{3,p1}(t) + \frac{E_{3,p2}}{E_4}\gamma_{3,p2}(t)  + \frac{t_{2\rightarrow 3, p2 \rightarrow p2}}{E_4}\gamma_{2,p2}(t) + \frac{t_{1 \rightarrow 3,p2 \rightarrow p2}}{E_4}\gamma_{1,p2}(t) \\
0 \\ 
0 \\ 
\end{pmatrix}_{E} + \nonumber \\
+E_5
\begin{pmatrix}
0 \\ 
0 \\ 
0 \\ 
0 \\ 
\frac{t_{1 \rightarrow 2,p1 \rightarrow p2}}{E_5}\gamma_{1,p1}(t) + \frac{t_{2,p1 \rightarrow p2}}{E_5}\gamma_{2,p1}(t) +  \frac{t_{3 \rightarrow 2,p1 \rightarrow p2}}{E_5}\gamma_{3,p1}(t) + \frac{t_{3 \rightarrow 2,p2 \rightarrow p2}}{E_5}\gamma_{3,p2}(t) + \frac{E_{2,p2}}{E_5}\gamma_{2,p2}(t)  +  \frac{t_{1 \rightarrow 2,p2 \rightarrow p2}}{E_5}\gamma_{1,p2}(t) \\
0 
\end{pmatrix}_E +
\end{eqnarray*}
\begin{eqnarray*}
\nonumber \\
+E_6 \begin{pmatrix}
0 \\ 
0 \\ 
0 \\ 
0 \\ 
0 \\ 
\frac{t_{1,p1 \rightarrow p2}}{E_6}\gamma_{1,p1}(t) + \frac{t_{2 \rightarrow 1,p1 \rightarrow p2}}{E_6}\gamma_{2,p1}(t) + \frac{t_{3 \rightarrow 1,p1 \rightarrow p2}}{E_6}\gamma_{3,p1}(t) +  \frac{t_{3 \rightarrow 1,p2 \rightarrow p2}}{E_6}\gamma_{3,p2}(t) + \frac{t_{2 \rightarrow 1,p2 \rightarrow p2}}{E_6}\gamma_{2,p2}(t) + \frac{E_{1,p2}}{E_6}\gamma_{1,p2}(t) \\
\end{pmatrix}_E  =\nonumber \\
\end{eqnarray*}
\normalsize
\begin{eqnarray}
=E_1(t) c_{E1,t}\ket{E_1,t}+E_2(t) c_{E2,t}\ket{E_2,t} + E_3(t) c_{E_3,t}\ket{E_3,t}+E_4(t) c_{E4,t}\ket{E_4,t}+E_5(t) c_{E_5,t}\ket{E_5,t} + E_6(t) c_{E_6,t}\ket{E_6,t} = \nonumber \\
=(E_1(t) \ket{E_1,t}\bra{E_1,t} + E_2(t) \ket{E_2,t}\bra{E_2,t} + E_3(t) \ket{E_3,t}\bra{E_3,t} + E_4(t) \ket{E_4,t}\bra{E_4,t} +  E_5(t) \ket{E_5,t}\bra{E_5,t}+ \nonumber \\ E_6(t) \ket{E_6,t}\bra{E_6,t} )\ket{\psi,t}. \nonumber \\
\end{eqnarray}
\normalsize
where $\ket{E_k,t}\bra{E_k,t}$ is projector on energy eigenstate $E_k$ and $\bra{E_k,t}\ket{E_l,t}=\delta_{k,l}$ and
\begin{eqnarray}
\hat{E}=
\begin{pmatrix}
E_1 & 0 & 0 & 0 & 0 & 0 \\
0 & E_2 & 0 & 0 & 0 & 0 \\
0 & 0 & E_3 & 0 & 0 & 0 \\
0 & 0 & 0 & E_4 & 0 & 0 \\
0 & 0 & 0 & 0 & E_5 & 0 \\
0 & 0 & 0 & 0 & 0 & E_6 \\
\end{pmatrix},  \ket{E_1,t}=
\begin{pmatrix}
1 \\
0 \\
0 \\
0 \\
0 \\
0 \\
\end{pmatrix},
\ket{E_1,t}\bra{E_1,t}=
\begin{pmatrix}
1 & 0 & 0 & 0 & 0 & 0 \\
0 & 0 & 0 & 0 & 0 & 0 \\
0 & 0 & 0 & 0 & 0 & 0 \\
0 & 0 & 0 & 0 & 0 & 0 \\
0 & 0 & 0 & 0 & 0 & 0 \\
0 & 0 & 0 & 0 & 0 & 0 \\
\end{pmatrix}, \nonumber \\
 \ket{E_2,t}=
\begin{pmatrix}
0 \\
1 \\
0 \\
0 \\
0 \\
0 \\
\end{pmatrix},
\ket{E_2,t}\bra{E_2,t}=
\begin{pmatrix}
0 & 0 & 0 & 0 & 0 & 0 \\
0 & 1 & 0 & 0 & 0 & 0 \\
0 & 0 & 0 & 0 & 0 & 0 \\
0 & 0 & 0 & 0 & 0 & 0 \\
0 & 0 & 0 & 0 & 0 & 0 \\
0 & 0 & 0 & 0 & 0 & 0 \\
\end{pmatrix}, .. , 
 \ket{E_6,t}=
\begin{pmatrix}
0 \\
0 \\
0 \\
0 \\
0 \\
1 \\
\end{pmatrix},
\ket{E_6,t}\bra{E_6,t}=
\begin{pmatrix}
0 & 0 & 0 & 0 & 0 & 0 \\
0 & 0 & 0 & 0 & 0 & 0 \\
0 & 0 & 0 & 0 & 0 & 0 \\
0 & 0 & 0 & 0 & 0 & 0 \\
0 & 0 & 0 & 0 & 0 & 0 \\
0 & 0 & 0 & 0 & 0 & 1 \\
\end{pmatrix}.
\end{eqnarray}
It is worth noticing that having knowledge on all eigenvalues $E_1(t), .., E_N(t)$ with time we can determine the eigenenergy occupancy with time from position occupancy in unique way.
From the above considerations the following relations takes place
\begin{eqnarray*}
\begin{pmatrix}
\frac{E_{1,p1}}{E_1} & \frac{t_{2 \rightarrow 1,p1 \rightarrow p1}}{E1} & \frac{t_{3 \rightarrow 1,p1 \rightarrow p1}}{E_1} & \frac{t_{3 \rightarrow 1, p2 \rightarrow p1}}{E_1} & \frac{t_{2 \rightarrow 1, p2 \rightarrow p1}}{E_1}  & \frac{t_{1,p2 \rightarrow p1}}{E_1} \\
\frac{t_{1 \rightarrow 2,p1 \rightarrow p1}}{E_2} & \frac{E_{2,p1}}{E_2}  & \frac{t_{3 \rightarrow 2, p1 \rightarrow p1}}{E_2} & \frac{t_{3 \rightarrow 2, p2 \rightarrow p1}}{E_2} & \frac{t_{2,p2 \rightarrow p1}}{E_2} & \frac{t_{1 \rightarrow 2,p2 \rightarrow p1}}{E_2} \\
\frac{t_{1 \rightarrow 3,p1 \rightarrow p1}}{E_3} & \frac{t_{2 \rightarrow 3, p1 \rightarrow p1}}{E_3} & \frac{E_{3,p1}}{E_3}  & \frac{t_{3,p2 \rightarrow p1}}{E_3} & \frac{t_{2 \rightarrow 3,p2 \rightarrow p1}}{E_3} & \frac{t_{1 \rightarrow 3,p2 \rightarrow p1} }{E_3} \\
\frac{t_{1 \rightarrow 3,p1 \rightarrow p2}}{E_4} & \frac{t_{2 \rightarrow 3, p1 \rightarrow p2}}{E_4} & \frac{t_{3,p1 \rightarrow p2}}{E_4} & \frac{E_{3,p2}}{E_4}  & \frac{t_{2\rightarrow 3, p2 \rightarrow p2}}{E_4} & \frac{t_{1 \rightarrow 3,p2 \rightarrow p2}}{E_4} \\
\frac{t_{1 \rightarrow 2,p1 \rightarrow p2}}{E_5} & \frac{t_{2,p1 \rightarrow p2}}{E_5} &  \frac{t_{3 \rightarrow 2,p1 \rightarrow p2}}{E_5} & \frac{t_{3 \rightarrow 2,p2 \rightarrow p2}}{E_5} & \frac{E_{2,p2}}{E_5}  &  \frac{t_{1 \rightarrow 2,p2 \rightarrow p2}}{E_5} \\
\frac{t_{1,p1 \rightarrow p2}}{E_6} & \frac{t_{2 \rightarrow 1,p1 \rightarrow p2}}{E_6} & \frac{t_{3 \rightarrow 1,p1 \rightarrow p2}}{E_6} &  \frac{t_{3 \rightarrow 1,p2 \rightarrow p2}}{E_6} & \frac{t_{2 \rightarrow 1,p2 \rightarrow p2}}{E_6} & \frac{E_{1,p2}}{E_6} \\
\end{pmatrix}_{E/x}
\begin{pmatrix}
\gamma_{E1-E2,p1}(t) \\
\gamma_{E3-E4,p1}(t) \\
\gamma_{E5-E6,p1}(t) \\
\gamma_{E5-E6,p2}(t) \\
\gamma_{E3-E4,p2}(t) \\
\gamma_{E1-E2,p2}(t)  \\
\end{pmatrix}_x=
\begin{pmatrix}
c_{E1,p1}(t) \\
c_{E2,p1}(t) \\
c_{E3,p1}(t) \\
c_{E3,p2}(t) \\
c_{E2,p2}(t) \\
c_{E1,p2}(t)  \\
\end{pmatrix}_E =\hat{A}(t)\hat{\gamma}. 
\end{eqnarray*}
By proper controlling matrix in position representation we can achieved desired occupancy of energetic levels with time expressed by $c_{E1,p1}(t),..,c_{E1,p2}(t)$ coefficients.
On another hand preimposing dependence of occupancy of energetic levels by quantum state expressed in $c_{E1,p1}(t)$,..,$c_{E1,p2}(t)$ with time one can achieve desired dependence of electrons positions $\gamma_{E1,p1}(t), .., \gamma_{E1,p2}(t)$  by using relation
$\ket{\psi,t}_x=\hat{\gamma}(t)=\hat{A}(t)^{-1}\hat{c}_E(t)=\hat{A}(t)^{-1}\ket{\psi,t}_E$.

\section{Case of electrostatic qubit interaction}
We consider most minimalist model of electrostatically interacting two position-based qubits that are double quantum dots A (with nodes 1 and 2  and named as U-upper qubit) and B (with nodes 1' and 2' and named as L-lower qubit) with local confinement potentials as given in the right side of Fig.2.
By introducing notation $\ket{1,0}_x=\ket{1},\ket{0,1}_x=\ket{2},\ket{1',0'}_x=\ket{1'},\ket{0',1'}_x=\ket{1'}$ the minimalistic Hamiltonian of the system of electrostatically interacting position based qubits can be written as
\begin{eqnarray}
\hat{H}=(t_{s21}(t)\ket{2}\bra{1}+t_{s12}(t)\ket{1}\bra{2})\hat{I}_{b})+(\hat{I}_a(t_{s2'1'}(t)\ket{2'}\bra{1'}+t_{s1'2'}(t)\ket{2'}\bra{1'})+ \nonumber \\
+(E_{p1}(t)\ket{1}\bra{1}+E_{p2}(t)\ket{2}\bra{2})\hat{I}_b+ \hat{I}_a(E_{p1'}(t)\ket{1'}\bra{1'}+E_{p2'}(t)\ket{2'}\bra{2'})+ \nonumber \\ 
+\frac{q^2}{d_{11'}}\ket{1,1'}\bra{1,1'}+\frac{q^2}{d_{22'}}\ket{2,2'}\bra{2,2'}+
\frac{q^2}{d_{12'}}\ket{1,2'}\bra{1,2'}+\frac{q^2}{d_{21'}}\ket{2,1'}\bra{2,1'}= \nonumber \\
H_{kinetic1}+H_{pot1}+H_{kinetic2}+H_{pot2}+H_{A-B}
\end{eqnarray} described by parameters
 $E_{p1}(t)$,$E_{p2}(t)$,$E_{p1'}(t)$,$E_{p2'}(t)$, $t_{s12}(t)$, $t_{s1'2'}(t)$ and distances between nodes k and l': $d_{11'}$,$d_{22'}$,$d_{21'}$,$d_{12'}$.
In such case q-state of the system is given as
\begin{equation}
\ket{\psi,t}=\gamma_1(t)\ket{1,0}_U\ket{1,0}_L+\gamma_2(t)\ket{1,0}_U\ket{0,1}_L+\gamma_3(t)\ket{0,1}_U\ket{1,0}_L+\gamma_4(t)\ket{0,1}_U\ket{0,1}_L,
\end{equation}
where normalization condiion gives $|\gamma_1(t)|^2+..|\gamma_4(t)|^2$. Probability of finding electron in upper system at node 1 is by action of projector $\hat{P}_{1U}=\bra{1,0}_U\bra{1,0}_L+\bra{1,0}_U\bra{0,1}_L$ on q-state $\hat{P}_{1U} \ket{\psi}$ so
it gives probability amplitude $|\gamma_1(t)+\gamma_3(t)|^2$ . On the other hand probability of finding electron from qubit A (U) at node 2 and electron from qubit B(L) at node 1 is obtained by projection $\hat{P}_{2U,1L}=\bra{0,1}_U\bra{1,0}_L$ acting on q-state
giving $(\bra{0,1}_U\bra{1,0}_L)\ket{\psi}$ that gives probability amplitude $|\gamma_3(t)|^2$.
Referring to picture from Fig.\ref{fig:central} we set distances between nodes as $d_{11'}=d_{22'}=d_1$,$d_{12'}=d_{21'}=\sqrt{(a+b)^2+d_1^2}$ and assume Coulomb electrostatic energy to
 be of the form $E_c(k,l)=\frac{q^2}{d_{kl'}}$ and hence we obtain the matrix  Hamiltonian given as $ \hat{H}(t)= $ 
\begin{equation}
\label{2bodies}
\begin{pmatrix}
E_{p1}(t)+E_{p1'}(t) + \frac{q^2}{d_1} & t_{s1'2'}(t) & t_{s12}(t) & 0 \\
t_{s1'2'}(t)^{*} & E_{p1}(t)+E_{p2'}(t)+\frac{q^2}{\sqrt{(d1)^2+(b+a)^2}} & 0 & t_{s12}(t) \\
t_{s12}^{*}(t) & 0 & E_{p2}(t)+E_{p1'}(t)+ \frac{q^2}{\sqrt{(d1)^2+(b+a)^2}} & t_{s1'2'}(t) \\
0 & t_{s12}^{*}(t) & t_{s1'2'}(t)^{*} & E_{p2}(t)+E_{p2'}(t)+ \frac{q^2}{d1} \\
\end{pmatrix}
\end{equation}
\normalsize
We can introduce notation $E_{c1}=\frac{q^2}{d_1}$ and $E_{c2}=\frac{q^2}{\sqrt{d_1^2+(b+a)^2}}$. In most general case of 2 qubit electrostatic interaction one of which has 4 different Coulomb terms on matrix diagonal $E_{c1}=\frac{q^2}{d_{11'}}$, $E_{c2}\frac{q^2}{d_{12'}}$, $E_{c3}=\frac{q^2}{d_{21'}}$, $E_{c4}=\frac{q^2}{d_{22'}}$ and $\ket{\psi,t}=\hat{U}(t,t_0)\ket{\psi,t_0}$.
We introduce $q_1=E_{p1}(t)+E_{p1'}(t)+E_{c11'}$,$q_2=E_{p1}(t)+E_{p2'}(t)+E_{c12'}$, $q_3=E_{p2}(t)+E_{p1'}(t)+E_{c21'}$,$q_4=E_{p2}(t)+E_{p2'}(t)+E_{c22'}$ and in such case by using formula 8 one can decompose 2 particle Hamiltonian \ref{2bodies} as
\begin{eqnarray}
\hat{H}=\Big[\frac{(q_1+q_2+q_3+q_4)}{4}\sigma_0 \times \sigma_0 +
\frac{(q_1-q_2+q_3-q_4)}{4}\sigma_0 \times \sigma_3 +
\frac{(q_1+q_2-q_3-q_4)}{4}\sigma_3 \times \sigma_0 + \nonumber \\
\frac{(q_1-q_2-q_3+q_4)}{4}\sigma_3 \times \sigma_3 +
+t_{sr1}(t)\sigma_0 \times \sigma_1 -t_{si1}(t) \sigma_0 \times \sigma_2+t_{sr2}(t)\sigma_1 \times \sigma_0 - t_{si2}(t) \sigma_2 \times \sigma_0 \Big.
\end{eqnarray}
A very similar procedure is for the case of 3 or N interacting particles so one deals with tensor product of 3 or N Pauli matrices.
In order to simplify representation of unitary matrix describing physical system of 2 particles evolution with time it is helpful to define  $Q_1(t)=\int_{t_0}^{t}(E_{p1}(t')+E_{p1'}(t')+E_{c11'})dt'$,$Q_2(t)=\int_{t_0}^{t}(E_{p1}(t')+E_{p2'}(t')+E_{c12'})dt'$, $Q_3(t)=\int_{t_0}^{t}(E_{p2}(t')+E_{p1'}(t')+E_{c21'})dt'$,$Q_4(t)=\int_{t_0}^{t}(E_{p2}(t')+E_{p2'}(t')+E_{c22'})dt'$ and $TR1(t)=\int_{t_0}^{t}dt't_{s1r}(t')$ , $TI1(t)=\int_{t_0}^{t}dt't_{s1i}(t')$. We consider the situation when there is no hopping between q-wells $t_{s2}=0$ so, the second particle is localized among two quantum wells and first particle can move freely among 2 q-wells.
We obtain the following unitary matrix evolution with time with following $\hat{U}(t,t_0)_{1,2}=\hat{U}(t,t_0)_{1,4}=0=\hat{U}(t,t_0)_{2,3}=\hat{U}_{3,4}$ and


\begin{eqnarray}
\hat{U}(t,t_0)_{1,1}= \nonumber \\  \frac{1}{2 \sqrt{
(Q_1(t)-Q_3(t))^2+4 \left(TR_1(t)^2+TI_1(t)^2\right)}} \Bigg[Q_1(t)
\left(-e^{i \hbar \sqrt{
(Q_1(t)-Q_3(t))^2+4 \left(TR_1(t)^2+TI_1(t)^2\right)
}}\right) \nonumber \\ +
\left(\sqrt{ |Q_1(t)-Q_3(t)|^2+4(TR_1(t)^2+TI_1(t)^2)}
+Q_3(t) \right) \times  \left(-e^{i \hbar \sqrt{
(Q_1(t)-Q_3(t))^2+4 \left(TR_1(t)^2+TI_1(t)^2\right)
}}\right)+ \nonumber \\ \sqrt{(Q_1(t)-Q_3(t))^2+4 \left(TR_1(t)^2+TI_1(t)^2\right)} +(Q_1(t)-Q_3(t)))
e^{-\frac{1}{2} i \hbar \left(\sqrt{|\int_{t_0}^{t}dt'(q_{1}(t')-q_{3}(t'))|^2+4
 \left(t_{s1r}^2+t_{si1}^2\right)}+(Q_1(t)+Q_3(t))\right)} \Bigg] \nonumber \\
\end{eqnarray}

\begin{eqnarray}
\hat{U}(t,t_0)_{1,3}= \frac{2 (TI_1(t)-iTR_1(t)) e^{-\frac{1}{2}(Q_1(t)+Q_3(t)) i \hbar } \sin \left(\frac{1}{2} \hbar
   \sqrt{|Q_1(t)-Q_3(t)|^2+4(TR_1(t)^2+TI_1(t)^2)}\right)}{\sqrt{|Q_1(t)-Q_3(t)|^2+4(TR_1(t)^2+TI_1(t)^2)}},
\end{eqnarray}
\\,
\begin{eqnarray}
\hat{U}(t,t_0)_{2,2}=\Bigg[e^{(\frac{1}{2} i \hbar \left(\sqrt{  (Q_2(t)-Q_4(t))^2 +4(TR_1(t)^2+TI_1(t)^2)}-(Q_2(t)+Q_4(t))\right))} \times \nonumber \\ \times
 \frac{\left(\sqrt{(Q_2(t)-Q_4(t))^2 +4(TR_1(t)^2+TI_1(t)^2)}-Q_2(t)+Q_4(t)\right) }{2
   \sqrt{(Q_2-Q_4)^2+4(TR_1(t)^2+TI_1(t)^2)}} \nonumber \\
   -e^{\left(\frac{1}{2} i \hbar
   \left(-\sqrt{(Q_2(t)-Q_4(t))^2+ 4(TR_1(t)^2+TI_1(t)^2)}-(Q_2(t)+Q_4(t))\right)\right)} \times \nonumber \\ \times
\frac{\left(-\sqrt{ (Q_2(t)-Q_4(t))^2+ 4(TR_1(t)^2+TI_1(t)^2) }-Q_2(t)+Q_4(t)\right) }{2\sqrt{(Q_2-Q_4)^2+4(TR_1(t)^2+TI_1(t)^2)}}
\end{eqnarray}
$ $
\\
\\
$ $
\begin{eqnarray}
\hat{U}(t,t_0)_{3,3}=\frac{\exp \left(-\frac{1}{2} i \hbar \left(\sqrt{(Q_1(t)^2-Q_3(t))^2+4
   \left(TR_1(t)^2+TI_1(t)^2\right)}+Q_1(t)+Q_3(t)\right)
   \right)}{2 \sqrt{(Q_1(t)^2-Q_3(t))^2+4
   \left( TR_1(t)^2+TI_1(t)^2 \right)}} \times \nonumber \\
\Bigg[Q_1(t) \left(-1+e^{i \hbar \sqrt{(Q_1(t)-Q_3(t))^2+4
   \left(TR_1(t)^2+TI_1(t)^2 \right)}}\right)+ \nonumber \\
   \left(\sqrt{(Q_1(t)-Q_3(t))^2+4
   \left(TR_1(t)^2+TI_1(t)^2\right)}-\text{q3}\right) e^{i \hbar
   \sqrt{(Q_1(t)-Q_3(t))^2+4
   \left(TR_1(t)^2+TI_1(t)^2\right)}}+\nonumber \\+\sqrt{(Q_1(t)-Q_3(t))^2+4 \left(TR_1(t)^2+TI_1(t)^2\right)}+Q_3(t)\Bigg]
\end{eqnarray}

\begin{eqnarray*}
 \hat{U}(t,t_0)_{4,4}=\frac{\exp \left(-\frac{1}{2} i \hbar \left(\sqrt{(Q_2(t)-Q_4(t))^2+4
   \left(TR_1(t)^2+TI_1(t)^2\right)}+Q_2(t)+Q_4(t)\right)
   \right)}{2 \sqrt{(Q_2(t)-Q_4(t))^2+4
   \left(TR_1(t)^2+TI_1(t)^2\right)}} \times \nonumber \\
\times \Bigg[Q_2(t) \left(-1+e^{i \hbar \sqrt{(Q_2(t)-Q_4(t))^2+4
   \left(TR_1(t)^2+TI_1(t)^2\right)}}\right)+\nonumber \\ +\left(\sqrt{(Q_2(t)-Q_4(t)^2)^2 +4
   \left( TR_1(t)^2+TI_1(t)^2\right)}-Q_4(t)\right) e^{i \hbar
   \sqrt{(Q_2(t)-Q_4(t))^2+ 4
   \left(TR_1(t)^2+TI_1(t)^2\right)}}+ \nonumber \\
   \sqrt{(Q_2(t)-Q_4(t))^2 +4 \left(TR_1(t)^2+TI_1(t)^2\right)}+Q_4(t)\Bigg]
\end{eqnarray*}
\begin{eqnarray}
\hat{U}(t,t_0)_{2,4}=\frac{2 (TI_1(t)-i TR_1(t)) e^{-\frac{1}{2} i \hbar
   (Q_2(t)+Q_4(t))} \sin \left(\frac{1}{2} \hbar
   \sqrt{(Q_2(t)-Q_4(t))^2+4
   \left(TR_1(t)^2+TI_1(t)^2\right)}\right)}{\sqrt{(Q_2(t)-Q_4(t))^2+4 \left(TR_1(t)^2+TI_1(t)^2\right)}}
\end{eqnarray}

\begin{figure}
\centering
\label{fig:spectra}
\includegraphics[scale=0.6]{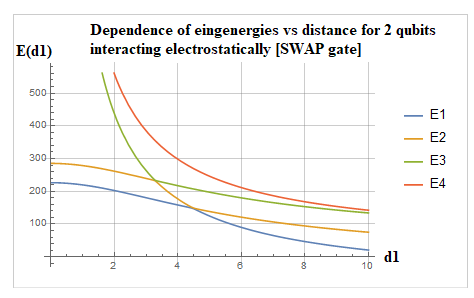}
\caption{Case of dependence of energy spectra on the distance $d_1$ for the case of 2 electrostatically interacting qubits from Fig.\,\ref{fig:central}.}
\end{figure}

The example of function dependence of eigenenergy spectra of 2 electrostatically interacting qubits on distance is given by Fig.\ref{fig:spectra}.

An important observation is that any element of matrix $\hat{H}(t')$ for $t' \in (t_0,t)$ denoted as $H_{k,l}(t')$ is transferred to element $\hat{U}_{k,l}(t,t_0)=e^{\frac{1}{\hbar i}\int_{t_0}^{t}dt'(H_{k,l}(t'))}$ of matrix $\hat{U}(t,t_0)$. 
We can easily generalize the presented reasoning for the system of N electrostatically coupled electrons confined by some local potentials. However we need to know the position dependent Hamiltonian eigenstate at the initial time $t_0$. In case $N>2$ finding such eigenstate is the numerical problem since
analytical solutions for roots of polynomials of one variable for higher order than 4 does not exist. Using numerical eigenstate at time instance $t_0$ we can compute the system quantum dynamics in analytical way.
This give us a strong and relatively simple mathematical tool giving full determination of quantum dynamical state at the any instance of time.
The act of measurement on position based qubit is represented by the operator $P_{Left}=\ket{1,0}_{E_1,E_2}\bra{1,0}_{E_1,E_2}$ and $P_{Right}=\ket{0,1}_{E_1,E_2}\bra{0,1}_{E_1,E_2}$.
\subsection{Simplified picture of symmetric Q-Swap gate}
\begin{figure}
\centering
\includegraphics[scale=0.8]{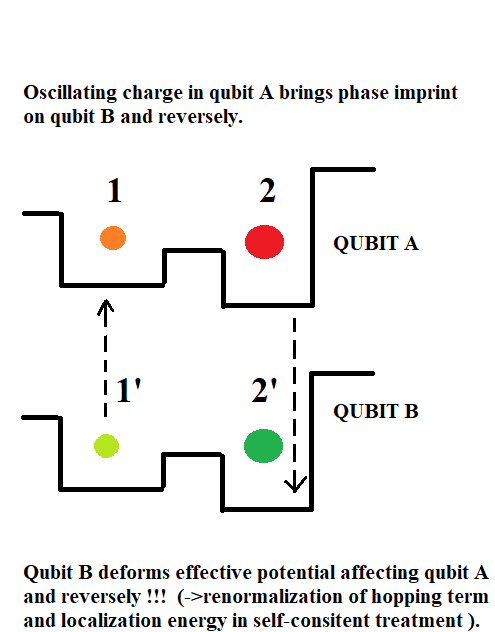}
\caption{Scheme of renormalization in the system of coupled qubits}
\label{renormalization}
\end{figure}

Now we need to find a system 4 eigenvalues and eigenstates(4 orthogonal 4-dimensional vectors)  so we are dealing with a matrix eigenvalue problem) what is the subject of classical algebra. Let us assume that 2 double quantum dot systems are symmetric and biased by the same voltages generating potential bottoms $V_s$ so we have $E_{p1}=E_{p2}=E_{p1'}=E_{p2'}=E_p=V_s$ and that $t_{s12}=t_{s1'2'}=t_s$. Denoting $E_c(1,1')=E_c(2,2')=E_{c1}$ and $E_c(1,2')=E_c(2,1')=E_{c2}$  we are obtaining 4 orthogonal Hamiltonian eigenvectors
\begin{eqnarray}
\ket{E_1}=
\begin{pmatrix}
-1 \\
0 \\
0 \\
+1
\end{pmatrix}
=-\ket{1,0}_U\ket{1,0}_L+\ket{0,1}_U\ket{0,1}_L \neq (a_1\ket{1,0}_U+a_2\ket{0,1}_U)(a_3\ket{1,0}_U+a_4\ket{0,1}_U),
\end{eqnarray}
\begin{eqnarray}
\ket{E_2}=
\begin{pmatrix}
1 \\
0 \\
0 \\
-1
\end{pmatrix}
=\ket{1,0}_U\ket{0,1}_L-\ket{0,1}_U\ket{1,0}_L \neq (a_1\ket{1,0}_U+a_2\ket{0,1}_U)(a_3\ket{1,0}_U+a_4\ket{0,1}_U).
\end{eqnarray}
We observe that two first energetic states are degenerated so the same quantum state corresponds to 2 different eigenenergies $E_1$ and $E_2$.
 This degeneracy is non-present if we come back to Schroedinger picture and  observe that localized energy and hopping terms for one particle are depending on another particle presence that will bring renormalization of wavevectors.
Situation is depicted in Fig.\ref{renormalization}. Degeneracy of eigenstates is lifted if we set $E_{p1}(|\psi(1')|^2,|\psi(2')|^2), E_{p2}(|\psi(1')|^2,|\psi(1')|^2)$, $E_{p1'}(|\psi(1)|^2,|\psi(2)|^2), E_{p2'}(|\psi(1)|^2,|\psi(1)|^2)$ and $t_{1 \rightarrow 2}(|\psi(1')|^2,|\psi(2')|^2)$, $t_{1' \rightarrow 2'}(|\psi(1')|^2,|\psi(2')|^2)$.

The same argument is for another
wavevectors as given below.
\begin{eqnarray}
\ket{E_{3(4)}}=
\begin{pmatrix}
1 \\
\mp\frac{4 t_s}{\pm(-E_{c1} + E_{c2}) + \sqrt{(E_{c1} - E_{c2})^2 + 16 t_s^2}} \\
\mp\frac{4 t_s}{\pm(-E_{c1} + E_{c2}) + \sqrt{(E_{c1} - E_{c2})^2 + 16 t_s^2}} \\
1
\end{pmatrix}
=\ket{1,0}_U\ket{1,0}_L+\ket{0,1}_U\ket{0,1}_L+c (\ket{1,0}_U\ket{0,1}_L +\ket{0,1}_U\ket{1,0}_L )= \nonumber \\
=(\ket{1,0}_U+\ket{0,1}_U)(\ket{1,0}_L+\ket{0,1}_L)+(c-1)(\ket{1,0}_U\ket{0,1}_L +\ket{0,1}_U\ket{1,0}_L) \nonumber \\
\neq (a_1\ket{1,0}_U+a_2\ket{0,1}_U)(a_3\ket{1,0}_U+a_4\ket{0,1}_U), \nonumber \\
\end{eqnarray}
where c=$\mp\frac{4 t_s}{\pm(-E_{c1} + E_{c2}) + \sqrt{(E_{c1} - E_{c2})^2 + 16 t_s^2}}$.
First two $\ket{E_1}$ and $\ket{E_2}$ energy eigenstates are always entangled, while  $\ket{E_3}$ and $\ket{E_4}$ eigenenergies are only partially entangled if $\mp\frac{4 t_s}{\pm(-E_{c1} + E_{c2}) + \sqrt{(E_{c1} - E_{c2})^2 + 16 t_s^2}} \neq 1$. If
$c=1=\mp\frac{4 t_s}{\pm(-E_{c1} + E_{c2}) + \sqrt{(E_{c1} - E_{c2})^2 + 16 t_s^2}}$ last two energy eigenstates are not entangled. The situation of c=1 takes place when $E_{c1}=E_{c2}$ so when two qubits are infinitely far away so when they are electrostatically decoupled. Situation of c=0 is interesting because it means that $\ket{E_3}$ and $\ket{E_4}$ are maximally entangled and it occurs when $t_s=0$ so when two electrons are maximally localized in each of the qubit so there is no hopping between left and right well.

The obtained eigenenergy states correspond to 4 eigenenergies
\begin{eqnarray}
E_1=E_{c1} + 2 V_s, E_2=E_{c2} + 2 V_s, E_1 > E_2 \nonumber \\
E_3= \frac{1}{2} ( (E_{c1} + E_{c2})  - \sqrt{(E_{c1} -E_{c2})^2 + 16 t_s^2} + 4 V_s)=  \nonumber \\
=\frac{1}{2} ( (q^2(\frac{1}{d_1}+\frac{1}{\sqrt{d_1^2+(a+b)^2}} ))  - \sqrt{(q^2(\frac{1}{d_1}-\frac{1}{\sqrt{d_1^2+(a+b)^2}} ))^2 + 16 t_s^2} + 4 V_s), \nonumber \\
E_4 =\frac{1}{2} ( (E_{c1} + E_{c2})  + \sqrt{(E_{c1} -E_{c2})^2 + 16 t_s^2} + 4 V_s)=  \nonumber \\
 \frac{1}{2} ( (q^2(\frac{1}{d_1}+\frac{1}{\sqrt{d_1^2+(a+b)^2}})+ \sqrt{(q^2(\frac{1}{d_1}-\frac{1}{\sqrt{d_1^2+(a+b)^2}}))^2 + 16 t_s^2} + 4 V_s), E_4 > E_3 .
\end{eqnarray}

We also notice that the eigenenergy states $\ket{E_1}$, $\ket{E_2}$ ,$\ket{E_3}$, $\ket{E_4}$ do not have its classical counterpart since upper electron exists at both positions 1 and 2 and lower electron exists at both positions at the same time. We observe that when distance between two systems of double quantum dots goes into infinity the energy difference between quantum state corresponding to $\ket{E_3}$ and $\ket{E_4}$ goes to zero. This makes those two entangled states degenerated.




Normalized 4 eigenvectors of 2 interacting qubits in SWAP Q-Gate configuration are of the following form

$ \ket{E_1}_n=
\frac{1}{\sqrt{\left(8\left(\frac{t_{sr1}-t_{sr2}}{\sqrt{(E_{c1}-E_{c2})^2+4
   (t_{sr1}-t_{sr2})^2}-E_{c1}+E_{c2}}\right)\right)^2+2
   }}
\begin{pmatrix}
-1,\\
-\frac{2(t_{sr1}-t_{sr2})}{\sqrt{(E_{c1}-E_{c2})^2+4 (t_{sr1}-t_{sr2})^2}-E_{c1}+E_{c2}}, \\
\frac{2(t_{sr1}-t_{sr2})}{\sqrt{(E_{c1}-E_{c2})^2+4(t_{sr1}-t_{sr2})^2}-E_{c1}+E_{c2}} \\
1
\end{pmatrix}=\frac{1}{\sqrt{\left(8\left(\frac{t_{sr1}-t_{sr2}}{\sqrt{(E_{c1}-E_{c2})^2+4
   (t_{sr1}-t_{sr2})^2}-E_{c1}+E_{c2}}\right)\right)^2+2
   }} \ket{E_1}
$

$ \ket{E_2}_{n}= -\frac{1}{\sqrt{\left(8\left(\frac{t_{sr1}-t_{sr2}}{\sqrt{(E_{c1}-E_{c2})^2+4
   (t_{sr1}-t_{sr2})^2}+E_{c1}-E_{c2}}\right)\right)^2+2
   }}
\begin{pmatrix}
-1 \\
\frac{2
   (t_{sr1}-t_{sr2})}{\sqrt{(E_{c1}-E_{c2})^2+4
   (t_{sr1}-t_{sr2})^2}+E_{c1}-E_{c2}} \\
-\frac{2(t_{sr1}-t_{sr2})}{\sqrt{(E_{c1}-E_{c2})^2+4
   (t_{sr1}-t_{sr2})^2}+E_{c1}-E_{c2}} \\,1
\end{pmatrix}=-
\frac{1}{\sqrt{\left(8\left(\frac{t_{sr1}-t_{sr2}}{\sqrt{(E_{c1}-E_{c2})^2+4
   (t_{sr1}-t_{sr2})^2}+E_{c1}-E_{c2}}\right)\right)^2+2
   }} \ket{E_2}
$

$\ket{E_3}_{n}= \frac{1}{\sqrt{\left(8\left(\frac{\text{tsr1}+\text{tsr2}}{\sqrt{(\text{Ec1}-\text{Ec2})^2+4
   (\text{tsr1}+\text{tsr2})^2}-\text{Ec1}+\text{Ec2}}\right)\right)^2+2
   }}
\begin{pmatrix}
1, \\
-\frac{2(t_{sr1}+t_{sr2})}{\sqrt{(E_{c1}-E_{c2})^2+4
   (t_{sr1}+t_{sr2})^2}-E_{c1}+E_{c2}}, \\
-\frac{2(t_{sr1}+t_{sr2})}{\sqrt{(E_{c1}-E_{c2})^2+4
   (t_{sr1}+t_{sr2})^2}-E_{c1}+E_{c2}}, \\
1
\end{pmatrix}=
\frac{1}{\sqrt{\left(8\left(\frac{\text{tsr1}+\text{tsr2}}{\sqrt{(\text{Ec1}-\text{Ec2})^2+4
   (\text{tsr1}+\text{tsr2})^2}-\text{Ec1}+\text{Ec2}}\right)\right)^2+2
   }} \ket{E_3}
$

$\ket{E_4}_{n}=  \frac{1}{\sqrt{\left(8\left(\frac{t_{sr1}+t_{sr2}}{\sqrt{(E_{c1}-E_{c2})^2+4
   (t_{sr1}+t_{sr2})^2}-E_{c2}+E_{c1}}\right)\right)^2+2
   }}
\begin{pmatrix}
1, \\
\frac{2(t_{sr1}+t_{sr2})}{\sqrt{(E_{c1}-E_{c2})^2+4(t_{sr1}+t_{sr2})^2}+E_{c1}-E_{c2}}, \\
\frac{2(t_{sr1}+t_{sr2})}{\sqrt{(E_{c1}-E_{c2})^2+4(t_{sr1}+t_{sr2})^2}+E_{c1}-E_{c2}},  \\
1
\end{pmatrix}=
\frac{1}{\sqrt{\left(8\left(\frac{t_{sr1}+t_{sr2}}{\sqrt{(E_{c1}-E_{c2})^2+4
   (t_{sr1}+t_{sr2})^2}-E_{c2}+E_{c1}}\right)\right)^2+2
   }} \ket{E_4}.
$ \newline
We are obtaining simplifications after assuming $t_{sr1}(t)=t_{sr2}(t)$ so we obtain
\begin{equation}
\ket{E_1}_{n}=\frac{1}{\sqrt{2}}
\begin{pmatrix}
-1 \\
0 \\
0 \\
1
\end{pmatrix},
\ket{E_2}_{n}=\frac{1}{\sqrt{2}}
\begin{pmatrix}
1 \\
0 \\
0 \\
-1
\end{pmatrix},
\end{equation}
\begin{equation}
\ket{E_3}_{n}=\sqrt{\frac{4t_s}{(E_{c2}-E_{c1}) + 8 t_s - \sqrt{ (E_{c1} - E_{c2})^2 + 16 t_s^2}}}
\begin{pmatrix}
1 \\
-\frac{4 t_s}{(-E_{c1} + E_{c2}) + \sqrt{(E_{c1} - E_{c2})^2 + 16 t_s^2}} \\
-\frac{4 t_s}{(-E_{c1} + E_{c2}) + \sqrt{(E_{c1} - E_{c2})^2 + 16 t_s^2}} \\
1
\end{pmatrix},
\end{equation}

\begin{equation}
\ket{E_4}_{n}=\sqrt{\frac{4t_s}{(E_{c1}-E_{c2}) + 8 t_s - \sqrt{ (E_{c1} - E_{c2})^2 + 16 t_s^2}}}
\begin{pmatrix}
1 \\
\frac{4 t_s}{(E_{c1} - E_{c2}) + \sqrt{(E_{c1} - E_{c2})^2 + 16 t_s^2}} \\
\frac{4 t_s}{(E_{c1} - E_{c2}) + \sqrt{(E_{c1} - E_{c2})^2 + 16 t_s^2}} \\
1
\end{pmatrix}.
\end{equation}

It is worth mentioning that if we want to bring two electrostatic qubits to the entangled state we need to cool down (or heat-up)  the system of interacting qubits to the energy $E_1$ (or to energy $E_2$).
Otherwise we might also wish to disentangle two electrostatically interacting qubits. In such way one of the scenario is to bring the quantum system either to energy $E_3$ or $E_4$ so only partial entanglement will be achieved.
Other scenario would be by bringing the occupancy of different energetic levels so net entanglement is reduced. One can use the entanglement witness in quantifing the existence of entanglement. One of the simplest q-state entanglement measurement is von Neumann entanglement entropy as it is expressed by formula \ref{entropyS} that requires the knowledge of q-system density matrix with time. Such matrices can be obtained analytically for the case of 2 electrostatically interacting qubits.

It is interesting to spot the dependence of eigenergies on distance between interacting qubits in the general case as it is depicted in Fig.6. Now we are moving towards description the procedure of cooling down or heating up in Q-Swap gate.
The proceudure was discussed previously in the case of single qubit. Now it is excercised in the case of 2-qubit electrostatic interaction. For the sake of simplicity we will change the occupancy of the energy level $E_1$ and energy level level $E_2$ and keep
the occupancy of other energy levels unchanged.  We can write the $\ket{E_2}\bra{E_1} $ as
\begin{eqnarray}
\ket{E_2}_n\bra{E_1}_n=\frac{1}{2}
\begin{pmatrix}
1 \\ 0 \\ 0 \\ -1
\end{pmatrix}
\begin{pmatrix}
-1 & 0 & 0 & 1
\end{pmatrix}=
\begin{pmatrix}
-1 & 0 & 0 & +1 \\
0 & 0 & 0 & 0 \\
0 & 0 & 0 & 0 \\
+1 & 0 & 0 & -1 \\
\end{pmatrix}, 
\ket{E_1}_n\bra{E_2}_n=\frac{1}{2}
\begin{pmatrix}
-1 \\ 0 \\ 0 \\ 1
\end{pmatrix}
\begin{pmatrix}
1 & 0 & 0 & -1
\end{pmatrix}=
\begin{pmatrix}
-1 & 0 & 0 & +1 \\
0 & 0 & 0 &  0 \\
0 & 0 & 0 & 0 \\
+1 & 0 & 0 & -1 \\
\end{pmatrix}. \nonumber \\
\end{eqnarray}
We are introducing $f_1$ and $f_2$ real valued functions of small magnitude $f(t)=f_1(t)=f_2(t), (|f_1|,|f_2|<<(E_1,E_2))$  and we are considering the following Hamiltonian having $H_0$ that is time-independent and other part dependent part as
\begin{eqnarray}
\hat{H}=\hat{H}_0+f_1(t)\ket{E_2}_n\bra{E_1}_n+f_2(t)\ket{E_1}_n\bra{E_2}_n=E_1\ket{E_1}\bra{E_1}+E_2\ket{E_2}\bra{E_2}+f_1(t)\ket{E_2}_n\bra{E_1}_n+f_2(t)\ket{E_1}_n\bra{E_2}_n= \nonumber
\end{eqnarray}
\begin{eqnarray}
=
\begin{pmatrix}
2E_{p}+ \frac{q^2}{d_1} & t_{s} & t_{s} & 0 \\
t_{s}^{*} & 2E_{p}+\frac{q^2}{\sqrt{(d1)^2+(b+a)^2}} & 0 & t_{s} \\
t_{s}^{*} & 0 & 2E_{p}+ \frac{q^2}{\sqrt{(d1)^2+(b+a)^2}} & t_{s} \\
0 & t_{s}^{*} & t_{s}^{*} & 2 E_{p}+ \frac{q^2}{d1} \\
\end{pmatrix}   \nonumber \\
+\frac{1}{2}\left(f_1
\begin{pmatrix}
-1 & 0 & 0 & 1 \\
0 & 0 & 0 & 0 \\
0 & 0 & 0 & 0 \\
1 & 0 & 0 & -1 \\
\end{pmatrix}
+f_2
\begin{pmatrix}
-1 & 0 & 0 & 1 \\
0 & 0 & 0 &  0 \\
0 & 0 & 0 & 0 \\
1 & 0 & 0 & -1 \\
\end{pmatrix}\right)=
\nonumber \\
=
\begin{pmatrix}
2E_{p}+ \frac{q^2}{d_1}-f(t) & t_{s} & t_{s} & f(t) \\
t_{s}^{*} & 2E_{p}+\frac{q^2}{\sqrt{(d1)^2+(b+a)^2}} & 0 & t_{s} \\
t_{s}^{*} & 0 & 2E_{p}+ \frac{q^2}{\sqrt{(d1)^2+(b+a)^2}} & t_{s} \\
f(t) & t_{s}^{*} & t_{s}^{*} & 2 E_{p}+ \frac{q^2}{d1}-f(t) \\
\end{pmatrix}=\hat{H}(t)_{E_1<->E_2,Q-Swap}.\nonumber \\
\end{eqnarray}

\normalsize
Initially we have established the following parameters of tight-binding model as $t_{s12}=t_{s1'2'}$. Changing $t_{s12}$ into $t_{s12}-\frac{f(t)}{2}$ and $t_{s1'2'}$ into $t_{s1'2'}+\frac{f(t)}{2}$ while keeping other parameters of tight-binding model unchanged will result
in the heating up (cooling down) of q-state of SWAP gate so population of energy level $E_1$ and $E_2$ are time-depenent, while populations of energy levels $E_3$ and $E_4$ are unchanged. Practically our results mean that we need to keep all our confiment potential bottoms constant, while changing barrier height between neighbouring q-dots in each of position based qubits.
In such way we have established the procedure of perturbative cooling (heating up) of q-state. Non-perturbative approach is absolutely possible but it requires full knowledge of time dependent eigenstates and eigenenergies (solutions of eigenenergies of 4th order polynomial are very lengthy in general case) and therefore corresponding expression are very lengthy. In similar fashion we can heat up or cool down two coupled Single Electron Lines \cite{SEL} as in Fig.1 or any other q-system having N interacting q-bodies that can be represented by the system of N-interacting position based qubits.
\subsection{Case of density matrix in case of 2 interacting particles in symmetric case}
We consider the simplifying matrix and highly symmetric matrix of the form
\small
\begin{eqnarray}
\hat{H}(t)= \nonumber \\
\begin{pmatrix}
2E_{p}(t)+ \frac{q^2}{d_1}=q_{11}+q_{22}  & t_{sr2}(t) & t_{sr1}(t) & 0 \\
t_{sr2}(t)  & 2E_{p}(t)+\frac{q^2}{\sqrt{(d_1)^2+(b+a)^2}}=q_{11}-q_{22} & 0 & t_{sr1}(t) \\
t_{sr1}(t) & 0 & 2E_{p}(t)+ \frac{q^2}{\sqrt{(d_1)^2+(b+a)^2}}=q_{11}-q_{22} & t_{sr2}(t) \\
0 & t_{sr1}(t) & t_{sr2}(t) & 2 E_{p}(t)+ \frac{q^2}{d1}=q_{11}+q_{22} \\
\end{pmatrix} =\nonumber \\
=\hat{\sigma}_0 \times \hat{\sigma}_0 q_{11} + \hat{\sigma}_{3} \times \hat{\sigma}_{3} q_{22} +t_{sr2}(t)  \hat{\sigma}_0 \times \hat{\sigma}_3 + t_{sr1}(t) \hat{\sigma}_3  \times \hat{\sigma}_0 \nonumber \\
\end{eqnarray}
\normalsize
that has only real value components $H_{k,l}$ with $q_{11}=E_p(t)+\frac{E_{c1}+E_{c2}}{2}=E_p(t)+\frac{1}{2}(\frac{q^2}{d_1}+\frac{q^2}{\sqrt{(d_1)^2+(b+a)^2}})$, $q_{22}=\frac{E_{c1}-E_{c2}}{2}=\frac{1}{2}(\frac{q^2}{d_1}-\frac{q^2}{\sqrt{(d_1)^2+(b+a)^2}})$ and $Q_{11}(t)=\int_{t_0}^{t}dt'q_{11}(t')$, $Q_{22}(t)=\int_{t_0}^{t}dt'q_{22}(t')$, $TR1(t)=\int_{t_0}^{t}dt't_{sr1}(t')$, $TR2(t)=\int_{t_0}^{t}dt't_{sr2}(t')$.  We obtain the density matrix

\begin{eqnarray}
\hat{U}(t)=
\begin{pmatrix}
U_{1,1}(t) & U_{1,2}(t) & U_{1,3}(t) & U_{1,4}(t) \\
U_{2,1}(t) & U_{2,2}(t) & U_{2,3}(t) & U_{2,4}(t) \\
U_{3,1}(t) & U_{3,2}(t) & U_{3,3}(t) & U_{3,4}(t) \\
U_{4,1}(t) & U_{4,2}(t) & U_{4,3}(t) & U_{4,4}(t) \\
\end{pmatrix},
\hat{\rho}(t)=\hat{U}(t,t_0)
\begin{pmatrix}
\rho_{1,1}(t_0) & \rho_{1,2}(t_0) & \rho_{1,3}(t_0) & \rho_{1,4}(t_0) \\
\rho_{2,1}(t_0) & \rho_{2,2}(t_0) & \rho_{2,3}(t_0) & \rho_{2,4}(t_0) \\
\rho_{3,1}(t_0) & \rho_{3,2}(t_0) & \rho_{3,3}(t_0) & \rho_{3,4}(t_0) \\
\rho_{4,1}(t_0) & \rho_{4,2}(t_0) & \rho_{4,3}(t_0) & \rho_{4,4}(t_0) \\
\end{pmatrix}\hat{U}^{-1}(t,t_0)\nonumber \\
\end{eqnarray}
with the following components of unitary matrix
\begin{eqnarray}
U_{1,1}(t)=\frac{e^{-i \hbar Q_{11}(t)}}{2}\Bigg[
-iQ_{22}(t)\times \nonumber \\ \times \left(\frac{\sin \left(\hbar
   \sqrt{|Q_{22}(t)|^2+(TR1(t)-TR2(t))^2}\right)}{\sqrt{|Q_{22}(t)|^2+(TR1(t)-TR2(t))^2}}+\frac{\sin \left(\hbar
   \sqrt{|Q_{22}(t)|^2+(TR1(t)+TR2(t))^2}\right)}{\sqrt{|Q_{22}(t)|^2+(TR1(t)+TR2(t))^2}}\right)+ \nonumber \\ +\cos \left(\hbar
   \sqrt{|Q_{22}(t)|^2+(TR1(t)-TR2(t))^2}\right)+\cos \left(\hbar \sqrt{|Q_{22}(t)|^2+(TR1(t)+TR2(t))^2}\right) \Bigg].
\end{eqnarray}
\begin{eqnarray}
U_{1,2}(t)=\frac{i e^{-i \hbar Q_{11}(t)} \Bigg(  (TR1(t)-TR2(t))  \sin \left(\hbar
   \sqrt{  |Q_{22}(t)|^2  +(TR1(t)-TR2(t))^2}\right) }{2 \sqrt{|Q_{22}(t)|^2+(TR1(t)-TR2(t))^2}  }
\nonumber \\
-\frac{(TR1(t)+TR2(t))  \sin \left(\hbar
   \sqrt{|Q_{22}(t)|^2+(TR1(t)+TR2(t))^2}\right) \Bigg)  }{2  \sqrt{|Q_{22}(t)|^2+(TR1(t)+TR2(t))^2}},
\end{eqnarray}
\begin{eqnarray}
U_{1,3}(t)=-i e^{-i \hbar Q_{11}(t)}\frac{ \Bigg[ (TR1(t)-TR2(t))  \sin \left(\hbar
   \sqrt{|Q_{22}(t)|^2+(TR1(t)-TR2(t))^2}\right)}{2 \sqrt{|Q_{22}(t)|^2+(TR1(t)-TR2(t))^2} } \nonumber \\
+  \frac{(TR1(t)+TR2(t))  \sin \left(\hbar
   \sqrt{|Q_{22}(t)|^2+|TR1(t)+TR2(t)|^2}\right)\Bigg]  }{2  \sqrt{|Q_{22}(t)|^2+(TR1(t)+TR2(t))^2}}. 
\end{eqnarray}

\begin{eqnarray}
U_{1,4}(t)=
\frac{1}{2} e^{-i \hbar Q_{11}(t)} \Bigg[i Q_{22}(t) \Bigg[\frac{\sin \left(\hbar
   \sqrt{ |Q_{22}(t)|^2+(TR1(t)-TR2(t))^2}\right)}{\sqrt{|Q_{22}(t)|^2+(TR1(t)-TR2(t))^2}}  \nonumber \\
   -\frac{\sin \left(\hbar
   \sqrt{|Q_{22}(t)|^2+(TR1(t)+TR2(t))^2}\right)}{\sqrt{|Q_{22}(t)|^2+(TR1(t)+TR2(t))^2}}\Bigg] \nonumber \\ -\cos \left(\hbar
   \sqrt{|Q_{22}(t)|^2+(TR1(t)-TR2(t))^2}\right)+\cos \left(\hbar \sqrt{|Q_{22}(t)|^2+(TR1(t)+TR2(t))^2}\right)\Bigg]
\end{eqnarray}

\begin{eqnarray}
U_{2,1}(t)=-\frac{i}{2}e^{-i \hbar Q_{11}(t)}
\frac{\Bigg[ (TR1(t)-TR2(t)) \sin \left(\hbar
   \sqrt{|Q_{22}(t)|^2+(TR1(t)-TR2(t))^2}\right)}{\sqrt{|Q_{22}(t)|^2+(TR1(t)-TR2(t))^2}} \nonumber \\
-\frac{ (TR1(t)+TR2(t)) \sin \left(\hbar
   \sqrt{|Q_{22}(t)|^2+(TR1(t)+TR2(t))^2}\right)\Bigg] }{\sqrt{|Q_{22}(t)|^2+(TR1(t)+TR2(t))^2}}
\end{eqnarray}

\begin{eqnarray}
U_{2,2}(t)=
\frac{1}{2} e^{-i \hbar Q_{11}(t)} \Bigg[ i Q_{22}(t) \Bigg[\frac{\sin \left(\hbar
   \sqrt{|Q_{22}(t)|^2+(TR1(t) -TR2(t))^2}\right)}{\sqrt{|Q_{22}(t)|^2+(TR1(t)-TR2(t))^2}}+\nonumber \\ \frac{\sin \left(\hbar
   \sqrt{|Q_{22}(t)|^2+(TR1(t)+TR2(t))^2}\right)}{\sqrt{|Q_{22}(t)|^2+(TR1(t)+TR2(t))^2}}\Bigg]+  \nonumber \\ +\cos \left(\hbar
   \sqrt{|Q_{22}(t)|^2+(TR1(t)-TR2(t))^2}\right)+\cos \left(\hbar \sqrt{|Q_{22}(t)|^2+(TR1(t)+TR2(t))^2}\right)\Bigg]
\end{eqnarray}

\begin{eqnarray}
U_{2,3}(t)=e^{-i \hbar Q_{11}(t)} 
\Bigg[\frac{  
 -Q_{22}(t)  \sin \left(\hbar
   \sqrt{|Q_{22}(t)|^2+(TR1(t)-TR2(t))^2}\right) 
}{2 \sqrt{|Q_{22}(t)|^2+(TR1(t)-TR2(t))^2} } + \nonumber \\ +
\frac{Q_{22}(t)  \sin \left(\hbar
   \sqrt{|Q_{22}(t)|^2+(TR1(t)+TR2(t))^2}\right)}{2  \sqrt{|Q_{22}(t)|^2+(TR1(t)+TR2(t))^2}}+\nonumber \\ +\frac{ i   \cos \left(\hbar \sqrt{|Q_{22}(t)|^2+(TR1(t)-TR2(t))^2}
   \right)-i   \cos \left(\hbar
   \sqrt{|Q_{22}(t)|^2+(TR1(t)+TR2(t))^2}\right)}{2  }\Bigg]
\end{eqnarray}

\begin{eqnarray}
U_{2,4}(t)=-\frac{i e^{-i \hbar Q_{11}(t) } \Bigg[(TR1(t)-TR2(t)) \sin \left(\hbar
   \sqrt{|Q_{22}(t)|^2+(TR1(t)-TR2(t))^2}\right)}{2 \sqrt{|Q_{22}(t)|^2+(TR1(t)-TR2(t))^2} }+\nonumber \\+\frac{(TR1(t)+TR2(t))  \sin \left(\hbar
   \sqrt{|Q_{22}(t)|^2+(TR1(t)-TR2(t))^2}\right)\Bigg]}{2  \sqrt{|Q_{22}(t)|^2+(TR1(t)+TR2(t))^2}}
\end{eqnarray}

\begin{eqnarray}
U_{3,1}(t)=-i e^{-i \hbar Q_{11}(t) }\frac{ \Bigg[(TR1(t)-TR2(t))  \sin \left(\hbar
   \sqrt{|Q_{22}(t)|^2+(TR1(t)-TR2(t))^2}\right)}{2 \sqrt{|Q_{22}(t)|^2+(TR1(t)-TR2(t))^2} }+\nonumber\\+\frac{(TR1(t)+TR2(t))  \sin \left(\hbar
   \sqrt{|Q_{22}(t)|^2+(TR1(t)+TR2(t))^2}\right)\Bigg]}{2  \sqrt{|Q_{22}(t)|^2+(TR1(t)+TR2(t))^2}}
\end{eqnarray}

\begin{eqnarray}
U_{3,2}(t)= e^{-i \hbar Q_{11}(t) } 
\Bigg[ \frac{ -Q_{22}(t)  \sin \left(\hbar
   \sqrt{|Q_{22}(t)|^2+(TR1(t)-TR2(t))^2}\right) }{2 \sqrt{|Q_{22}(t)|^2+(TR1(t)-TR2(t))^2}  }+ \nonumber \\ \frac{Q_{22}(t)  \sin \left(\hbar
   \sqrt{|Q_{22}(t)|^2+(TR1(t)+TR2(t))^2}\right)}{2  \sqrt{|Q_{22}(t)|^2+(TR1(t)+TR2(t))^2}}+\nonumber \\ +\frac{i   \cos \left(\hbar
   \sqrt{|Q_{22}(t)|^2+(TR1(t)-TR2(t))^2}\right)-i   \cos \left(\hbar
   \sqrt{|Q_{22}(t)|^2+(TR1(t)+TR2(t))^2}\right)\Bigg]}{2  }
\end{eqnarray}

\begin{eqnarray}
U_{3,3}(t)=\frac{1}{2} e^{-i \hbar Q_{11}(t)} \Bigg[ i Q_{22}(t) \Bigg[\frac{\sin \left(\hbar
   \sqrt{|Q_{22}(t)|^2+(TR1(t)-TR2(t))^2}\right)}{\sqrt{|Q_{22}(t)|^2+(TR1(t)-TR2(t))^2}}\nonumber \\ +\frac{\sin \left(\hbar
   \sqrt{|Q_{22}(t)|^2+(TR1(t)+TR2(t))^2}\right)}{\sqrt{|Q_{22}(t)|^2+(TR1(t)+TR2(t))^2}}\Bigg]+\nonumber \\ +\cos \left(\hbar
   \sqrt{|Q_{22}(t)|^2+(TR1(t)-TR2(t))^2}\right)+\cos \left(\hbar \sqrt{|Q_{22}(t)|^2+(TR1(t)+TR2(t))^2}\right)\Bigg]
\end{eqnarray}

\begin{eqnarray}
U_{3,4}(t)=(\sin (\hbar Q_{11}(t))+i \cos (\hbar Q_{11}(t) ))\frac{ \Bigg[(TR1(t)-TR2(t))  \sin \left(\hbar
   \sqrt{|Q_{22}(t)|^2+(TR1(t)-TR2(t))^2}\right)}{2 \sqrt{|Q_{22}(t)|^2 +(TR1(t)-TR2(t))^2} }+\nonumber \\   -\frac{(TR1(t)+TR2(t))  \sin \left(\hbar
   \sqrt{|Q_{22}(t)|^2+(TR1(t)+TR2(t))^2}\right)\Bigg]}{2  \sqrt{|Q_{22}(t)|^2+(TR1(t)+TR2(t))^2}}
\end{eqnarray}

\begin{eqnarray}
U_{4,1}(t)=\frac{1}{2} e^{-i \hbar Q_{11}(t)} \Bigg[i Q_{22}(t) \Bigg[\frac{\sin \left(\hbar
   \sqrt{|Q_{22}(t)|^2+(TR1(t)-TR2(t))^2}\right)}{\sqrt{|Q_{22}(t)|^2+(TR1(t)-TR2(t))^2}}\nonumber \\ -\frac{\sin \left(\hbar
   \sqrt{|Q_{22}(t)|^2+(TR1(t)+TR2(t))^2}\right)}{\sqrt{|Q_{22}(t)|^2+(TR1(t)+TR2(t))^2}}\Bigg]+\nonumber \\  -\cos \left(\hbar
   \sqrt{|Q_{22}(t)|^2+(TR1(t)-TR2(t))^2}\right)+\cos \left(\hbar \sqrt{|Q_{22}(t)|^2+(TR1(t)+TR2(t))^2}\right)\Bigg]
\end{eqnarray}

\begin{eqnarray}
U_{4,2}(t)=-\frac{i e^{-i \hbar Q_{11}(t) } \Bigg[(TR1(t)-TR2(t))  \sin \left(\hbar
   \sqrt{|Q_{22}(t)|^2+(TR1(t)-TR2(t))^2}\right)}{2 \sqrt{|Q_{22}(t)|^2+(TR1(t)-TR2(t))^2} }+\nonumber \\+\frac{(TR1(t)+TR2(t))  \sin \left(\hbar
   \sqrt{|Q_{22}(t)|^2+(TR1(t)+TR2(t))^2}\right)\Bigg]}{2  \sqrt{|Q_{22}(t)|^2+(TR1(t)+TR2(t))^2}}
\end{eqnarray}

\begin{eqnarray}
U_{4,3}(t)=i e^{-i \hbar Q_{11}(t)}\Bigg[ \frac{  (TR1(t)-TR2(t))  \sin \left(\hbar
   \sqrt{|Q_{22}(t)|^2+(TR1(t)-TR2(t))^2}\right)}{2 \sqrt{|Q_{22}(t)|^2+(TR1(t)-TR2(t))^2} }+\nonumber \\ -\frac{(TR1(t)+TR2(t))  \sin \left(\hbar
   \sqrt{|Q_{22}(t)|^2+(TR1(t)+TR2(t))^2}\right)}{2  \sqrt{|Q_{22}(t)|^2+(TR1(t)+TR2(t))^2}}\Bigg]
\end{eqnarray}

\begin{eqnarray}
U_{4,4}(t)=\frac{1}{2} e^{-i \hbar Q_{11}(t)} \Bigg[-i Q_{22}(t) \Bigg[\frac{\sin \left(\hbar
   \sqrt{|Q_{22}(t)|^2+(TR1(t)-TR2(t))^2}\right)}{\sqrt{|Q_{22}(t)|^2+(TR1(t)-TR2(t))^2}}+ \nonumber \\
   \frac{\sin \left(\hbar
   \sqrt{|Q_{22}(t)|^2+(TR1(t)+TR2(t))^2}\right)}{\sqrt{|Q_{22}(t)|^2+(TR1(t)+TR2(t))^2}}\Bigg]+ \nonumber \\ + \cos \left(\hbar
   \sqrt{|Q_{22}(t)|^2+(TR1(t)-TR2(t))^2}\right)+\cos \left(\hbar \sqrt{|Q_{22}(t)|^2+(TR1(t)+TR2(t))^2}\right)\Bigg]
\end{eqnarray}
We set the quantum state to be $\ket{\psi,t_0}=\ket{E_1}$ at time $t_0$ so it is maximally entangled and its density matrix is $\rho(t_0)=\ket{\psi,t_0}\bra{\psi,t_0}=\ket{E_1}\bra{E_1}=\frac{1}{2}
\begin{pmatrix}
+1 & 0 & 0 & -1 \\
0 & 0 & 0 & 0  \\
0 & 0 & 0 & 0  \\
-1 & 0 & 0 & 1 \\
\end{pmatrix}
$.
 Finally we obtain the following density matrix
\begin{eqnarray}
\rho_{1,1}(t)=\frac{(TR1(t)-TR2(t))^2 \cos \left(2 \hbar \sqrt{|Q_{22}(t)|^2+(TR1(t)-TR2(t))^2}\right)+2 |Q_{22}(t)|^2+(TR1(t)-TR2(t))^2}{4
   \left(|Q_{22}(t)|^2+(TR1(t)-TR2(t))^2\right)}\nonumber \\
\end{eqnarray}

\begin{eqnarray}
\rho_{1,2}(t)=\frac{(TR1(t)-TR2(t)) \Bigg[-i \sqrt{|Q_{22}(t)|^2+(TR1(t)-TR2(t))^2} \sin \left(2 \hbar \sqrt{|Q_{22}(t)|^2+(TR1(t)-TR2(t))^2}\right)}{4 \left(|Q_{22}(t)|^2+(TR1(t)-TR2(t))^2\right)}+ \nonumber \\ +\frac{Q_{22}(t) \cos
   \left(2 \hbar \sqrt{|Q_{22}(t)|^2+(TR1(t)-TR2(t))^2}\right)-Q_{22}(t)\Bigg]}{4 \left(|Q_{22}(t)|^2+(TR1(t)-TR2(t))^2\right)}\nonumber \\
\end{eqnarray}

\begin{eqnarray}
\rho_{1,3}(t)=-(TR1(t)-TR2(t))\frac{ \Bigg[-i \sqrt{|Q_{22}(t)|^2+(TR1(t)-TR2(t))^2} \sin \left(2 \hbar \sqrt{|Q_{22}(t)|^2+(TR1(t)-TR2(t))^2}\right)}{4 \left(|Q_{22}(t)|^2+(TR1(t)-TR2(t))^2\right)}+ \nonumber \\ +\frac{Q_{22}(t) \cos
   \left(2 \hbar \sqrt{|Q_{22}(t)|^2+(TR1(t)-TR2(t))^2}\right)-Q_{22}(t)\Bigg]}{4 \left(|Q_{22}(t)|^2+(TR1(t)-TR2(t))^2\right)}\nonumber \\
\end{eqnarray}

\begin{eqnarray}
\rho_{1,4}(t)=-\frac{(TR1(t)-TR2(t))^2 \cos \left(2 \hbar \sqrt{|Q_{22}(t)|^2+(TR1(t)-TR2(t))^2}\right)+2 |Q_{22}(t)|^2+(TR1(t)-TR2(t))^2}{4
   \left(|Q_{22}(t)|^2+(TR1(t)-TR2(t))^2\right)}\nonumber \\
\end{eqnarray}

\begin{eqnarray}
\rho_{2,1}(t)=
\frac{(TR1(t)-TR2(t)) \Bigg(i \sqrt{|Q_{22}(t)|^2+(TR1(t)-TR2(t))^2} \sin \left(2 \hbar \sqrt{|Q_{22}(t)|^2+(TR1(t)-TR2(t))^2}\right)}{4 \left(|Q_{22}(t)|^2+(TR1(t)-TR2(t))^2\right)}+\nonumber \\+\frac{Q_{22}(t) \cos
   \left(2 \hbar \sqrt{|Q_{22}(t)|^2+(TR1(t)-TR2(t))^2}\right)-Q_{22}(t)\Bigg]}{4 \left(|Q_{22}(t)|^2+(TR1(t)-TR2(t))^2\right)}\nonumber \\
\end{eqnarray}

\begin{eqnarray}
\rho_{2,2}(t)=\frac{(TR1(t)-TR2(t))^2 \sin ^2\left(\hbar \sqrt{|Q_{22}(t)|^2+(TR1(t)-TR2(t))^2}\right)}{2 \left(|Q_{22}(t)|^2+(TR1(t)-TR2(t))^2\right)}
\end{eqnarray}

\begin{eqnarray}
\rho_{2,3}(t)=-\frac{(TR1(t)-TR2(t))^2 \sin ^2\left(\hbar \sqrt{|Q_{22}(t)|^2+(TR1(t)-TR2(t))^2}\right)}{2 \left(|Q_{22}(t)|^2+(TR1(t)-TR2(t))^2\right)}
\end{eqnarray}

\begin{eqnarray}
\rho_{2,4}(t)=-\frac{(TR1(t)-TR2(t)) \Bigg[ i \sqrt{|Q_{22}(t)|^2+(TR1(t)-TR2(t))^2} \sin \left(2 \hbar \sqrt{|Q_{22}(t)|^2+(TR1(t)-TR2(t))^2}\right)}{4 \left(|Q_{22}(t)|^2+(TR1(t)-TR2(t))^2\right)}+\nonumber \\+\frac{Q_{22}(t)\cos
   \left(2 \hbar \sqrt{|Q_{22}(t)|^2+(TR1(t)-TR2(t))^2}\right)-Q_{22}(t)\Bigg]}{4 \left(|Q_{22}(t)|^2+(TR1(t)-TR2(t))^2\right)}\nonumber \\
\end{eqnarray}

\begin{eqnarray}
\rho_{3,1}(t)=-(TR1(t)-TR2(t))\frac{ \Bigg[i \sqrt{|Q_{22}(t)|^2+(TR1(t)-TR2(t))^2} \sin \left(2 \hbar \sqrt{|Q_{22}(t)|^2+(TR1(t)-TR2(t))^2}\right)}{4 \left(|Q_{22}(t)|^2+(TR1(t)-TR2(t))^2\right)}+ \nonumber \\ +\frac{Q_{22}(t) \cos
   \left(2 \hbar \sqrt{|Q_{22}(t)|^2+(TR1(t)-TR2(t))^2}\right)-Q_{22}(t)\Bigg]}{4 \left(|Q_{22}(t)|^2+(TR1(t)-TR2(t))^2\right)}\nonumber \\
\end{eqnarray}

\begin{eqnarray}
\rho_{3,2}(t)=-\frac{(TR1(t)-TR2(t))^2 \sin ^2\left(\hbar \sqrt{|Q_{22}(t)|^2+(TR1(t)-TR2(t))^2}\right)}{2 \left(|Q_{22}(t)|^2+(TR1(t)-TR2(t))^2\right)}
\end{eqnarray}

\begin{eqnarray}
\rho_{3,3}(t)=\frac{(TR1(t)-TR2(t))^2 \sin ^2\left(\hbar \sqrt{|Q_{22}(t)|^2+(TR1(t)-TR2(t))^2}\right)}{2 \left(|Q_{22}(t)|^2+(TR1(t)-TR2(t))^2\right)}
\end{eqnarray}

\begin{eqnarray}
\rho_{3,4}(t)=\frac{(TR1(t)-TR2(t)) \Bigg(i \sqrt{|Q_{22}(t)|^2+(TR1(t)-TR2(t))^2} \sin \left(2 \hbar \sqrt{|Q_{22}(t)|^2+(TR1(t)-TR2(t))^2}\right)}{4 \left(|Q_{22}(t)|^2+(TR1(t)-TR2(t))^2\right)}+ \nonumber \\ \frac{Q_{22}(t) \cos
   \left(2 \hbar \sqrt{|Q_{22}(t)|^2+(TR1(t)-TR2(t))^2}\right)-Q_{22}(t)\Bigg]}{4 \left(|Q_{22}(t)|^2+(TR1(t)-TR2(t))^2\right)}\nonumber \\
\end{eqnarray}

\begin{eqnarray}
\rho_{4,1}(t)=-\frac{(TR1(t)-TR2(t))^2 \cos \left(2 \hbar \sqrt{|Q_{22}(t)|^2+(TR1(t)-TR2(t))^2}\right)+2|Q_{22}(t)|^2+(TR1(t)-TR2(t))^2}{4
   \left(|Q_{22}(t)|^2+(TR1(t)-TR2(t))^2\right)}\nonumber \\
\end{eqnarray}

\begin{eqnarray}
\rho_{4,2}(t)=-(TR1(t)-TR2(t))\frac{ \Bigg[ -i \sqrt{|Q_{22}(t)|^2+(TR1(t)-TR2(t))^2} \sin \left(2 \hbar \sqrt{|Q_{22}(t)|^2+(TR1(t)-TR2(t))^2}\right)}{4 \left(|Q_{22}(t)|^2+(TR1(t)-TR2(t))^2\right)}+ \nonumber \\ + \frac{Q_{22}(t) \cos
   \left(2 \hbar \sqrt{|Q_{22}(t)|^2+(TR1(t)-TR2(t))^2}\right)-Q_{22}(t)\Bigg]}{4 \left(|Q_{22}(t)|^2+(TR1(t)-TR2(t))^2\right)}\nonumber \\
\end{eqnarray}

\begin{eqnarray}
\rho_{4,3}(t)=(TR1(t)-TR2(t))\frac{ \Bigg[-i \sqrt{|Q_{22}(t)|^2+(TR1(t)-TR2(t))^2} \sin \left(2 \hbar \sqrt{|Q_{22}(t)|^2+(TR1(t)-TR2(t))^2}\right)}{4 \left(|Q_{22}(t)|^2+(TR1(t)-TR2(t))^2\right)}+\nonumber \\ \frac{Q_{22}(t) \cos
   \left(2 \hbar \sqrt{|Q_{22}(t)|^2+(TR1(t)-TR2(t))^2}\right)-Q_{22}(t)\Bigg]}{4 \left(|Q_{22}(t)|^2+(TR1(t)-TR2(t))^2\right)}\nonumber \\
\end{eqnarray}

\begin{eqnarray}
\rho_{4,4}(t)=\frac{(TR1(t)-TR2(t))^2 \cos \left(2 \hbar \sqrt{|Q_{22}(t)|^2+(TR1(t)-TR2(t))^2}\right)+2|Q_{22}(t)|^2+(TR1(t)-TR2(t))^2}{4
   \left(|Q_{22}(t)|^2+(TR1(t)-TR2(t))^2\right)}\nonumber \\
\end{eqnarray}

It turns out that $\rho^n(t)=\rho(t)$ so one deals with a pure quantum state.
Now we are obtaining reduced matrices describing the state of particle B  from 2 particle density matrix.

\begin{eqnarray}
\rho_{B}(t)=
\begin{pmatrix}
\rho_{11}(t)+\rho_{22}(t) & \rho_{13}(t)+\rho_{24}(t) \\
\rho_{31}(t)+\rho_{42}(t) & \rho_{33}(t)+\rho_{44}(t) \\
\end{pmatrix}= \nonumber \\
\begin{pmatrix}
\frac{1}{2} & \frac{Q_{22}(t) (TR1(t)-TR2(t)) \sin ^2\left(\hbar \sqrt{|Q_{22}(t)|^2+(TR1(t)-TR2(t))^2}\right)}{|Q_{22}(t)|^2+(TR1(t)-TR2(t))^2} \\
\frac{ Q_{22}(t) (TR1(t)-TR2(t)) \sin ^2\left(\hbar \sqrt{|Q_{22}(t)|^2+(TR1(t)-TR2(t))^2}\right)}{|Q_{22}(t)|^2+(TR1(t)-TR2(t))^2} & \frac{1}{2} \\
\end{pmatrix}. \nonumber \\
\end{eqnarray}
Consequently we can compute entanglement entropy.
At first we evaluate
\begin{eqnarray}
Log(\rho_{B}(t))=
\begin{pmatrix}
a & b \\
c & d \\
\end{pmatrix},
\end{eqnarray}
\small
\begin{eqnarray*}
a=\frac{1}{2} \Bigg[\log \Bigg[ Q_{22}(t) (TR1(t)-TR2(t)) \cos \left(2 \hbar \sqrt{|Q_{22}(t)|^2+(TR1(t)-TR2(t))^2}\right)+\nonumber \\ +|Q_{22}(t)|^2+Q_{22}(t)
   (TR2(t)-TR1(t))+(TR1(t)-TR2(t))^2 \Bigg] \nonumber \\
   -2\log \Bigg[
   |Q_{22}(t)|^2+(TR1(t)-TR2(t))^2
   \Bigg] \nonumber \\
   +\log \Bigg[\Bigg[Q_{22}(t) (TR2(t)-TR1(t)) \cos \left(2 \hbar
   \sqrt{|Q_{22}(t)|^2+(TR1(t)-TR2(t))^2}\right)+ \nonumber \\
   |Q_{22}(t)|^2+Q_{22}(t)
   (TR1(t)-TR2(t))+(TR1(t)-TR2(t))^2\Bigg] 
   -\log (4)\Bigg]
\end{eqnarray*}
\normalsize
$b=-\tanh ^{-1}\left(\frac{Q_{22}(t)(TR1(t)-TR2(t)) \left(\cos \left(2 \hbar
   \sqrt{|Q_{22}(t)|^2+(TR1(t)-TR2(t))^2}\right)-1\right)}{|Q_{22}(t)|^2+(TR1(t)-TR2(t))^2}\right)=c$

\small
\begin{eqnarray}
d=\frac{1}{2}  \Bigg[\log\Bigg[ Q_{22}(t) (TR1(t)-TR2(t)) \cos \left(2 \hbar
\sqrt{| Q_{22}(t)|^2+(TR1(t)-TR2(t))^2}\right)+\nonumber \\ | Q_{22}(t)|^2+Q_{22}(t)
   (TR2(t)-TR1(t))+(TR1(t)-TR2(t))^2\Bigg] \nonumber \\  -2\log\Bigg[|Q_{22}(t)|^2+(TR1(t)-TR2(t))^2\Bigg]\Bigg] \nonumber \\
+\log \Bigg[] Q_{22}(t)(TR2(t)-TR1(t)) \cos \left(2 \hbar
   \sqrt{|Q_{22}(t)|^2+(TR1(t)-TR2(t))^2}\right)+ \nonumber \\ |Q_{22}(t)|^2+Q_{22}(t)
   (TR1(t)-TR2(t))+(TR1(t)-TR2(t))^2\Bigg] 
-\log (4)\Bigg]
\end{eqnarray}
\normalsize
and we obtain the formula when we start from $TR1(t_0)=TR2(t_0)$ as

\begin{eqnarray}
S_{B}(t)=Tr[\rho_{B}(t) Log[\rho_{B}(t)]]= \nonumber \\
=Tr \Bigg[
\begin{pmatrix}
\frac{1}{2} & \frac{Q_{22}(t) (TR1(t)-TR2(t)) \sin ^2\left(\hbar \sqrt{|Q_{22}(t)|^2+(TR1(t)-TR2(t))^2}\right)}{|Q_{22}(t)|^2+(TR1(t)-TR2(t))^2} \\
\frac{ Q_{22}(t) (TR1(t)-TR2(t)) \sin ^2\left(\hbar \sqrt{|Q_{22}(t)|^2+(TR1(t)-TR2(t))^2}\right)}{|Q_{22}(t)|^2+(TR1(t)-TR2(t))^2} & \frac{1}{2} \\
\end{pmatrix} \times \nonumber \\
Log \Bigg[
\begin{pmatrix}
\frac{1}{2} & \frac{Q_{22}(t) (TR1(t)-TR2(t)) \sin ^2\left(\hbar \sqrt{|Q_{22}(t)|^2+(TR1(t)-TR2(t))^2}\right)}{|Q_{22}(t)|^2+(TR1(t)-TR2(t))^2} \\
\frac{ Q_{22}(t) (TR1(t)-TR2(t)) \sin ^2\left(\hbar \sqrt{|Q_{22}(t)|^2+(TR1(t)-TR2(t))^2}\right)}{|Q_{22}(t)|^2+(TR1(t)-TR2(t))^2} & \frac{1}{2} \\
\end{pmatrix}
\Bigg] \Bigg]=\nonumber \\
=-\log (4)\frac{1}{2}+ 
\frac{1}{2} \Bigg[\log \Bigg[Q_{22}(t) (TR1(t)-TR2(t)) \cos \left(2 \hbar \sqrt{|Q_{22}(t)|^2+(TR1(t)-TR2(t))^2}\right)+\nonumber \\ +|Q_{22}(t)|^2+Q_{22}(t)
   (TR2(t)-TR1(t))+(TR1(t)-TR2(t))^2\Bigg]+
   \nonumber \\ +\log \Bigg[Q_{22}(t)(TR2(t)-TR1(t)) \cos \left(2 \hbar
   \sqrt{|Q_{22}(t)|^2+(TR1(t)-TR2(t))^2}\right) \nonumber \\ +|Q_{22}(t)|^2+Q_{22}(t)
   (TR1(t)-TR2(t))+(TR1(t)-TR2(t))^2\Bigg]\nonumber \\ -2\log\Bigg[
   |Q_{22}(t)|^2+(TR1(t)-TR2(t))^2
   \Bigg] \nonumber \\ +\frac{4 Q_{22}(t) (TR2(t)-TR1(t)) \sin ^2\left(\hbar
   \sqrt{|Q_{22}(t)|^2+(TR1(t)-TR2(t))^2}\right) }{|Q_{22}(t)|^2+(TR1(t)-TR2(t))^2} \times \nonumber \\
\times   \tanh ^{-1}\left(\frac{Q_{22}(t) (TR1(t)-TR2(t)) \left(\cos \left(2 \hbar
   \sqrt{|Q_{22}(t)|^2+(TR1(t)-TR2(t))^2}\right)-1\right)}{|Q_{22}(t)|^2+(TR1(t)-TR2(t))^2}\right)\Bigg]
   \nonumber \\
\end{eqnarray}
\normalsize

The results obtained allows for monitoring of entanglement entropy with time.

\begin{figure}
\centering
\includegraphics[scale=0.95]{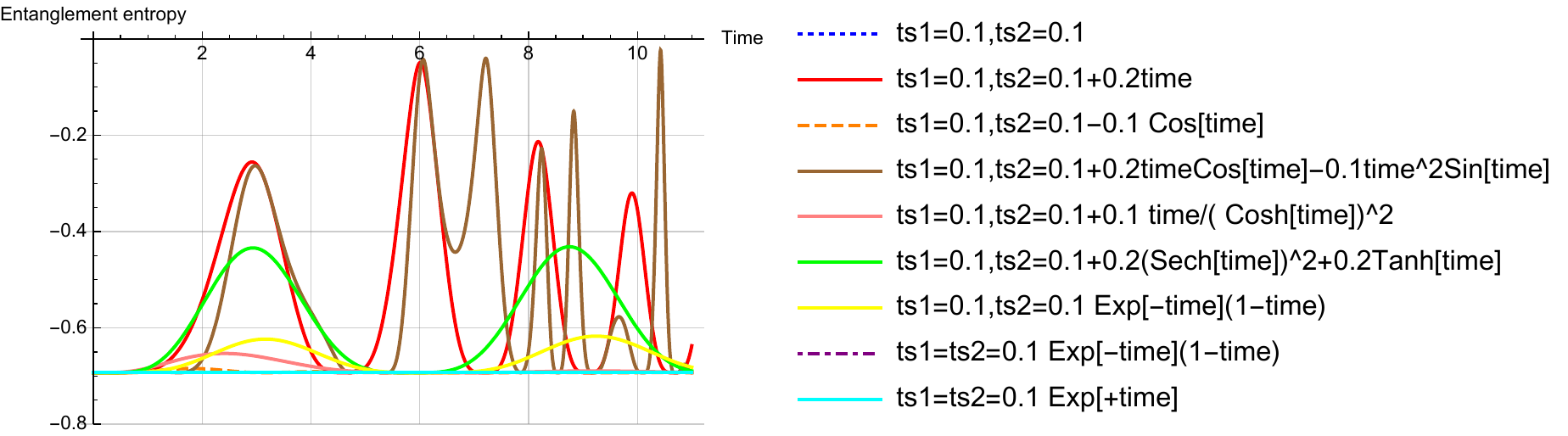} 
\caption{Entanglement entropy with time for 2 interacting particles for different functions of hopping constant with time.}
\end{figure}

\normalsize
\section{Case of 2 coupled Single Electron Lines}
\begin{figure}[h]
\centering
\includegraphics[width=0.5\linewidth]{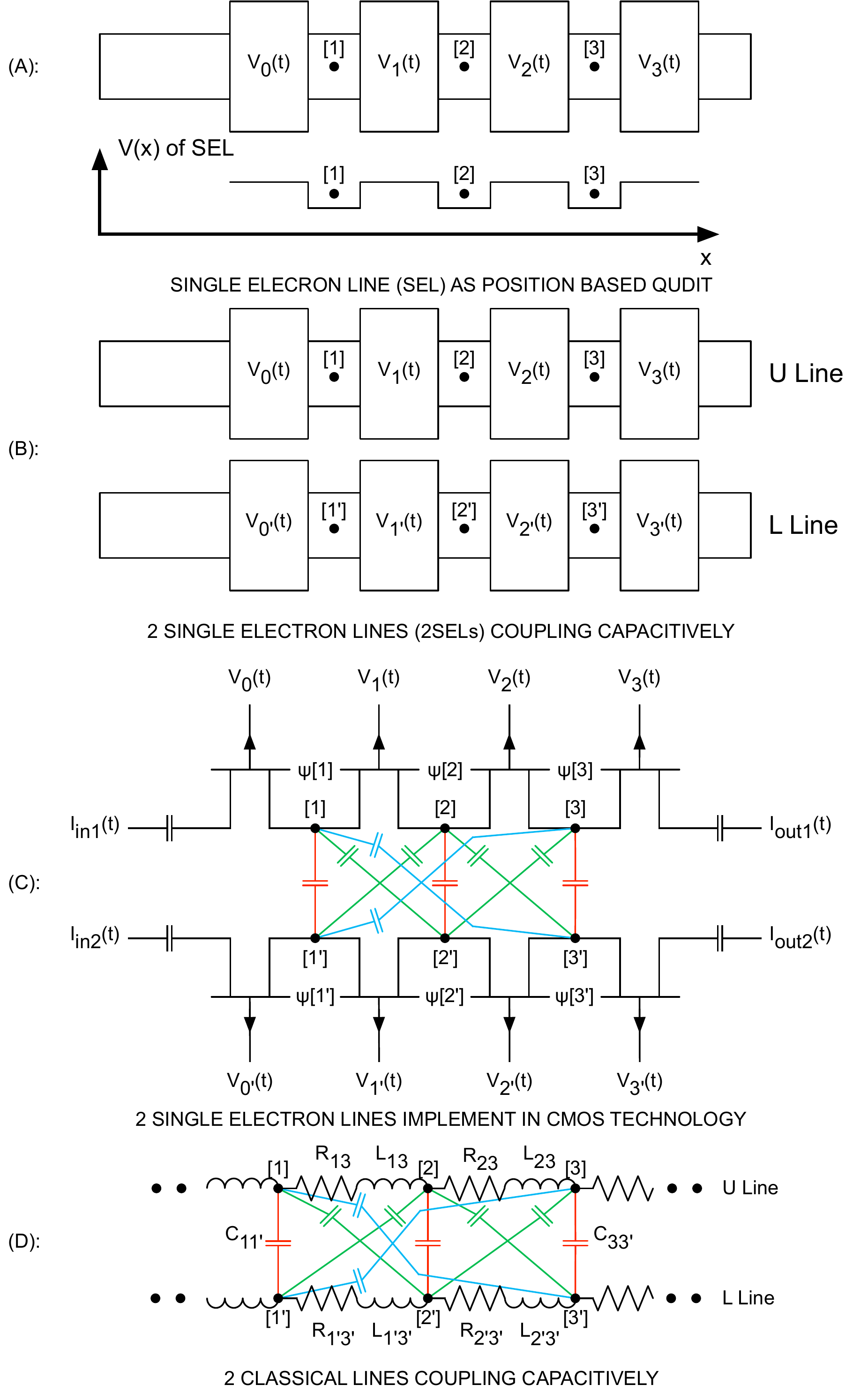}
\caption{Nanometer CMOS structure \cite{Pomorski_spie}, effective potential and circuit representation of: (A) electrostatic position-dependent qubit \cite{Pomorski_spie} (the quantum dot dimensions are 80$\times$80\,nm$^2$ in 22FDX FDSOI CMOS technology); (B)--(C) two electrostatic position-dependent qubits representing two inductively interacting lines (upper "U" and lower "L" quantum systems) in minimalistic way (more rigorously they shall be named as MOS transistor single-electron lines). Presented systems are subjected to the external voltage biasing that controls the local potential landscape in which electrons are confined.  Classical limit is expressed by circuit D.}
\label{PositionDependentQubit}
\end{figure}

We follow the reasoning described in \cite{SEL}.
At first, we consider a physical system of an electron confined in a potential with two minima (position-dependent
qubit with presence of electron at node 1 and 2) or three
minima (position dependent qubit with presence of electron
at nodes 1, 2 and 3), as depicted in Fig.\,\ref{PositionDependentQubit}(A),
which was also considered by Fujisawa \cite{Fujisawa} and Petta \cite{Petta} and which forms a position-dependent qubit (or qudit). We can write the Hamiltonian in the second quantization as
\begin{equation}
\hat{H}=\sum_{i,j}t_{i \rightarrow j}\hat{a}^{\dag}_{i}\hat{a}_{j}+\sum_{i} E_{p}(i)\hat{a}^{\dag}_{i}\hat{a}_{i}+\sum_{i,j,k,l}\hat{a}^{\dag}_{i}\hat{a}^{\dag}_{j}\hat{a}_{i}\hat{a}_{j}V_{i,j},
\end{equation}
where $\hat{a}^{\dag}_{i}$ is a fermionic creator operator at $i$-th point in the space lattice and $\hat{a}_{j}$ is fermionic annihilator operator at $j$-th point of the lattice. The hopping term $t_{i \rightarrow j}$ describes hopping from $i$-th to $j$-th lattice point and is a measure of kinetic energy. The potential $V_{i,j}$ represents particle-particle interaction and term $E_{p}(i)$ incorporates potential energy. In this approach we neglect the presence of a spin.
It is convenient to write a system Hamiltonian of position based qubit in spectral form as
\begin{eqnarray}
\label{eq:1}
\hat{H}(t) = E_{p1}(t) \ket{1,0} \bra{1,0} +E_{p2}(t)\ket{0,1} \bra{0,1} + \nonumber \\
t_{1 \rightarrow 2}(t) \ket{0,1} \bra{1,0}+t_{2 \rightarrow 1}(t) \ket{1,0} \bra{0,1}= \nonumber \\
=\frac{1}{2}(\hat{\sigma}_0+\hat{\sigma}_3)E_{p1}(t)+\frac{1}{2}(\hat{\sigma}_0-\hat{\sigma}_3)E_{p2}(t)+ \nonumber \\
\frac{1}{2}(\hat{\sigma}_1-i\hat{\sigma}_2)t_{2 \rightarrow 1}(t)+\frac{1}{2}(i\hat{\sigma}_2-\hat{\sigma}_1)t_{1 \rightarrow 2}(t) 
\end{eqnarray}
where Pauli matrices are $\hat{\sigma}_0, .. , \hat{\sigma}_3$ while system quantum state is given as $\ket{\psi(t)}=\alpha(t) \ket{1,0} + \beta(t) \ket{0,1}$ with $|\alpha|^2 + |\beta|^2 =1$ and is expressed in Wannier function eigenbases $\ket{1,0}=w_L(x)$ and $\ket{0,1}=w_R(x)$ which underlines the presence of electron on the left/right side as equivalent to picture from Schr\"odinger equation \cite{Pomorski_spie}. We obtain two energy eigenstates 
\begin{eqnarray*}
\ket{E_{1(2)}}=
\begin{pmatrix}
\frac{ (E_{p2}-E_{p1}) \pm \sqrt{4t_{1 \rightarrow 2}t_{2 \rightarrow1} + |E_{p1}-E_{p2}|^2 } }{2t_{1 \rightarrow 2}} \\
1
\end{pmatrix}=\nonumber \\
\frac{ (E_{p2}-E_{p1}) \pm \sqrt{4t_{1 \rightarrow 2}t_{2 \rightarrow1} + |E_{p1}-E_{p2}|^2 } }{2t_{1 \rightarrow 2}}\ket{1,0}+\ket{0,1}.
\end{eqnarray*}
and energy eigenvalues 
%
\begin{eqnarray}
E_{1(2)} = \frac{1}{2}(E_{p1} + E_{p2} \pm \sqrt{4t_{1 \rightarrow 2}t_{2 \rightarrow1} + |E_{p1}-E_{p2}|^2 })= \nonumber \\
\frac{1}{2}(E_{p1} + E_{p2} \pm 2|t_{1 \rightarrow 2}|\sqrt{1 + |\frac{E_{p1}-E_{p2}}{2t_{1 \rightarrow 2}t_{2 \rightarrow 1}}|^2 }) \approx \nonumber \\
\frac{1}{2}(E_{p1} + E_{p2} \pm 2|t_{1 \rightarrow 2}|(1 +\frac{1}{2} |\frac{E_{p1}-E_{p2}}{2t_{1 \rightarrow 2}t_{2 \rightarrow 1}}|^2 )) \approx \nonumber \\  \frac{1}{2}(E_{p1} + E_{p2}) \pm |t_{1 \rightarrow 2}|.
\end{eqnarray}
The last approximation is obtained in the limit of $t_{1 \rightarrow 2} \gg E_{p1}, E_{p2}$ (classical limit when system energy becomes big and $|t|$ has the interpretation of kinetic energy) what is the case depicted in the middle Fig.3 when $|t| \rightarrow +\infty$.
Since Schroedinger formalism can be also applied to the position based qubit that has discrete eigenenergy spectra, one expects that value $E_p$ and $t_s$ takes discrete values. It is even more pronounced when one is using formula being prescription for $E_p$ and $t_s$ parameters as
\begin{equation}
    E_p(i)=\int_{-\infty}^{+\infty}dx\psi_i^{*}(x)\hat{H}_0\psi_i(x),
\end{equation}
where $\psi(x)_i$ is wavefunction of electron localized at i-th node (i-th quantum well) and $\hat{H}$ is effective Hamiltonian. In similar fashion we can define hopping constant from node i-th to node j-th as energy participating in energy transport from one quantum well into the neighbouring quantum well so we define
\begin{equation}
    t_{s, i \rightarrow j}=\int_{-\infty}^{+\infty}dx\psi_i^{*}(x)\hat{H}_0\psi_j(x),
\end{equation}
Another interesting fact is the transition from Schroedinger picture to the tight-binding picture that can be done by $\ket{\psi}=\int_{-\infty}^{\infty}\psi(x)dx \ket{x} \approx \sum_{k=-\infty}^{k=+\infty}\Delta x \psi(k) \ket{k \Delta x}$, where $\Delta x$ is the distance between nodes. Having momentum operator defined as $\frac{\hbar}{\Delta x \sqrt{-1}}(-\ket{k+1}\bra{k}+\ket{k}\bra{k+1})=\frac{\hbar}{\Delta x \sqrt{-1}}\frac{d}{dx}_k$. We obtain the second derivative by Euler formula $(\frac{d^2}{dx^2})_k=\frac{1}{(\Delta x)^2}(\ket{k+1}\bra{k}+\ket{k}\bra{k+1}-2\ket{k}\bra{k})$. Now we can recover the Schroedinger equation and we observe that $t_{s,i \rightarrow i+1}=\frac{\hbar^2}{2m \Delta x}w$, where w is positive and integer. Therefore $t_{s,i \rightarrow i+1}$ has the positive discrete values. We also observe that the potential in the Schroedinger equation can be connected with $E_p(i)-2t_{s,i \rightarrow i+1}=V_p(i)$ at i-th node. Since kinetic energy is discrete and potential energy in Schroedigner equation is continuous one obtains discrete $E_p$.
The eigenstate depends in the tight binding model depends on an external vector potential source acting on the qubit by means of $t_{1 \rightarrow 2}=|t_{1 \rightarrow 2}|e^{i \alpha}=t_{2 \rightarrow 1}^{*}$. Since every energy eigenstate is spanned by $\ket{0,1}$ and $\ket{1,0}$, we will obtain oscillations of occupancy between two wells \cite{SEL},\cite{Panos},\cite{Pomorski_spie}. It is worth-mentioning that the act of measurement will affect the qubit quantum state. Since we are dealing with a position-based qubit, we can make measurement of the electron position with the use an external single-electron device (SED) in close proximity to the qubit. This will require the use of projection operators that represent eigenenergy measurement as $\ket{E_{0(1)}}\bra{E_{0(1)}}$ or, for example, measurement of the electron position at left side so we use the projector $\ket{1,0}\bra{0,1}$.
We can extend the model for the case of three (and more) coupled wells. In such a case, we obtain the system Hamiltonian for a position based qubit: 
\begin{eqnarray}
  \label{simple_equation1}
  \hat{H} = \sum_{s}E_{ps}\ket{\textbf{s}}\bra{\textbf{s}}+\sum_{l,s,s \neq l}t_{s \rightarrow l}\ket{\textbf{l}}\bra{\textbf{s}},
  \end{eqnarray}
where $\ket{\textbf{1}}=\ket{1,0,0}, \ket{\textbf{2}}=\ket{0,1,0}, \ket{\textbf{3}}=\ket{0,0,1} $
%
and its Hamiltonian matrix

\begin{eqnarray}
H(t)=
\begin{pmatrix}
  E_{p1}(t) & t_{2 \rightarrow 1}(t) & t_{3 \rightarrow 1}(t) \\
  t_{1 \rightarrow 2}(t) & E_{p2}(t) & t_{3 \rightarrow 2}(t) \\
  t_{1 \rightarrow 3}(t) & t_{2 \rightarrow 3}(t) & E_{p3}(t)
\end{pmatrix}
\end{eqnarray}

and quantum state $\ket{\psi}$ (with a normalization condition $|\alpha|^2+|\beta|^2+|\gamma|^2=1$) is given as
\begin{eqnarray}
\ket{\psi} =
\begin{pmatrix}
  \alpha(t) \\
  \beta(t) \\
  \gamma(t)
\end{pmatrix}
= \alpha(t) \ket{1,0,0} + \beta(t) \ket{0,1,0} + \gamma(t) \ket{0,0,1}. \nonumber \\
\end{eqnarray}
Coefficients $\alpha(t)$, $\beta(t)$ and $\gamma(t)$ describe oscillations of occupancy of one electron at wells 1, 2 and 3.
The problem of qubit equations of motion can be formulated by having
$\ket{\psi} = c_1(0)e^{-\frac{i}{\hbar}t E_1}\ket{E_1}+c_2(0)e^{-\frac{i}{\hbar}t E_2}\ket{E_2}+c_3(0)e^{-\frac{i}{\hbar}t E_3}\ket{E_3}$, where $|c_1(0)|^2$,$|c_2(0)|^2$ and $|c_3(0)|^2$ are probabilities of occupancy of $E_1$, $E_2$ and $E_3$ energetic levels. Energy levels are roots of 3rd order polynomial
\begin{eqnarray*}
(-E_{p1} E_{p2} E_{p3} + E_{p3} t_{12}^2 + E_{p1} t_{23}^2+E_{p2} t_{13}^2-2t_{s12}t_{s13}t_{s23})\nonumber \\ + (E_{p1} E_{p2} + E_{p1} E_{p3}  + E_{p2} E_{p3} - t_{12}^2 -
t_{23}^2-t_{13}^2)E  \nonumber \\ -(E_{p1}+E_{p2}+E_{p3}) E^2 + E^3=0,
\end{eqnarray*}
where $\ket{E_1},\ket{E_2},\ket{E_3}$ are 3-dimensional Hamiltonian eigenvectors.

By introducing two electrostatically interacting qudits, we are dealing with the Hamiltonian of the upper and lower lines as well as with their Coulomb electrostatic interactions. We are obtaining the Hamiltonian in spectral representation acting on the product of Hilbert spaces in the form of $\hat{H} = \hat{H}_U \times I_{L}+I_{U} \times \hat{H}_L+\hat{H}_{U-L}$
%
%
where $H_u$ and $H_l$ are Hamiltonians of separated upper and lower qudits, $H_{l-u}$ is a two-line Coulomb interaction and $I_{u(l)}=\ket{1,0,0}_{u(l)}\bra{1,0,0}_{u(l)}+\ket{0,1,0}_{u(l)}\bra{0,1,0}_{u(l)}+\ket{0,0,1}_{u(l)}\bra{0,0,1}_{u(l)}$. The electrostatic interaction is encoded in $\textcolor{black}{E_c(1,1')=E_c(2,2')=E_c(3,3')=\frac{e^2}{4\pi \epsilon_0 \epsilon d}=q_1}$ (red capacitors of Fig.1) and \textcolor{black}{$q_2=E_c(2,1')=E_c(2,3')=E_c(1,2')=E_c(3,2')=\frac{e^2}{4\pi \epsilon_0 \epsilon \sqrt{d^2+(a+b)^2}}$}
and electrostatic energy of green capacitors of Fig.1. is
\begin{equation}
\textcolor{black}{E_c(1,3')=E_c(3,1')=q_2=\frac{e^2}{4\pi \epsilon_0 \epsilon \sqrt{d^2+4(a+b)^2}}},
\end{equation}
where $a$, $b$ and $d$ are geometric parameters of the system, e is electron charge and $\epsilon$ is a relative dielectric constant of the material; $\epsilon_0$ corresponds to the dielectric constant of vacuum. The very last Hamiltonian corresponds to the following quantum state $\ket{\psi(t)}$ ($|\gamma_1(t)|^2+..|\gamma_9(t)|^2=1$) given as
\begin{eqnarray}
\label{stateSEL}
\ket{\psi(t)}=\gamma_1(t)\ket{1,0,0}_u\ket{1,0,0}_l+\gamma_2(t)\ket{1,0,0}_u\ket{0,1,0}_l \nonumber \\
+\gamma_3(t)\ket{1,0,0}_u\ket{0,0,1}_l +\gamma_4(t)\ket{0,1,0}_u\ket{1,0,0}_l \nonumber \\
+\gamma_5(t)\ket{0,1,0}_u\ket{0,1,0}_l + \gamma_6(t)\ket{0,1,0}_u\ket{0,0,1}_l \nonumber \\
+ \gamma_7(t)\ket{0,0,1}_u\ket{0,0,1}_l +\gamma_8(t)\ket{0,0,1}_u\ket{0,1,0}_l \nonumber \\
+\gamma_9(t)\ket{0,0,1}_u\ket{0,0,1}_l, \nonumber \\
\end{eqnarray}
where $|\gamma_1(t)|^2$ is the probability of finding two electrons at nodes 1 and 1' at time $t$ (since $\gamma_1$ spans $\ket{1,0,0}_u\ket{1,0,0}_l$), etc. The Hamiltonian has nine eigenenergy solutions that are parametrized by geometric factors and hopping constants $t_{k,m}$ as well as energies $E_p(k)$ for the case of `u' or 'l' system. Formally, we can treat $E_{p}(k)=t_{k \rightarrow k} \equiv t_{k,k} \equiv t_k \in \textbf{R}$ as a hopping from $k$-th lattice point to the same lattice point $k$. We obtain the following Hamiltonian
\vspace{-3mm}
\begin{eqnarray}
\hat{H}=
\begin{pmatrix}
\xi_{1,1'} & \textcolor{red}{t_{ 1' \rightarrow 2' }} & \textcolor{orange}{t_{1' \rightarrow 3'}} & \textcolor{green}{t_{1 \rightarrow 2}} & 0 & 0 & \textcolor{blue}{t_{1 \rightarrow 3}} & 0 & 0 \\ 
\textcolor{red}{t_{ 2' \rightarrow 1' }} & \xi_{1,2'} & \textcolor{gray}{t_{2' \rightarrow 3'}} & 0 & \textcolor{green}{t_{1 \rightarrow 2}} & 0 & 0 & \textcolor{blue}{t_{1 \rightarrow 3}} & 0 \\ 
\textcolor{orange}{t_{3' \rightarrow 1'}} & \textcolor{gray}{t_{3' \rightarrow 2'}} & \xi_{1,3'} & 0 & 0 & \textcolor{green}{t_{1 \rightarrow 2}} & 0 & 0 & \textcolor{blue}{t_{1 \rightarrow 3}} \\
\textcolor{green}{t_{2 \rightarrow 1}} & 0 & 0 & \xi_{2,1'} & \textcolor{red}{ t_{1' \rightarrow 2'} } & \textcolor{orange}{t_{1' \rightarrow 3'}} & \textcolor{yellow}{t_{2 \rightarrow 3}} & 0 & 0 \\
0 & \textcolor{green}{t_{2 \rightarrow 1}} & 0 & \textcolor{red}{t_{2' \rightarrow 1'}} & \xi_{2,2'} & \textcolor{gray}{t_{2' \rightarrow 3'}} & 0 & \textcolor{yellow}{t_{2 \rightarrow 3}} & 0 \\
0 & 0 & \textcolor{green}{t_{2 \rightarrow 1}} & \textcolor{orange}{t_{3' \rightarrow 1'}} & \textcolor{gray}{t_{3' \rightarrow 2'}} & \xi_{2,3'} & 0 & 0 & \textcolor{yellow}{t_{2 \rightarrow 3}} \\
\textcolor{blue}{t_{3 \rightarrow 1}} & 0 & 0 & \textcolor{yellow}{t_{3 \rightarrow 2}} & 0 & 0 & \xi_{3,1'} & \textcolor{red}{t_{1' \rightarrow 2'}} & \textcolor{orange}{t_{1' \rightarrow 3'}} \\
0 & \textcolor{blue}{t_{3 \rightarrow 1}} & 0 & 0 & \textcolor{yellow}{t_{3 \rightarrow 2}} & 0 & \textcolor{red}{t_{2' \rightarrow 1'}} & \xi_{3,2'} & \textcolor{gray}{t_{2' \rightarrow 3'}} \\
0 & 0 & \textcolor{blue}{t_{3 \rightarrow 1}} & 0 & 0 & \textcolor{yellow}{t_{3 \rightarrow 2}} & \textcolor{orange}{t_{3' \rightarrow 1'}} & \textcolor{gray}{t_{3' \rightarrow 2'}} & \xi_{3,3'} \\
\end{pmatrix}
=
\begin{pmatrix}
H(1)_{1',3'} & \textcolor{green}{H_{1,2}} & \textcolor{blue}{H_{1,3}} \\
\textcolor{green}{H(1)_{2,1}} & H(2)_{1',3'} & \textcolor{yellow}{H_{2,3}} \\
\textcolor{blue}{H_{3,1}} & \textcolor{yellow}{H_{3,2}} & H(3)_{1,3'} \\
\end{pmatrix}
\end{eqnarray}
\vspace{-4mm}
\normalsize

\noindent with diagonal elements $( [\xi_{1,1'}, \xi_{1,2'}, \xi_{1,3'} ]$ , $ [\xi_{2,1'}, \xi_{2,2'}, \xi_{2,3'}],$ $ [\xi_{3,1'}, \xi_{3,2'}, \xi_{3,3'} ])$ set to
$([(E_{p1}+E_{p1'}+E_c(1,1'))$, $ (E_{p1}+E_{p2'}+E_c(1,2'))$ , $ (E_{p1}+E_{p3'} + E_c(1,3'))]$, $[((E_{p1}+E_{p1'}+E_c(1,1'))$, $ (E_{p2}+E_{p2'}+E_c(2,2'))$ , $ (E_{p2}+E_{p3'} + E_c(2,3'))]$, $[((E_{p3}+E_{p1'}+E_c(3,1')$, $ (E_{p3}+E_{p2'}+E_c(3,2'))$, $ (E_{p3}+E_{p3'} + E_c(3,3'))])$. In the absence of magnetic field, we have $t_{k \rightarrow m}=t_{m \rightarrow k}=t_{k,l}=t_{m,k} \in \textbf{R}$ and in the case of nonzero magnetic field $t_{k,m}=t_{m,k}^{*}\in \textbf{C}$.
It is straightforward to determine the matrix of two lines with $N$ wells [=3 in this work] each following the mathematical structure of two interacting lines with three wells in each line. Matrices $H_{1,2},H_{2,3},H_{1,3}$ are diagonal of size $N \times N$ with all the same terms on the diagonal. At the same time, matrices $H(1)_{1',N'}$,..,$H(N)_{1',N'} $ have only different diagonal terms corresponding to $((\xi_{1,N'}, .. , \xi_{1,N'} )$, .., $((\xi_{N,N'}, .. , \xi_{N,N'} )$ elements. In simplified considerations we can set $t_{1 \rightarrow N}=t_{N \rightarrow 1}$ and $t_{1' \rightarrow N'}=t_{N' \rightarrow 1'}$ to zero since a probability for the wavefunction transfer from 1st to $N$-th lattice point is generally proportional to $\approx \exp(-s N),$ where $s$ is some constant. It shall be underlined that in the most general case of two capacitevly coupled symmetric SELs with three wells each (being parallel to each other), we have six (all different $E_{p}(k)$ and $E_{p}(l')$) plus six (all different $t_{k \rightarrow s}$, $t_{k' \rightarrow s'}$) plus three geometric parameters ($d$, $a$ and $b$) as well as a dielectric constant hidden in the effective charge of interacting electrons $q$. Therefore, the model Hamiltonian has 12+4 real-valued parameters (4 depends on the material and geomtry of 2 SELs). They can be extracted from a particular transistor implementation of two SELs (Fig.\,\ref{PositionDependentQubit}C). There are two main physically important regimes when $t$ $\ll$ $E_p$ and when $t$ $\gg$ $E_p$. They correspond to the case of electron tunneling from one quantum well into another (electron is not in highly excited state ) and the case when electron wavepacket can move freely between neighbouring wells (electron is in highly excited state).

\section{Analytical and Numerical Modeling of Capacitively Coupled SELs}

\subsection{Analytical Results}

The greatest simplification of matrix (8) is when we set all $t_{k' \rightarrow m'}=t_{o \rightarrow m}$ = $|t|$, and all $E_{p}(k)=E_{p}(m')=E_p$ for $N$=3. Let us first consider the case of two insulating lines (all wells on each line are completely decoupled so there is no electron tunneling between the barriers and the barrier energies are high) where there are trapped electrons so $|t|=0$ (electrons are confined in quantum wells and cannot move towards neighbouring wells). In such a case, we deal with a diagonal matrix that has three different eigenvalues on its diagonal and has three different eigenenergy values
\begin{eqnarray}
\label{InsulatorEnergy}
\hat{E}=
\left\{
  \begin{array}{lr}
E_1= q_1 = E_p + \frac{e^2}{4 \pi \epsilon \epsilon_0 d}, \\
E_2= q_2 =E_p + \frac{e^2}{4 \pi \epsilon \epsilon_0 \sqrt{|d|^2+(a+b)^2}}, \\
E_3= q_3 =E_p + \frac{e^2}{4 \pi \epsilon \epsilon_0 \sqrt{|d|^2+4(a+b)^2}},
  \end{array}
\right.
\end{eqnarray}

so $E_3<E_2<E_1$. In the limit of infinite distance between SELs, we have nine degenerate eigenergies. They are set to $E_{pk}$ which corresponds to six decoupled quantum systems (the first electron is delocalized into three upper wells, while the second electron is delocalized into three lowers wells).

Let us also consider the case of ideal metal where electrons are completely delocalized. In such a case, all $t_{k(k')} \gg E_{pl(s)}$ which brings Hamiltonian diagonal terms to be negligible in comparison with other terms. In such a case, we can set all diagonal terms to be zero which is an equivalent to the case of infinitely spaced SELs lines. It simply means that in the case of ideal metals, two lines are not `seeing' each other.

Let us now turn to the case where processes associated with hopping between wells have similar values of energy to the energies denoted as $E_{pk(l')}$. In such a case, the Hamiltonian matrix can be parametrized only by three real value numbers due to symmetries depicted in Fig.\,\ref{PositionDependentQubit}B (we divide the matrix by a constant number $|t|$) so
\begin{eqnarray}
\left\{
  \begin{array}{lr}
q_{1_1}=\frac{2E_p+\frac{e^2}{d}}{|t|}, \nonumber \\
q_{1_2}=\frac{2E_p+\frac{e^2}{\sqrt{d^2+(a+b)^2}}}{|t|}, \nonumber \\
q_{1_3}=\frac{2E_p+\frac{e^2}{\sqrt{d^2+4(a+b)^2}}}{|t|}. \nonumber
  \end{array}
\right.
\end{eqnarray}
For a fixed $|t|$, we change the distance $d$ and observe that $q_{1_1}$ can be arbitrary large, while $q_{1_2}$ and $q_{1_3}$ have finite values for $d$=0. Going into the limit of infinite distance $d$, we observe that all $q_{1_1}$, $q_{1_2}$ and $q_{1_3}$ approach a finite value $\frac{2E_p}{|t|}$. We obtain the simplified Hamiltonian matrix that is a Hermitian conjugate and has a property $H_{k,k}=H_{N-k+1,N-k+1}$. It is in the form
\begin{equation}
\label{Matrix}
\hat{H}=
\begin{pmatrix}
q_{1_1} & 1 & 0 & 1 & 0 & 0 & 0 & 0 & 0 \\
1 & q_{1_2} & 1 & 0 & 1 & 0 & 0 & 0 & 0 \\
0 & 1 & q_{1_3} & 0 & 0 & 1 & 0 & 0 & 0 \\
1 & 0 & 0 & q_{1_2} & 1 & 0 & 1 & 0 & 0 \\
0 & 1 & 0 & 1 & q_{1_1} & 1 & 0 & 1 & 0 \\
0 & 0 & 1 & 0 & 1 & q_{1_2} & 0 & 0 & 1 \\
0 & 0 & 0 & 1 & 0 & 0 & q_{1_3} & 1 & 0 \\
0 & 0 & 0 & 0 & 1 & 0 & 1 & q_{1_2} & 1 \\
0 & 0 & 0 & 0 & 0 & 1 & 0 & 1 & q_{1_1} \\
\end{pmatrix}
\end{equation}

We can analytically find nine energy eigenvalues and they correspond to the entangled states. We have
\begin{equation}
\label{eigenenergies}
\left\{
  \begin{array}{lr}
E_1=q_{1_1}, \\
E_2=q_{1_2}, \\
E_3=\frac{1}{2}( q_{1_1}+q_{1_2} -\sqrt{8+(q_{1_1}-q_{1_2})^2} ), \\
E_4=\frac{1}{2}( q_{1_1}+q_{1_2} + \sqrt{8+(q_{1_1}-q_{1_2})^2} ), \\
E_5=\frac{1}{2}( q_{1_2}-q_{1_3} - \sqrt{8+(q_{1_2}-q_{1_3})^2} ), \\
E_6=\frac{1}{2}( q_{1_2}-q_{1_3} + \sqrt{8+(q_{1_2}-q_{1_3})^2} ).
  \end{array}
\right.
\end{equation}
%
%
The last 3 energy eigenvalues are the most involving analytically and are the roots of a 3rd order polynomial
\begin{eqnarray}
     (2q_{1_1}+6q_{1_3}-q_{1_1} q_{1_2} q_{1_3}) + 
(-8+q_{1_1} q_{1_2}+ \nonumber \\ q_{1_1} q_{1_3}+q_{1_2} q_{1_3})E_k
      -(q_{1_1}+q_{1_2}+q_{1_3})E_k^2+E_k^3=0. \nonumber \\
\end{eqnarray}
We omit writing direct and very lengthy formulas since the solutions of a 3rd-order polynomial are commonly known. The eigenvectors have the structure given in Appendix~\ref{app:a}.

We can readily recognize that all nine energy eigenvectors are entangled. In particular first two eigenenergy states (given also in formula \ref{ent1}) are linear combination of position dependent states,
\begin{eqnarray}
\ket{E_1}=\ket{1,0,0}_U\ket{1,0,0}_L -\ket{0,1,0}_U\ket{0,1,0}_L+ \nonumber \\
\ket{0,0,1}_U\ket{0,0,1}_L, \nonumber \\
 \ket{E_2}=\ket{1,0,0}_U\ket{0,1,0}_L-\ket{0,1,0}_U\ket{1,0,0}_L \nonumber \\ - \ket{0,1,0}_U\ket{0,0,1}_L +
\ket{0,0,1}_U\ket{0,1,0}_L,
\end{eqnarray}
 so they have no equivalence in the classical picture of two charged balls in channels that are repelling each other.

\begin{figure}[htb]
\includegraphics[width=0.3\linewidth,height=2.5cm]{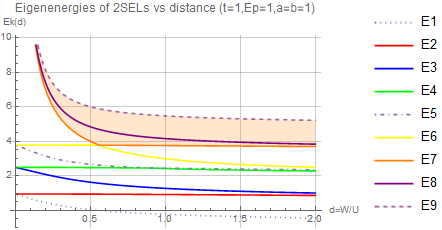}\\
\includegraphics[width=0.3\linewidth,height=2.5cm]{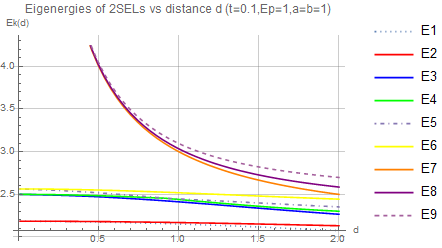}\\
\includegraphics[width=0.3\linewidth,height=2.5cm]{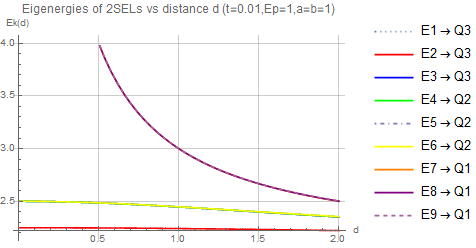}
\caption{Cases of: (a) metal $(t=1,E_p=1)$; (b) semiconductor $(t=0.1,E_p=1)$; and (c) insulator $(t=0.01,E_p=1)$ state of 2-SELs given by eigenenergy spectra as function of distance $d$ between two lines ($a=b=1,e=1$).} 
\label{MottTransition0}
\end{figure}

\begin{figure}[htb]
\includegraphics*[width=0.3\columnwidth]{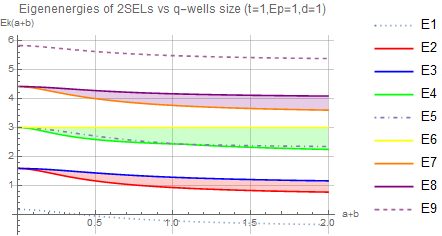}\\
\includegraphics*[width=0.3\columnwidth,height=2.5cm]{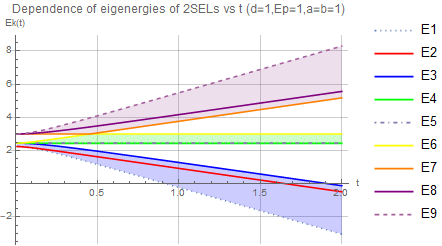}\\
\includegraphics*[width=0.3\columnwidth,height=2.5cm]{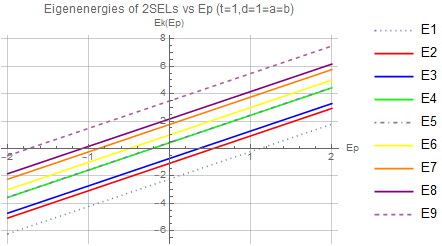}
\caption{Dependence of eigenenergy spectra vs. (a) quantum well size $a+b$, (b) hopping term $|t|$, and (c) chemical potential $E_p$ parameter.}
\label{MottTransition1}
\end{figure}

\subsection{Numerical Results for Case of Capacitively Coupled SETs}

At first, we are analyzing available spectrum of eigenenergies as in the case of insulator-to-metal phase transition \cite{Spalek}, which can be implemented in a tight-binding model by a systematic increase of the hopping term from small to large values, while at the same time keeping all other parameters constant, as depicted in Fig.\,\ref{MottTransition1}. Described tight-binding model can minimic a metal ($t$=1), semiconductor ($t$=0.1) or insulator state ($t$=0.01), as given in Fig.\ref{MottTransition0}. We can recognized 2-SELs eigenergy spectra dependence on distance between the two lines. Characteristic narrowing of bands is observed when one moves from large towards small distance $d$ between SELs (what can be related to the ratio of $W/U$ in the Hubbard model) and it is one of the signs of transition from metallic to insulator regime (Mott-insulator phase transition \cite{Spalek}). One of the plots referring to $t=0.01$ describes Anderson localization of electrons and, in such a case, energy eigenspectra are determined by formula (\ref{InsulatorEnergy}) and hopping terms $t$ can be completely neglected since electrons are localized in the quantum-well potential minima. 

\begin{figure} 
\centering
\includegraphics[width=0.25\linewidth]{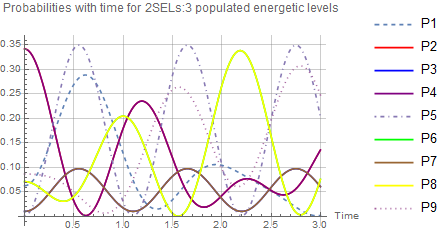}
\includegraphics[width=0.25\linewidth]{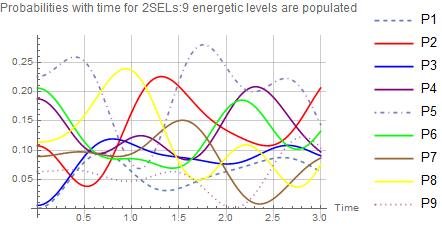}
\includegraphics[width=0.25\linewidth]{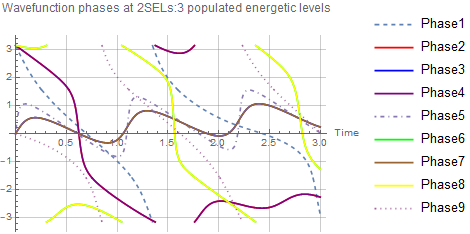}
\includegraphics[width=0.25\linewidth]{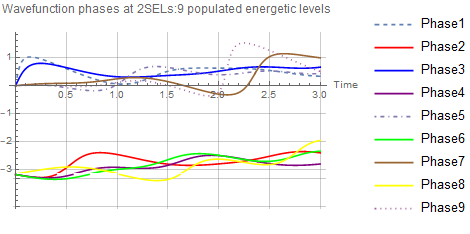}
\vspace{-4mm}
\caption{Quantum state of two SELs over time: Upper (Lower) plots populate 3 (9) energy levels as given by Scenario I (Scenario II). The probabilities of finding both electrons simultaneously at the input $p_1(t)=|\gamma_1(t)|^2$ and output $p_9(t)=|\gamma_1(t)|^2$ is shown with time as well as evolution of phases $\phi_1(t),..,\phi_9(t)$ of $\gamma_1(t)=|\gamma_1(t)|e^{\phi_1(t)}$, .., $\gamma_9(t)=|\gamma_9(t)|e^{\phi_9(t)}$ corresponding to equation (\ref{stateSEL}).}
\label{OccupancyOscillations}
\end{figure}

Bottom plots of Fig.\,\ref{MottTransition1}. describe the ability of tunneling eigenenergy spectra with respect to quantum well lengths ($a+b$), $E_p$ and $t$ parameters. The last two parameters can be directly controlled by an applied voltage as earlier shown in Fig.\ref{PositionDependentQubit}, where eight voltage signals are used for controlling the effective tight-binding Hamiltonian. It is informative to notice that change of the quantum well length, expressed by $a+b$, does not affect the eigenenergy of 2-SELs significantly. The observed change affects the ratio of electrostatic to kinetic energy and thus is similar to the change in energy eigenspectra generated by different distances $d$. We can spot narrowing of the bands when moving from the situation of lower to higher electrostatic energy of interacting electron and again it is typical for metal-insulator phase transition. Change of ratio kinetic to electrostatic energy can be obtained by keeping quantum well size constant, constant distance between 2 SELs and by change of hopping constant $t$ that is the measure of electron ability in conducting electric or heat current. Again one observes the narrowing of bands when we reduce $t$ so the dominant energy of electron is due to the electron-electron interaction. The last plot of Fig.\ref{MottTransition1} describes our ability of tunneling eigenenergy spectra of system in linear way just by change of $E_p$ parameter. In very real way we can recognize the ability of tunning the chemical potential (equivalent to Fermi energy at temperatures T=0K) by controlling voltages given in Fig.\ref{PositionDependentQubit}. in our artificial lattice system. Due to controllability of energy eigenspectra by controlling voltages from Fig.\ref{PositionDependentQubit} one can recognize 2 SELs system as the first stage of implementation of programmable quantum matter. In general case considered 2-SELs Hamiltonian consists 12 different $E_p$ parameters and 6 different $t$ parameters that can be controlled electrostatically (18 parameters under electrostatic control) by 2-SELS controlling voltages $V_0(t), .., V_3(t), V_{0'}(t), .., V_{3'}(t) $ depicted in Fig.1. \newline \newline The numerical modeling of electron transport across coupled SELs is about solving a set of nine coupled recurrent equations of motion as it is in the case of time-dependent 2 SELs Hamiltonian.
In this work we consider time-independent Hamiltonian implying constant occupation of energetic levels. Therefore the quantum state can be written in the form
$\ket{\psi(t')}$ $=\alpha_1 e^{\frac{\hbar}{i}E_1 t'}\ket{E_1}+$..$+\alpha_9 e^{\frac{\hbar}{i}E_9 t'}\ket{E_9}$, so the probability of occupancy of energetic level $E_1$ is $|\alpha_1|^2=|\bra{E_1} \ket{\psi(t)}|^2=p_{E1}=constant$, etc. Since we have obtained analytical form of all states $\ket{E_k}$ and eigenenergies $E_k$ we have analytical form of quantum state dynamics $\ket{\psi(t')}$ with time. From obtained analytical solutions presented in Appendix~\ref{app:a} we recognize that every eigenenergy state is the linear combination of positon-based states $\ket{k} \bigotimes \ket{l'}$ what will imply that quantum state can never be fully localized at two nodes k and l' as it is pointed by analytically obtained eigenstates of the 2-SELs Hamiltonian that are given in Appendix~\ref{app:a}.
In the conducted numerical simulations we visualize analytical solutions. We set $\hbar=1$ and $\alpha_1=..=\alpha_8=\frac{1}{9}$, $\alpha_9=\sqrt{1-\frac{8}{81}}$ (Scenario I that has populated all 9 energetic levels) or $\alpha_1=\alpha_2=\frac{1}{2}$,$\alpha_9=\frac{\sqrt{2}}{2}$,$\alpha_3=..=\alpha_8$ (Scenario II that has populated 3 energetic levels) that will correspond to top or bottom plots of Fig.\ref{OccupancyOscillations}. We can recognize that probability of occupancy of (1,1') from Fig.1. (when two electrons are at input of 2-SELs) is given by $|(\bra{1,0,0}\bigotimes\bra{1,0,0})\ket{\psi(t)}|^2=|\gamma_1(t)|^2=p_1(t)$ (two electrons as SELs inputs) can be compared with occupancy of (3,3') given by $p_9(t)=|\gamma_9(t)|^2=|(\bra{0,0,1}\bigotimes\bra{0,0,1})\ket{\psi(t)}|^2$ (2 electrons at SELs outputs) as depicted in Fig.\,\ref{OccupancyOscillations}. It is relatively easy to identify probability of finding first electron at input as the sum of $p_1(t)+p_2(t)+p_3(t)$.

Various symmetries can be traced in the Scenario II (9 populated energy levels) given by Fig.\ref{OccupancyOscillations}. as between probability $p_2(t)$ and $p_8(t)$ or in the upper part of Fig.
\ref{OccupancyOscillations} in the Scenario I (3 populated energy levels) when $p_2(t)=p_8(t)$ or $\phi_2(t)=phase(\gamma_2(t))=\phi_8(t)$. The same symmetry relations applies to the case of probability $p_4(t)$ and $p_6(t)$ as well as $\phi_4(\gamma_4(t))$ and $\phi_6(\gamma_6(t))$. These symmetries has its origin in the fact that 2 SELs system is symmetric along x axes what can be recoginezed in symmetries of simplified Hamiltonian matrix \ref{Matrix}. It shall be underlined that in the most general case when system matrix has no symmetries the energy eigenspectra might have less monotonic behaviour.

\subsection{Act of Measurement and Dynamics of Quantum State}

The quantum system dynamics over time is expressed by the equation of motion $\hat{H}(t')\ket{\psi(t')}=i\hbar \frac{d}{dt'}\ket{\psi(t')}$ that can be represented in discrete time step by relation
\begin{equation}
  \frac{dt'}{i \hbar}\hat{H}(t')\ket{\psi(t')}+\ket{\psi(t')}=\ket{\psi(t'+dt')}.
\end{equation}
It leads to the following equations of motion for quantum state expressed by equation (\ref{stateSEL}) as follows
\begin{eqnarray}
\vec{\gamma}(t'+dt')=
\left\{
  \begin{array}{lr}
\gamma_1(t')+dt'\sum_{k=1}^{9} \hat{H}_{1,k}(t')\gamma_k(t')=\\ f_1(\vec{\gamma}(t'),dt')[\hat{H}(t')], \\
.. \\
\gamma_9(t')+dt'\sum_{k=1}^{9} \hat{H}_{9,k}(t')\gamma_k(t')=\\ f_9(\vec{\gamma}(t'),dt')[\hat{H}(t')] \\
  \end{array}
\right\}= \nonumber\\ =\vec{f}(\vec{\gamma}(t'),dt')[\hat{H}(t')]=\vec{f}(\vec{\gamma}(t'),dt')_{[\hat{H}(t')]}. \nonumber \\
\end{eqnarray}
%
Symbol $[.]$ denotes functional dependence of $\vec{f}(\vec{\gamma}(t'),dt')$ on Hamiltonian $\hat{H}(t')$. The measurement can be represented by projection operators $\hat{\Pi}(t')$ equivalent to the matrix that acts on the quantum state over time. The lack of measurement can simply mean that the state projects on itself so the projection is the identity operation ($\hat{\Pi}(t')=\hat{I}_{9 \times 9}$). Otherwise, the quantum state is projected on its subset and hence the projection operator can change in a non-continuous way over time. We can formally write the quantum state dynamics with respect to time during the occurrence of measurement process (interaction of external physical system with the considered quantum system) as
%
\begin{eqnarray}
\vec{\gamma}(t'+dt')= \nonumber \\
\frac{\hat{\Pi}(t'+dt')(\vec{f}(\vec{\gamma}(t'),dt'))}{(\hat{\Pi}(t'+dt')\vec{f}(\vec{\gamma}(t'),dt'))^{\dag}(\hat{\Pi}(t'+dt')\vec{f}(\vec{\gamma}(t'),dt'))}. 
\end{eqnarray}
%
Let us refer to some example by assuming that a particle in the upper SELs was detected by the upper output detector (Fig.\,\ref{PositionDependentQubit}b). In such a case, the following projector $\hat{\Pi}(t,t+\Delta t) $ is different from the identity in time interval $(t,t+\Delta t)$ with $1_{1_{t,t+\Delta t}}=1$ set to 1 in this time interval and 0 otherwise. The projector acts on the quantum state (diagonal matrix is given by diag symbol). It is given as
\begin{eqnarray}
   \hat{\Pi}(t,t+\Delta t)= (1-1_{t,t+\Delta t})(\hat{I}_U \times \hat{I}_L)+ \nonumber \\ 1_{t,t+\Delta t}(\ket{0,0,1}_U\bra{0,0,1}_U \times \hat{I}_L)= \nonumber \\
   (1-1_{t,t+\Delta t})(\hat{I}_U \times \hat{I}_L)+ \nonumber \\
1_{t,t+\Delta t}( \ket{0,0,1}_U\bra{0,0,1}_U \times (\ket{1,0,0}_L\bra{1,0,0}_L+ \nonumber \\ \ket{0,1,0}_L\bra{0,1,0}_L+ \ket{0,0,1}_L\bra{0,0,1}_L)) = \nonumber \\
  =(1-1_{t,t+\Delta t})\hat{I}_{9 \times 9}
 +1_{t,t+\Delta t} diag(0,0,1) \times \hat{I}_{3 \times 3}
 \\ \nonumber  =diag((1-1_{t,t+\Delta t}),(1-1_{t,t+\Delta t}),(1-1_{t,t+\Delta t}), \\ \nonumber
(1-1_{t,t+\Delta t}),(1-1_{t,t+\Delta t}),(1-1_{t,t+\Delta t}),1,1,1)
\end{eqnarray}

\section{Correlations for the case of 2  electrostatically interacting qubits}
We define correlation function as $C_e(a,b)=\frac{N_{+,+}+N_{-,-}-N_{+,-}-N_{-,+}}{N_{+,+}+N_{-,-}+N_{+,-}+N_{-,+}}$, where $N_{+,+}$ represents presene of 2 electrons at points 2 and 2', $N_{-,-}$ represents presence of electrons at points 1 and 1',
$N_{+,-}$ is corresponding to presence of electrons at point 2 and 1' and $N_{-,+}$ is corresponding to presence of electrons at point 1 and 2'. It is convenient to introduce the operator
\begin{equation}
N_{+,+}+N_{-,-}-N_{+,-}-N_{-,+}=
\bra{\psi,t}
\begin{pmatrix}
1 & 0 & 0 & 0 \\
0 & -1 & 0 & 0 \\
0 & 0 & -1 & 0 \\
0 & 0 & 0 & 1 \\
\end{pmatrix}
\ket{\psi,t} =\bra{\psi,t_0}U(t,t_0)^{-1} \sigma_3 \times \sigma_3  U(t,t_0)\ket{\psi,t_0}
\end{equation}
Consequently we obtain

$ \ket{E_1}_n=
\frac{1}{\sqrt{\left(8\left(\frac{t_{sr1}-t_{sr2}}{\sqrt{(E_{c1}-E_{c2})^2+4
   (t_{sr1}-t_{sr2})^2}-E_{c1}+E_{c2}}\right)\right)^2+2
   }}
\begin{pmatrix}
-1,\\
-\frac{2(t_{sr1}-t_{sr2})}{\sqrt{(E_{c1}-E_{c2})^2+4 (t_{sr1}-t_{sr2})^2}-E_{c1}+E_{c2}}, \\
\frac{2(t_{sr1}-t_{sr2})}{\sqrt{(E_{c1}-E_{c2})^2+4(t_{sr1}-t_{sr2})^2}-E_{c1}+E_{c2}} \\
1
\end{pmatrix}=\frac{1}{\sqrt{\left(8\left(\frac{t_{sr1}-t_{sr2}}{\sqrt{(E_{c1}-E_{c2})^2+4
   (t_{sr1}-t_{sr2})^2}-E_{c1}+E_{c2}}\right)\right)^2+2
   }} \ket{E_1}
$

$ \ket{E_1}_n=
\frac{1}{\sqrt{\left(8\left(\frac{t_{sr1}-t_{sr2}}{\sqrt{(E_{c1}-E_{c2})^2+4
   (t_{sr1}-t_{sr2})^2}-E_{c1}+E_{c2}}\right)\right)^2+2
   }}
\begin{pmatrix}
-1,\\
-\frac{2(t_{sr1}-t_{sr2})}{\sqrt{(E_{c1}-E_{c2})^2+4 (t_{sr1}-t_{sr2})^2}-E_{c1}+E_{c2}}, \\
\frac{2(t_{sr1}-t_{sr2})}{\sqrt{(E_{c1}-E_{c2})^2+4(t_{sr1}-t_{sr2})^2}-E_{c1}+E_{c2}} \\
1
\end{pmatrix}=\frac{1}{\sqrt{\left(8\left(\frac{t_{sr1}-t_{sr2}}{\sqrt{(E_{c1}-E_{c2})^2+4
   (t_{sr1}-t_{sr2})^2}-E_{c1}+E_{c2}}\right)\right)^2+2
   }} \ket{E_1}
$

$ \ket{E_2}_{n}= -\frac{1}{\sqrt{\left(8\left(\frac{t_{sr1}-t_{sr2}}{\sqrt{(E_{c1}-E_{c2})^2+4
   (t_{sr1}-t_{sr2})^2}+E_{c1}-E_{c2}}\right)\right)^2+2
   }}
\begin{pmatrix}
-1 \\
\frac{2
   (t_{sr1}-t_{sr2})}{\sqrt{(E_{c1}-E_{c2})^2+4
   (t_{sr1}-t_{sr2})^2}+E_{c1}-E_{c2}} \\
-\frac{2(t_{sr1}-t_{sr2})}{\sqrt{(E_{c1}-E_{c2})^2+4
   (t_{sr1}-t_{sr2})^2}+E_{c1}-E_{c2}} \\,1
\end{pmatrix}=-
\frac{1}{\sqrt{\left(8\left(\frac{t_{sr1}-t_{sr2}}{\sqrt{(E_{c1}-E_{c2})^2+4
   (t_{sr1}-t_{sr2})^2}+E_{c1}-E_{c2}}\right)\right)^2+2
   }} \ket{E_2}
$

$\ket{E_3}_{n}= \frac{1}{\sqrt{\left(8\left(\frac{\text{tsr1}+\text{tsr2}}{\sqrt{(\text{Ec1}-\text{Ec2})^2+4
   (\text{tsr1}+\text{tsr2})^2}-\text{Ec1}+\text{Ec2}}\right)\right)^2+2
   }}
\begin{pmatrix}
1, \\
-\frac{2(t_{sr1}+t_{sr2})}{\sqrt{(E_{c1}-E_{c2})^2+4
   (t_{sr1}+t_{sr2})^2}-E_{c1}+E_{c2}}, \\
-\frac{2(t_{sr1}+t_{sr2})}{\sqrt{(E_{c1}-E_{c2})^2+4
   (t_{sr1}+t_{sr2})^2}-E_{c1}+E_{c2}}, \\
1
\end{pmatrix}=
\frac{1}{\sqrt{\left(8\left(\frac{\text{tsr1}+\text{tsr2}}{\sqrt{(\text{Ec1}-\text{Ec2})^2+4
   (\text{tsr1}+\text{tsr2})^2}-\text{Ec1}+\text{Ec2}}\right)\right)^2+2
   }} \ket{E_3}
$

$\ket{E_4}_{n}=  \frac{1}{\sqrt{\left(8\left(\frac{t_{sr1}+t_{sr2}}{\sqrt{(E_{c1}-E_{c2})^2+4
   (t_{sr1}+t_{sr2})^2}-E_{c2}+E_{c1}}\right)\right)^2+2
   }}
\begin{pmatrix}
1, \\
\frac{2(t_{sr1}+t_{sr2})}{\sqrt{(E_{c1}-E_{c2})^2+4(t_{sr1}+t_{sr2})^2}+E_{c1}-E_{c2}}, \\
\frac{2(t_{sr1}+t_{sr2})}{\sqrt{(E_{c1}-E_{c2})^2+4(t_{sr1}+t_{sr2})^2}+E_{c1}-E_{c2}},  \\
1
\end{pmatrix}=
\frac{1}{\sqrt{\left(8\left(\frac{t_{sr1}+t_{sr2}}{\sqrt{(E_{c1}-E_{c2})^2+4
   (t_{sr1}+t_{sr2})^2}-E_{c2}+E_{c1}}\right)\right)^2+2
   }} \ket{E_4}.
$

\subsection{Correlation Function for Classical and Quantum Approaches for Single-Electron Lines}

\begin{figure}
   \centering
     \includegraphics[scale=0.6]{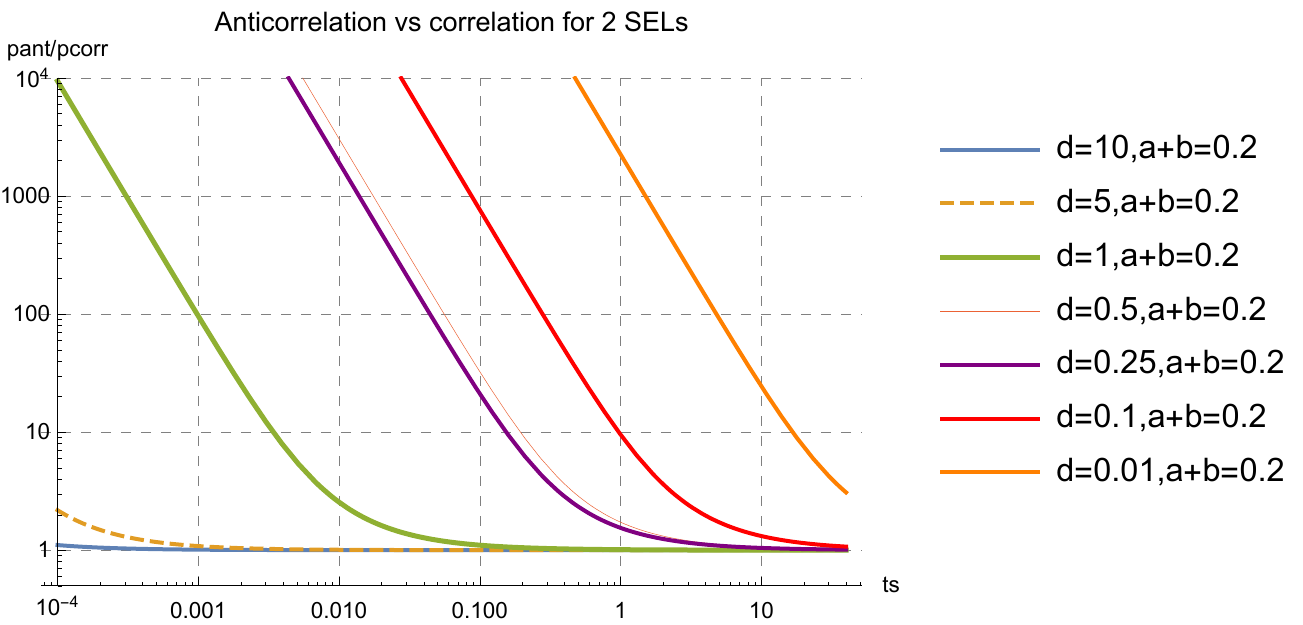}
     \includegraphics[scale=0.6]{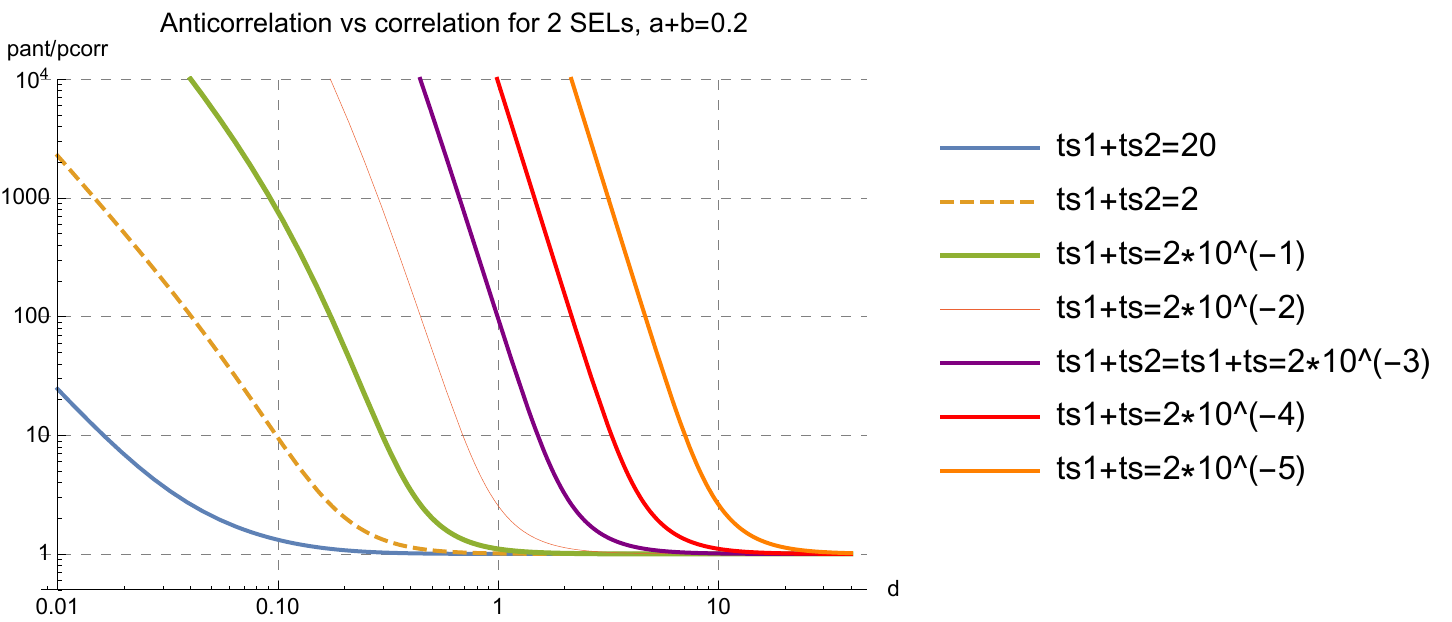}
    \includegraphics[scale=0.6]{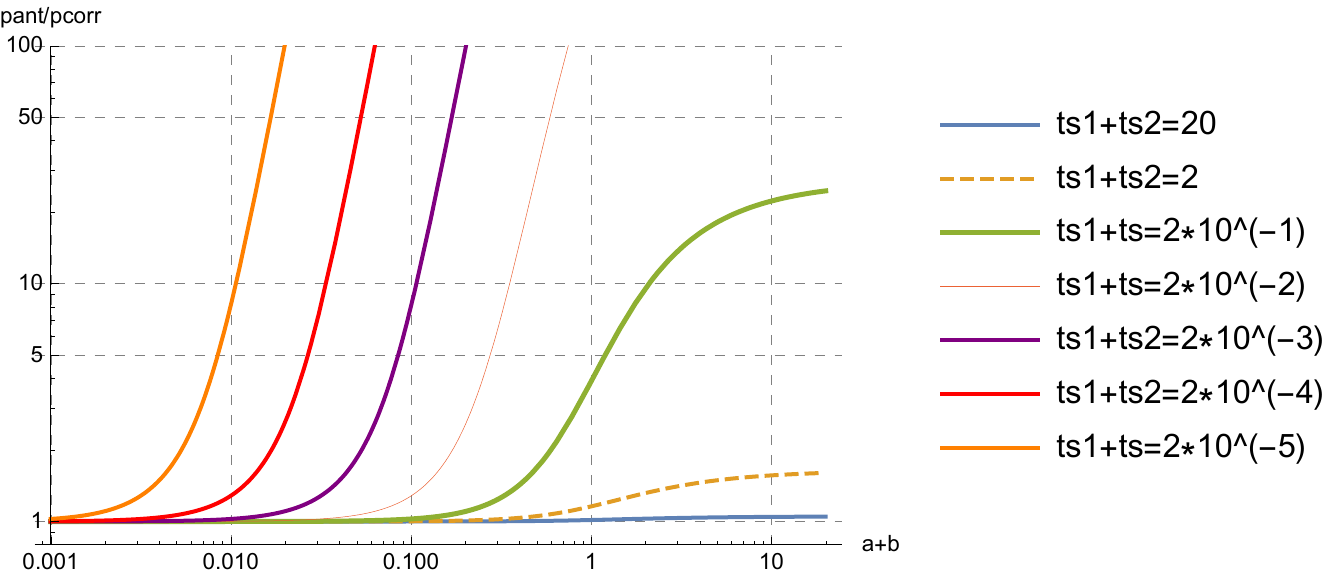}
 \caption{The ratio of probabilities in 2 SELs ground state between correlated and anticorrelated  quantum state components is very strongly depending on hopping constants by term $t_{s1}+t_{s2}$ and very strongly depends on size of quantum wells denoted by a and distance between two two neighbours a.}
  \label{fig:pcorranticorr}
\end{figure}

In this work, two capacitively coupled single-electron lines (SEL) are treated by the tight-binding model with the use of three nodes for each line to describe the electron occupancy. It should be highlighted that the most simplistic approach towards the two SELs can be attempted
with the use of two nodes for each line. In such a case, it is possible to introduce a correlation function for both quantum and classical treatments of the system under consideration. Let us start from the quantum approach. The Hamiltonian of the system having flat bottoms of potentials can be written as
\begin{equation}
H=\left(
\begin{array}{cccc}
 E_{c1}+2 E_{p} & e^{i \beta } t_{s2} & e^{i \alpha }
   t_{s1} & 0 \\
 e^{-i \beta } t_{s2} & E_{c2}+2 E_{p} & 0 & e^{i \alpha }
  t_{s1} \\
 e^{-i \alpha } t_{s1} & 0 & E_{c2}+2 E_{p} & e^{i \beta }
   t_{s2} \\
 0 & e^{-i \alpha } t_{s1} & e^{-i \beta } t_{s2} & E_{c1}+2 E_{p} \\
\end{array}
\right),
\end{equation}
where $E_{c1}=\frac{q^2}{d}$ and $E_{c2}=\frac{q^2}{\sqrt{d^2+a^2}}$, so $E_{c1}-E_{c2}=\frac{q^2}{d}-\frac{q^2}{\sqrt{d^2+a^2}}>0$. The hopping terms are parametrized by $t_{s1}$ and $t_{s2}$. 
This last Hamiltonian refers to the quantum state describing the occupancy of four nodes at upper U=(1,2) or lower line L=(1',2') by two spatially separated electrons
\begin{eqnarray}
\ket{\psi}=\gamma_1(t)\ket{1}\ket{1'}+\gamma_2(t)\ket{1}\ket{2'}+ 
\gamma_3(t)\ket{2}\ket{1'}+ 
\gamma_4(t)\ket{2}\ket{2'}.
\end{eqnarray}
Normalization condition requires $|\gamma_1|^2+\cdots+|\gamma_4|^2=1$. We have four eigenenergies
\begin{eqnarray}
E_1=\frac{1}{2}(E_{c1}\!+\!E_{c2}\!+\!4E_p\!-\!\sqrt{(E_{c1}\!-\!E_{c2}\!)^2+\!4(t_{s1}\!-\!t_{s2})^2)}), \nonumber \\
E_2=\frac{1}{2}(E_{c1}\!+\!E_{c2}\!+\!4E_p\!+\!\sqrt{(E_{c1}\!-\!E_{c2}\!)^2+\!4(t_{s1}\!-\!t_{s2})^2)}),\nonumber \\
E_3=\frac{1}{2}(E_{c1}\!+\!E_{c2}\!+\!4E_p\!-\!\sqrt{(E_{c1}\!-\!E_{c2}\!)^2+\!4(t_{s1}\!+\!t_{s2})^2)}), \nonumber \\
E_4=\frac{1}{2}(E_{c1}\!+\!E_{c2}\!+\!4E_p\!+\!\sqrt{(E_{c1}\!-\!E_{c2}\!)^2+\!4(t_{s1}\!+\!t_{s2})^2)}), \nonumber \\
\label{eigenergiessSEL}
\end{eqnarray}
fulfilling $E_1<E_2$, $E_3<E_4$ as corresponding to four eigenenergy states
\begin{eqnarray}
\ket{E_1}=
\begin{pmatrix}
-e^{i (\alpha +\beta )}, \\
-\frac{2 e^{i \alpha }(t_{s1}-t_{s2})}{\sqrt{(E_{c1}-E_{c2})^2+4(t_{s1}-t_{s2})^2}-E_{c1}+E_{c2}}, \\
\frac{2 e^{i \beta} (t_{s1}-t_{s2})}{\sqrt{(E_{c1}-E_{c2})^2+4(t_{s1}-t_{s2})^2}-E_{c1}+E_{c2}},\\
1
\end{pmatrix}, \nonumber \\
\ket{E_2}=
\begin{pmatrix}
-e^{i (\alpha +\beta )}, \\
\frac{2 e^{i \alpha }(t_{s1}-t_{s2})}{\sqrt{(E_{c1}-E_{c2})^2+4
   (t_{s1}-t_{s2})^2}+E_{c1}-E_{c2}}, \\
-\frac{2 e^{i \beta} (t_{s1}-t_{s2})}{\sqrt{(E_{c1}-E_{c2})^2+4(t_{s1}-t_{s2})^2}+E_{c1}-E_{c2}}, \\
1
\end{pmatrix}, \nonumber \\
\ket{E_3}=
\begin{pmatrix}
e^{i (\alpha +\beta )}, \\
-\frac{2 e^{i \alpha }(t_{s1}+t_{s2})}{\sqrt{(E_{c1}-E_{c2})^2+4(t_{s1}+t_{s2})^2}-E_{c2}+E_{c1}}, \\
-\frac{2 e^{i \beta} (t_{s1}+t_{s2})}{\sqrt{(E_{c1}-E_{c2})^2+4(t_{s1}+t_{s2})^2}-E_{c2}+E_{c1}}, \\
1
\end{pmatrix}, \nonumber \\
\ket{E_4}=
\begin{pmatrix}
e^{i (\alpha +\beta )}, \\
\frac{2 e^{i \alpha }(t_{s1}+t_{s2})}{\sqrt{(E_{c1}-E_{c2})^2+4(t_{s1}+t_{s2})^2}+E_{c1}-E_{c2}}, \\
\frac{2 e^{i \beta} (t_{ts1}+ t_{s2})}{\sqrt{(E_{c1}-E_{c2})^2+4t_{s1}+t_{s2})^2}+E_{c1}-E_{c2}}, \\
1
\end{pmatrix}.
\end{eqnarray}
with ground state
\begin{equation}
E_3=E_g=\frac{1}{2}(E_{c1}\!+\!E_{c2}\!+\!4E_p\!-\!\sqrt{(E_{c1}\!-\!E_{c2}\!)^2+\!4(t_{s1}\!+\!t_{s2})^2)}).
\end{equation}
We observe that in the ground state, the probability of occurrence of two particles at the maximum distance $p_{1,2'}=p_{2,1'}=p_{anticorr}$ to the probability of two particles occurrence at the minimum distance $p_{1,1'}=p_{2,2'}=p_{corr}$ is given by the formula:
\begin{eqnarray}
\frac{p_{acorr}}{p_{corr}}=
=\Bigg[\frac{\sqrt{(E_{c1}-E_{c2})^2+4(t_{s1}+t_{s2})^2}-(E_{c1}-E_{c2})}{2(t_{s1}+t_{s2})}\Bigg]^2=
\nonumber \\
\Bigg[\frac{\sqrt{q^2(\sqrt{d^2+(a+b)^2}-d)^2+4(t_{s1}+t_{s2})^2}}{2(t_{s1}+t_{s2})d\sqrt{d^2+(a+b)^2}} 
-\frac{q^2(\sqrt{d^2+(a+b)^2}-d)}{2(t_{s1}+t_{s2})d\sqrt{d^2+(a+b)^2}}\Bigg]^2
\nonumber \\
\label{pacorrpcorrration}
\end{eqnarray}
It is worth mentioning that ground state of 2 coupled SELs brings electrons partly into anticorrelated position (2 electrons at maximum distance) and correlated positions (2 electrons at minimum distance) what simply means that anticorrelation is not greatly pronounced in quantum case at it is the case of classical picture. One can refer to the following dependence of ratio between probabilties for the state to be anticorrelated or correlated
state as depicted by Fig.\ref{fig:pcorranticorr}.

The quantum state in case of time-independent Hamiltonian can be expressed as
\begin{eqnarray}
\ket{\psi}\!=\!\sqrt{p_{E1}}e^{\phi_{E10}i}e^{\frac{1}{\hbar i}E_1 t}\ket{E_1}\!+\!\sqrt{p_{E2}}e^{\phi_{E20}i}e^{\frac{1}{\hbar i}E_2 t}\ket{E_2}\!+ \nonumber \\
\sqrt{p_{E3}}e^{\phi_{E30}i}e^{\frac{1}{\hbar i}E_3 t}\ket{E_3}\!+\!\sqrt{p_{E4}}e^{\phi_{E40}i}e^{\frac{1}{\hbar i}E_4 t}\ket{E_4} \nonumber \\.
\end{eqnarray}

Having $E_{c1}=\frac{q^2}{d}$ and $E_{c2}=\frac{q^2}{\sqrt{d^2+a^2}}$ so $E_{c1}-E_{c2}=\frac{q^2}{d}-\frac{q^2}{\sqrt{d^2+a^2}}>0$ and hopping terms $t_{s1}, t_{s2}$ we obtain Hamiltonian and a correlation function $C$.

\begin{figure}
\includegraphics[scale=1.1]{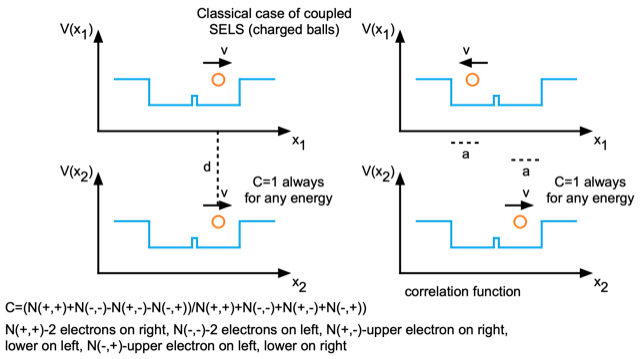} 
\caption{Case of electrostatically coupled charged particles confined by local potentials and electrostatically interacting. Concept of correlation/anticorrelation in their positions.}
\label{fig:boat2}
\centering
\includegraphics[scale=0.7]{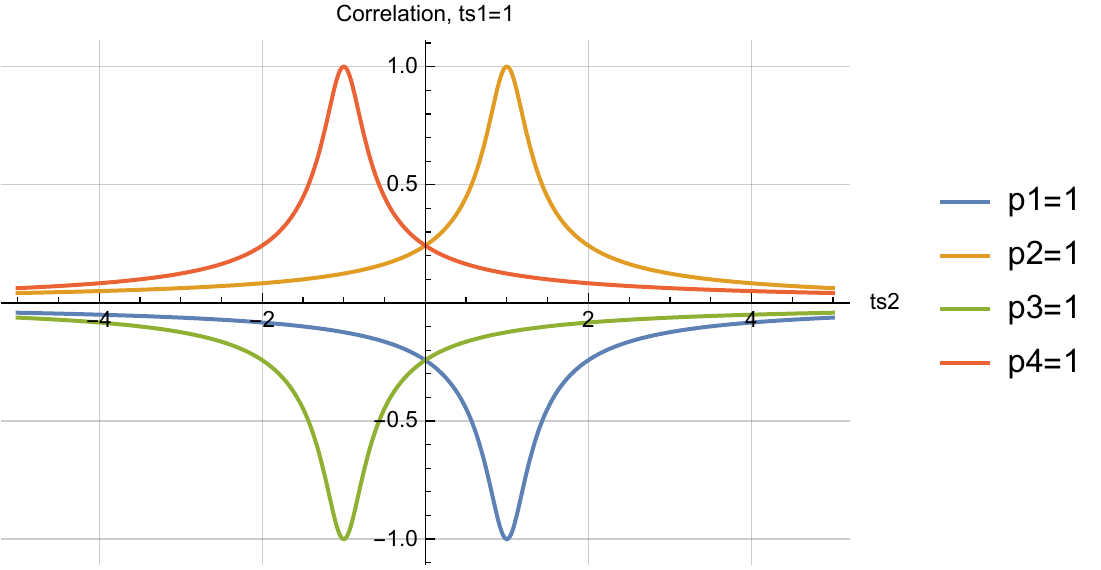}\includegraphics[scale=0.7]{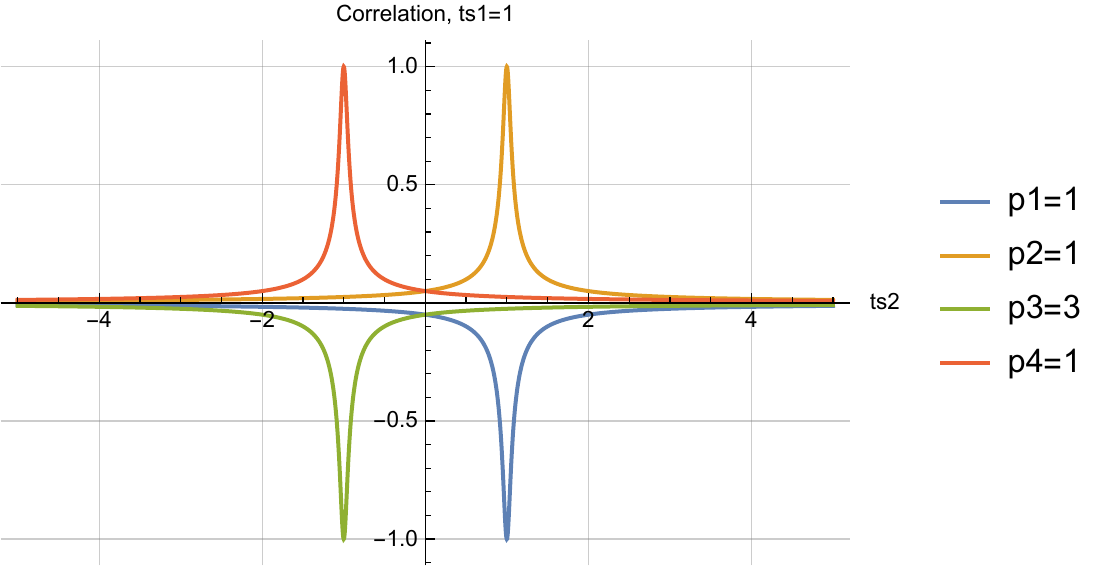}
\caption{Four main fundamental configurations named as anticorrelation and correlation for system of coupled SEL depicted in Fig.\,\ref{fig:boat2}. The correlation function $C$ that are grasped by formula (\ref{CC}) corresponding to the full occupancy of one among four eigenenergies.}
\label{fig:boat1}
\includegraphics[scale=0.6]{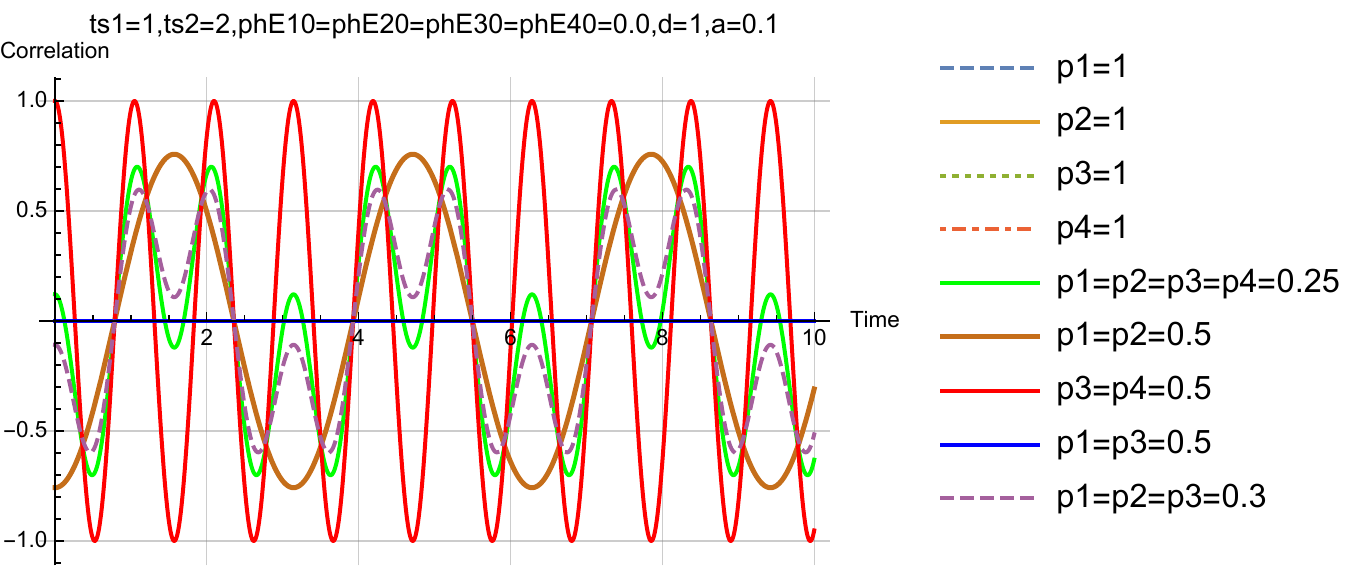} 
\caption{Correlation function $C$ with time for time-independent Hamiltonian corresponding to full and partial occupancy of 4 eigenergies of 2-SEL system.}
\label{fig:boat1a}
\end{figure}

\begin{figure}
\centering
\includegraphics[scale=0.6]{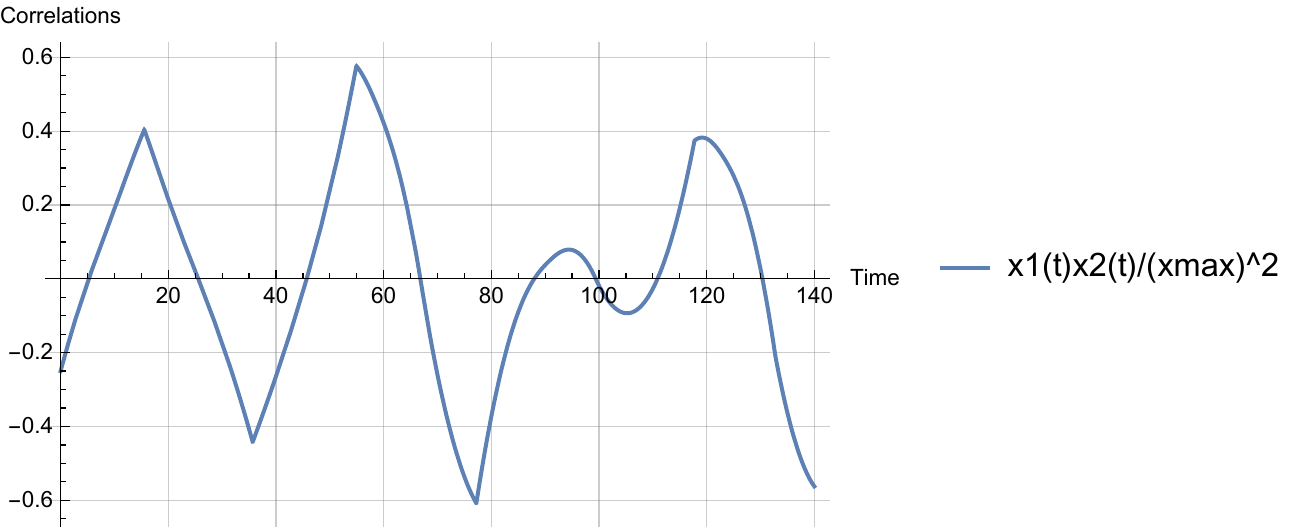} 
\includegraphics[scale=0.6]{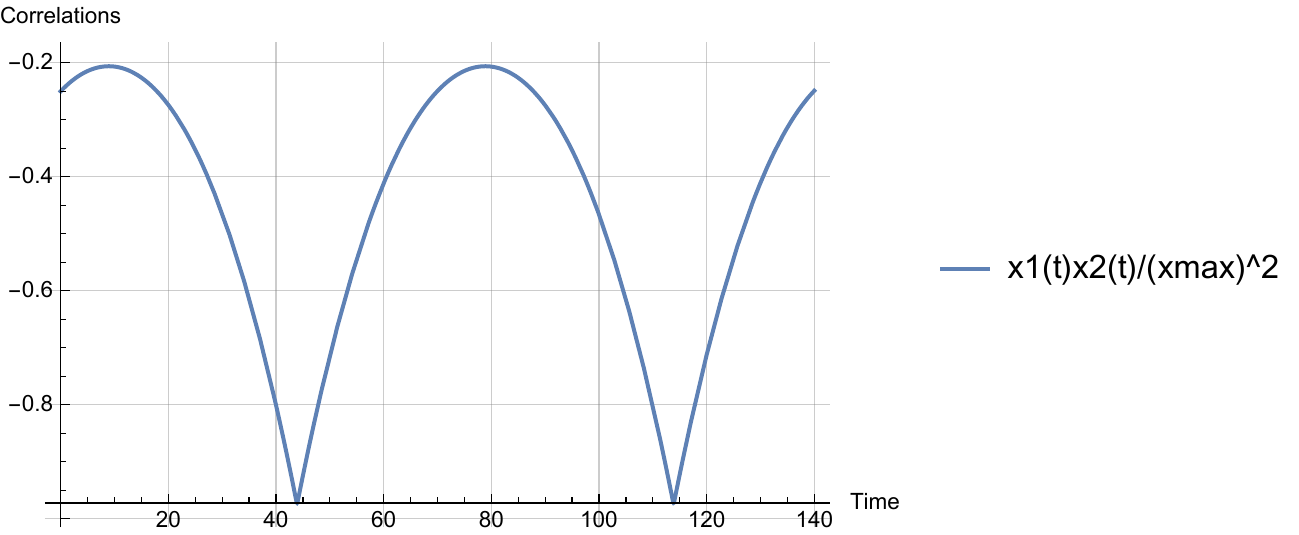} 
\includegraphics[scale=0.6]{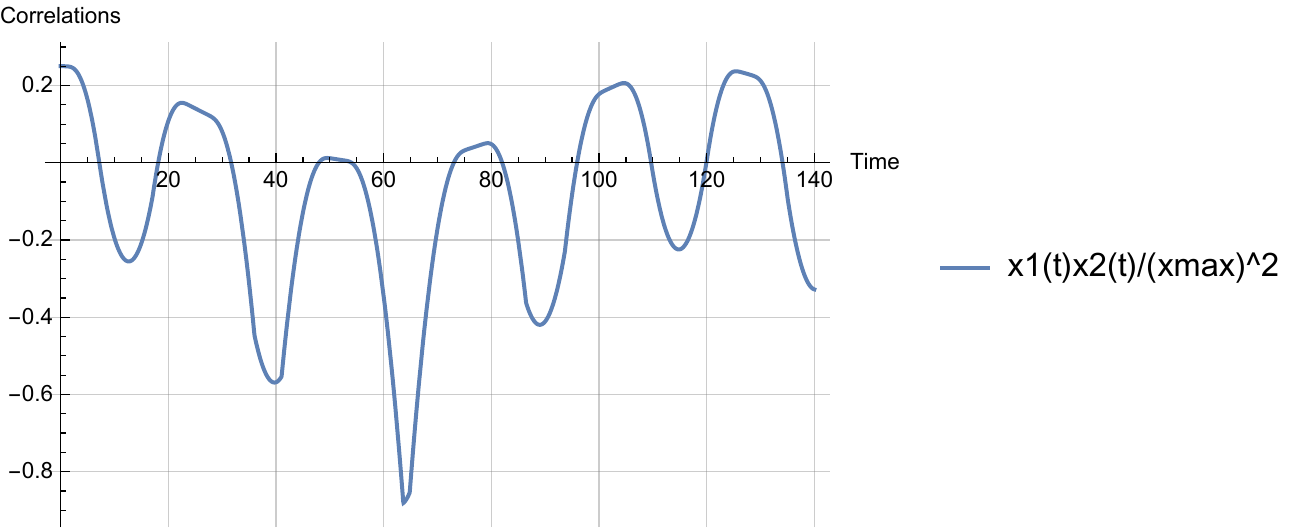} 
\caption{Varying dependence of classical correlation function [$C=\frac{x_1(t) x_2(t)}{x_{max}^2}$] over time. Upper case refers to 2-SELs with particles of significantly different speeds at anticorrelated positions at initial time; middle figure describes two perfectly anticorrelated particles (\ref{app:b}); third case refers to the proceeding figure \ref{fig:2cSEL1b}.}
\label{CorrelationNoWM}
\end{figure}

We refer to the physical situation depicted in Fig.\,\ref{fig:boat2} and utilize the correlation function $C$ to capture as to what extent the two electrons are in a correlated state being both either on the left or on the right side that is corresponding to terms $N_{\rm -,-}, N_{+,+}$, or in an anticorrelated state (expressed by terms $N_{+,-}$ and $N_{-,+}$). Such function is commonly used in spin systems and is a measure of non-classical correlations. Using a tight-binding model describing two electrostatically coupled SELs and using the same correlation function applicable in the test of Bell theory of entangled spins \cite{WikipediaBellTheorem}, we obtain the correlation function $C$ given by formula:

$ $ \newline \newline
\begin{eqnarray}
C=\frac{N_{+,+}+N_{-,-}-N_{-,+}-N_{+,-}}{N_{+,+}+N_{-,-}+N_{-,+}+N_{+,-}}= \nonumber \\
4 [\frac{\sqrt{p_{E1}} \sqrt{p_{E2}} (t_{s1}-t_{s2}) \cos [-t \sqrt{(E_{c1}-E_{c2})^2+ 
4 (t_{s1}-t_{s2})^2}+\phi_{E10}-\phi_{E20}]}{\sqrt{(E_{c1}-E_{c2})^2+4 (t_{s1}-t_{s2})^2}} \nonumber \\
+\frac{\sqrt{p_{E3}} \sqrt{p_{E4}}
   (t_{s1}+t_{s2}) \cos [-t \sqrt{(E_{c1}-E_{c2})^2+4 (t_{s1}+t_{s2})^2}+\phi_{E30}-\phi_{E40}]}{\sqrt{(E_{c1}-E_{c2})^2+4
   (t_{s1}+t_{s2})^2}}] \nonumber \\
-(E_{c1}-E_{c2}) [\frac{p_{E1}-p_{E2}}{\sqrt{(E_{c1}-E_{c2})^2+4
   (t_{s1}-t_{s2})^2}}+\frac{p_{E3}-p_{E4}}{\sqrt{(E_{c1}-E_{c2})^2+4 (t_{s1}+t_{s2})^2}}]
  \label{CC}
\end{eqnarray}

Classical intuition points out that when the kinetic energy of electrons goes to zero they shall be anticorrelated due to the presence of the repulsive Coulomb force. On the other hand, when the kinetic energy is dominant, the Coulomb interaction does not matter so much and the correlation function shall be zero or positive. Four fundamental solutions for the correlation function corresponding to the occupancy of four eigenenergies are given by Fig.\,\ref{fig:boat1}. Indeed, when only the ground state is occupied so $p_1=1$, then $C<1$, as depicted in Fig.\,\ref{fig:boat1a}.
It is remarkable to observe that $C=0$ if $p_1=p_3=0.5$. 
We also observe that if the two qubits are electrostatically decoupled then  $C=0$ does not need to be applied. However, for certain cases, the weaker the Coulomb interaction the sharper the peaks in the 2-SEL correlation function $C$, as depicted by Fig.\ref{fig:boat1a}. 

\begin{figure}
\centering
  \includegraphics[scale=0.8]{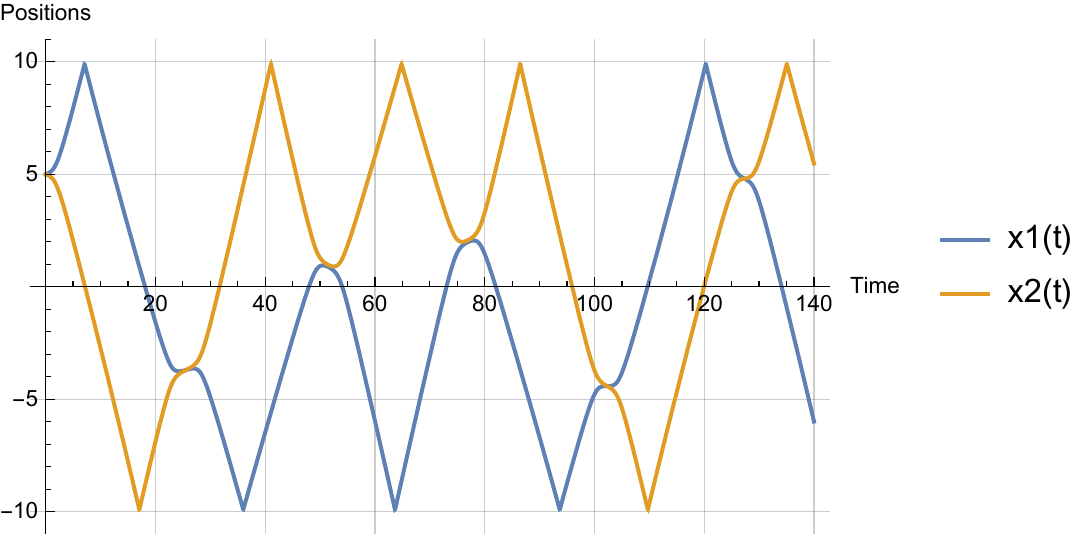} 
  \includegraphics[scale=0.8]{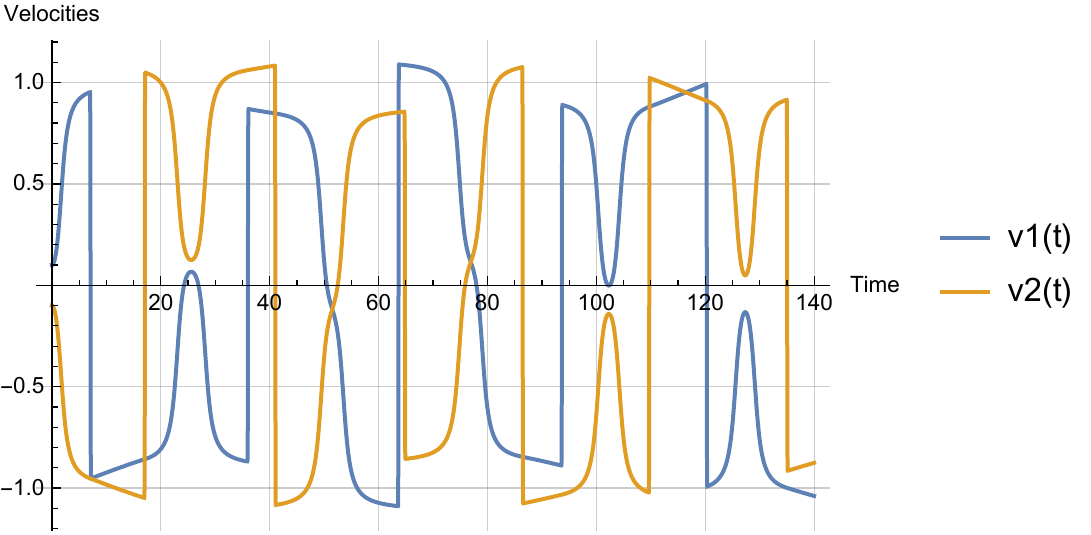}
  \caption{Evolution of positions $x_{i}(t)$ and velocities $v_{i}(t)$ for the system of 2 electrostatically coupled SELs in classical picture.}
  \label{fig:2cSEL1b}
 \end{figure}
\begin{figure}
\centering
\includegraphics[scale=0.8]{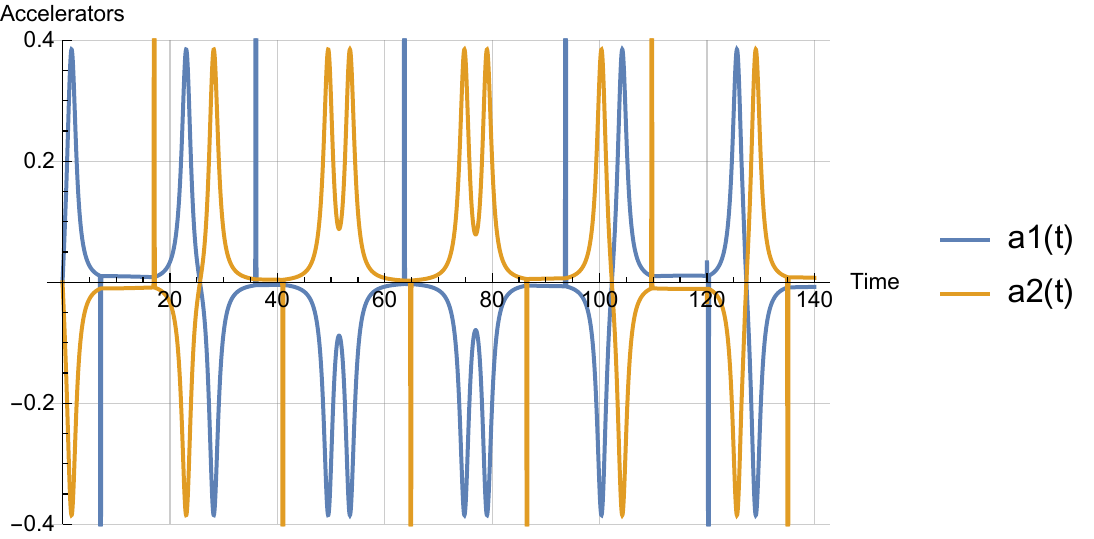} 
 \caption{Acceleration for the system of 2 coupled oscillators from Fig.\ref{fig:2cSEL1b} confined by local potential with coordinates $x_1,x_2 \in (-x_{max},x_{max})$ with $x_{max}=10$ . }
 \label{fig:2cSEL1}
\end{figure}


Now we turn towards the classical description of the two coupled single-electron lines using Newtonian dynamics as we expect qualitative changes in the correlation function due to the unique differences between the quantum and classical pictures. The confinement potential is approximated as a step function and presence of Poyting vector is neglected in the space as Hamiltonian system is time-independent, and system Hamiltonian corresponds to the classical mechanical energy that is preserved if we omit radiation emission for two particles subjected to acceleration and deceleration during different moments of motion that can be periodic or aperiodic. We have the minimalistic classical Hamiltonian for 2SELs given as
\begin{eqnarray}
    \hat{H}=\frac{1}{2m_1}p_1(t)^2+\frac{1}{2m_2}p_2(t)^2+\frac{q^2}{\sqrt{d^2+(x_1(t)-x_2(t))^2}} \nonumber \\
    +V_0\Theta(x_1(t)-x_{max1})+ V_0\Theta(-x_1(t)-x_{max1})+ \nonumber \\
    V_0\Theta(x_2(t)-x_{max2})+V_0\Theta(-x_2(t)-x_{max2}) \nonumber \\
    +V_{b1}\Theta(x_1(t)-x_{b1})+V_{b1}\Theta(-x_1(t)-x_{b1})+ \nonumber \\
    V_{b2}\Theta(x_2(t)-x_{b2})+V_{b2}\Theta(-x_2(t)-x_{b2}). \nonumber \\
    \label{HC}
\end{eqnarray}
We simplify the situation by having two symmetric masses $m_1=m_2=m$ and same charges $q$, and having $x_{max1}=x_{max2}=x_{max}$. We set $x_{b1}=x_{b2} \rightarrow 0$.
There are always two possible grounds states of the classically interacting electrons in 2 SELs configuration corresponding to the same energy when charged particles of same charge are confined in local potential that corresponds to two positions of particle that are at maximum distance $x_2(t)=\mp x_{min}=constans, \frac{dx_1}{dt}(t)=0,\frac{d^2x_1}{dt^2}(t)=0,
x_2(t)=\pm x_{min}, \frac{dx_1}{dt}(t)=0,\frac{d^2x_1}{dt^2}(t)=0 $.
Classical ground state is maximally anticorrelated.
On the contrary the same situation in quantum picture has only one ground state and this state is not maximally anticorrelated and is partly correlated what is expressed by formula \ref{pacorrpcorrration}.
Moreover, in the classical picture of 2 SELs, one can observe the emergence of deterministic chaos that is heavily pronounced in the classical system, as depicted in Figs.\,\ref{fig:2cSEL1b}--\ref{fig:2cSEL1}. Now we are moving towards a description of classical 2-SEL system in case of perfect correlated or anticorrelated electrons. From the classical Hamiltonian we determine the equations of motion of the two electrons assuming the existence of the antisymmetric case $\pm x(t)=x_1(t)=-x_2(t)$ at all instances of motion for the system symmetric around $x=0$.
We assume that the distance between electrons $\sqrt{d^2+x(t)^2} \approx d$.
We have
\begin{eqnarray}
m v^2(t)+\frac{q^2}{\sqrt{d^2+x^2}}=E_c>0, \nonumber
\frac{d^2x}{dt}=\frac{x q^2}{(\sqrt{d^2+x^2})^3}.
\end{eqnarray}
In simplified case $d \gg x$ and thus we can write
\begin{eqnarray}
m\frac{d^2x(t)}{dt^2}=x \frac{ q^2}{d^{\frac{3}{2}}}.
\end{eqnarray}
and it has solutions
for each electron position
\begin{equation}
x(t)=\frac{\sqrt{m} v_0 d^{3/4}}{q} \sinh(\frac{q}{\sqrt{m}d^{3/4}}t).
\end{equation}
We notice that
$x_{max}=\frac{\sqrt{m} v_0 d^{3/4}}{q} sinh( \frac{q}{\sqrt{m}d^{3/4}}\frac{T}{4} )$ and the period of oscillations is
\begin{equation}
T=4 \frac{\sqrt{m}d^{3/4}}{q}Arc_{\sinh}(\frac{q x_{max}}{\sqrt{m} v_0 d^{3/4}})
\end{equation}
if $x_1(t=0)=x_2(t=0)=0$ and when $\frac{dx_1}{dt}(t=0)=-\frac{x_2}{dt}(t=0)=v_0 \neq 0$, which is a definition of perfect anticorrelation.
Collision with walls is occurring at $\frac{T}{4}$ time while the total size of classical well is $2 x_{max}$.
We observe that $x_1(t)=\frac{\sqrt{m} v_0 d^{3/4}}{q} \sinh(\frac{q}{\sqrt{m}d^{3/4}}t)=-x_2(t)$ for $t \in [0,\frac{T}{4}]$.
Correlation function $C$ is given analytically
\begin{equation}
C(t)=-\frac{1}{x_{max}^2}\frac{ m v_0^2 d^{3/2}}{q^2}( \sinh(\frac{q}{\sqrt{m}d^{3/4}}t))^2<0
\label{CC1}
\end{equation}
and is negative for any energy $E_c$ (function of $q$, $d$, $v_0$) of the system.
Such situation occurs only in some subsets of the classical case since in the quantum case the sign of function $C$ depends on the occupancy of the energetic levels. In the classical treatment of 2-SELs there exist the case of
perfectly correlated electrons at any distance that is independent on the system energy if we are above the ground state.

It is possible to specify such situation when at $t=0$ we have $x_{1}(t=0)=x_{2}(t=0)=0$ and when $\frac{dx_{1}}{dt}(t=0)=\frac{x_{2}}{dt}(t=0)=v_0$. In such a case, the Coulomb force will act perpendicular to the direction of motion and will play no role
in the electron movement. Electron movement will be correlated and with constant speed over time, with periodic reflections from the potential walls.
The correlation function will have the form
\begin{equation}
C(t)=\frac{1}{x_{max}^2}(v_0)^2 t^2
\label{CC2}
\end{equation}
within time $t \in [0,T/4]$. A perfect correlation of electrons in the classical situation can occur for any energy (if kinetic energy is larger than zero) of the system $E_c>0$. It is one of the key differences from the quantum situation when the positive value of correlation function can occur only for certain system eigenenergies as given by formula (\ref{CC}).

It should be underlined that the perfectly correlated electrons generate higher overall magnetic field energy as it is the case of two electric currents of the same sign (correlated electron movement in one direction) generated by each electron. In the case of anticorrelated electrons we are dealing with electric currents of opposite sign that are generating magnetic field in the opposite directions, thus decreasing the overall magnetic field. Therefore, thermal equilibrium of 2-SEL will favor anticorrelation of two electrons. It shall be underlined that, in accordance with the classical thermodynamics that applies to the case of two electrons treated classically, the movement of electron with certain acceleration will cause the occurrence of non-zero Poynting vector into the space and thus electron's energy will be emitted in the form of electromagnetic radiation. In such way one can introduce effective dissipative term
to the movement of electrons and it will cause the system mechanical energy to eventually vanish. After sufficiently long time the electrons will stop their oscillatory movement and they will move into ground state that is perfectly anticorrelated and corresponds to the case when
$x_2=x_1=\pm x_{max}$ and $\frac{d}{dt}x_1=\frac{d}{dt}x_2$ and when $\frac{d^2}{dt^2}x_1=0=\frac{d^2}{dt^2}x_2=0$. It is also worth mentioning that the ground state of two classical electrons in 2-SELs is different from the quantum ground state of 2-SELs.

A case described by two perfectly anticorrelated electrons at any distance in the classical treatment that is independent of the system energy. Such situation does not take place in the quantum case as treated by the tight-binding model given by formula (\ref{CC}) that has discrete spectra of energies as specified by (\ref{eigenergiessSEL}).

We can write the equations of motion of two electrons assuming the existence of antisymmetric case $x_1(t)=-x_2(t)$ at all instances of motion for the system symmetric around $x=0$.
From the equation
\begin{eqnarray}
mv^2(t)+\frac{q^2}{\sqrt{d^2+4x(t)^2}}=E_c={\rm const}>0, \nonumber
\end{eqnarray}
we obtain the equation
$\sqrt{\frac{q^4}{(E_c-mv^2(t))^2}-d^2}=2 x(t)$ and consequently we obtain the equation of motion
\begin{eqnarray}
m\frac{dv}{dt}= 
\frac{1}{2} \sqrt{\frac{q^4}{(E_c-m v^2(t))^2}-d^2} \cdot q^2 (E_c\!-\!m v^2(t))^3 \frac{1}{q^6}= \nonumber \\ \frac{1}{2q^4}\sqrt{q^4(E_{c}-mv^2(t))^4-d^2(E_{c}-mv^2(t))^6}. 
\end{eqnarray}
Finally we obtain the equation
\begin{equation}
\frac{dv}{\sqrt{q^4(E_{c}\!-\!mv^2(t))^4-d^2(E_{c}\!-\!mv^2(t))^6}} = dt\frac{1}{2 m q^4} 
\end{equation}
We introduce a new variable $u=\frac{d}{q}(E_{c}-mv^2)$. We have $du=-2\frac{m d}{q} v dv$. We also notice that $\sqrt{(\frac{E_c}{m}-\frac{q}{m d}u)}=v$. The last expressions imply
\begin{eqnarray}
dv=-\frac{q}{ 2 md }\frac{du}{v}= 
-\frac{q}{ 2 md }\frac{du}{\sqrt{(\frac{E_c}{m}-\frac{q}{m d}u)}}=-\frac{\sqrt{q}}{ 2 \sqrt{ md} }\frac{du}{\sqrt{(\frac{E_c d}{q}-u)}}.
\end{eqnarray}
The last expression allows us to write integral
\begin{eqnarray}
\int \frac{dv}{\sqrt{q^4(E_{c}-mv^2))^4-d^2(E_{c}-mv^2)^6}}=
\frac{ d^2}{q^4} \int \frac{du}{\sqrt{(\frac{E_c d}{q}-u)}} \frac{1}{u^2 \sqrt{1-u^2}}= \nonumber \\
=s_1 \int \frac{du}{\sqrt{(s-u)}} \frac{1}{u^2 \sqrt{1-u^2}}. 
\end{eqnarray}
Setting $s_1=\frac{ d^2}{q^4}$ and $s=\frac{E_c d}{q}$ we obtain the integral $s_1 \int \frac{du}{\sqrt{(s-u)}} \frac{1}{u^2 \sqrt{1-u^2}}$ that has a solution as three types of elliptic functions given in Appendix~\ref{app:b}.

\subsection{Classical Weak Measurement on 2-SEL System}

Measurement on a given physical system is about introducing an interaction of it with an external physical system that acts as a probe. If this interaction is strong (weak) we are dealing with a strong (weak) measurement.
We shall introduce an external charged particle at a certain distance that can move only in parallel to the system being probed and then we apply Newtonian equations of motion. For the sake of simplicity, we consider only interaction of the probe that is moving electron across one line with nearest charged particle, as depicted in Fig.\,\ref{fig:wm}. At a first level of approximation, the movement of external electron is the perturbation to the physical system of two electrons (2-SELs).


\begin{figure}[b!]
\includegraphics[scale=0.6]{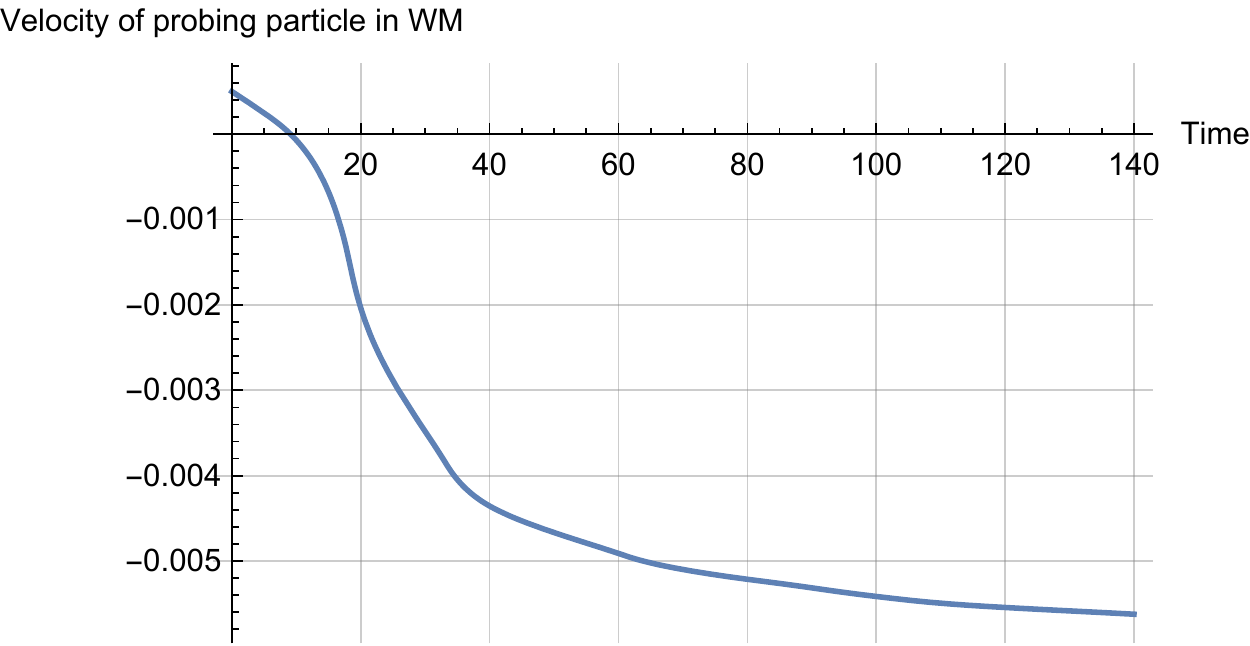}
\includegraphics[scale=0.6]{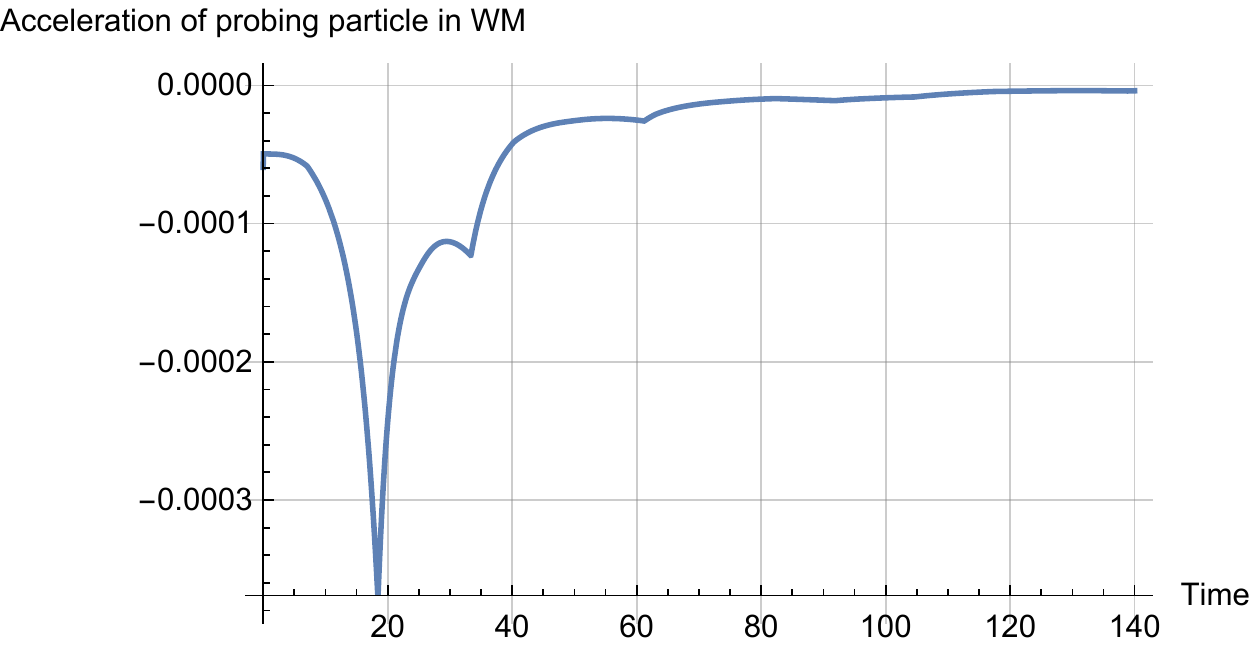}
\caption{Case of the classical measurement with electron used for probing of 2-SELs.}
\includegraphics[scale=0.6]{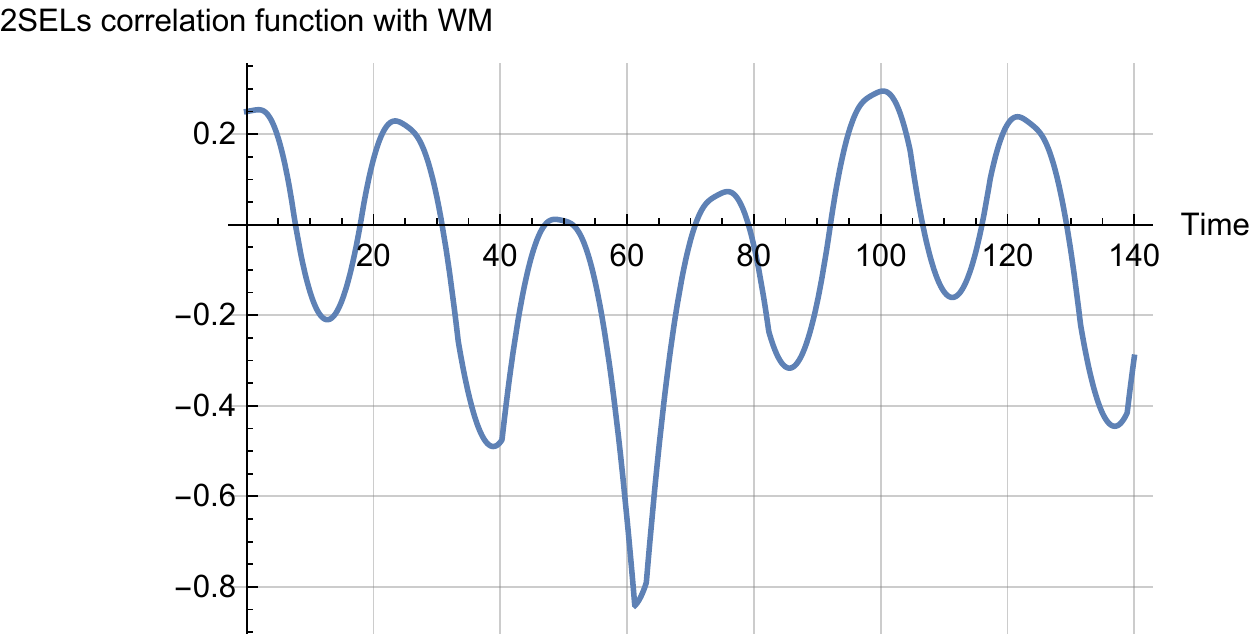}
\caption{Correlation function for 2-SELs under classical weak measurement from external probing charged particle. One shall refer to the bottom plot of Fig.\,\ref{CorrelationNoWM} and to Fig.\ref{fig:wm}.}
\end{figure}

\begin{figure}[b!]
\includegraphics[width=1.0\linewidth]{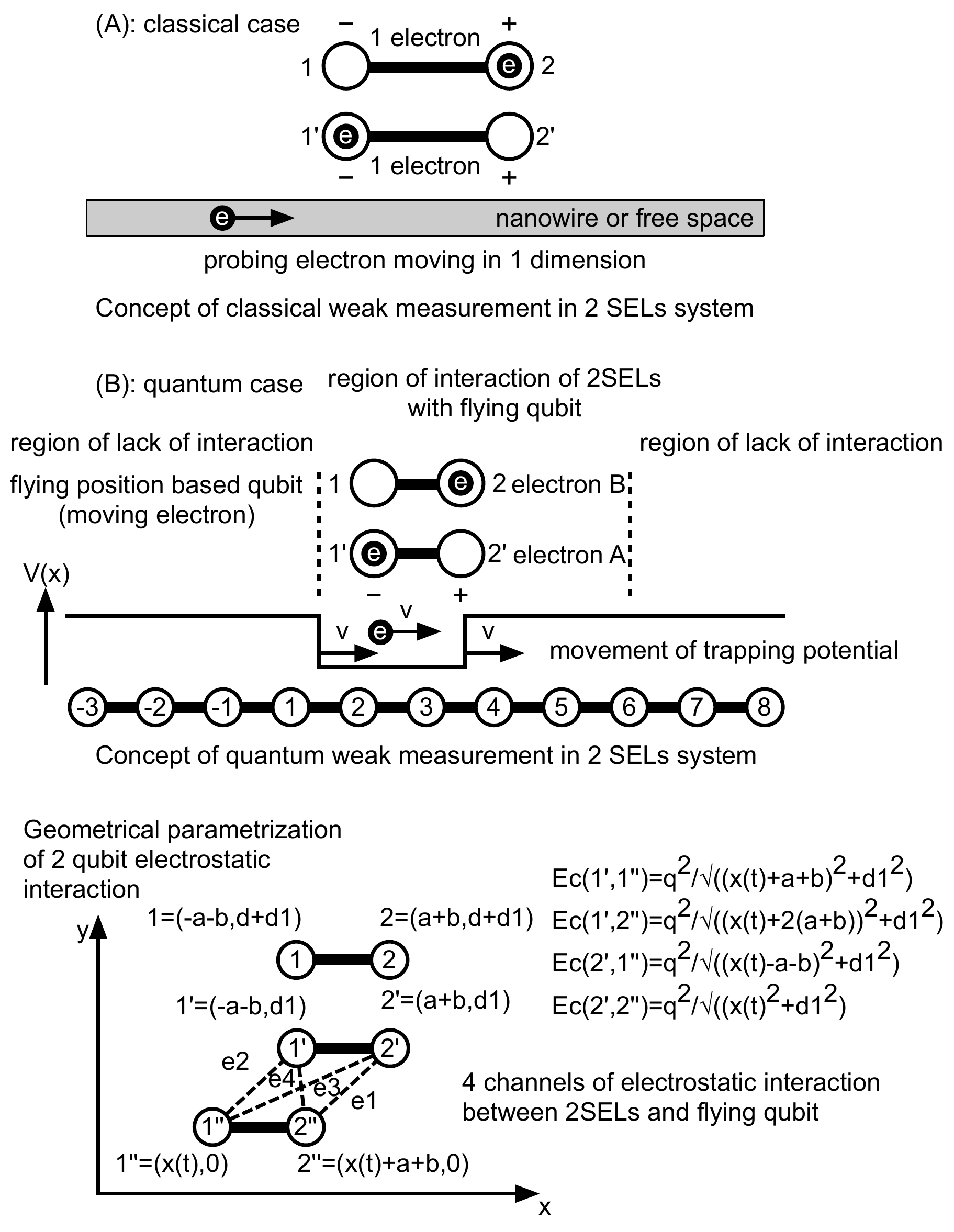}
\caption{Concept of classical and quantum weak measurements in a double single-electron line system. All simulations were conducted for the classical case.}
\label{fig:wm}
\end{figure}

%

\subsection{Weak Quantum Measurement on 2-SEL System}

We consider an interaction of two single-electron lines (2-SEL) that incorporate qubits A and B with an external line along which there is a movement of position-based qubit $C$. The CMOS structures have the capability to impose a constrained `movement' of a virtual qubit along single-electron lines. This way, the moving qubit becomes effectively a flying qubit, which is a term usually reserved for polarized photons participating in quantum information processing.
At a very far distance, there is no interaction between the flying qubit and 2-SELs. In such a case one can have a tensor of two density matrices being a density matrix of 2-SELs denoted by $\rho_{AB}$ and the external flying qubit.
We have a three-body quantum density matrix given as
\begin{eqnarray}
\hat{\rho}_{ABC}=\hat{\rho}_C \times \hat{\rho}_{AB}=
\begin{pmatrix}
\rho_{C}[1,1] \hat{\rho}_{AB} &  \rho_{C}[1,2] \rho_{AB} \\
\rho_{C}[2,1] \hat{\rho}_{AB} &  \rho_{C}[2,2] \rho_{AB} \\
\end{pmatrix}= 
\begin{pmatrix}
\hat{A}_1 & \hat{B}_1 \\
\hat{C}_1 & \hat{D}_1 \\
\end{pmatrix}.
\end{eqnarray}
We immediately recognize that we can obtain the density matrix of particle C by tracing out the existence of density matrix AB
\begin{equation}
\hat{\rho}_C=\sum_{i_A=\{1,2\},j_B=\{1',2'\}} \bra{i_A,j_B}\hat{\rho}_{ABC}\ket{i_A,j_B}.
\end{equation} In similar way we obtain the density matrix for 2-SEL system
\begin{equation}
\hat{\rho}_{AB}=\sum_{k_C=\{1,2\}} \bra{k_C}\hat{\rho}_{ABC}\ket{k_C}.
\label{detailedDM}
\end{equation}

The last expressions can be expressed by formula
\begin{eqnarray}
\hat{\rho}_C=
\begin{pmatrix}
Tr(\hat{A}_1) & Tr(\hat{B}_1) \\
Tr(\hat{C}_1) & Tr(\hat{D}_1) \\
\end{pmatrix},
\hat{\rho}_{AB}=\hat{A}_1+\hat{D}_1.
\end{eqnarray}
System of 2-SELs with the flying qubit can be reagarded as non-dissipative system and thus one can write the following equations of motion
\begin{equation}
\rho(t)= e^{\frac{1}{-i \hbar}H_0 t} e^{\frac{1}{i \hbar}\int_{0}^{t}\hat{H}(t')dt'}\rho(t)e^{\frac{1}{-i \hbar}\int_{0}^{t}\hat{H}(t')dt'}
e^{\frac{1}{-i \hbar}H_0 t},
\end{equation}
where $H_0$ is a time-independent Hamiltonian of isolated 2-SELs and isolated external qubit, while $H(t')$ stands for electrostatic interaction between the flying qubit and 2-SELs. We have the total system Hamiltonian having time-independent and time-dependent components
\begin{eqnarray}
\hat{H}(t)=\hat{H}_0+\hat{H}_1(t)
=(\hat{I}_C \times \hat{H}_{AB}+\hat{H}_{C} \times \hat{I}_{AB})_0 + \hat{H}_{AC}(t) \times \hat{I}_{B},
\label{DHC}
\end{eqnarray}
where $\hat{I}_{AB}$ and $\hat{I}_{C}$ are identity matrices acting on the 2-SELs and flying qubit, while $\hat{H}_{AB}$ is 2-SEL Hamiltonian. $\hat{H}_C$ is the flying qubit Hamiltonian and $\hat{H}_{AC}(t)$ is the interaction Hamiltonian between A line and flying qubit C (note: for the sake of simplicity we neglect the interaction between B line and C qubit). The detailed structure of those Hamiltonians are given in Appendix C.

Defining 2-SEL correlation function previously defined by formula (\ref{CC}), so $C=C_{AB}$, incorporated into three-body system takes form as $C_{AB,C}=\hat{I}_C \times \hat{C}_{AB}$ and, consequently, we obtain the following time dependence of correlation function given as
\begin{equation}
C(t)=Tr(C_{AB,C}\rho(t)).
\end{equation}
Details of the calculations can be found in Appendix C. Finally, we obtain the formula for correlation function of the 2-SEL system interacting weakly with the flying qubit in the form as

\vspace{-3mm}

\begin{eqnarray}
C(t)=\frac{(E_{c1}-E_{c2})^2-4 \cos \left(\frac{t \sqrt{(E_{c1}-E_{c2})^2+16}}{\hbar}\right) \left(\cos \left(\int_{0}^{t}dt'\frac{E_{c11''}(t')-E_{c1''2}(t')}{\hbar}\right)+\cos
  \left(\int_{0}^{t}dt'\frac{E_{c2''1}(t')-E_{c2pp2}(t')}{\hbar}\right)-2\right)}{(E_{c1}-E_{c2})^2+16} \nonumber \\ +\frac{4 \cos \left(\int_{0}^{t}dt'\frac{E_{c11''}(t')-E_{c1''2}(t')}{\hbar}\right)+4 \cos
  \left(\int_{0}^{t}dt'\frac{E_{c2''1}(t')-E_{c2''2}(t')}{\hbar}\right)+8}{(E_{c1}-E_{c2})^2+16}\nonumber \\
\end{eqnarray}
%
where
\begin{eqnarray}
E_{c11''}(t)=\frac{q^2}{\sqrt{(x(t)+a+b)^2+d_1^2}}, \nonumber \\
E_{c12''}(t)=\frac{q^2}{\sqrt{(x(t)+2(a+b))^2+d_1^2}}, \nonumber \\
E_{c21''}(t)=\frac{q^2}{\sqrt{(x(t)-(a+b))^2+d_1^2}}, \nonumber \\
E_{c22''}(t)=\frac{q^2}{\sqrt{(x(t))^2+d_1^2}}.
\end{eqnarray}
The movement of the flying qubit can be described, for example, by a constat velocity $v=v_0$, so $x(t)=x_0+v_0 t$. In case of a time-dependent flying qubit, $x(t)=x(t_0)+\int_{t_0}^{t}v_f(t')dt'$, where $v_f(t)$ is an instantaneous speed of the flying qubit. The only assumption for this model is that particle at time $t=0$ is at a far distance from 2-SELs.


\section{Entangling two qubits by means of RF fields}
\begin{figure}
    \centering
\includegraphics[width=3.0in]{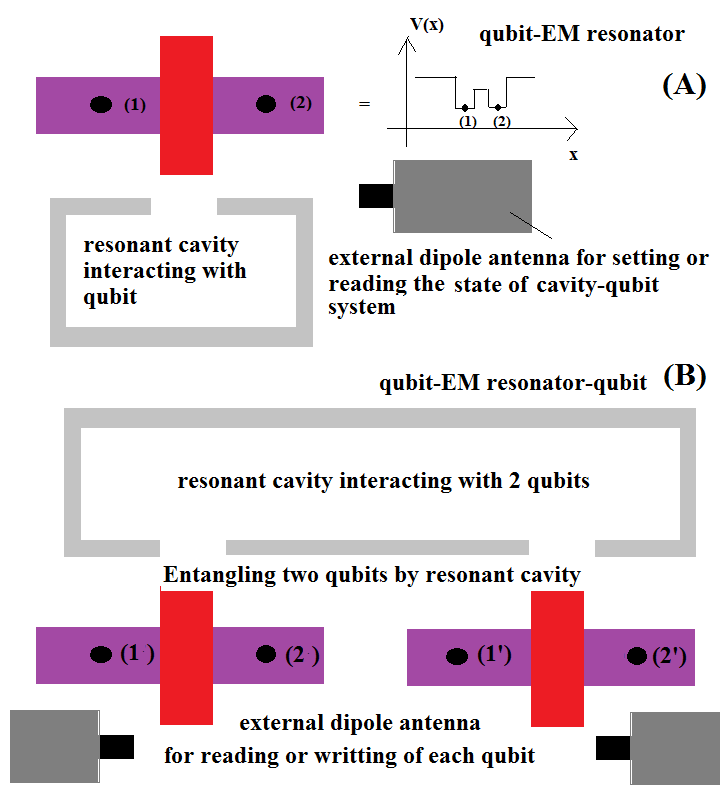} 
\caption{Position based qubit in RF field (A) and position based qubits placed at high distance interlinked by waveguide (B). }
\label{waveguide}
\end{figure}

We are placing electrostatic position-based qubit in external electromagnetic cavity. We assume that electromagnetic cavity maintains quantum coherence what is possible in case of cavities with small dissipation (high-quality factor) as it is the case of superconducting cavity. At the same time we are assuming quantum coherence of semiconductor position based qubit. What is more we assume coherehent interaction of electromagnetic radiation with electron trapped in positon based qubit. Such system is depicted in Fig.\ref{waveguide}. We expect that during this interaction it will be possible to entangle electromagnetic radiation with position based qubit. This entanglement will be essential in quantum information processing.
Before moving to the detailed picture of qubit-radiation interaction let us review non-local realism in quantum mechanics.
\subsection{Non-local realism in quantum mechanical picture}
Quantum mechanics gives only probabilistic description of physical processes what does not support classical determinism but only stochastic determinism. Given particle can be localized in certain area of space as when it is in the potential minimum that is around certain point or can be distributed over big area as it is the case of conductive electron in metal. Once the measurement is conducted on the particle its position can be determined very exactly but at the prize that particle momentum is highly perturbed and essentialy infomation about particle momentum is lost. In that way one cannot fully determine both position and momentum of the particle what is expressed in the non-commutation relation between momentum and position and it leads to the Heisenberg principle. The phenomena that one cannot determine position and momentum of the particle is commonly known from wave mechanics. Under the circumstance of particle being localized or delocalized the particles interact what affects the probability distribution. In very real sense quantum particle is like classical particle under very high noise so it is pointless to talk about the individual particle position but it makes sense to talk about probability of finding particle in given ensemble of particles. We use to say that canonical ensemble is attached to the individual behaviour of particle. Thus dealing with conglomerate of particles we
are dealing with statistical ensemble [of single particle] attached to another statistical ensemble of environment in which the given particle is placed. Such reasoning indeed draws analogies of statistical mechanics with quantum mechanics. At some point one can say that there is no big difference between quantum mechanical or classical particle under the impact of external potential. Local principle holds for both classical and quantum pictures and two particles interact if they are close one to another.  Coulomb electrostatic energy has the same formula both in classical and in quantum picture. However first main difference is the fact that quantum particle can be subjected to the self-interference as it is the case of two slit experiment when given wave (quantum particle) appears in certain regions with higher probability (higher wave intensity) and in other regions with lower probability. Self-interference requires that wavefunction of given particle is coherent what is strongly dependent on the environment. Self interference has classical counterpart in the theory of waves as given electromagnetic wave can interfere with itself. There is however the effect that has no classical counterpart in quantum picture and is named as entanglement that is the manifestation of non-local correlation. In classical physics it is however not suprising that when two particles are interacting the change of state of one particle brings the change of state of another particle. However the surprising aspect is when two particles being  at very high distances are essentially no interacting and change of the state of one of particles is affecting the state of another particles in immediate way. Such event is called spooky action on the distance and is the example of non-local correlation that can only occur in quantum theory and is the manifestation of particle entanglement. In this work we will describe the entanglement between waveguide and position based qubits as well as entanglement between two far position based qubits mediated by waveguide.Most common picture of entanglement is illustrated by the Bell states.

\subsection{Interaction of radiation with position dependent qubit}
We are referring to the situation of placement of position based qubit in external radiofrequency field
of electromagnetic cavity or waveguide as depicted in Fig.\ref{waveguide}.
Because of simplicity we are going to use Jaynes-Cumming Hamiltonian \cite{Jaynes} that describes the interactiton
atom with cavity by means of electromagnetic field [more precise name can be tight-binding Jaynes-Cumming Hamiltonian or Hubbard Jaynes-Cumming Hamiltonian]. In the simplest approach
the cavity Hamiltonian describing waveguide without dissipation is represented as
\begin{equation}
H_{cavity} = \hbar \omega_c (\frac{1}{2}+\hat{a}^{\dag}\hat{a}) = E_{\phi1} \ket{E_{\phi1}}\bra{E_{\phi1}} +E_{\phi2} \ket{E_{\phi2}}\bra{E_{\phi2}},  
\end{equation}
where $\hat{a}^{\dag}$ ($\hat{a}$) is the photon creation (annihilation) operator and number of photons in cavity is given as $n=\hat{a}^{\dag}\hat{a}$. At the same we can represent the two level qubit
system
\begin{equation}
H_{qubit} = E_g \ket{g}\bra{g} +E_e \ket{e}\bra{e}. 
\end{equation}
The interaction Hamilonian is of the following form
\begin{equation}
H_{qubit-cavity} = g(\hat{a}^{\dag}\sigma_{-}+\hat{a}\sigma_{+}),   
\end{equation}
where $\sigma_{-}=\sigma_{1}-i \sigma_{2}$, $\sigma_{+}=\sigma_{1}+i \sigma_{2}$.
The qubity-cavity interaction has the electric-dipole nature so quasiclassicaly we can write
\begin{equation}
H_{qubit-cavity} = \hat{d} \cdot \hat{E} = g (\sigma_{-} + \sigma_{+})(\hat{a}+\hat{a}^{\dag})\approx g(\hat{a}^{\dag}\sigma_{-}+\hat{a}\sigma_{+}).
\end{equation}
Here we have neglected the terms $g (\sigma_{-}\hat{a}+ \sigma_{+}\hat{a}^{\dag})$ and our approach is known as rotating phase. Constant g is depending on the distance between waveguide and position-dependent qubit as depicted Fig.2.
During photon emission from qubit the energy level is lowered and reversely during photon absorption the energy level of qubit is raised what is seen in the term $\hat{a}\sigma_{+}$. The system Hamiltonian is given as
$H=H_{cavity}+H_{qubit}+H_{qubit-cavity}$.
It is not hard to construct the Hilbert space for Jaynes-Cumming Hamiltonian.
Essentially we are considering the tensor product of qubit space and cavity space.
\begin{eqnarray}
\ket{\psi}=\gamma_1\ket{\phi_1}\ket{0}+\gamma_2\ket{\phi_1}\ket{1}+\gamma_3\ket{\phi_2}\ket{0}+\gamma_4\ket{\phi_2}\ket{1}=
\begin{pmatrix}
\gamma_1  \\
\gamma_2  \\
\gamma_3  \\
\gamma_4  \\
\end{pmatrix},\nonumber \\
1=\braket{\psi}{\psi} =|\gamma_1|^2+..+|\gamma_4|^2.
\end{eqnarray}
Here $\ket{0}=\ket{g}$ and $\ket{1}=\ket{e}$ stands for $E_g$ and $E_e$ energetic state of position based qubit, while $\ket{\phi_1}$ and $\ket{\phi_2}$ stands for cavity with 1 and 2 photons.
We have the following matrices $H_{qubit}+H_{cavity}$, $H_{qubit-cavity}$
\begin{eqnarray}
H_{qubit}+H_{cavity}=\nonumber \\
=
\begin{pmatrix}
E_g+E_{ph1} & 0 & 0 & 0 \\
0 & E_e+E_{ph1} & 0 & 0 \\
0 & 0 & E_g+E_{ph2} & 0 \\
0 & 0 & 0 & E_e+E_{ph2} \\
\end{pmatrix}=  
E_g\ket{E_g}\bra{E_g}+E_e\ket{E_e}\bra{E_e}+E_{\phi_{1(2)}}\ket{E_{\phi_{1(2)}}}\bra{E_{\phi_{1(2)}}}. 
\end{eqnarray}
\begin{eqnarray}
H_{qubit-cavity}=
\begin{pmatrix}
0 & 0 & 0 & 0 \\
0 & 0 & g_1 & 0 \\
0 & g_1 & 0 & 0 \\
0 & 0 & 0 & 0 \\
\end{pmatrix}= 
=g_1(\ket{E_{\phi_1}, E_e}\bra{E_{\phi_2}, E_g}+\ket{E_{\phi_2}, E_g}\bra{E_{\phi_1}, E_e}),
\end{eqnarray}
what implies
\begin{eqnarray}
H_{qubit}+H_{cavity}+H_{qubit-cavity} 
=
\begin{pmatrix}
E_g+E_{ph1} & 0 & 0 & 0 \\
0 & E_e+E_{ph1} & g_1 & 0 \\
0 & g_1 & E_g+E_{ph2} & 0 \\
0 & 0 & 0 & E_e+E_{ph2} \\
\end{pmatrix}.
\end{eqnarray}
The last Hamiltonian gives the eigenstates
\begin{eqnarray}
|E_1>=
\begin{pmatrix}
1 \\
0 \\
0 \\
0 \\
\end{pmatrix}, 
|E_2>=
\begin{pmatrix}
0 \\
0 \\
0 \\
1 \\
\end{pmatrix}, 
\nonumber \\
|E_3>=
\begin{pmatrix}
0 \\
\frac{(E_e-E_g)-(E_{ph_2}-E_{ph_1})-\sqrt{((E_e-E_g)-(E_{ph2}-E_{ph1}))^2+4|g_1|^2}}{2g_1} \\
1 \\
0 \\
\end{pmatrix}=, \nonumber \\
\frac{(E_e-E_g)-(E_{ph_2}-E_{ph_1})-\sqrt{((E_e-E_g)-(E_{ph2}-E_{ph1}))^2+4|g_1|^2}}{2g_1}|\phi_1>|E_e> 
+ |\phi_2>|E_g>,    \nonumber \\
|E_4>=
\begin{pmatrix}
0 \\
\frac{(E_e-E_g)-(E_{ph_2}-E_{ph_1})+\sqrt{((E_e-E_g)-(E_{ph2}-E_{ph1}))^2+4|g_1|^2}}{2g_1} \\
1 \\
0 \\
\end{pmatrix},
\end{eqnarray}
\normalsize
and one obtains eigenenergies of the form
\begin{eqnarray}
E_1=E_g+E_{ph1},
\nonumber \\
E_2=E_e+E_{ph2}, \nonumber \\
 E_3=\frac{1}{2}(E_g+E_e+E_{ph1}+E_{ph2} \nonumber \\
 -\sqrt{((E_e-E_g)-(E_{ph2}-E_{ph1}))^2+4|g_1|^2}, \nonumber \\
 E_4=\frac{1}{2}(E_g+E_e+E_{ph1}+E_{ph2} \nonumber \\
 +\sqrt{((E_e-E_g)-(E_{ph2}-E_{ph1}))^2+4|g_1|^2}, \nonumber \\
\end{eqnarray}
We recognize that state correponding to eigenenergies $E_3$ and $E_4$ are entangled states of matter and radiation while states correponding to eigenenergies $E_1$ and $E_2$ are non-entangled states of matter and radiation.
In particular if state  $E_3$ is subjected to the measurment of number of photons and value 1 was encountered than it implies that position based qubit is in the excited state correspondig to the energy $E_e$ . Otherwise if the number of photon is encountered to be 2 than the state of qubit is enconuntered to be $E_g$.
\subsection{Case of 2 qubits interaction via waveguide on the distance and teleportation on the distance}
We have the following Hamiltonian for 2 qubits interacting with waveguide in the case when qubit 1 is relatively far from qubit 2. If waveguide has L lenght and c is speed of signal propagation along waveguide we have $\Delta t=L/c$ and Hamiltonian is of the form:
\begin{eqnarray}
H=(E_{\phi1}|\phi_1><\phi_1|+E_{\phi2}|\phi_2><\phi_2|)I_{qubit1}I_{qubit2}+ \nonumber \\
+I_{cavity}(E_{g1}|g1><g1|+E_{e1}|e1><e1|)I_{qubit2}+ \nonumber \\ + I_{cavity}(E_{g1}|g1><g1|+E_{e1}|e1><e1|)I_{qubit2}+ \nonumber \\
 + I_{cavity}I_{qubit1}(E_{g2}|g2><g2|+E_{e2}|e2><e2|) + \nonumber \\  +g_1f_1(t)[(|\phi_1><\phi_2|)(|e1><g1|)+ \nonumber \\
 +(|\phi_2><\phi_1|)(|g1><e1|)]I_{qubit2}+ \nonumber \\ +g_2f_1(t+\Delta t)[(|\phi_1><\phi_2|)I_{qubit1}(|e2><g2|)+
 \nonumber \\
 +(|\phi_2><\phi_1|)I_{qubit1}(|g2><e2|)].
\end{eqnarray}
It is formally 3 interating body system (qubit1)-(waveguide)-(qubit2) in which qubit 1 cannot directly interact with qubit 2 and the quantum state has the form
\begin{eqnarray}
|\psi(t)>=\alpha_1(t)|\phi_1>|g1>|g2>+\alpha_2(t)|\phi_1>|g1>|e2>+ \nonumber \\
+\alpha_3(t)|\phi_1>|e1>|g2>+\alpha_4(t)|\phi_1>|e1>|e2>+ \nonumber \\
\alpha_5(t)|\phi_2>|g1>|g2>+\alpha_6(t)|\phi_2>|g1>|e2>+ \nonumber \\
\alpha_7(t)|\phi_2>|e1>|g2>+\alpha_8(t)|\phi_2>|e1>|e2>,
\end{eqnarray}
The normalization condition is fullfilled $|\alpha_1(t)|^2+..|\alpha_8(t)|^2=1$. The system matrix is of the structure given below
\tiny
\begin{eqnarray}
\tiny{H}=
\begin{pmatrix}
E_{g1}+E_{g2}+E_{\phi1} & 0 & 0 & 0 & 0  & 0 & 0 & 0 \\
0                                          & E_{g1}+E_{e2}+E_{\phi1}  & 0 & 0 & f_1(t)e^{-i d_2 t} g_2  & 0 & 0 & 0 \\
0                                         & 0                                             & E_{e1}+E_{g2}+E_{\phi1} & 0 & f_1(t)g_1e^{-i d_1 t}  & 0 & 0 & 0 \\
0 & 0 & 0 & E_{e1}+E_{e2}+E_{\phi1} & 0 & g_1 f_1(t)e^{-i d_1 t}  & g_2f_1(t)e^{-i d_2 t} & 0  \\
0 & f_1(t)e^{i d_2 t} g_2 & f_1 (t)e^{i d_1 t} g_1  &  0 & E_{g1}+E_{g2}+E_{\phi2} & 0 & 0 & 0 \\
0 & 0 & 0 & g_1 f_1(t)e^{i d_1 t} & 0  & E_{g1}+E_{e2}+E_{\phi2} & 0 & 0 \\
0 & 0 & 0 & f_1(t)e^{i d_2 t} g_2 & 0  & 0 & E_{e1}+E_{g2}+E_{\phi2} & 0 \\
0 & 0 & 0 & 0 & 0  & 0 & 0 & E_{e1}+E_{e2}+E_{\phi2} \\
\end{pmatrix}
\end{eqnarray}
\normalsize
This matrix can be simplified. We can preassume that $g_1f_1(t)=g f(t)e^{id_1(t)}$ and $g_2f_2(t)=g f(t)e^{id_2(t)}$  
and we can divide all matrix by this value. Second simplification is by $E_g=E_{g1}=E_{g2}=E_{\phi1}=E_{\phi2}-E_{\phi1}=E_{e1}-E_{g1}=E_{e2}-E_{g2}$. In such case we obtain 
\begin{eqnarray}\hat{H}=
\begin{pmatrix}
3E_g & 0    & 0    & 0 & 0  & 0 & 0 & 0 \\
0    & 4E_g & 0    & 0 & g_2 e^{-i d_2 (t)}  & 0 & 0 & 0 \\
0    & 0    & 4E_g & 0 & g_1 e^{-i d_1 (t)}  & 0 & 0 & 0 \\
0    & 0    & 0    & 5E_g & 0 & g_1 e^{-id_1t}  & g_2 e^{-id_2(t)} & 0  \\
0 & g_2 e^{i d_2 (t)} & g_1 e^{i d_1 (t)}  &  0 & 4E_g & 0 & 0 & 0 \\
0 & 0 & 0 & g_1 e^{i d_1 (t)} & 0  & 5E_g & 0 & 0 \\
0 & 0 & 0 & g_2 e^{i d_2 (t)} & 0  & 0 & 5E_g & 0 \\
0 & 0 & 0 & 0 & 0  & 0 & 0 & 6E_g \\
\end{pmatrix}.
\end{eqnarray}
\normalsize
It shall be underlined that is $g_1$ and $g_2$ are proportional to the electric field in the resonator cavity so they are depending on frequency of oscillations and amplitude of electric field in resonator cavity. If we are dealing with 2 or more qubits we assume that they are at coupled to EM field in different way and that they catch oscillating EM field at different phase what is expressed by phase factors $e^{i d_1 (t)}$ , $e^{i d_1 (t)}$ .
The last Hamiltonian matrix has the following energy eigenvalues $3E_g,4E_g,5E_g,6E_g,4E_g-\sqrt{g_1^2+g_2^2},5E_g-\sqrt{g_1^2+g_2^2}$, $4E_g+\sqrt{g_1^2+g_2^2}$, $5E_g+\sqrt{g_1^2+g_2^2}$.
In general case $g_1$ is depending on how waveguide with hole is close to the position dependent qubit. Otherwise position dependent qubit must be placed in resonant cavity.
We assume $E_p=E_{p1}=E_{p2}=E_{p1'}=E_{p2'}$ and we have found the following eingestates
\begin{eqnarray}
|E_1>=
\begin{pmatrix}
1 \\
0 \\
0 \\
0 \\
0 \\
0 \\
0 \\
0 \\
\end{pmatrix}=|\phi_1>|g_1>|g_2>
=\frac{1}{2}\ket{\phi_1}(|x_1>-|x_2>)(|x_{1'}>-|x_{2'}>), \nonumber \\
\end{eqnarray}
\begin{eqnarray}
|E_2>=
\begin{pmatrix}
0 \\
-(g_1/g_2)e^{i(-d_2+d_1)} \\
+1 \\
0 \\
0 \\
0 \\
0 \\
0 \\
\end{pmatrix}=
-(\frac{g_1}{g_2}e^{i(-d_2+d_1)}|\phi_1>|g_1>|e_2>+|\phi_1>|e_1>|g_2> \nonumber \\
|E_3>= 
\begin{pmatrix}
0 \\
0 \\
0 \\
0 \\
0 \\
-\frac{g_2}{g_1}e^{i(-d_2+d_1)} \\
1 \\
0 \\
\end{pmatrix}= 
-\frac{g_2}{g_1}e^{i(-d_2+d_1)}|\phi_2>|g1>|e_2>+|\phi_2>|e1>|g_2>,
\end{eqnarray}
\begin{eqnarray}
|E_4>=
\begin{pmatrix}
0 \\
0 \\
0 \\
0 \\
0 \\
0 \\
0 \\
1 \\
\end{pmatrix}=+|\phi_2>|e_1>|e_2>, 
\end{eqnarray}
\begin{eqnarray}
|E_5>=
\begin{pmatrix}
0 \\
-\frac{g_2e^{id_2}}{\sqrt{g_1^2+g_2^2}} \\
-\frac{g_1e^{id_1}}{\sqrt{g_1^2+g_2^2}} \\
0 \\
1 \\
0 \\
0 \\
0 \\
\end{pmatrix}=\nonumber \\
-\frac{g_2e^{id_2}}{\sqrt{g_1^2+g_2^2}}|\phi_1>|g_1>|e_2> - \frac{g_1e^{id_1}}{\sqrt{g_1^2+g_2^2}}|\phi_1>|e_1>|g_2>
+\frac{1}{\sqrt{2}}|\phi_1>|e_1>|e_2>,
\end{eqnarray}
\begin{eqnarray}
|E_6>=
\begin{pmatrix}
0 \\
0 \\
0 \\
-\frac{e^{2id_2}\sqrt{g_1^2+g_2^2}}{g_2} \\
0 \\
1 \\
1 \\
0 \\
\end{pmatrix}=-\frac{e^{2id_2}\sqrt{g_1^2+g_2^2}}{g_2}|\phi_1>|e_1>|e_2>+ 
|\phi_2>|e_1>|g_2>+|\phi_2>|g_1>|e_2>,
\nonumber \\
|E_7>= 
\begin{pmatrix}
0 \\
e^{-id_2}\frac{g_2}{\sqrt{g_1^2+g_2^2}} \\
e^{-id_1}\frac{g_1}{\sqrt{g_1^2+g_2^2}} \\
0 \\
1 \\
0 \\
0 \\
0 \\
\end{pmatrix} = \nonumber \\
=e^{-id_2}\frac{g_2}{\sqrt{g_1^2+g_2^2}}(|\phi_1>|e_1>|g_2>+ 
e^{-id_1}\frac{g_1}{\sqrt{g_1^2+g_2^2}}(|\phi_1>|g_1>|e_2>+ 
|\phi_2>|g_1>|g_2>, \nonumber \\
\end{eqnarray}
\begin{eqnarray}
|E_8> =
\begin{pmatrix}
0 \\
0 \\
0 \\
e^{-i(2d_1+d_2)}\frac{\sqrt{g_1^2+g_2^2}}{g_2} \\
0 \\
e^{-id_2+id_1}\frac{g_1}{g_2} \\
1 \\
0 \\
\end{pmatrix} =\nonumber \\ e^{-i(2d_1+d_2)}\frac{\sqrt{g_1^2+g_2^2}}{g_2}|\phi_1>|e1>|e2>+ 
e^{-id_2+id_1}\frac{g_1}{g_2}|\phi_2>|e_1>|g_2>+|\phi_2>|g_1>|e_2>
\nonumber \\
\end{eqnarray},
6 eigenstates among 8 Eigenstates (except E1 and E8) are entangled in energy bases.

It is noticeable to underline that all 8 energy eigenstates are entangled in position based representation especially when all $E_p$ values corresponding to nodes in 2 different qubits are different.

\section{Analytic extensions of topology of chain of coupled quantum dots}
\begin{figure}
\centering
\includegraphics[scale=0.4]{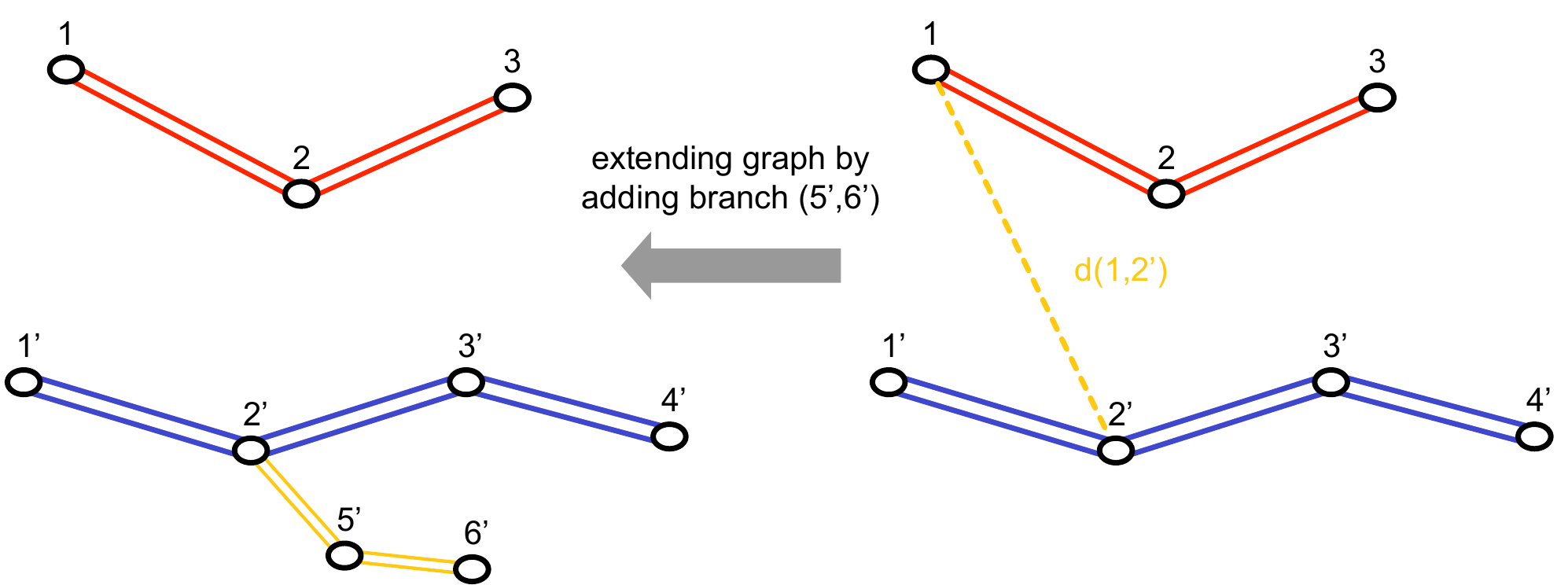} 
\caption{Example of arbitrary extension of network of electrostatically coupled quantum dots with reference to technological scheme depicted in Fig.\,\ref{fig:central}}
\label{fig:graph}
\end{figure}

Since we have electrostatic control of interaction between quantum dots we can turn on coupling between two chains of quantum dots as it is depicted in Fig.\ref{fig:graph},
where Coulomb electrostatic interaction occurs between m and n' node of two separated chain and is given by $E_c(m,n')=f(m,n')=\frac{q^2}{d_{m,n'}}$. The quantum state of right-system is given as
\begin{eqnarray}
\ket{\psi}=\gamma_{1,1'}(t)\ket{1}\ket{1'}+\gamma_{1,2'}(t)\ket{1}\ket{2'}+ \gamma_{1,3'}(t)\ket{1}\ket{3'}+\gamma_{1,4'}(t)\ket{1}\ket{4'}
+\gamma_{2,1'}(t)\ket{2}\ket{1'}+\gamma_{2,2'}(t)\ket{2}\ket{2'}+\nonumber \\+ \gamma_{2,3'}(t)\ket{2}\ket{3'}+\gamma_{2,4'}(t)\ket{2}\ket{4'}
+\gamma_{3,1'}(t)\ket{3}\ket{1'}+\gamma_{3,2'}(t)\ket{3}\ket{2'}+\gamma_{3,3'}(t)\ket{3}\ket{3'}+\gamma_{3,4'}(t)\ket{3}\ket{4'}.  \nonumber \\
\end{eqnarray}
where $\sum_{k,l'}|\gamma_{k,l'}|^2=1$.
After extension by 2 elements the quantum state of left system is given as
\begin{eqnarray}
\ket{\psi}=\gamma_{1,1'}(t)\ket{1}\ket{1'}+\gamma_{1,2'}(t)\ket{1}\ket{2'}+\gamma_{1,3'}(t)\ket{1}\ket{3'}+\gamma_{1,4'}(t)\ket{1}\ket{4'}
+\gamma_{2,1'}(t)\ket{2}\ket{1'}+\gamma_{2,2'}(t)\ket{2}\ket{2'}+ \nonumber \\ +\gamma_{2,3'}(t)\ket{2}\ket{3'}+\gamma_{2,4'}(t)\ket{2}\ket{4'} 
+\gamma_{3,1'}(t)\ket{3}\ket{1'}+\gamma_{3,2'}(t)\ket{3}\ket{2'}+\gamma_{3,3'}(t)\ket{3}\ket{3'}+\gamma_{3,4'}(t)\ket{3}\ket{4'}.  \nonumber \\
+\gamma_{1,5'}(t)\ket{1}\ket{5'} +\gamma_{2,5'}(t)\ket{2}\ket{5'}+ \gamma_{3,5'}(t) \ket{3}\ket{5'}
+\gamma_{1,6'}(t)\ket{1}\ket{6'}+\gamma_{2,6'}(t)\ket{2}\ket{6'}+\gamma_{3,6'}(t)\ket{3}\ket{6'}. \nonumber \\
\end{eqnarray}
where again  $\sum_{s,w'}|\gamma_{s,w'}|^2=1$.
The Hamiltonian of the system before extension is
\begin{equation}
\hat{H}=
\begin{pmatrix}
H_1 & H_2 \\
H_3 & H_4 \\
\end{pmatrix}
\end{equation}
and after extension into system depicted in Fig.\,\ref{fig:graph} (left side) is
$ $
\begin{equation}
\label{extension}
\hat{H}(t)_{ext}=
\begin{pmatrix}
H_1 & H_2  & H_{e1}  \\
H_3 & H_4  & H_{e2} \\
H_{e5} & H_{e4}  & H_{e3} \\
\end{pmatrix}
\end{equation}
with matrix subcomponents
$ \hat{H}_1(t)= $
\begin{eqnarray*}
\begin{pmatrix}
 E_{p1}+E_{p1'}+ \frac{q^2}{d_{1,1'}} & t_{s1',2'} & 0  & 0 & t_{s12} & 0 \\ 
 t_{s2',1'}                & E_{p1}+E_{p2'}+\frac{q^2}{d_{1,2'}} & t_{s2',3'}  & 0 & 0 & t_{s12} \\ 
 0                        & t_{s3',2'} & E_{p1}+E_{p3'}+ \frac{q^2}{d_{1,3'}}   & t_{s3',4'} & 0 & 0 \\ 
 0                        & 0 & t_{s4',3'}  & E_{p1}+E_{p4'}+\frac{q^2}{d_{1,4'}} & t_{s1',2'}  & 0 \\ 
 t_{s2,1}                        & 0 & 0          & 0 & E_{p2}+E_{p1'}+ \frac{q^2}{d_{1',2}} & t_{2',3'} \\ 
 0                        & t_{s2,1} & 0          & 0 & t_{s2',1'} &  E_{p2}+E_{p2'}+\frac{q^2}{d_{2,2'}} \\ 
 \end{pmatrix},
\end{eqnarray*}
\normalsize

\begin{eqnarray*}
\hat{H}_2= 
\begin{pmatrix}
0 & 0 & 0  & 0 & 0 & 0 \\ 
0  & 0 & 0 & 0 & 0 & 0 \\ 
t_{s1,2}  & 0 & 0  & 0 & 0 & 0 \\ 
0 & t_{s1,2} & 0  & 0 & 0 & 0 \\ 
0 & 0 &  t_{s2,3} & 0 & 0  & 0 \\ 
t_{s2',3'} & 0 & 0  &  t_{s2,3} & 0 &  0 \\ 
 \end{pmatrix}, 
\hat{H}_3= 
\begin{pmatrix}
0 & 0 & t_{s2,1}  & 0 & 0 & t_{s3',2'} \\ 
0 & 0 & 0  & t_{s2,1} & 0 & 0 \\ 
0 & 0 & 0  & 0 & t_{s3,2} & 0 \\ 
0 & 0 & 0  & 0 & 0 & t_{s3,2} \\ 
0 & 0 & 0  & 0 & 0 & 0 \\ 
0 & 0 & 0  & 0 & 0 &  0 \\ 
 \end{pmatrix}, \nonumber \\
\end{eqnarray*}

$\hat{H}_4(t)= $
\tiny
\begin{eqnarray*}
\begin{pmatrix}
 E_{p2}(t)+E_{p3'}(t)+ \frac{q^2}{d_{2,3'}} & t_{s3',4'}(t) & 0  & 0 & 0 & 0 \\ 
 t_{s4',3'}(t)            & E_{p2}(t)+E_{p4'}(t)+ \frac{q^2}{d_{2,3'}} & 0 & 0 & 0 & 0 \\ 
 0                        & 0 & E_{p3}(t)+E_{p1'}(t)+ \frac{q^2}{d_{2,3'}}  & t_{s1',2'}(t) & 0 & 0 \\ 
 0                        & 0 & t_{s2',1'}(t)  & E_{p3}(t)+E_{p2'}(t)+\frac{q^2}{d_{3,2'}} & t_{s2',3'}(t) & 0 \\ 
 0                        & 0 & 0          & t_{3',2'}(t) & E_{p3}(t)+E_{p3'}(t)+\frac{q^2}{d_{3,3'}} & t_{s3',4'}(t) \\ 
 0                        & 0 & 0          & 0 & t_{s4',3'}(t) &  E_{p3}(t)+E_{p4'}(t)+\frac{q^2}{d_{3,4'}} \\ 
 \end{pmatrix},
\end{eqnarray*}
\normalsize
\normalsize
We can determine inductive step of quantum dot graph extension by adding matrices $\hat{H}_{e1},..,\hat{H}_{e5}$ to the formula \ref{extension} in the form as given

\begin{eqnarray}
 \hat{H}_{e1}=  
 \begin{pmatrix}
 0 & 0 & 0 & 0 & 0 &0 \\
 t_{s2',5'} & 0 & 0 & 0 & 0 &0 \\
 0 & 0 & 0 & 0 & 0 &0 \\
 0 & 0 & 0 & 0 & 0 &0 \\
 0 & 0 & 0 & 0 & 0 &0 \\
 0 & t_{s2',5'} & 0 & 0 & 0 &0 \\
 \end{pmatrix},
 =\hat{H}_{e5}^{\dag}, 
 \hat{H}_{e2}= 
 \begin{pmatrix}
 0 & 0 & 0 & 0 & 0 &0 \\
 0 & 0 & 0 & 0 & 0 &0 \\
 0 & 0 & 0 & 0 & 0 &0 \\
 0 & 0 & t_{s2',5'} & 0 & 0 &0 \\
 0 & 0 & 0 & 0 & 0 &0 \\
 0 & 0 & 0 & 0 & 0 &0 \\
 \end{pmatrix}
 =\hat{H}_{e4}^{\dag},
\end{eqnarray}


\begin{eqnarray*}
\hat{H}_{e3}=
 \begin{pmatrix}
E_{p1}+E_{p5'}+ \frac{q^2}{d_{1,5'}} & t_{s1,2} & 0 & t_{s5',6'} & 0 &0 \\
 t_{s2,1} & E_{p2}+E_{p5'}+\frac{q^2}{d_{2,5'}} & t_{s2,3} & 0 & t_{s5',6'} &0 \\
 0 & t_{s3,2} & E_{p3}+E_{p5'}+\frac{q^2}{d_{3,5'}} & 0 & 0 & t_{s5',6'} \\
 t_{s6',5'} & 0 & 0 & E_{p1}+E_{p6'}+\frac{q^2}{d_{1,6'}} & t_{s1,2} &0 \\
 0 & t_{s6',5'} & 0 & t_{s2,1} & E_{p2}+E_{p6'}+\frac{q^2}{d_{2,6'}} &t_{s2,3} \\
 0 & 0 & t_{s6',5'} & 0 & t_{s3,2} & E_{p3}+E_{p6'}+\frac{q^2}{d_{3,6'}} \\
 \end{pmatrix}.
\end{eqnarray*}

\normalsize
Similarly to before having the knowledge of quantum state at $t_0$ we can evaluate the state at time $t$ by computing $\exp(\int_{t_0}^{t}\frac{1}{\hbar i}\hat{H}_{ext}(t)dt')\ket{\psi,t_0)}=\ket{\psi,t)}$ what bases on the same method already presented before in Equation 8. 
We can also perform the procedure of heating up or cooling down of the quantum state in the way as it was described before.
\normalsize
\section{Electrostatic interaction of Josephson junction qubit with semiconductor electrostatic qubit}
The state of Josephson junction is well described by Bogoliubov-de Gennes (BdGe) equation \cite{PSSB2012} pointing the correlation between electron and holes as
\begin{equation}
\begin{pmatrix}
H_0 & \Delta(x) \\
\Delta(x)^{*} & -H_0^{\dag} \\
\end{pmatrix}
\begin{pmatrix}
u_n(x) \\
v_n(x) \\
\end{pmatrix}
=E_n
\begin{pmatrix}
u_n(x) \\
v_n(x) \\
\end{pmatrix} ,
\end{equation}
where $H_0=-\frac{\hbar^2}{2m}\frac{d^2}{dx^2}$ is free electron Hamiltonian with self-consistency relation  $\Delta(x)=\sum_n(1-2f(E_n))u_n(x)v_n(x)^{*}$, where $\Delta(x)$ is the superconducting order parameter and $f(E_n)=\frac{1}{1+e^{-\frac{E_n}{k_bT}}}$ is Fermi-Dirac distribution function and $u_n(x)$ and $v_n(x)$ are electron and hole wavefunctions. In case of bulk superconductor with constant superconducting order parameter we obtain $E_n=\pm \sqrt{|H_0|^2+|\Delta|^2}$.
In later considerations we are going to omit the self-consistency relation assuming the depedence of superconducting order parameter as step-like function. It shall be underlined that BdGe equation is mean field
equation that is dervied basing on BCS theory of superconductivity. It it thus naturally valid for the case of many particles.
Semiconductor single electron line with 2 nodes can be regarded as electrostatic position dependent qubit and can be described by
$H_{semi}=t_{s1,2}\ket{1}\bra{2}+t_{s2,1}\ket{2}\bra{1}+E_{p1}\ket{1}\bra{1}+E_{p2}\ket{2}\bra{2}, $

\begin{figure}
\centering
\label{fig:JJvssemi}
\includegraphics[scale=0.5]{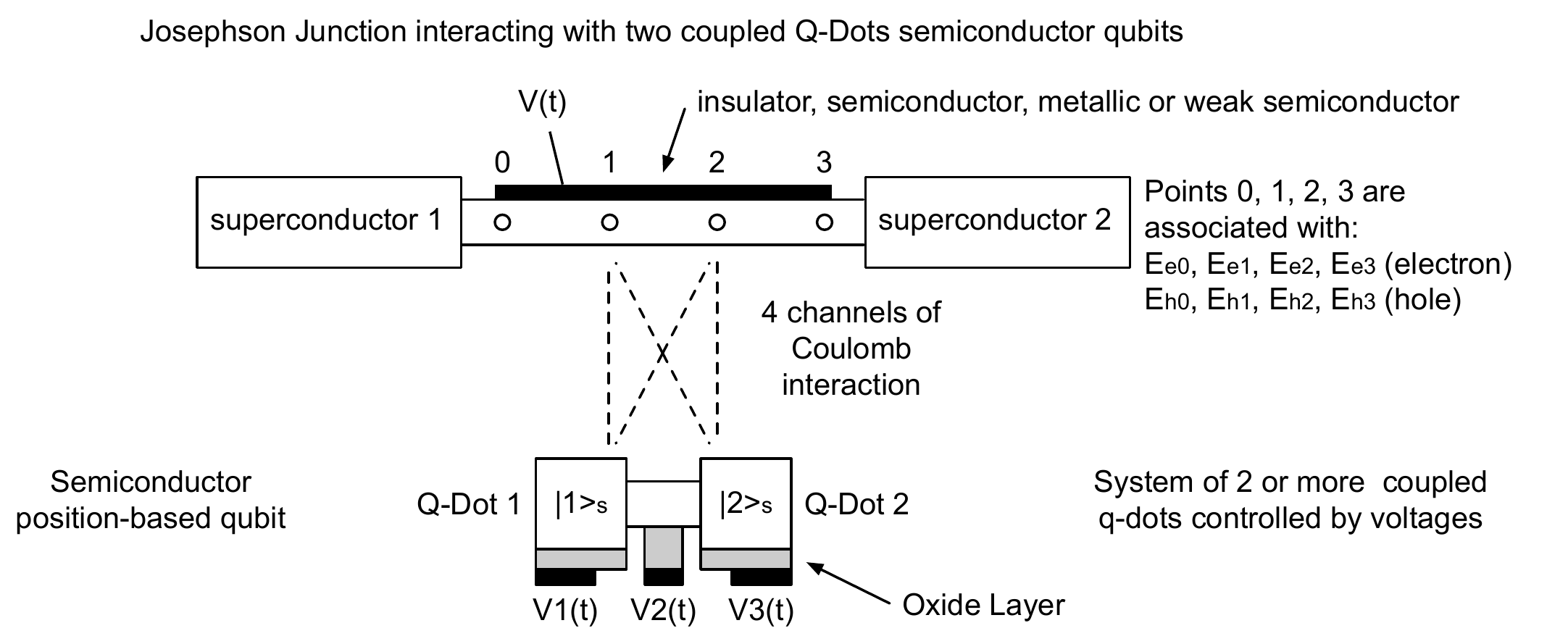} 
\caption{Superconducting Josephson junction interacting with semiconductor position based qubit in minimalistic tight-binding approach, where tight-binding BdGe equation describing Josephson junction is coupled electrostatically to tight-binding model of semiconductor position based qubit.}
\end{figure}

We refer to the physical situation depicted in Fig.8.
We can express coupling of 2 systems assuming 4 nodes for electron or hole and 2 nodes for electron confined in semiconductor so we have eigenvector having 16 components ($\ket{0}_e\ket{1}_s$, $\ket{0}_e\ket{2}_s$,$\ket{1}_e\ket{1}_s$, $\ket{1}_e\ket{2}_s$,$\ket{2}_e\ket{1}_s$, $\ket{2}_e\ket{2}_s $, $ \ket{3}_e\ket{1}_s, \ket{2}_e\ket{2}_s $ ), ($\ket{0}_h\ket{1}_s$, $\ket{0}_h\ket{2}_s$,$\ket{1}_h\ket{1}_s$, $\ket{1}_h\ket{2}_s$,$\ket{2}_h\ket{1}_s$, $\ket{2}_h\ket{2}_s $, $ \ket{3}_h\ket{1}_s, \ket{2}_h\ket{2}_s $ )
where s refers to semiconductor qubit whose quantum state is superposition of $
\ket{1}_s$ and $ \ket{2}_s$ and states $\ket{0}_e$, .., $\ket{3}_e$, $\ket{0}_h$, .., $\ket{3}_h$ characterizes the state of electron and hole respectively in ABS [Andreev Bound State when electron moving in normal (non-superconducting) region between superconducors is reflected as hole when it comes into superconducint area and when hole moving in normal region is reflected as electron when it meets superconductor etc .. ] of Josephson junction. This time the quantum state of the system can be written as
\begin{eqnarray}
\ket{\psi,t}= 
\gamma_1(t)\ket{0}_e\ket{1}_s +
\gamma_2(t)\ket{0}_e\ket{2}_s +\gamma_3(t)\ket{1}_e\ket{1}_s+ 
\gamma_4(t)\ket{1}_e\ket{2}_s+\gamma_5(t)\ket{2}_e\ket{1}_s+ 
\gamma_6(t)\ket{2}_e\ket{2}_s+
\gamma_7(t)\ket{2}_e\ket{1}_s \nonumber \\ +\gamma_8(t)\ket{2}_e\ket{2}_s+ 
\gamma_9(t)\ket{0}_h\ket{1}_s+\gamma_{10}(t)\ket{0}_h\ket{2}_s+\gamma_{11}(t)\ket{1}_h\ket{1}_s 
+\gamma_{12}(t)\ket{1}_h\ket{2}_s+\gamma_{13}(t)\ket{2}_h\ket{1}_s+\gamma_{14}(t)\ket{2}_h\ket{2}_s+ 
\nonumber \\ \gamma_{15}(t)\ket{2}_e\ket{1}_s+  
\gamma_{16}(t)\ket{2}_h\ket{2}_s. \nonumber \\
\end{eqnarray}
Normalization condition implies $|\gamma_1(t)|^2+|\gamma_2(t)|^2+..+|\gamma_{16}(t)|^2=1$ at any instance of time t. Such system has 16 eigenenergies. The probability of find electron at node 1 under any presence of electron in semiconductor qubit at node 1 or 2 is obtained by appling projection of $\bra{1}_e\bra{1}_s+\bra{1}_e\bra{2}_s$ so $|\bra{1}_e\bra{1}_s+\bra{1}_e\bra{2}_s \ket{\psi,t}|^2$ is probability of finding electron at node 1 in Josephson junction.
We obtain the following structures of matrices corresponding to $H_0$ part of BdGe equation in the forma as
\begin{eqnarray}
\hat{H}_{0[e]}=
\begin{pmatrix}
E_{p1} + E_{e0}                    & t_s                                             & t_{e(1,0)}                                           & 0                                                           & t_{e(2,0)}                                          & 0                                            & t_{e(3,0)}         & 0                                   \\
t_s^{*}                                    & E_{p2} +  E_{e0}                    & 0                                                          & t_{e(1,0)}                                            & 0                                                         & t_{e(2,0)}                             & 0                       & t_{e(3,0)}                     \\
t_{e(1,0)}^{*}                        & 0                                                & E_{p1} + \frac{q^2}{a} +  E_{e1}   & t_s                                                        & t_{e(2,1)}                                          & 0                                            & t_{e(3,1)}         & 0                                   \\
0                                               & t_{e(1,0)}^{*}                         & t_s^{*}                                                & E_{p2} +  E_{e1} + \frac{q^2}{b}   & 0                                                         & t_{e(2,1)}                             & 0                       & t_{e(3,1)}                    \\
t_{e(2,0)}^{*}                        & 0                                                & t_{e(2,1)}^{*}                                    & 0                                                           & E_{p1} +E_{e2} + \frac{q^2}{b}   & t_s                                         & t_{e(3,2)}        & 0                                   \\
0                                               &  t_{e(2,0)}^{*}                        & 0                                                           & t_{e(2,1)}^{*}                                    & t_s^{*}                                              & E_{p2} + E_{e2}+ \frac{q^2}{a} & 0                       & t_{e(3,2)}                    \\
t_{e(3,0)}^{*}                        & 0                                                &  t_{e(3,1)}^{*}                                   & 0                                                           & t_{e(3,2)}^{*}                                  & 0                                            & E_{p1} + E_{3e}       & t_s                                \\
0                                               & t_{e(3,0)}^{*}                         & 0                                                           &  t_{e(3,1)}^{*}                                   & 0                                                         & t_{e(3,2)}^{*}                     & t_s^{*}            & E_{p2} + E_{3e}           \\
\end{pmatrix} \nonumber \\
\end{eqnarray}
\normalsize
Parameters $E_{p1}$, $E_{p2}$, $t_s$ correspond to semiconductor position based qubit and distance between semiconductor qubit and Josephson junction is given by a and b. Other parameters $E_{e0},E_{e1}, E_{e2},E_{e3}$ , $E_{h0},E_{h1}, E_{h2},E_{h3}$
describes localization energy of electron and hole at nodes 0, 1, 2 and 3 of Josephson junction.
In analogical way we can write
\begin{eqnarray}
\hat{H}_{0[h]}=
\begin{pmatrix}
E_{p1} + E_{h0}                    & t_s                                             & t_{h(1,0)}                                           & 0                                                           & t_{h(2,0)}                                          & 0                                                         & t_{h(3,0)}                 & 0                                   \\
t_s^{*}                                    & E_{p2} +  E_{h0}                    & 0                                                          & t_{h(1,0)}                                            & 0                                                         & t_{h(2,0)}                                          & 0                                & t_{h(3,0)}                     \\
t_{h(1,0)}^{*}                        & 0                                                & E_{p1} - \frac{q^2}{a} +  E_{h1}   & t_s                                                        & t_{h(2,1)}                                          & 0                                                         & t_{h(3,1)}                 & 0                                   \\
0                                               & t_{h(1,0)}^{*}                         & t_s^{*}                                                & E_{p2} +  E_{h1} - \frac{q^2}{b}   & 0                                                         & t_{h(2,1)}                                          & 0                                & t_{h(3,1)}                    \\
t_{h(2,0)}^{*}                        & 0                                                & t_{h(2,1)}^{*}                                    & 0                                                           & E_{p1} +E_{h2} - \frac{q^2}{b}   & t_s                                                      & t_{h(3,2)}                 & 0                                   \\
0                                               &  t_{h(2,0)}^{*}                        & 0                                                           & t_{h(2,1)}^{*}                                    & t_s^{*}                                              & E_{p2} + E_{h2}- \frac{q^2}{a}   & 0                                & t_{h(3,2)}                    \\
t_{h(3,0)}^{*}                        & 0                                                &  t_{h(3,1)}^{*}                                   & 0                                                           & t_{h(3,2)}^{*}                                  & 0                                                         & E_{p1} + E_{3h}       & t_s                                \\
0                                               & t_{h(3,0)}^{*}                         & 0                                                           &  t_{h(3,1)}^{*}                                   & 0                                                         & t_{h(3,2)}^{*}                                  & t_s^{*}                     & E_{p2} + E_{3h}           \\
\end{pmatrix} \nonumber \\
\end{eqnarray}
\normalsize
and two other matrices
$\hat{\Delta}_1=diag(\Delta(0),\Delta(0),\Delta(1),\Delta(1),\Delta(2),\Delta(2),\Delta(3),\Delta(3)),
\hat{\Delta}_2=\hat{\Delta}_1^{\dag}$.
Finally we obtain the following structure of tight-binding Bogoliubov-de Gennes equations including the interaction of semiconductor qubit with Josephson junction described in the minimalistic way in the form
\begin{equation}
\hat{H}_{eff}=
\begin{pmatrix}
\hat{H}_{0[e]} & \hat{\Delta}_1 \\
\hat{\Delta}_2 & \hat{H}_{0[h]}
\end{pmatrix}.
\end{equation}

Similarly as before, having knowledge of quantum state at $t_0$ we can evaluate the state at time $t$ by computing $\exp(\int_{t_0}^{t}\frac{1}{\hbar i}\hat{H}_{ext}(t)dt')\ket{\psi,t_0)}=\ket{\psi,t)}$ which bases on the same method already presented before in Eq.\,(8).
We can also perform the procedure of heating up or cooling down of the quantum state in the way as it was described before or we can regulate the population of pointed energetic level(s).

\begin{figure}
\centering
\label{fig:Spectrumd}
\includegraphics[scale=0.7]{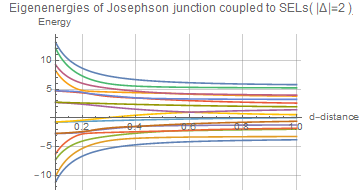}
\caption{Eigenenergies of semiconductor qubit coupled to Josephson junction in dependence on distance in tight-binding minimalitic approach.}
\end{figure}

\begin{figure}
\centering
\label{fig:SpectrumDelta}
\includegraphics[scale=0.7]{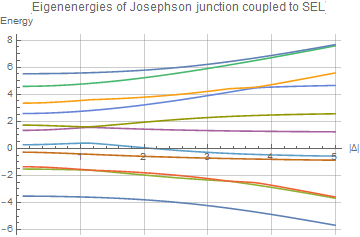}
\caption{Eigenenergies of semiconductor qubit coupled to Josephson junction in dependence on superconducting order parameter in minimalitic approach.}
\end{figure}

In most minimalistic tight-binding model of Josephson junction Sc-I-Sc (Superconductor-Insulator-Superconductor) we set $\Delta(1)=\Delta(2)=0$ what corresponds to the simplest form of Andreev Bound State in Tunneling Josephson junction. However in weak-links and in the Field Induced Josephson junctions all diagonal elements are non-zero and $|\Delta|$ has maximum at $\Delta(0)$ and $\Delta(3)$ that can be considered as superconducting state of bulk superconductors. Quite naturally, Field Induced Josephson junction \cite{PSSB2012} can have special profile of dependence of superconducting order parameter $\Delta(x)$ on position x with presence of built-in magnetic fields in area of junction. It will also have special complex-valued hopping constants for electron and hole in area of superconductor that
will incorporate the profile of magnetic field present across Josephson junction.
Specified Hamiltonian describing electrostatic interface between superconducting Josephson junction and semiconductor position-based qubit has the following parameters describing the state of position based semiconductor qubit $E_{p1}$, $E_{p2}$ , $t_s=t_{sr}+i t_{is}$ (4 real valued time dependent functions),
and parameters describing the state of Josephson junction $E_{e0}$, $E_{e1}$, $E_{e2}$,$E_{e3}$, $E_{h0}$,$E_{h1}$, $E_{h2}$,$E_{h3}$ , $\Delta(0)$, $\Delta(1)$, $\Delta(2)$, $\Delta(3)$, $t_{e(1,0)}$, $t_{e(2,1)}$, $t_{e(2,3)}$, $t_{e(3,0)}$, $t_{h(1,0)}$, $t_{h(2,1)}$, $t_{h(2,3)}$, $t_{h(3,0)}$  as well as geometrical parameters describing electrostatic interaction between semiconductor JJ and semiconductor qubit by a and b. It is worth mentioning that electrostatic interaction taken into account is only between nodes 1-1s, 1-2s,2-1s,2-2s what means 4 channels for Coulomb interaction and simplifies the model greatly so one can find analytical solutions as well. The assumption with four channels of electrostatic interaction is physically justifiable if one assumes that $\Delta(0) \neq 0, \Delta(3) \neq 0$ and $(\Delta(1), \Delta(2)) \rightarrow 0$. Therefore formally we have omitted the following channels of electrostatic interaction $0-1s,3-1s, 0-2s,3-2s$. It is commonly known that superconducting state especially with strong superconductivity as in case of bulk superconductor is not supporting and shielding itself from the external and internal electrostatic field of certain strength as it naturally protects its ground superconducting macroscopic state.
Having established the mathematical structure describing the electrostatic interaction between semiconductor position-based qubit and Josephson junction we can move into first analytical and numerical calculations. First simplification is that $\Delta(1)=\Delta(2)=0$ and $\Delta=\Delta(0)=\Delta(3) \in R$ so it means that there is no net electric current flowing via Josephson junction since the electric current flow imposes the condition of phase difference among superconducting order parameter $\Delta(0)$ and $\Delta(3)$ and in such case superconducting order parameter is complex valued scalar. Also it implies that there is no magnetic field in our system since magnetic field brings phase imprint between $\Delta(0)$ and $\Delta(3)$. Second simplification is that $E_{p1}=E_{p2}=E_p, t_s \in R$. Third simplification is that $E_{e0}=E_{e1}=E_{e2}=E_{e3}=-E_{h0}=-E_{h1}=-E_{h2}=-E_{h3}=V$ so it implies electron-hole symmetry in area of ABS that is the middle of Josephson junction.  In such way all hole eigenenergies are corresponding to electron eigenenergies with $-$ sign. Last assumption is that electron or hole hopping in the area of ABS in between nearest neighbours is such that $t_{e(k,k+1)} \neq 0$ and $t_{h(k,k+1)} \neq 0$ and is 0 otherwise. One can name such feature of transport in Josephson junction as diffusive and not ballistic what brings the mathematical simplifications. Having established such facts we can move into analytical and numerical calculations. The Hamiltonian of physical system has such structure that allows analytic determination of all eigenenergies since Hamiltonian matrix has many symmetries. In particular we can obtain the spectrum of eigenenergies in dependence on the distance a as depicted in Fig.\,\ref{fig:Spectrumd}
and spectrum of eigeneneries in dependence of superconducting order parameter as given in Fig.\,\ref{fig:SpectrumDelta}. One can recognize certain similarities with Fig.\,\ref{fig:spectra}. It simply means that increase of superconducting order parameter strength brings similar effect as increase of distance between interaction of semiconductor position based qubit and Josephson junction.

\begin{figure}
\centering
\label{fig:SpectrumV}
\includegraphics[scale=0.7]{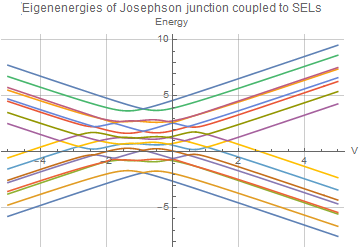}  
\caption{Tunnning the spectrum of eigenenergies in electrostatic qubit interacting with Josephson junction while we are changing the chemical potential of insulator region in Josephson junction at all nodes 0, 1, 2 and 3 in the same time.}
\end{figure}

One of the most interesting feature is tuning the landscape of eigenenergies by applying small voltage (below the size $2 e \Delta$) to non-superconducting region of Josephson junction. In such case one obtains the features as described in Fig.\,\ref{fig:SpectrumV}. In the described considerations the spin degree-of-freedom was omitted in case of Josephson junction as well as in case of semiconductor position based qubit. However they could be easily included but it would increase the size of matrix describing interaction between superconductor Josephson junction and semiconductor electrostatic qubit from 16 by 16 to the size 8*4=32 so one obtains matrix 32 by 32. Adding strong spin-orbit interaction to the Hamiltonian of Josephson junction under the presence of magnetic field allows to describe topological Josephson junction. In such way we can obtain the effective 32 by 32 Hamiltonian for interaction between semiconductor position based qubit and topological Josephson junction in minimalistic way.
It shall be also underlined that so far we have used BdGe formalism that is suitable for mean field theory domain. However, in our case we have considered very special interactions between individual (electrons, holes) present in area of Josephson junction and specific individual electron present in area of semiconductor qubit. Usage of BdGe formalism is therefore first level of possible approximation and further more detailed study can be
attempted in determination of microscopic processes present interacting Josephson junction with semicondutor qubit in more detailed way. It is sufficient to mention that in our case superconductors shall have relatively small size so we are dealing with relatively small number of electrons and holes in non-superconducting area. More detailed considerations are however beyond the scope of this work and requires Density Functional Theory (DFT) methods, etc.

\section{Conclusions}

The obtained results have meaning in the development of single-electron electrostatic quantum neural networks, quantum gates, such as CNOT, SWAP, Toffoli and Fredkin gates as well as any other types of quantum gates with $N$ inputs and $M$ outputs. Single-electron semiconductor devices can be attractive from point of view of power consumption and they can approach similar performance as Rapid Single Quantum Flux superconducting circuits \cite{Pomorski_spie} having much smaller dimensions than superconducting circuits. In conducted computations the spin degree-of-freedom was neglected. However it can be added in straightforward way doubling the size of Hilbert space.  The obtained results allow us to obtain the entanglement of qubit A (for example) using biparticle Von Neumann entropy $S(t)_A$ of qubit A in two electrostatically interacting qubits with time as given by formula
\begin{equation}
\label{entropyS}
S(t)=-Tr[\hat{\rho_A(t)}(\log(\hat{\rho_A}(t)))],
\end{equation}
where Tr[.] is matrix trace operator and $\rho_A$ is the reduced density matrix of A qubit after presence of B qubit was traced out. The obtained results can be mapped to Schr\"odinger formalism \cite{Xu} in order to obtain higher accuracy and resolution in the description of quantum state dynamics. One can use the results in the determination of quantum transport in the single electron devices or arbitrary topology, which can be helpful in optimization of device functionality and sequence of controlling sequences shaping the electron confinement potential. Topological phase transitions as described by \cite{QPT}, \cite{Choi}, \cite{Belzig} are expected to take place in arrays of coupled electrostatic qubits due to the similarity of tight-binding applied in semiconductor coupled quantum well model to Josephson model in Cooper pair box superconducting qubits. All results are straightforward to be generalized for electrons and holes confined in net of coupled quantum dots (which changes only sign of electrostatic energy so $q^2 \rightarrow -q^2$) under the assumption that recombination processes do not occur. What is more the interaction between electrostatic position based qubit and Josephson junction was formulated and solved in tight-binding model. In a quite straightforward way one obtains the electrostatically coupled networks of graphs interacting with single Josephson junction in analytical way.
It will be important in the development of interface between semiconductor CMOS quantum computer and already developed superconducting computer.

\section{Acknowledgment}

This work was supported by Science Foundation Ireland under Grant 14/RP/I2921. We would like to thank to professor Andrew Mitchell (UCD) for his long discussions on methodology of approach and to Erik Staszewski (erik.staszewski@ucd.ie) and to David Watt\'e (david.watte@ucdconnect.ie) for his assistance in graphical design of figures.

\section{Appendix A}
\label{app:a}

The simplified Hamiltonian, given by equation (\ref{Matrix}) for two electrostatically interacting single-electron lines (Fig.\,\ref{PositionDependentQubit}) has eigevalues pointed by formulas (10)--(12) and has following eigenvectors
\begin{eqnarray}
\label{Matrix1}
\hat{H}=
\begin{pmatrix}
q_{1_1} & 1 & 0 & 1 & 0 & 0 & 0 & 0 & 0 \\
1 & q_{1_2} & 1 & 0 & 1 & 0 & 0 & 0 & 0 \\
0 & 1 & q_{1_3} & 0 & 0 & 1 & 0 & 0 & 0 \\
1 & 0 & 0 & q_{1_2} & 1 & 0 & 1 & 0 & 0 \\
0 & 1 & 0 & 1 & q_{1_1} & 1 & 0 & 1 & 0 \\
0 & 0 & 1 & 0 & 1 & q_{1_2} & 0 & 0 & 1 \\
0 & 0 & 0 & 1 & 0 & 0 & q_{1_3} & 1 & 0 \\
0 & 0 & 0 & 0 & 1 & 0 & 1 & q_{1_2} & 1 \\
0 & 0 & 0 & 0 & 0 & 1 & 0 & 1 & q_{1_1} \\
\end{pmatrix},  \\
\end{eqnarray}
\begin{eqnarray}
\label{ent1}
 \ket{E_1}=
 \begin{pmatrix}
 1, \\
 0, \\
 0, \\
 0, \\
 -1, \\
 0, \\
 0, \\
 0, \\
 1 \\
 \end{pmatrix},
 \ket{E_2}=
 \begin{pmatrix}
 0, \\
 1, \\
 0, \\
 -1, \\
 0, \\
 -1, \\
 0, \\
 1, \\
 0 \\
 \end{pmatrix}, \nonumber \\
\label{ent4}
  \ket{E_{3(4)}}=
 \begin{pmatrix}
 -1, \\
 \frac{1}{4}( q_{1_1} - q_{1_2} \pm \sqrt{8+(q_{1_1}-q_{1_2})^2} ), \\
 0, \\
 \frac{1}{4}( q_{1_1} - q_{1_2} \pm \sqrt{8+(q_{1_1}-q_{1_2})^2} ), \\
 0, \\
 -\frac{1}{4}( q_{1_1} - q_{1_2} \pm \sqrt{8+(q_{1_1}-q_{1_2})^2} ), \\
 0, \\
 -\frac{1}{4}( q_{1_1} - q_{1_2} \pm \sqrt{8+(q_{1_1}-q_{1_2})^2} ), \\
 1 \\
 \end{pmatrix},
 \end{eqnarray}
 \begin{eqnarray}
\label{ent5}
   \ket{E_{5(6)}}=
 \begin{pmatrix}
 -1, \\
 \frac{1}{4}( q_{1_2} - q_{1_3} \pm \sqrt{8+(q_{1_2}-q_{1_3})^2} ), \\
 0, \\
 \frac{1}{4}( q_{1_2} - q_{1_3} \pm \sqrt{8+(q_{1_2}-q_{1_3})^2} ), \\
 0, \\
 -\frac{1}{4}( q_{1_2} - q_{1_3} \pm \sqrt{8+(q_{1_2}-q_{1_3})^2} ), \\
 0, \\
 -\frac{1}{4}( q_{1_2} - q_{1_3} \pm \sqrt{8+(q_{1_2}-q_{1_3})^2} ), \\
 1 \\
 \end{pmatrix}, \nonumber \\
\end{eqnarray}
\begin{eqnarray}
\label{ent7}
 \ket{E_{k=(7 .. 9)}}= 
 \begin{pmatrix}
 1, \\
(E_{k=(7 .. 9)} - q_{1_1})/2, \\
\frac{(-E_{k=(7 .. 9)} + q_{1_1}) ( -2 + E_{k=(7 .. 9)}^2 + q_{1_1} q_{1_2} - E_{k=(7 .. 9)} (q_{1_1} + q_{1_2}))}{ 2 (-3 E_{k=(7 .. 9)} + q_{1_1} + 2 q_{1_3})}, \\
(E_{k=(7 .. 9)} - q_{1_1})/2, \\
2, \\
(E_{k=(7 .. 9)} - q_{1_1})/2 , \\
2, \\
\frac{(-E_{k=(7 .. 9)} + q1_1)(-2 + E_{k=(7 .. 9)}^2 + q_{1_1} q_{1_2} -E_{k=(7 .. 9)} (q_{1_1} + q_{1_2}))} {2 (-3 E_{k=(7 .. 9)} + q_{1_1} + 2 q_{1_3})}
\end{pmatrix}.\nonumber \\
 \end{eqnarray}
It is important to recognize that in the case of electrons partly or wholly localized at the nodes of 2-SEL system, such that all hoping constants $t_{s1,kl}$ and $t_{s2,r'u'}$ are zero, we have no quantum entanglement between 2-SELs if it populates one energetic level and its Hamiltonian becomes diagonal. It brings the following energy eigenstates:
 \begin{eqnarray}
 \ket{E_1}=
 \begin{pmatrix}
 1 \\
 0 \\
 0 \\
 0 \\
 0 \\
 0 \\
 0 \\
 0 \\
 0 \\
 \end{pmatrix},
 ..
 \ket{E_9}=
 \begin{pmatrix}
 0 \\
 0 \\
 0 \\
 0 \\
 0 \\
 0 \\
 0 \\
 0 \\
 1 \\
 \end{pmatrix},
 \end{eqnarray}
 and Hamiltonian of system simulating two electrostatically charged insulators has the following structure
 \begin{eqnarray}
\hat{H}=
\begin{pmatrix}
q_{1_1} & 0 & 0 & 0 & 0 & 0 & 0 & 0 & 0 \\
0 & q_{1_2} & 0 & 0 & 0 & 0 & 0 & 0 & 0 \\
0 & 0 & q_{1_3} & 0 & 0 & 0 & 0 & 0 & 0 \\
0 & 0 & 0 & q_{1_2} & 0 & 0 & 0 & 0 & 0 \\
0 & 0 & 0 & 0 & q_{1_1} & 0 & 0 & 0 & 0 \\
0 & 0 & 0 & 0 & 0 & q_{1_2} & 0 & 0 & 0 \\
0 & 0 & 0 & 0 & 0 & 0 & q_{1_3} & 0 & 0 \\
0 & 0 & 0 & 0 & 0 & 0 & 0 & q_{1_2} & 0 \\
0 & 0 & 0 & 0 & 0 & 0 & 0 & 0 & q_{1_1} \\
\end{pmatrix},  \nonumber \\
\end{eqnarray}
what brings following eigenenergy values
\begin{eqnarray}
E_1=q_{1_1},  E_2=q_{1_2},  E_3=q_{1_3},
E_4=q_{1_2},  E_5=q_{1_1},  \nonumber \\
E_6=q_{1_2},  E_7=q_{1_3},  E_8=q_{1_2},
E_9=q_{1_1}.  \nonumber \\
\end{eqnarray}
\section{Appendix B: Details of Analytical Solution for Case of Coupled SELs}
\label{app:b}
We continue derivation of the equation of motion imposed by classical picture of 2-SELs
and from Hamiltonian \ref{HC} we obtain the following expression for velocity of interacting particles with positions $x_1(t)=-x_2(t)$ and velocity vs time as
\begin{eqnarray}
\int \frac{dv}{\sqrt{q^4(E_{c}-mv^2))^4-d^2(E_{c}-mv^2)^6}}=
\frac{ d^2}{q^4} \int \frac{du}{\sqrt{(\frac{E_c d}{q}-u)}} \frac{1}{u^2 \sqrt{1-u^2}} 
=s_1 \int \frac{du}{\sqrt{(s-u)}} \frac{1}{u^2 \sqrt{1-u^2}}
. 
\end{eqnarray}
Setting $s_1=\frac{ d^2}{q^4}$ and $s=\frac{E_c d}{q}$, we obtain the integral $s_1 \int \frac{du}{\sqrt{(s-u)}} \frac{1}{u^2 \sqrt{1-u^2}}$ that has the solution as
\begin{eqnarray} s_1 \int \frac{du}{\sqrt{(s-u)}} \frac{1}{u^2 \sqrt{1-u^2}}=
	\frac{s_1}{s \sqrt{1-u^2}} \Bigg[ \frac{\left(u^2-1\right) \sqrt{s-u}}{u} + \nonumber \\ + \frac{i (s-1) \sqrt{s-u} \sqrt{\frac{u-1}{s-1}} \left({\rm Elliptic_E}\left(i \sinh ^{-1}\left(\sqrt{\frac{u-s}{s+1}}\right),\frac{s+1}{s-1}\right)-{\rm Elliptic_F}\left(i \sinh
   ^{-1}\left(\sqrt{\frac{u-s}{s+1}}\right),\frac{s+1}{s-1}\right)\right)}{\sqrt{\frac{u-s}{u+1}}} + \nonumber \\
	+\frac{i s \sqrt{s-u} \sqrt{\frac{u-1}{s-1}} {\rm Elliptic_F}\left(i \sinh ^{-1}\left(\sqrt{\frac{u-s}{s+1}}\right),\frac{s+1}{s-1}\right)}{\sqrt{\frac{u-s}{u+1}}} \nonumber \\
-\frac{\left(\sqrt{s-1}+\sqrt{s+1}\right) \left(\sqrt{s-1}-\sqrt{s-u}\right)^2 \sqrt{\frac{\sqrt{s-1} \left(\sqrt{s+1}-\sqrt{s-u}\right)}{\left(\sqrt{s-1}+\sqrt{s+1}\right)
   \left(\sqrt{s-1}-\sqrt{s-u}\right)}} \sqrt{\frac{\sqrt{s-1} \left(\sqrt{s-u}+\sqrt{s+1}\right)}{\left(\sqrt{s-1}-\sqrt{s+1}\right)
   \left(\sqrt{s-u}-\sqrt{s-1}\right)}}}{\sqrt{s} \left(s-\sqrt{s-1} \sqrt{s+1}-1\right)} \times \nonumber \\
	\times \sqrt{\frac{\sqrt{s-1} \sqrt{s-u}-\sqrt{s+1} \sqrt{s-u}+s-\sqrt{s-1} \sqrt{s+1}-1}{\left(\sqrt{s-1}+\sqrt{s+1}\right) \left(\sqrt{s-1}-\sqrt{s-u}\right)}} \times \nonumber \\ \times \Bigg[ \left(\sqrt{s-1}+\sqrt{s}\right){\rm Elliptic_F}\left(\sin ^{-1}\left(\sqrt{\frac{\left(\sqrt{s-1}-\sqrt{s+1}\right) \left(\sqrt{s-1}+\sqrt{s-u}\right)}{\left(\sqrt{s-1}+\sqrt{s+1}\right)
   \left(\sqrt{s-1}-\sqrt{s-u}\right)}}\right),\frac{\left(\sqrt{s-1}+\sqrt{s+1}\right)^2}{\left(\sqrt{s-1}-\sqrt{s+1}\right)^2}\right)  \nonumber \\
	-2 \sqrt{s-1} \times \nonumber \\ \times {\rm Elliptic_{Pi}} \Big[ \frac{\left(\sqrt{s-1}-\sqrt{s}\right) \left(\sqrt{s-1}+\sqrt{s+1}\right)}{\left(\sqrt{s-1}+\sqrt{s}\right) \left(\sqrt{s-1}-\sqrt{s+1}\right)},\sin
   ^{-1}\left(\sqrt{\frac{\left(\sqrt{s-1}-\sqrt{s+1}\right) \left(\sqrt{s-1}+\sqrt{s-u}\right)}{\left(\sqrt{s-1}+\sqrt{s+1}\right)
   \left(\sqrt{s-1}-\sqrt{s-u}\right)}}\right), \nonumber \\ \frac{\left(\sqrt{s-1}+\sqrt{s+1}\right)^2}{\left(\sqrt{s-1}-\sqrt{s+1}\right)^2} \Big]
 \nonumber \\ -\frac{\left(\sqrt{s-1}+\sqrt{s+1}\right) \left(\sqrt{s-1}-\sqrt{s-u}\right)^2 \sqrt{\frac{\sqrt{s-1} \left(\sqrt{s+1}-\sqrt{s-u}\right)}{\left(\sqrt{s-1}+\sqrt{s+1}\right)
   \left(\sqrt{s-1}-\sqrt{s-u}\right)}} \sqrt{\frac{\sqrt{s-1} \left(\sqrt{s-u}+\sqrt{s+1}\right)}{\left(\sqrt{s-1}-\sqrt{s+1}\right)
   \left(\sqrt{s-u}-\sqrt{s-1}\right)}}}{\sqrt{s} \left(-s+\sqrt{s-1} \sqrt{s+1}+1\right)} \times \nonumber \\ \times
\sqrt{\frac{\sqrt{s-1} \sqrt{s-u}-\sqrt{s+1} \sqrt{s-u}+s-\sqrt{s-1} \sqrt{s+1}-1}{\left(\sqrt{s-1}+\sqrt{s+1}\right) \left(\sqrt{s-1}-\sqrt{s-u}\right)}} \times \nonumber \\
	\times \Big[ \left(\sqrt{s-1}-\sqrt{s}\right) {\rm Elliptic_F}\left(\sin ^{-1}\left(\sqrt{\frac{\left(\sqrt{s-1}-\sqrt{s+1}\right) \left(\sqrt{s-1}+\sqrt{s-u}\right)}{\left(\sqrt{s-1}+\sqrt{s+1}\right)
   \left(\sqrt{s-1}-\sqrt{s-u}\right)}}\right),\frac{\left(\sqrt{s-1}+\sqrt{s+1}\right)^2}{\left(\sqrt{s-1}-\sqrt{s+1}\right)^2}\right) \Big] \nonumber
	\\ - 2 \sqrt{s-1}{\rm Elliptic_{Pi}} \Big[ \frac{\left(\sqrt{s-1}+\sqrt{s}\right) \left(\sqrt{s-1}+\sqrt{s+1}\right)}{\left(\sqrt{s-1}-\sqrt{s}\right) \left(\sqrt{s-1}-\sqrt{s+1}\right)}, \nonumber \\
\sin ^{-1}\left(\sqrt{\frac{\left(\sqrt{s-1}-\sqrt{s+1}\right) \left(\sqrt{s-u}+\sqrt{s-1}\right)}{\left(\sqrt{s-1}+\sqrt{s+1}\right) \left(\sqrt{s-1}-\sqrt{s-u}\right)}}\right), 
 \frac{\left(\sqrt{s-1}+\sqrt{s+1}\right)^2}{\left(\sqrt{s-1}-\sqrt{s+1}\right)^2} \Big] \Bigg],
 \label{long}
\end{eqnarray}
where $\rm Elliptic_F[.,.]$
is the elliptic integral of the first kind, $\rm Elliptic_E[.,.]$ is the elliptic integral of the second kind and $\rm Elliptic_{Pi}[.,.]$ is the complete elliptic integral of the third kind as in accordance with nomenclature used by Mathematica symbolic software \cite{Mathematica}.

\section{Appendix C: Details of Anticorrelation Function Calculation for the Case of Weak Measurement Performed on the 2-SELs}
\label{app:c}

We refer to the Hamiltonian of 2-SEL system coupled to flying qubit given by equation (\ref{DHC}) and we recognize that the time-dependent Hamiltonian $\hat{H}_{AC}(t)$ and evolution operator based on it is as follows
\begin{equation}
e^{\frac{1}{\hbar i} \int_{0}^{t} (\hat{H}_{AC}(t')\times \hat{I}_B)dt'}=
\end{equation}
\tiny
\begin{equation*}
\begin{pmatrix}
e^{\frac{1}{\hbar i}\int_{0}^{t}E_{c1''1}(t')dt'} & 0 & 0 & 0 & 0 & 0 & 0 & 0 \\
0 & e^{\frac{1}{\hbar i}\int_{0}^{t}E_{c1''1}dt'} & 0 & 0 & 0 & 0 & 0 & 0 \\
0 & 0 & e^{\frac{1}{\hbar i}\int_{0}^{t}E_{c2''1}(t')dt'} & 0 & 0 & 0 & 0 & 0 \\
0 & 0 & 0  & e^{\frac{1}{\hbar i}\int_{0}^{t}E_{c2''1}(t')dt'} & 0 & 0 & 0 & 0 \\
0 & 0 & 0  & 0 & e^{\frac{1}{\hbar i}\int_{0}^{t}E_{c1''2}(t')dt'} & 0 & 0 & 0 \\
0 & 0 & 0  & 0 &  0 & e^{\frac{1}{\hbar i}\int_{0}^{t}E_{c1''2}(t')dt'} & 0 & 0 \\
0 & 0 & 0  & 0 & 0 & 0 & e^{\frac{1}{\hbar i}\int_{0}^{t}E_{c2''2}(t')dt'} & 0 \\
0 & 0 & 0  & 0 & 0 & 0 & 0 & e^{\frac{1}{\hbar i}\int_{0}^{t}E_{c2''2}dt'} \\
\end{pmatrix}.
\end{equation*}
\normalsize

Now we are defining the correlation function for 2-SELs in case of the system interaction with the external flying qubit given by the matrix
\begin{eqnarray}
C_{AB,C}=\hat{I}_C \times \hat{C}_{AB}= 
\begin{pmatrix}
1 & 0 \\
0 & 1 \\
\end{pmatrix} \times
\begin{pmatrix}
1 & 0 & 0 & 0 \\
0 & -1 & 0 & 0 \\
0 & 0 & -1 & 0 \\
0 & 0 & 0 & 1 \\
\end{pmatrix}= 
\begin{pmatrix}
+1 & 0 & 0 & 0 & 0 & 0 & 0 & 0 \\
0 & -1 & 0 & 0 & 0 & 0 & 0 & 0 \\
0 & 0 & -1 & 0 & 0 & 0 & 0 & 0 \\
0 & 0 & 0  & +1 & 0 & 0 & 0 & 0 \\
0 & 0 & 0  & 0 & +1 & 0 & 0 & 0 \\
0 & 0 & 0  & 0 &  0 & -1 & 0 & 0 \\
0 & 0 & 0  & 0 & 0 & 0 & -1 & 0 \\
0 & 0 & 0  & 0 & 0 & 0 & 0 & +1 \\
\end{pmatrix}.
\end{eqnarray}

Now we construct Hamiltonian for non-interacting C and AB physical systems given as
\begin{eqnarray*}
\hat{H}=\hat{I}_C \times \hat{H}_{AB} + \hat{H}_C \times \hat{I}_{AB}= \nonumber \\
=
\begin{pmatrix}
\hat{H}_{AB} & \hat{0}_{4 \times 4} \\
\hat{0}_{4 \times 4} & \hat{H}_{AB}
\end{pmatrix} + 
\begin{pmatrix}
\hat{H}_{C}[1,1] & 0                & 0                & 0                    & \hat{H}_{C}[1,2]  & 0 & 0 & 0\\
0                & \hat{H}_{C}[1,1] & 0                & 0                    & 0                 & \hat{H}_{C}[1,2] & 0 & 0\\
0                & 0                & \hat{H}_{C}[1,1] & 0                    &                0  & 0 & \hat{H}_{C}[1,2] & 0 \\
0                & 0                & 0                & \hat{H}_{C}[1,1] & 0 & 0 & 0 & \hat{H}_{C}[1,2] \\
\hat{H}_{C}[2,1] & 0                & 0                & 0 & \hat{H}_{C}[2,2] & 0 & 0 & 0\\
0                & \hat{H}_{C}[2,1] & 0                & 0 & 0 & \hat{H}_{C}[2,2] & 0 & 0\\
0                & 0                & \hat{H}_{C}[2,1] & 0 & 0 & 0 & \hat{H}_{C}[2,2] & 0 \\
0                & 0                & 0                & \hat{H}_{C}[2,1] & 0 & 0 & 0 & \hat{H}_{C}[2,2] \\
\end{pmatrix}
=
\end{eqnarray*}
\tiny
\begin{eqnarray*}
=
\begin{pmatrix}
E_{p1}+E_{p1'}+E_{c1} & t_{s1'2'}             & t_{s12}               & 0                     & 0 & 0 & 0 & 0 \\
t_{s1'2'}^{*}         & E_{p1}+E_{p2'}+E_{c2} & 0                     & t_{s12}               & 0 & 0 & 0 & 0 \\
t_{s12}^{*}           & 0                     & E_{p2}+E_{p1'}+E_{c2} & t_{s1'2'}             & 0 & 0 & 0 & 0 \\
0                     & t_{s12}^{*}           & t_{s1'2'}^{*}         & E_{p2}+E_{p2'}+E_{c1} & 0 & 0 & 0 & 0 \\
0                     & 0                     & 0                     & 0                     & E_{p1}+E_{p1'}+E_{c1} & t_{s1'2'} & t_{s12} & 0  \\
0                     & 0                     & 0                     & 0                     & t_{s1'2'}^{*} & E_{p1}+E_{p2'}+E_{c2} & 0 & t_{s12} \\
0                     & 0                     & 0                     & 0                     & t_{s12}^{*} & 0 & E_{p2}+E_{p1'}+E_{c2} & t_{s1'2'} \\
0                     & 0                     & 0                     & 0                     & 0 & t_{s12} & t_{s1'2'}^{*} & E_{p2}+E_{p2'}+E_{c1} \\
\end{pmatrix} +
 \nonumber \\ +
\begin{pmatrix}
E_{p1''} & 0        & 0 & 0 & t_{s1''2''} & 0 & 0 & 0 \\
0        & E_{p1''} & 0 & 0 & 0 & t_{s1''2''} & 0 & 0 \\
0        & 0        & E_{p1''} & 0 & 0 & 0 & t_{s1''2''} & 0 \\
0        & 0        & 0 & E_{p1''} & 0 & 0 & 0 & t_{s1''2''} \\
t_{s1''2''}^{*}        & 0        & 0 & 0 & E_{p2''} & 0 & 0 & 0 \\
0        & t_{s1''2''}^{*}        & 0 & 0 & 0 & E_{p2''} & 0 & 0 \\
0        & 0        & t_{s1''2''}^{*} & 0 & 0 & 0 & E_{p2''} & 0 \\
0        & 0        & 0 & t_{s1''2''}^{*} & 0 & 0 & 0 & E_{p2''} \\
\end{pmatrix} = \nonumber \\
\begin{pmatrix}
E_{p1}+E_{p1'}+E_{c1} & t_{s1'2'}             & t_{s12}               & 0                     & t_{s1''2''} & 0           & 0            & 0 \\
t_{s1'2'}^{*}         & E_{p1}+E_{p2'}+E_{c2} & 0                     & t_{s12}               & 0           & t_{s1''2''} & 0           & 0 \\
t_{s12}^{*}           & 0                     & E_{p2}+E_{p1'}+E_{c2} & t_{s1'2'}             & 0           & 0           & t_{s1''2''} & 0 \\
0                     & t_{s12}^{*}           & t_{s1'2'}^{*}         & E_{p2}+E_{p2'}+E_{c1} & 0           & 0           & 0           & t_{s1''2''} \\
t_{s1''2''}^{*}       & 0                     & 0                     & 0                     & E_{p1}+E_{p1'}+E_{c1} & t_{s1'2'} & t_{s12} & 0  \\
0                     & t_{s1''2''}^{*}       & 0                     & 0                     & t_{s1'2'}^{*} & E_{p1}+E_{p2'}+E_{c2} & 0 & t_{s12} \\
0                     & 0                     & t_{s1''2''}^{*}       & 0                     & t_{s12}^{*} & 0 & E_{p2}+E_{p1'}+E_{c2} & t_{s1'2'} \\
0                     & 0                     & 0                     & t_{s1''2''}^{*}       & 0 & t_{s12} & t_{s1'2'}^{*} & E_{p2}+E_{p2'}+E_{c1} \\
\end{pmatrix} + \nonumber \\
+diag(E_{p1''},E_{p1''},E_{p1''},E_{p1''},E_{p2''},E_{p2''},E_{p2''},E_{p2''}).
\end{eqnarray*}
\normalsize
We recognize that diagonal elements of $\hat{I}_C \times \hat{H}_{AB} + \hat{H}_C \times \hat{I}_{AB}$ are
\begin{eqnarray*}
(E_{p1}+E_{p1'}+E_{c1}+E_{p1''},E_{p1}+E_{p2'}+E_{c2}+E_{p1''},E_{p2}+E_{p1'}+E_{c2}+E_{p1''}, E_{p2}+E_{p2'}+E_{c1}+E_{p1''}, \nonumber \\
E_{p1}+E_{p1'}+E_{c1}+E_{p2''},E_{p1}+E_{p2'}+E_{c2}+E_{p2''},E_{p2}+E_{p1'}+E_{c2}+E_{p2''},E_{p2}+E_{p2'}+E_{c1}+E_{p2''}).
\end{eqnarray*}
Now we consider the interaction between qubits C and A denoted by $H_{CA}$ and it will be incoroporated into global Hamiltonian $\hat{H}_{CA} \times \hat{I}_{B}$ that has the following diagonal matrix representation
\begin{equation}
\hat{H}_{CA}=
\begin{pmatrix}
E_{c1''1}(t) & 0            & 0            & 0 \\
0         & E_{c1''2}(t) & 0            & 0 \\
0         & 0            & E_{c2''1}(t) & 0 \\
0         & 0            & 0            & E_{c2''2}(t) \\
\end{pmatrix}
\end{equation}
and consequently
\begin{equation}
\hat{H}_{CA}\times \hat{I}_{B}=
\begin{pmatrix}
E_{c1''1}(t)     & 0            & 0            & 0 & 0 & 0 & 0 & 0 \\
0                & E_{c1''1}(t) & 0            & 0 & 0 & 0 & 0 & 0 \\
0                & 0            & E_{c1''2}(t) & 0 & 0 & 0 & 0 & 0 \\
0                & 0            & 0            & E_{c1''2}(t) & 0 & 0 & 0 & 0 \\
0                & 0            & 0            & 0 & E_{c2''1}(t) & 0 & 0 & 0 \\
0                & 0            & 0            & 0 & 0 & E_{c2''1}(t) & 0 & 0 \\
0                & 0            & 0            & 0 & 0 & 0 & E_{c2''2}(t) & 0 \\
0                & 0            & 0            & 0 & 0 & 0 & 0 & E_{c2''2}(t) \\
\end{pmatrix}.
\end{equation}
We have the total Hamiltonian for the flying qubit interacting with 2-SELs given as
\begin{equation}
\hat{H}=\hat{I}_C \times \hat{H}_{AB} + \hat{H}_C \times \hat{I}_{AB}+\hat{H}(t)_{CA}\times \hat{I}_{B} .
\end{equation}
We recognize that the diagonal terms of total matrix are given as a following sequence 
$ $
\newline
$(E_{p1}+E_{p1'}+E_{c1}+E_{p1''}+E_{c1''1}(t),E_{p1}+E_{p2'}+E_{c2}+E_{p1''}+E_{c1''1}(t),$ $E_{p2}+E_{p1'}+E_{c2}+E_{p1''}+E_{c1''2}(t), E_{p2}+E_{p2'}+E_{c1}+E_{p1''}+E_{c1''2}(t),$ \newline  
$E_{p1}+E_{p1'}+E_{c1}+E_{p2''}+ E_{c2''1}(t),E_{p1}+E_{p2'}+E_{c2}+E_{p2''}+$ $E_{c2''1}(t),E_{p2}+E_{p1'}+E_{c2}+E_{p2''}+ E_{c2''2}(t),E_{p2}+E_{p2'}+E_{c1}+E_{p2''}+ E_{c2''2}(t))$.
\normalsize
\newline
Setting $E_{p1}=E_{p1'}=E_{p1''}=E_{p2}=E_{p2'}=E_{p2''}=E_{p}$, we obtain diagonal terms as
\begin{eqnarray}
(E_{c1}+3E_p+E_{c1''1}(t),E_{c2}+3E_p+E_{c1''1}(t),E_{c2}+3E_{p}+E_{c1''2}(t), 3E_{p}+E_{c1}+E_{c1''2}(t), \nonumber \\
E_{c1}+3 E_{p}+ E_{c2''1}(t),3E_{p}+E_{c2}+E_{c2''1}(t),3E_{p}+E_{c2}+E_{c2''2}(t),3E_{p}+E_{c1}+ E_{c2''2}(t)).
\end{eqnarray}
Substracting element $3E_p+E_{c1}$ we obtain
\begin{eqnarray}
E_{c1''1}(t),E_{c2}-E_{c1}+E_{c1''1}(t),E_{c2}-E_{c1}+E_{c1''2}(t),E_{c1''2}(t), \nonumber \\
E_{c2''1}(t),E_{c2}-E_{c1}+E_{c2''1}(t),E_{c2}-E_{c1}+E_{c2''2}(t),E_{c2''2}(t)).
\end{eqnarray}
Now we are constructing the density matrix for the case of non-interacting qubit C with 2-SELs denoted as AB system. We assume that qubit C is
in the ground state and that symmetric 2-SELs line is populated at energy $E_1$ or $E_2$. In such a case, the density matrices are as follows
\begin{equation}
\hat{\rho}_C=
\begin{pmatrix}
+\frac{1}{2} & -\frac{1}{2} \\
-\frac{1}{2} & +\frac{1}{2} \\
\end{pmatrix},
\hat{\rho}_{AB}=
\begin{pmatrix}
+\frac{1}{2} & 0 & 0 & -\frac{1}{2} \\
0 & 0 & 0 & 0 \\
0 & 0 & 0 & 0 \\
-\frac{1}{2} & 0 & 0 & +\frac{1}{2} \\
\end{pmatrix}
\end{equation}
Therefore, the density matrix of non-interacting qubit C with 2-SELs line denoted as AB system is given as
\begin{equation}
\hat{\rho}_{ABC}=
\begin{pmatrix}
+\frac{1}{2} & -\frac{1}{2} \\
-\frac{1}{2} & +\frac{1}{2} \\
\end{pmatrix} \times
\begin{pmatrix}
+\frac{1}{2} & 0 & 0 & -\frac{1}{2} \\
0 & 0 & 0 & 0 \\
0 & 0 & 0 & 0 \\
-\frac{1}{2} & 0 & 0 & +\frac{1}{2} \\
\end{pmatrix}=
\begin{pmatrix}
+\frac{1}{4} & 0 & 0 & -\frac{1}{4} & -\frac{1}{4} & 0 & 0 & +\frac{1}{4} \\
0 & 0 & 0 & 0 & 0 & 0 & 0 & 0 \\
0 & 0 & 0 & 0 & 0 & 0 & 0 & 0 \\
-\frac{1}{4} & 0 & 0 & +\frac{1}{4} & +\frac{1}{4} & 0 & 0 & -\frac{1}{4} \\
-\frac{1}{4} & 0 & 0 & +\frac{1}{4} & +\frac{1}{4} & 0 & 0 & -\frac{1}{4} \\
0 & 0 & 0 & 0 & 0 & 0 & 0 & 0 \\
0 & 0 & 0 & 0 & 0 & 0 & 0 & 0 \\
+\frac{1}{4} & 0 & 0 & -\frac{1}{4} & -\frac{1}{4} & 0 & 0 & +\frac{1}{4} \\
\end{pmatrix}.
\end{equation}
 The density matrix follows the equation of motion

\begin{equation}
\rho(t)=e^{\frac{1}{i \hbar}\int_0^{t}H(t')dt'}
\begin{pmatrix}
+\frac{1}{4} & 0 & 0 & -\frac{1}{4} & -\frac{1}{4} & 0 & 0 & +\frac{1}{4} \\
0 & 0 & 0 & 0 & 0 & 0 & 0 & 0 \\
0 & 0 & 0 & 0 & 0 & 0 & 0 & 0 \\
-\frac{1}{4} & 0 & 0 & +\frac{1}{4} & +\frac{1}{4} & 0 & 0 & -\frac{1}{4} \\
-\frac{1}{4} & 0 & 0 & +\frac{1}{4} & +\frac{1}{4} & 0 & 0 & -\frac{1}{4} \\
0 & 0 & 0 & 0 & 0 & 0 & 0 & 0 \\
0 & 0 & 0 & 0 & 0 & 0 & 0 & 0 \\
+\frac{1}{4} & 0 & 0 & -\frac{1}{4} & -\frac{1}{4} & 0 & 0 & +\frac{1}{4} \\
\end{pmatrix}
e^{-\frac{1}{i \hbar}\int_0^{t}H(t')dt'}
.
\end{equation}

Since the structure of the Hamiltonian matrix $\hat{H}(t)=\hat{I}_C \times \hat{H}_{AB}+\hat{H}_C \times \hat{I}_{AB} + \hat{H}_{CA}(t) \times \hat{I}_B$ describing the interaction of three electrons confined to the flying position-based qubit C and 2-SEL system is known at all instances of time in the analytical way as well as the operators $e^{\pm \frac{1}{i \hbar}\int_0^{t}H(t')dt'}$ are known in the analytical way, the structure of the density matrix is known in the analytical way. This implies our full knowledge of the qubit C state and 2-SELs system at any instance of time thanks to the formula (\ref{detailedDM}). Such reasoning opens the perspective of analytical approach towards quantum $N$-body electron (hole) system confined to the three disconnected graphs of quantum dots of any topology in the 3D space subjected to the steering mechanism from voltage polarization applied to CMOS gates, as depicted in Fig.\ref{fig:central}. It is thus the subject of the future more detailed studies with use of both analytical and numerical tools. It also opens the perspective on new experiments and new technological novelties in the area of cryogenic CMOS single-electron device electronics that have both importance in the implementation of quantum computer as well as in the development of classical single electron electronics.


\begin{thebibliography}{9}
\bibitem{Dirk}
D. Leipold, Controlled Rabi Oscillations as foundation for entangled quantum aperture logic, Seminar
at UC Berkley Quantum Labs, 25th July 2018
\bibitem{Panos} P.Giounanlis, E.Blokhina, K.Pomorski, D.R.Leipold, R.B.Staszewski, Modeling of Semiconductor Electrostatic Qubits Realized Through Coupled Quantum Dots,
10.1109/ACCESS.2019.2909489,IEEE Access, 2019
\bibitem{Pomorski_spie}  Krzysztof Pomorski, Panagiotis Giounanlis, Elena Blokhina, Dirk Leipold, Pawel Peczkowski, Robert Bogdan Staszewski, From two types of electrostatic position-dependent semiconductor qubits to quantum universal gates and hybrid semiconductor-superconducting quantum computer, Proc. SPIE 11054, Superconductivity and Particle Accelerators 2018, 110540M, 2019
\bibitem{Fujisawa} T. Fujisawa, T. Hayashi, HD Cheong, YH Jeong, and Y. Hirayama.
Rotation and phase-shift operations for a charge qubit in a double
quantum dot. Physica E: Low-dimensional Systems and Nanostructures,
21(2-4):10461052, 2004.
\bibitem{Petta} K. D. Petersson, J. R. Petta, H. Lu, and A. C. Gossard. Quantum
coherence in a one-electron semiconductor charge qubit. Phys. Rev. Lett.,
105:246804, 2010.
\bibitem{Spalek}
Jozef Spalek, Wstep do fizyki materii skondensowanej, PWN, 2015.
\bibitem{Pomorski_compel} K.Pomorski, H.Akaike, A.Fujimaki, and K.Rusek. Relaxation method
in description of ram memory cell in rsfq computer, COMPEL, 38(1):395414, 2019.
\bibitem{Choi}
M.S.Choi, J.Yi, M.Y.Choi, J.Choi, and S.I.Lee. Quantum phase
transitions in josephson-junction chains. Phys. Rev. B, 57:R716R719, 1998.
\bibitem{QPT}
S.Sachdev. Quantum phase transitions. Cambridge Univ. Press, 2011.
\bibitem{Xu} H. Q. Xu. Method of calculations for electron transport in multiterminal quantum systems based on real-space lattice models. Phys. Rev. B, 66:165305.
\bibitem{Belzig}
D. Maile, S. Andergassen, and W. Belzig. Quantum phase transition
with dissipative frustration. Phys. Rev. B, 97, 2018.
\bibitem{PSSB2012}
K.Pomorski, P.Prokopow, Possible existence of field-induced Josephson junctions, Vol.249, No. 9, Physica Status Solidi B, 2012
\bibitem{SEL}
Krzysztof Pomorski, Panagiotis Giounanlis, Elena Blokhina, Dirk Leipold, Robert Bogdan Staszewski,
Analytic view on Coupled Single-Electron Lines, Semiconductor Science and Technology, $iopscience.iop.org/10.1088/1361-6641/ab4f40$ ,2019
\bibitem{Jaynes}
E. T. Jaynes and F. W. Cummings, "Comparison of quantum and semiclassical radiation theories with application to the beam maser," in Proceedings of the IEEE, vol. 51, no. 1, pp. 89-109, Jan. 1963.
doi: 10.1109/PROC.1963.1664
\bibitem{Mathematica}
Wolfram Mathematica:  $http://www.wolfram.com/mathematica/$
\bibitem{WikipediaBellTheorem}
Wikipedia:Bell theorem
\end{thebibliography}

\end{document}